\documentclass[12pt]{article} 

\usepackage{url,hyperref,lineno,microtype,subcaption}
\usepackage[onehalfspacing]{setspace}

\usepackage{amsmath,amsfonts,amssymb,graphicx,amsthm,geometry}
\usepackage{stackrel,stmaryrd,mathabx,etoolbox,framed,accents}
\usepackage[all]{xy}

\usepackage{titletoc}

\def\be{\begin{equation}}
\def\ee{\end{equation}}
\def\ba{\begin{eqnarray}}
\def\ea{\end{eqnarray}}

\makeatletter
\newsavebox{\@brx}
\newcommand{\llangle}[1][]{\savebox{\@brx}{\(\m@th{#1\langle}\)}%
  \mathopen{\copy\@brx\kern-0.5\wd\@brx\usebox{\@brx}}}
\newcommand{\rrangle}[1][]{\savebox{\@brx}{\(\m@th{#1\rangle}\)}%
  \mathclose{\copy\@brx\kern-0.5\wd\@brx\usebox{\@brx}}}
\makeatother

\newgeometry{vmargin={25mm}, hmargin={22mm,22mm}}

\begin{document}

\title{Geometric tool kit for higher spin gravity (part III):\\ An introduction to the general theory of\\
	 connections on fibre bundles 
} 

\author{Xavier Bekaert}

\date{Institut Denis Poisson, Unit\'e Mixte de Recherche $7013$ du CNRS\\
Universit\'e de Tours, Universit\'e d'Orl\'eans\\
Parc de Grandmount, 37200 Tours, France\\
\vspace{2mm}
{\tt xavier.bekaert@univ-tours.fr}
}

\maketitle

\vspace{5mm}

\begin{abstract}
	 These notes provides an introduction to a wide variety of notions of connection that appear in the contemporary mathematical literature. Besides the familiar Ehresmann connections on principal bundles and Koszul connections on vector bundles, there is a large zoo of perhaps less familiar notions, most notably Cartan connections on principal bundles and their avatars on tractor bundles, which may deserve more attention for further elucidating higher-spin interactions and symmetries.	 
	This review is motivated both by the previous (explicit or implicit) use of these notions of connection in higher-spin gravity and by their potential future applications in this context. Much of the material covered here is also relevant to advanced mathematical topics in gravity, such as the rigorous treatment of conformal geometry (which is of interest for holography) and may help fill gaps left by standard textbooks (most of which typically do not address Cartan connections, tractor calculus, and related topics).
\end{abstract}

\thispagestyle{empty}

\clearpage


\pagestyle{empty}

\setcounter{tocdepth}{2}

\startcontents[main]

\printcontents[main]{}{1}{\section*{Contents}}

\clearpage


\setcounter{page}{1}
\pagestyle{plain}

\section{Introduction}

The Erlangen programme, formulated in 1872 by Klein, successfully unified most of the extensions of Euclidean geometry known at his time (\textit{e.g.} hyperbolic, projective, affine, etc) under the common theme of transitive actions of a Lie group on a manifold (in modern language).
However, another generalisation of Euclidean geometry had been put forward and fell out of the scope of the Erlangen programme: Riemannian geometry, which was formulated in 1854 by Riemann in his G\" ottingen lecture.
It was Cartan who succeeded to unify the visions of Klein and Riemann in a single approach during the 1920s. The perspective of Cartan on differential geometries (\textit{e.g.} Riemannian geometry, conformal geometry, etc) was to view them as (curved) ``deformations'' of the (flat) ``model'' geometries offered by Klein's programme.\footnote{Cartan presented his bird view of differential geometry in his book \cite{Cartan} but it was his student Ehresmann who formalised Cartan's ideas on connections in the modern language of fibre bundles \cite{Ehresmann:1950}. Nevertheless, Cartan connections quickly fell into oblivion due the modern focus on principal connections on $G$-structures by differential geometers, pushed forward by influential textbooks such as \cite{Kob63} (see, however, \cite{Kob72} for a notable exception). The classical textbook on Cartan connections that revived this subject is \cite{Sharpe} (see Sharpe's preface for some comments on the ``troubled history'' of Cartan geometries). A nice pedagogical review for theoretical physicists is \cite{Wise:2006sm}. A more modern and advanced textbook is \cite{CrampinSaunders}.} 

The aim of this text is to provide a self-contained overview of the vast landscape of connections\footnote{As a pun word and witty tribute to their seminal works, the trio of French mathematicians who built the foundations of the modern theory of connections (\textit{i.e.} \'Elie Cartan, Charles Ehresmann and Jean-Louis Koszul) is referred here as the ``French Connection''. Let us note that each of them gave his name to a general type of connection. As a genealogical remark, one may observe that Charles Ehresmann and Jean-Louis Koszul were the students of, respectively, \'Elie Cartan and Henri Cartan (father and son).}
(by \'Elie Cartan, Charles Ehresmann, Jean-Louis Koszul, etc) on the
 various standard types of fibre bundles (principal, vector, associated, etc).\footnote{This review is intended for theoretical physicists; accordingly, the abstract, geometric, and global definitions of connections are complemented by concrete expressions in local coordinates.} The long term goal is to look for their higher-derivative generalisations  (in the sense advocated in \cite{Bekaert:2023cmi}) suitable for higher-spin gravity (along the lines initiated in \cite{Bekaert:2022dlx}), but this goes beyond the scope of the present review. Nevertheless, to prepare the ground for such a future generalisation we deliberately complement the geometric definition of each notion of connections with an equivalent algebraic definition allowing for a natural higher-derivative generalisation.

Higher-spin gravity\footnote{Many pedagogical reviews on higher-spin gravity of various levels are available by now: advanced ones \cite{Reviews} as well as introductory ones \cite{Introductions}. Two books of conference proceedings also offer a panorama of this research area \cite{Proceedings}.} is formulated, in the frame-like approach, in terms of a Cartan-like gauge connection: a differential one-form on the spacetime manifold taking values in the higher-spin algebra. The nature and geometrical intepretation of this Cartan-like connection appearing in the available fully non-linear theories remains somewhat mysterious. Among other things, this problem is related to the fact that, despite some superficial similarities, a Cartan connection is \textit{not} the same thing as a principal connection. In more physical terms, a gravity theory is not a Yang-Mills theory, even when gravity is formulated \`a la Cartan. These subtleties are well-known for spins lower than two but they become conceptual obstacles for spins higher than two. 

In fact, the Cartan-like connection takes values in a higher-spin algebra. The latter is a Lie algebra, typically  infinite-dimensional, obtained from (a suitable quotient of) the universal enveloping algebra $U(\mathfrak{g})$ corresponding to the isometry (or conformal) algebra $\mathfrak{g}$ underlying the gravity theory that it extends. While Cartan connections provide a covariant way to identify the tangent space of the orthonormal (or conformal) frame bundle with the Lie algebra $\mathfrak{g}$, the higher-derivative analogue of this geometric picture, relevant for higher-spin gravity, remains elusive. These subtleties indicate that it is somewhat misleading to think about a Cartan-like gauge connection in fully non-linear higher-spin gravity as a sort of principal connection on an infinite-dimensional principal bundle. Hopefully, an exhaustive and systematic study of the theory of connections might help to come up with a generalisation that would help our conceptual and technical understanding of higher-spin gravity.

\subsection*{Plan}

The most general notion of connection is the definition by
Ehresmmann of connections on general fibre bundles as  horizontal distributions on their total spaces. This definition is reviewed in Section \ref{fibrations1}. 

In the particular case of principal bundles, two notions of connections stand out: principal connections and Cartan connections, which are reviewed at length in Section \ref{principalb} with an emphasis on the Lie algebroid viewpoint.  

Cartan geometries are the curved version of Klein geometries. In Section \ref{modelgeo}, the latter are reviewed as particular cases of the former (as well as geometries on their own). In particular, flat Cartan connection one-forms coincide with Maurer-Cartan one-forms in the case of Klein geometries. Various important examples of Klein geometries are briefly presented: affine, conformal, Euclidean, hyperbolic, spherical, \textit{etc}. 
The Klein geometries relevant for three-dimensional gravity theories are also presented (in Subsection \ref{H^2}).

Koszul connection on vector bundles are also known as linear connections. Their definition is reviewed in Section \ref{vectb}. The algebraic definition of a Koszul connection in the particular case of the tangent bundle produces the celebrated definition of an affine connection.

Principal bundles and vector bundles are deeply related via the notion of associated bundles, such as the profound bound between frame and tangent bundles. This deep relationship implies that a connection on one bundle can induce a connection on another. Some of the various possibilities are reviewed in Section \ref{associatedbundles}. Important examples of associated bundles are tractor bundles (reviewed in Subsection \ref{tractbdles}) which appear frequently in conformal geometry.  

To conclude, the relations between principal bundles (most notably, reductions) and the links between the corresponding Ehresmann and Cartan connections are investigated in Section \ref{relprincbundles}. 

Some additional material, interesting in its own right but not essential to the main discussion, has been moved to the appendix. In particular,
homogeneous bundles and induced representations are reviewed in Appendix \ref{inducedreps}. An abstract definition of higher-spin algebras is provided (in Subsection \ref{homvectbundles}).
Furthermore, the fundamentals of conformal/projective geometry relevant for holography of low-dimensional (usual and higher-spin) gravity theories are introduced (in Subsection \ref{GeneralisedVermamodules}). 
Various examples of standard geometric constructions are briefly reviewed in the language of $G$-structures in Appendix \ref{Gstructures}. The distinct conventions and jargons of conformal field theorists and geometers, about primary fields and densities, are compared in Appendix \ref{densitiesweights}. Finally, the soldering tools are reviewed in Appendix \ref{solderring}.

\subsection*{Prerequisites}

Some frightening words of modern mathematics literature will be used repeatedly throughout the notes because they provide such a convenient tool, for organising material and summarising many concepts in few words, that we decided to take the risk of using them, despite the fact it might scare some physicist colleagues.
The definitions of the following terms will not be reviewed here because they were already introduced in the part II of this series. The precise location of the corresponding definitions is indicated.

\begin{itemize}
	\item Module (see \textit{e.g.} \cite[Section 2.1]{Bekaert:2023jvl} or \cite[Section 1.1]{Sardanashvily2})
	\item Short exact sequence (see \textit{e.g.} \cite[Section 2.1]{Bekaert:2023jvl} or \cite[Section 1.1)]{Sardanashvily2})
	\item Lie algebroid  (see \textit{e.g.} \cite[Section 3.2.1]{Bekaert:2023jvl} or \cite[Definition 3.3.1]{Mackenzie2})
	\item Action algebroid  (see \textit{e.g.} \cite[Section 3.2.2]{Bekaert:2023jvl} or \cite[Example 3.3.7]{Mackenzie2})
\end{itemize}

The present lecture notes are the third part of a series. Connections on principal and vector bundles (and, more generally, on Lie algebroids) were already introduced in the second part \cite{Bekaert:2023jvl} but they will be reviewed here in a self-contained manner

\pagebreak

 \section{Ehresmmann connections on general fibre bundles}\label{fibrations1}

Let $E$ be a fibre bundle over $M$ defined by the fibration $\pi:E\twoheadrightarrow M$ of the total space $E$ over its base $M$.
In a local trivialisation, the coordinates of a point $p\in E$ read as ($x^\mu$, $y^i$) where $x^\mu$ denotes the coordinates of the base point $\pi(p)\in M$ while $y^i$ denotes coordinates on the typical fibre, \textit{i.e.} $\pi:(x^\mu,y^i)\mapsto x^\mu\,$.
Let 
\be
E_m=\pi^{-1}(m)=\{p\in E\mid \pi(p)=m\}
\ee
denote the fibre above the base point $m\in M$. Fibres will always be implicitly assumed connected.

A \textbf{fibre bundle automorphism} is a diffeomorphism of the total space $E$ that maps fibres to fibres. In particular, it induces a diffeomorphism of the base $M$. In coordinates, it corresponds to a transformation of the form
\be
(x^\mu,y^i)\,\to\,\Big(\,x^{\prime\mu}(x)\,,\,y^{\prime i}(x,y)\,\Big)\,.
\ee
The collection of all (local)  sections $\sigma:M\hookrightarrow E$ of the fibration $\pi:E\twoheadrightarrow M$ (\textit{i.e.} maps such that $\pi\circ\sigma=id_M$) will be denoted $\Gamma(\pi)$, or $\Gamma(E)$ when there is no ambiguity, and called the \textbf{sheaf of sections} (with a slight abuse of terminology).

 \subsection{Pullback bundle of a fibre bundle}

Let $F:N\to M$ be a map from a manifold $N$ to a manifold $M$, which reads $F:z^\alpha\mapsto x^\mu$ in a local trivialisation, where $z^\alpha$ denotes some coordinates on $N$. Consider the fibre bundle $\pi:E\twoheadrightarrow M$ as above.

One may define the fibre bundle over $N$ whose fibre at a point $n\in N$ is defined to be the fibre at $F(n)\in M$ of the fibre bundle $\pi:E\twoheadrightarrow M$. The fibre bundle defined in this way is called the \textbf{pullback bundle of the fibration} $\pi$ \textbf{by the map} $F$. Its total space is denoted $F^*E$ and its fibration is denoted $F^*\pi:F^*E\twoheadrightarrow N$.
This means that the fibre above the base point $n\in N$ is
\be\label{defibrepullback}
F_n^*E=E_{{}_{F(n)}}=\pi^{-1}\big(F(n)\big)=\{\,p\in E\mid \pi(p)=F(n)\,\}
\ee
More concretely, the coordinates of a point $q\in F^*E$ read in a local trivialisation as ($z^\alpha$, $y^i$) where $y^i$ are local coordinates on the fibre of the bundle $E$. The fibration of the pullback bundle reads as $F^*\pi:(z^\alpha,y^i)\mapsto z^\alpha$. 
There is also a canonical morphism of fibre bundles, $\pi^*F:F^*E\to E$, which maps the fibre $F_n^*E$ to the fibre $E_{{}_{F(n)}}$ in an obvious way, according to the definition \eqref{defibrepullback}.

This canonical construction of the pullback bundle can be summarised in the commutative diagram
\be
\begin{array}
	[c]{ccc}
	F^*E&\stackrel{\pi^*F}{\to}&E\\
	&&\\
	F^*\pi\downarrow\quad&&\downarrow \pi\\
	&&\\
	N &\stackrel{F}{\to}& M
\end{array}
\ee
Both vertical arrows are fibrations while the upper horizontal arrow is the morphism of fibre bundles induced from the lower horizontal arrow (which is a morphism of manifolds). 

\subsection{Tangent structures of a fibre bundle}

With respect to a local trivialisation, vector fields on $E$ will read as
\be
{Y} = Y^{\mu}(x,y)\,\partial_{\mu}+Y^{i}(x,y)\,\partial_i\,.
\label{loccoords}
\ee
At the level of tangent bundles, the pushforward by $\pi$ is the vector bundle morphism $\pi_*:TE\twoheadrightarrow TM$
from the tangent bundle of $E$ onto the tangent bundle of $M$. 
In coordinates it reads as
\be
\pi_{*p}\,:\,T_pE\twoheadrightarrow T_{\pi(p)}M\,:\, (\,Y^{\mu},Y^i)\mapsto Y^\mu\,,
\label{pist}
\ee
where only the mapping in the fibres (of the tangent bundles) has been given.\footnote{For more details and rigorous proofs of the material reviewed in this subsection, see \textit{e.g.} \cite[Chapter 3.2]{Saunders}.}

\subsubsection{Vertical and projectable vector fields}

The kernel of $\pi_*$ is a vector sub-bundle over $E$ of the tangent bundle $TE$. It is called the \textbf{vertical distribution} and it is denoted $VE$ ($=\text{Ker}\,\pi_*$).
From \eqref{pist}, it is clear that a vertical tangent vector is such that all base components vanish, \textit{i.e.} $Y^{\mu}=0$.
Therefore, the coordinates on the \textbf{vertical space} $V_pE$ are the components $Y^i$.
The sections of the vertical distribution $VE$ are called \textbf{vertical vector fields} and read in local coordinates as
\be
{Y}_V = Y^{i}(x,y)\,\partial_i\,.
\label{vertvectf}
\ee
The space $ \Gamma(VE)$ of vertical vector fields will be called the \textbf{vertical sheaf}.
The vertical sheaf is endowed with a structure of Lie algebra via the Lie bracket of vector fields on $E$ (and even of Lie algebroid over $E$ via the canonical embedding $i:VE\hookrightarrow TE$ as anchor). Indeed, the vertical distribution is involutive:
the Lie bracket of two vertical vector fields is also vertical. In fact, the vertical distribution defines the \textbf{vertical foliation} whose leaves are simply the fibres of the bundle $E$ over $M$.
The collection $\{\partial_i\}$ of vertical vector fields is
a coordinate basis of the vertical sheaf $ \Gamma(VE)$ seen as ${C}^\infty(E)$-module,
as is clear from \eqref{vertvectf}.

At the level of tangent sheaves, the pushforward by $\pi$ is the linear map 
\be\label{pushpi}
\pi_*:\Pi \Gamma(TE)\twoheadrightarrow  \Gamma(TM)
\ee
where the subspace 
\be
\Pi \Gamma(TE)\,:=\,(\pi_*)^{-1} \Gamma(TM)\,\subset\, \Gamma(TE)
\ee
is spanned by the vector fields on $E$ which are projectable on $M$ by $\pi$. In coordinates, a vector field on $E$ projectable on $M$ reads as
\be
{Y}_\Pi = Y^{\mu}(x)\,\partial_{\mu}+Y^{i}(x,y)\,\partial_i\,,
\label{Pidecomp}
\ee
since $\pi_*{Y}_\Pi = Y^{\mu}(x)\,\partial_{\mu}$ is a well-defined vector field on $M$.
The kernel of the pushforward \eqref{pushpi} is the vertical sheaf $ \Gamma(VE)$.
There is a short exact sequence of ${C}^\infty(M)$-modules
\be
0\to  \Gamma(VE)\stackrel{i}{\hookrightarrow}\Pi \Gamma(TE)\stackrel{\pi_*}{\twoheadrightarrow} \Gamma(TM)\to 0\,,
\label{shortexactEM}
\ee
where all arrows are ${C}^\infty(M)$-linear maps.

\subsubsection{Projectable vector fields as infinitesimal automorphisms}

A vector field ${Y}$ on $E$ is projectable on $M$ if and only if its Lie bracket along any vertical vector field is also vertical:
\be
{Y}\subset \Pi \Gamma(TE)\quad \Longleftrightarrow \quad
{\mathcal L}_{ \Gamma(VE)}{Y}=[ \Gamma(VE),{Y}]\subset  \Gamma(VE)\,.
\label{projectable}
\ee
This is geometrically clear since it is equivalent to say that a vector field is projectable if and only if its infinitesimal variation along any vertical flow is purely vertical. Notice that another way to intepret the condition \eqref{projectable} is by writing it as ${\mathcal L}_{{Y}} \Gamma(VE)=[{Y}, \Gamma(VE)]\subset  \Gamma(VE)$, thence a projectable vector field preserves the vertical sheaf (\textit{i.e.} the Lie derivative of a vertical vector field along a projectable vector field is a vertical vector field)
and thus generates a flow of automorphisms of the fibre bundle (\textit{i.e.} diffeomorphisms of $E$ mapping fibres on fibres).
In other words, the projectable vector fields are the infinitesimal automorphisms of a fibre bundle. 

The vertical sheaf $ \Gamma(VE)$ is thus an ideal of the Lie algebra of projectable vector fields $\Pi\Gamma(TE)$.
In fact, as Lie algebras one may characterise $\Pi\Gamma(TE)$ as the normaliser of the vertical sheaf $ \Gamma(VE)$ inside the tangent sheaf $\Gamma(TE)$. 
The Lie algebra of vector fields on $M$ is isomorphic to the corresponding quotient:
\be\label{TMPiTEVE}
\Gamma(TM)\cong \Pi\Gamma(TE)\,/\, \Gamma(VE)\,.
\ee
This is manifest in local coordinates, \textit{cf.} equation \eqref{Pidecomp}, and was already expressed in the short exact sequence \eqref{shortexactEM}.

\vspace{3mm}
\noindent\textbf{Remark:} From now on, the Lie algebras of vector fields on $M$ will either be denoted $ \Gamma(TM)$ or $\mathfrak{X}(M)$ (when the emphasis is on the Lie algebra structure).
\vspace{3mm}

The pullback by $\pi$ provides an embedding $\pi^*:{C}^\infty(M)\hookrightarrow {C}^\infty(E):f\mapsto f\circ \pi$ of the commutative algebra of functions of the base space $M$ inside the commutative algebra of functions of the total space $E$.
The image $\pi^*{C}^\infty(M)\subset {C}^\infty(E)$ can be identified with the subspace of functions on $E$ that are annihilated by any vertical vector field, \textit{i.e.} that are constant along the fibre.
In coordinates, this is obvious since this subspace correspond to functions which do not depend on the fibre coordinates $y^i$.

The vertical distribution is canonical for a fibre bundle, this provides a notion of verticality of tangent vectors. However, their horizontality is not canonical and requires the introduction of an Ehresman connection (\textit{i.e.} a horizontal distribution, defined in the next subsection). However, for the dual concepts the situation is reverse: there is a canonical notion of \textbf{horizontal cotangent vectors} as the one-forms on $T_pE$ annihilating $V_pE$. Accordingly, a differential one-form is horizontal if it is a section of the horizontal cotangent bundle denoted Ann\,$VE\,\subset\,T^*E\,$. However, there is no canonical notion of verticality for cotangent vectors.

\subsection{French connection (part I)\,: Charles Ehresmann}

The short exact sequence of linear maps
\be
0\to V_pE\stackrel{i|_p}{\hookrightarrow}T_pE\stackrel{\pi_*|_p}{\twoheadrightarrow}T_{\pi(p)}M\to 0\,
\label{shortexactTM}
\ee
expresses that the tangent space at the base point $\pi(p)\in M$ is isomorphic to the 
quotient of the tangent space at $p\in E$ by the vertical subspace: $T_{\pi(p)}M\cong T_pE/V_pE$.
A linear splitting of the short exact sequences \eqref{shortexactTM} for all points $p\in E$,
\be
0\leftarrow V_pE\stackrel{\omega|_p}{\twoheadleftarrow}T_pE\stackrel{\gamma|_p}{\hookleftarrow}T_{\pi(p)}M\leftarrow 0\,,
\label{splitTM}
\ee
is called an \textbf{Ehresmann connection} on the fibre bundle $\pi:E\twoheadrightarrow M$ \cite{Ehresmann:1950} where, by definition, $\omega|_p\circ i|_p=id_{V_pE}$ and $\pi_*|_p\circ\gamma|_p=id_{T_{\pi(p)}M}$.
The splitting is linear in the sense that each arrow in \eqref{splitTM} is a morphism of vector spaces.
If the fibres of $E$ are endowed with an extra (\textit{e.g.} group) structure, then special instances of Ehresmann connections 
(\textit{e.g.} principal connections) arise when the field of morphisms has a compatible vertical dependence (\textit{e.g.} equivariant).

\subsubsection{Horizontal lift and vertical projector}

The short exact sequences \eqref{shortexactTM} holds in each fibre so there is a short exact sequence of vector bundles
\be
0\to VE\stackrel{i}{\hookrightarrow}TE\stackrel{\pi_*}{\twoheadrightarrow}TM\to 0\,.
\label{shortexactvectTM}
\ee
However, the sequence \eqref{shortexactvectTM}
is not uniform because the first two terms (\textit{i.e.} $VE$ and $TE$) are vector bundles over the fibre bundle $E$ while
the third term (\textit{i.e.} $TM$) is a vector bundler over the base manifold $M$.
Nevertheless, there is also a canonical short exact sequence of morphisms of vector bundles over $E$:
\be
0\to VE\stackrel{i}{\hookrightarrow}TE\stackrel{\pi_*}{\twoheadrightarrow}TM\underset{M}{\times}E\to 0\,
\label{shortexactvectTEM}
\ee
where $TM\times_M E$ is the fibrewise product of the two fibre bundles over $M$. Local coordinates on  $TM\times_M E$ are  $(x^\mu,y^i,Y^\mu)$ where we made manifest that it can also be seen as a vector bundle over $E$ with fibre coordinates $Y^\mu$.
An Ehresmann connection provides a linear splitting of \eqref{shortexactvectTEM} 
\be
0\leftarrow  VE\stackrel{\omega}{\twoheadleftarrow} TE\stackrel{\gamma}{\hookleftarrow}TM\times_M E\leftarrow 0\,.
\label{splitT(E)}
\ee
where each arrow is a morphism of vector bundles over $E$.

The field $\gamma$ on $E$ of sections 
$\gamma|_p:T_{\pi(p)}M\hookrightarrow T_pE$
is called the \textbf{horizontal lift} of tangent vectors on the base manifold $M$ to tangent vectors on the total space $E$. 
In local coordinates, it reads as 
\be
\gamma: X^\mu\partial_\mu\mapsto X^{\mu}\,\nabla_\mu\,,\qquad\text{where}\quad \nabla_\mu\,:=\,\partial_{\mu}\,-\,\omega_\mu^i(x,y)\,\partial_i
\label{lincon}
\ee
is the horizontal lift of the coordinate basis $\partial_\mu$, which contains an extra vertical term.
For any tangent vector on $M$, the pushforward by the fibration $\pi$ of its horizontal lift  
gives back the original vector field on $M$.
The image $\gamma|_p\big(\,T_{\pi(p)}M\,\big)$ is the horizontal lift at $p$ of all tangent vectors at the point $\pi(p)$ on the base, it is called the \textbf{horizontal space} at $p$ and is denoted $H_pE$. The coordinate basis of the horizontal space is $\{\nabla_\mu\}$.
The image $\gamma\big(\,TM\,\big)$ is formed by all horizontal lifts of all base tangent vectors, it is a vector sub-bundle of $TE$ denoted $HE$ and called
the \textbf{horizontal distribution} on $E$.
The splitting \eqref{splitTM} means that the tangent bundle is the direct sum of the vertical and horizontal distributions:
$TE=VE\oplus HE$. Equivalently, the vertical and horizontal distributions are transversal in the sense that $VE\,\cap\, HE$ is the zero section of $TE$.

In \eqref{splitTM}, the field $\omega$ on $E$ of retractions $\omega|_p:T_pE\hookrightarrow V_pE$ is a vertical-valued differential one-form on $E$ called the \textbf{connection one-form} or \textbf{vertical projector}. It maps tangent vectors on $E$ to vertical vectors. 
In local coordinates, it reads as 
\be
\omega{(x,y)}: Y^{\mu}\partial_{\mu}+Y^{i}\partial_i\mapsto \Big(Y^\mu\,\omega_\mu^i(x,y)+Y^{i}\Big)\,\partial_i\,,
\ee
or, equivalently,
\be
\omega=\Big(\omega_\mu^i(x,y)\,dx^\mu\,+\,dy^{i}\Big)\otimes\partial_i\,.
\ee
The kernel of the connection one-form is the horizontal distribution: $\text{Ker}\,\omega=HE$. In other words, a tangent vector at $p$ is horizontal if and only if it is sent to zero by the one-form $\omega|_p$\,. The connection one-form can be seen as a field of vertical projections therefore its complement ${\mathcal H}:=id_{TE}-\omega$ defines a field of projections onto the horizontal subspaces called the \textbf{horizontal projector}.
In local coordinates, it reads as 
\be
{\mathcal H}:Y^{\mu}\partial_{\mu}+Y^{i}\partial_i\mapsto Y^\mu\,\nabla_\mu\,.
\ee

A \textbf{horizontal section} is a section $\sigma:M\hookrightarrow E$ such that the image of its pushforward 
\be
\sigma_*:TM\hookrightarrow TE:X^\mu\partial_\mu\mapsto X^{\mu}\big(\partial_\mu+\partial_\mu \sigma^i(x)\,\partial_i\big)
\ee
is horizontal, \textit{i.e.} $\sigma_*(TM)\subset HE$. The composition of the pushforward by a section with the connection one-form reads
\be\label{covderEhr}
\omega\circ\sigma_*:TM\to VE
:X^\mu\partial_\mu\mapsto X^\mu\,\Big(\,\partial_\mu \sigma^i(x)+\omega_\mu^i\big(\,x,\sigma(x)\,\big)\,\Big)\partial_i\,,
\ee
or, equivalently,
\be
\sigma^*\omega=\partial_i\otimes\Big(\,d\sigma^i(x)+\omega^i\big(\,x,\sigma(x)\,\big)\,\Big)\,,
\ee
where $\omega^i:=\omega_\mu^i dx^\mu$.
Therefore, a horizontal section is a solution of the system of first-order partial differential equations $\partial_\mu \sigma^i(x)+\omega_\mu^i\big(\,x,\sigma(x)\,\big)=0$.
The Ehresmann connection is said \textbf{flat} if the horizontal distribution $HE$ is involutive, in which case the leaves are  horizontal sections and the corresponding foliation is called the \textbf{horizontal foliation} defined by a flat Ehresmann connection.

The short exact sequence \eqref{shortexactEM} of Lie algebras and ${C}^\infty(M)$-modules
expresses that the Lie algebra $\Gamma(TM)$ of vector fields on $M$ is isomorphic to the quotient of the Lie algebras $\Pi\Gamma(TE)$ of
vector fields on $E$ projectable on $M$ by the $\Gamma(VE)$ of vertical vector fields on $E$.
In other words, the Lie algebra of projectable
vector fields is an extension of the Lie algebra of base vector fields by the Lie algebra of vertical vector fields.
An Ehresmann connection provides a splitting of the short exact sequence \eqref{shortexactEM} of ${C}^\infty(M)$-modules
\be
0\leftarrow  \Gamma(VE)\stackrel{\omega}{\twoheadleftarrow}\Pi\Gamma(TE)\stackrel{\gamma}{\hookleftarrow}\Gamma(TM)\leftarrow 0\,.
\label{splitT(M)}
\ee
More concretely, the horizontal lift 
\be
\gamma\,:\, \Gamma(TM)\hookrightarrow\Pi \Gamma(TE)\,:\,{X}=X^\mu(x)\,\partial_\mu\mapsto \gamma({{X}})=X^\mu(x)\nabla_\mu
\ee 
is a
${C}^\infty(M)$-linear map, sending vector field on the base $M$ to vector fields on $E$ projectable on $M$, and such that,
for any vector field ${X}$ on $M$, the pushforward by the fibration $\pi$ of its horizontal lift $\gamma({{X}})$ 
gives back the original vector field on $M$: $\pi_*\big(\gamma({{X}})\big)={X}$ (since $\pi_*\circ\gamma=id_{ \Gamma(TM)}$  by definition of splitting).

\subsubsection{Curvature two-form}\label{curv2formsect}

A horizontal lift is flat
if and only if it is a Lie algebra morphism $\gamma:\Gamma(TM)\hookrightarrow \Gamma(TE)$ from the tangent sheaf of vector fields on $M$ to the tangent sheaf of vector fields on $E$.
More generally, the \textbf{curvature of the horizontal lift} $\gamma$ is the vertical-valued antisymmetric ${C}^\infty(M)$-bilinear map
\be\label{Kurvature}
K\,:\, \Gamma(TM)\times  \Gamma(TM)\to  \Gamma(VE)\,:\,({X}_1,{X}_2)\,\mapsto\, 
\big[\,\gamma({{X}_1})\,,\,\gamma({{X}_2})\,\big]-\gamma\big(\,[{X}_1,{X}_2]\,\big)
\ee 
which measures the failure of the horizontal lift $\gamma$ to be a Lie algebra morphism. This is to be contrasted with the pushforward $\pi_*$ by the fibration which is always a Lie algebra morphism. The verticality of the curvature (\textit{i.e.} $\pi_*\circ K=0$) follows from the properties that the pushforward $\pi_*$ by the fibration is a Lie algebra morphism (\textit{i.e.} $\pi_*[{Y}_1,{Y}_2]=[\pi_*{Y}_1,\pi_*{Y}_2]$) and that the horizontal lift $\gamma$ is a section of $\pi_*$ (\textit{i.e.} $\pi_*\circ \gamma=id_{ \Gamma(TM)}$). The ${C}^\infty(M)$-bilinearity  of the curvature follows from the Leibniz rule for the Lie bracket.
More geometrically, the antisymmetric ${C}^\infty(M)$-bilinear map \eqref{Kurvature} can be thought as a vertical-valued differential two-form 
$K:\wedge^2TM\to VE$ on $M$.

A horizontal differential form on $E$ 
 annihilates any vertical vector field. In local coordinates, a horizontal differential $q$-form reads as 
\be
\alpha=\frac1{q!}\alpha_{\mu_1\ldots\mu_q}(x,y)dx^{\mu_1}\wedge\ldots\wedge dx^{\mu_q}\,.
\ee
The \textbf{curvature two-form} of $\omega$ is the vertical-valued horizontal\footnote{The horizontality can be shown by making use of formula \eqref{domegaformula} and checking that (i) if ${X}_1$ is a horizontal vector field and ${X}_2$ a vertical one (hence $\omega({X}_1)=0$, $\omega({X}_2)={X}_2$ and $[{X}_1,{X}_2]$ is vertical), then $(d\omega)({X}_1,{X}_2)=0$, and (ii) if ${X}_1$ and ${X}_2$ are vertical (\textit{i.e.} $\omega({X}_1)={X}_1$ and $\omega({X}_2)={X}_2$), then $(d\omega)({X}_1,{X}_2)=[{X}_1,{X}_2]$.} differential two-form on $E$, defined by
\be\label{curv2formEhresm}
\Omega=d\omega+\frac12\,[\omega,\omega]\,,
\ee
where the Lie bracket on the right-hand side is the Lie bracket of vector fields on $E$.
More explicitly, the curvature two-form $\Omega$ is the antisymmetric ${C}^\infty(E)$-bilinear map
\ba\label{curv2form}
\Omega&:& \Gamma(TE)\times \Gamma(TE)\to  \Gamma(VE)\\
&:&({X}_1,{X}_2)\mapsto {\mathcal L}_{{X}_1}\,\omega({{X}_2})\,-\,{\mathcal L}_{{X}_2}\,\omega({{X}_1})\,
-\,\omega\big(\,[{X}_1,{X}_2]\,\big)\,+\,\big[\,\omega({{X}_1})\,,\,\omega({{X}_2})\,\big]\nonumber\,,
\ea 
as can be obtained from \eqref{curv2formEhresm} by making use of the formula
\be\label{domegaformula}
(d\omega)({X}_1,{X}_2)\,=\,{\mathcal L}_{{X}_1}\,\omega({X}_2)\,-\,{\mathcal L}_{{X}_2}\,\omega({X}_1)\,-\,\omega\big(\,[{X}_1,{X}_2]\,\big)\,,
\ee
where ${X}_1$, ${X}_2$ are vector fields on $E$ and $[{X}_1,{X}_2]$ denotes their Lie bracket. 
The horizontality allows to express the curvature two-form in a compact way in terms of the horizontal projector ${\mathcal H}$ as 
\be
\Omega\,=\,(d\omega)\circ{\mathcal H}\,=\,-\,\omega\circ[\,,\,]\circ{\mathcal H}\,,
\ee
meaning that
\be
\Omega({X}_1,{X}_2)=(d\omega)({\mathcal H}{X}_1,{\mathcal H}{X}_2)=-\omega\big(\,[{\mathcal H}{X}_1,{\mathcal H}{X}_2]\,\big)\,.
\ee
The right-hand-side makes manifest that the curvature two-form (i) is horizontal, (ii) is vertical-valued and (iii) vanishes identically if and only if the horizontal distribution is involutive (\textit{i.e.} if and only if the Lie bracket of horizontal vector fields is always horizontal).

\pagebreak

 \section{Cartan and Ehresmann connections on principal bundles}\label{principalb}

 \subsection{Principal bundles via group actions}

Consider a free (and proper\footnote{See \textit{e.g.} \cite[Section 4.2]{Sharpe} for a definition of this technical assumption.}) right action of a Lie group $H$ on a manifold $P$. The $H$-orbit through the point $p\in P$ is denoted as
$pH$.
The orbit space $P/H$ is a quotient manifold
and the canonical projection on $H$-orbits,
\be\label{piprincipbundle}
\pi:P\twoheadrightarrow P/H:p\mapsto pH\,,
\ee
endows $P$ with a structure of principal $H$-bundle over $P/H$ (see \textit{e.g.} \cite[Theorem 4.2.4]{Sharpe} or \cite[Theorem 24.1.1]{Dubrovin}). Each orbit is a manifold (\textit{cf.} \cite[Lemma 4.2.13]{Sharpe}) on which the right $H$-action is regular (\textit{i.e.} free and transitive).
A principal $H$-bundle $P$ over $P/H$ can be denoted in a symbolic way that mimicks a short exact sequence as follows: 
\be
H\lefttorightarrow P \twoheadrightarrow P/H\,,
\ee
where the first arrow stands for the free right action of the structure group $H$ on the total space $P$ of the principal bundle.

\vspace{3mm}
\noindent{\small\textbf{Example (Ray bundle)\,:} The simplest example of principal bundle corresponds to the case when the structure group is the one-dimensional additive Lie group $H=\mathbb R$. Geometrically, the orbits are parametrised open curves (\textit{i.e.} maps from $\mathbb R$ to $P$) hence a principal $\mathbb R$-bundle $P$ identifies with a congruence of parametrised open curves on the total space $P$. Similarly, a principal $U(1)$-bundle identifies with a congruence of parametrised closed curves.\footnote{\label{strictlyspeaking}To be more precise, a technical assumption is that the $\mathbb R$-action (corresponding to the parametrisation of curves) on $P$ should be proper for both congruences, otherwise the quotient $P/\mathbb{R}$ might not be a manifold.}
	For this reason, principal $\mathbb R$-bundles are sometimes called \textbf{ray bundles}. They should be distinguished from \textbf{lines bundles} which are vector bundles of rank one (while ray bundles can be thought as affine bundles of rank one).}
\vspace{3mm}

One will later introduce the standard global definitions of connections, \textit{etc}, but in practice all our considerations will be local in essence so one will actually be interested mostly in the infinitesimal right action.
Therefore, the Lie algebra $\mathfrak{h}$ of $H$ will play a crucial role.
Consequently, it will be useful to introduce a basis  for $\mathfrak{h}$, which will be denoted $\{\texttt{T}_i\}$ with $C^i_{jk}$ the corresponding structure constants: $[\,\texttt{T}_j,\texttt{T}_k]=C^i_{jk}\texttt{T}_i$.

 \subsection{Tangent structures of a principal bundle}

One can summarise the main messages of this section as follows. There are three canonical algebroids over the total space $P$ of a principal bundle: the tangent bundle $TP$, the vertical distribution $VP$ and the fundamental action algebroid $P\rtimes\mathfrak{h}$, where there is a canonical isomorphism $VP\cong P\rtimes\mathfrak{h}$ between the latter two.
Quotienting these three Lie algebroids by the action of the structure group $H$, one obtains three canonical algebroids over the base space $P/H$: the Atiyah algebroid $\frac{TP}{H}$, the vertical subalgebroid $\frac{VP}{H}$ and the adjoint algebroid $P\times_{Ad_H}\mathfrak{h}$, where there is a canonical isomorphism $\frac{VP}{H}\cong P\times_{Ad_H}\mathfrak{h}$ between the latter two.  This is summarised in Table \ref{Lielagebroidsprincipbundle}.

\begin{table}
	\begin{center}
		\begin{tabular}{
				|c|c|}
			\hline
			Lie algebroids over total space $P$ & Lie algebroids over base space $P/H$ \\ 
			\hline\hline
			Tangent algebroid $TP$ & Atiyah algebroid $\frac{TP}{H}$\\
			$\cup$&$\cup$\\
			Vertical distribution $VP$ & Vertical subalgebroid $\frac{VP}{H}$\\
			$\cong$&$\cong$\\
			Fundamental action algebroid & Adjoint algebroid\\
			$P\rtimes\mathfrak{h}$&$P\times_{Ad_H}\mathfrak{h}$\\\hline
		\end{tabular}
	\end{center}
	\caption{Canonical Lie algebroids for a principal bundle $H\lefttorightarrow P \twoheadrightarrow P/H$}\label{Lielagebroidsprincipbundle}
\end{table}

\subsubsection{Two algebroids over the total space: the vertical distribution and the fundamental action algebroid}

Consider a principal $H$-bundle $P$, \textit{i.e.} $H\lefttorightarrow P \twoheadrightarrow P/H$.
At the level of tangent bundles, the pushforward by the canonical projection \eqref{piprincipbundle} 
is the vector bundle morphism 
\be
\pi_*:TP\twoheadrightarrow T\frac{P}{H}\,.
\ee
The leaves of the vertical distribution $VP$ are the orbits $pH$.

The right action of $H$ on $P$ is an antihomomorphism of groups
\be\label{rightactionHP}
R_\bullet\,:\,H\hookrightarrow{Diff}(P)\,:\,h\mapsto R_h\,,
\ee 
where the diffeomorphism $R_h:p\mapsto ph$ stands for the vertical translation by $h\in H$. The action is free, that is to say for any $p\in P$: $ph=p\Leftrightarrow h=e$\,.
The infinitesimal version of the free action of the Lie group $H$ on the principal bundle $P$
is the free action of the Lie algebra of $\mathfrak{h}$ on $P$. More precisely, it is an injective Lie algebra morphism 
\be
\#\,:\,\mathfrak{h}\hookrightarrow  \Gamma(VP)\,:\,y\mapsto y^\#=\left.\frac{d}{dt}(R_{\exp ty})^*\,\right|_{t=0}\,,
\label{fund}
\ee
called the \textbf{fundamental action of the Lie algebra} $\mathfrak{h}$ \textbf{on the principal bundle} $P$,
that sends (see \textit{e.g.} \cite[Proposition I.4.1]{Kob63}) a Lie algebra element to a nowhere (or everywhere) vanishing vertical vector field on $P$, \textit{i.e.}
\be
y^\#|_p\neq0\,,\, \forall p\in P\,\quad\,\Leftrightarrow\,\quad\, y\neq 0
\ee
or, equivalently, 
\be
\exists p\in P\,:\,y^\#|_p=0\,\quad\,\Leftrightarrow\,\quad\, y= 0\,.
\ee
The nowhere (or everywhere) vanishing vertical vector fields on $P$ that span the image $\mathfrak{h}^\#$ are called the \textbf{fundamental vector fields} of the principal bundle $P$.
They span a Lie algebra isomorphic to $\mathfrak{h}$.
The fundamental action intertwines two $\mathfrak{h}$-modules: it sends the adjoint module, $\mathfrak{h}$, onto the module 
of fundamental vector fields, $\mathfrak{h}^\#$. In fact, the fundamental actions relates the adjoint representation of $\mathfrak{h}$ on itself to the Lie derivative as follows: 
\be
\#\circ ad = {\mathcal L}\circ \#\,.
\ee
More explicitly, $(ad_yz)^\#={\mathcal L}_{y^\#}z^\#$ for any $y$ and $z$ in $\mathfrak{h}$. Therefore,
\be
{\mathcal L}_{\texttt{T}_j^\#}\texttt{T}_k^\#=C^i_{jk}\texttt{T}_i^\#\,.
\label{strctss}
\ee
In fact, the free action of $\mathfrak{h}$ on $P$ defines a free action Lie algebroid
\be\label{fundactanchor}
\#_\bullet\,:\,P\rtimes\mathfrak{h}\hookrightarrow TP\,:\,(p,y)\mapsto y^\#|_p\,.
\ee
which will be called the \textbf{fundamental action algebroid} whose total space is denoted $P\rtimes_\#\mathfrak{h}$.
Since the fundamental action is \textit{free}, it provides an $H$-equivariant field of \textit{isomorphisms}
\be
\#_p\,:\,\mathfrak{h}\stackrel{\sim}{\to} V_pP\,:\,y\mapsto y^\#|_p
\label{diese}
\ee
such that
each vertical space $V_pP$ is isomorphic to the Lie algebra $\mathfrak{h}$.
In fact, the restriction of the codomain of the anchor \eqref{fundactanchor} to the vertical distribution $VP\subset TP$ defines an isomorphism of Lie algebroids over $P$
\be\label{fundiso}
\#_\bullet\,:\,P\rtimes\mathfrak{h}\stackrel{\sim}{\to}VP\,:\,(p,y)\mapsto y^\#|_p\,.
\ee
between the fundamental action algebroid $P\rtimes_\#\mathfrak{h}$ and the vertical distribution $VP$, which will be called the \textbf{fundamental isomorphism}.\footnote{See \textit{e.g.} \cite[Proposition 5.1.3]{Hamilton} for a proof of the isomorphism at the level of vector bundles. The extension to Lie algebroids is straightforward (one should check that it is a Lie algebra isomorphism) but note that, strictly speaking, it is an anti-isomorphism because the map reverses the sign of the Lie bracket. his subtlety (only due to sign convention mismatch related to left vs right action conventions) will be left implicit, in order to avoid overloading the discussion (see \textit{e.g.} \cite[Appendix B]{Mackenzie} for a careful discussion on these conventions).}
The inverse of the fundamental isomorphism will be called the \textbf{canonical isomorphism}. It reads
\be\label{fundiso'}
\iota\,:\,VP\stackrel{\sim}{\to}P\rtimes\mathfrak{h}\,:\,y^\#\mapsto \big(\,\tau_P(y)\,,\,y\,\big)\,,
\ee
where $\tau_P:TP\twoheadrightarrow P$ is the fibration of the tangent bundle over $P$.
The fundamental vector fields $\{\texttt{T}_i^\#\}\subset \mathfrak{h}^\#$ corresponding to a basis $\{\texttt{T}_i\}\subset\mathfrak{h}$ of the Lie algebra 
provide a basis\footnote{Note that this basis is holonomic if and only if $\mathfrak{h}$ is Abelian.} of the vertical sheaf $ \Gamma(VP)\cong {C}^\infty(P)\rtimes\mathfrak{h}$ isomorphic to the space of $\mathfrak{h}$-valued functions on $P$. Let $y^i$ be the coordinates on $\mathfrak{h}$, \textit{i.e.} $y=y^i\texttt{T}_i\in \mathfrak{h}$ then vertical vector fields read as 
\be
{Y}_V = Y^{i}(x,y)\,\texttt{T}_i^\#\,.
\label{vertvectfpr}
\ee

\vspace{3mm}
\noindent{\small\textbf{Example (Ray bundle)\,:} Let $M$ be a manifold endowed with a complete and nowhere-vanishing vector field ${X}\in \Gamma(TM)$ without periodic integral curves. The integral curve 
	\be
	C_m\,:\,\mathbb{R}\hookrightarrow M\,:\,t\mapsto C_m(t)
	\ee
	of the vector field ${X}$, passing through the point $m\in M$, is the unique solution to the differential equation
	\be
	\frac{dC_m(t)}{dt}={X}|_{C_m(t)}
	\ee 
	with $t\in\mathbb{R}$ and initial condition $C_m(0)=m$ while the left-hand side stands for the directional derivative along $C_m$. 
	Since ${X}$ is assumed to be complete, its integral curves exist for all values of the parameter $t$ and, consequently, the flow 
	\be
	\exp(\bullet{X})\,:\,\mathbb{R}\to Diff(M)\,:\,\lambda\mapsto\exp(t{X})
	\ee
	is global, thus inducing a well-defined right-action of the additive Lie group $\mathbb{R}$ on $M$. Since ${X}$ is also assumed to be nowhere-vanishing, this action is free (\textit{i.e.} the flow is injective). 
	The integral curve 
	\be
	C_m(\mathbb{R})=\{C_m(t)\in M\mid t\in \mathbb{R} \}\subset M
	\ee
	through $m\in M$ is thus the orbit of $m$ under the $\mathbb R$-action. Without loss of generality, such a principal $\mathbb R$-bundle $M$ can be assumed to be trivial (see \textit{e.g.} \cite[Proposition 16.14.5]{Dieudonne1970}), \textit{i.e.} it is isomorphic to $C_M\times \mathbb R$ where $C_M=M/\mathbb{R}$.}
	
\vspace{3mm}These observatins can be summarised as follows (strictly speaking, see the technical caveat in Footnote \ref{strictlyspeaking} about item 3):

\vspace{5mm}
\begin{framed}
	\begin{center}
		\textbf{Congruence of curves as a principal bundle}
	\end{center}
	
	\noindent
	For a smooth manifold, the following notions are equivalent:
	\begin{enumerate}
		\item a congruence $C_\bullet$ of parameterised open curves on $M$ whose images form a one-dimensional foliation of $M$, \textit{i.e.} a collection of embeddings $C_m:\mathbb{R}\hookrightarrow M$ of the real line $\mathbb R$ into the manifold $M$, 
		\item a complete nowhere-vanishing vector field ${X}$ on $M$ without periodic integral curves,
		\item a principal $\mathbb{R}$-bundle $M$  with fundamental vector field ${X}$,
		\item a free proper action $\mathbb{R}\lefttorightarrow M$ of the additive group $\mathbb{R}$ on $M$,
		\item an injective flow on $M$, \textit{i.e.} a group morphism $\exp(\bullet{X})\,:\,\mathbb{R}\hookrightarrow Diff(M)\,:\,t\mapsto\exp(\,t{X})$\,.
	\end{enumerate}
	\vspace{3mm}\end{framed}

The relation between these geometrical structures is as follows: the congruence $C_\bullet$ is the family of integral curves of the vector field ${X}$, as well as the flow lines of the injective flow and it is also the vertical foliation of the manifold $M$.

\subsubsection{Two algebroids over the base space: the adjoint and Atiyah algebroids}

Since vertical vector fields on $P$ are derivations of the commutative algebra ${C}^\infty(P)$ of functions on $P$, \textit{i.e.}
$ \Gamma(VP)\subset  \Gamma(TP)=\mathfrak{der}\big({C}^\infty(P)\big)$, the fundamental action \eqref{fund} can also be seen as a representation of
$\mathfrak{h}$ on the structure algebra ${C}^\infty(P)$ of the principal bundle $P$.
In fact, the free $H$-action on the principal bundle $P$ induces a faithful representation of $H$ on the structure algebra ${C}^\infty(P)$ thus,
infinitesimally, the fundamental $\mathfrak{h}$-action induces a faithful representation of $\mathfrak{h}$ on ${C}^\infty(P)$
\be
{\mathcal L}\circ \#\,:\,\mathfrak{h}\hookrightarrow Der\big(\,{C}^\infty(P)\,\big)\,:\,y\mapsto {\mathcal L}_{y^\#}\,.
\ee
The structure algebra analogue of the canonical projection $\pi:P\twoheadrightarrow P/H$ is the canonical embedding $\pi^*:{C}^\infty(P/H)\hookrightarrow {C}^\infty(P)$. In this way, the structure algebra ${C}^\infty(P/H)$ of the base $P/H$ is canonically embedded into the structure algebra ${C}^\infty(P)$
of the principal bundle $P$ as the space 
\be
{C}^\infty(P)^H\cong \pi^*\big({C}^\infty(P/H)
\ee
of $H$-\textbf{invariant functions} on $P$.
Infinitesimally, the invariance condition is that the Lie derivative along a fundamental vector field should vanish:\footnote{Note that when the Lie group $H$ is connected (as always assumed here), the infinitesimal $\mathfrak{h}$-invariance condition is equivalent to the finite $H$-invariance condition. This property holds more generally (see \cite{Kosmann:1979} for details), in particular it is true for the sections of associated vector bundles.\label{Hinv}}
\be
f\in C^\infty(P)\quad\text{and}\quad{\mathcal L}_{\mathfrak{h}^\#}f=0\quad\Longleftrightarrow\quad f\in{C}^\infty(P)^H \cong \pi^*{C}^\infty(P/H)\,.
\label{CP/H}
\ee

The $H$-action on the principal bundle $P$ also induces an $H$-action on its tangent bundle $TP$. The space of orbits of tangent vectors is the quotient denoted by $\frac{TP}{H}$.
The quotient $\frac{TP}{H}$ is the total space of a vector bundle over $P/H$ called the \textbf{Atiyah bundle of the principal} $H$-\textbf{bundle} $P$. Similarly, the space of orbits of vertical vectors is the quotient denoted by $\frac{VP}{H}$. It is the total space of a vector sub-bundle $\frac{VP}{H}\subset\frac{TP}{H}$ of the Atiyah bundle, which will be called the \textbf{vertical sub-bundle of the Atiyah bundle}. 
The short exact sequence \eqref{shortexactvectTM} of vector bundles reads in the present case
\be
0\to VP\stackrel{i}{\hookrightarrow}TP\stackrel{\pi_*}{\twoheadrightarrow}T\frac{P}{H}\to 0\,.
\label{shortexactvectTP/H}
\ee
The drawback of this short exact sequence \eqref{shortexactvectTP/H} is that it is \textit{not} uniform (because the first two terms, \textit{i.e.} $VP$ and $TP$, are vector bundles over the total space $P$ while
the third term, \textit{i.e.} $T\frac{P}{H}$, is a vector bundler over the base manifold $P/H$).
Nevertheless, by passing to the quotient by the action of $H$ one obtains a short exact sequence of morphisms of Lie algebroids over $P/H$:
\be
0\to \frac{VP}{H}\stackrel{i}{\hookrightarrow}\frac{TP}{H}\stackrel{\pi_*}{\twoheadrightarrow}T\frac{P}{H}\to 0\,,
\label{shortexactvectcan}
\ee
where, strictly speaking, $i$ and $\pi_*$ stand for the corresponding map induced on the quotient by the $H$-action.
This short exact sequence \eqref{shortexactvectcan} of Lie algebroid mrophisms will be called the \textbf{canonical sequence of the principal bundle}.
The Atiyah bundle $\frac{TP}{H}$ endowed with a structure of transitive Lie algebroid over $P/H$
via the anchor $\pi_*$ in \eqref{shortexactvectcan} is called the \textbf{Atiyah (or gauge) algebroid of the principal} $H$-\textbf{bundle} $P$.
The vertical sub-bundle $\frac{VP}{H}\subset\frac{TP}{H}$ endowed with a structure of Lie algebra bundle over $P/H$
via the trivial anchor is called the \textbf{vertical subalgebroid of the principal} $H$-\textbf{bundle} $P$.

The canonical isomorphism \eqref{fundiso'} between the vertical distribution $VP$ and the fundamental action algebroid $P\rtimes_\#\mathfrak{h}$ 
implies that the short exact sequence \eqref{shortexactvectTP/H} can be written equivalently as
\be
0\to P\rtimes\mathfrak{h}\stackrel{\#}{\hookrightarrow}TP\stackrel{\pi_*}{\twoheadrightarrow}T\frac{P}{H}\to 0\,,
\ee
where $\#$ is the anchor of the fundamental action algebroid $P\rtimes\mathfrak{h}$.
The fundamental isomorphism is equivariant, hence it descends to the quotient and translates into the following isomorphism of Lie algebroids over $P/H$ \cite[Proposition A.3.2]{Mackenzie}
\be\label{isofundact}
\#:P\times_{Ad_H}\mathfrak{h}\stackrel{\sim}{\rightarrow}\frac{VP}{H}\,,
\ee
between the adjoint algebroid and the vertical subalgebroid.
Similarly, its inverse 
\be\label{fundisofundact}
\iota:\frac{VP}{H}\stackrel{\sim}{\rightarrow}P\times_{Ad_H}\mathfrak{h}\,,
\ee
is equivalent to the canonical isomorphism \eqref{fundiso'}.
In this way, one obtains the short exact sequence of morphisms of Lie algebroids over $P/H$
\be
0\to P\times_{Ad_H}\mathfrak{h}\stackrel{\#}{\hookrightarrow}\frac{TP}{H}\stackrel{\pi_*}{\twoheadrightarrow}T\frac{P}{H}\to 0\,
\label{shortexactvectAt}
\ee
where, strictly speaking, $\#$ and $\pi_*$ stand for the corresponding map induced on the quotient by the $H$-action.
The short exact sequence \eqref{shortexactvectAt} of Lie algebroid morphisms is called the \textbf{Atiyah sequence of the principal} $H$-\textbf{bundle} $P$ \cite[Definition A.3.6]{Mackenzie}.

The Lie algebra $\Gamma\big(\frac{TP}{H}\big)$ of sections of the Atiyah bundle is called the \textbf{Atiyah algebra of the principal bundle}. A section of the Atiyah bundle $\frac{TP}{H}$ is an 
$H$-\textbf{invariant vector field} on $P$, \textit{i.e.} a vector field mapped to itself by the right $H$-action.
In other words, one has the following isomorphism of Lie algebras: $\Gamma\big(\frac{TP}{H}\big)\cong\mathfrak{X}(P)^H$.
Infinitesimally, the invariance condition is that the Lie derivative along a fundamental vector field should vanish\footnote{When the Lie group $H$ is connected, this infinitesimal $\mathfrak{h}$-invariance condition is equivalent to the finite $H$-invariance condition (\textit{cf.} the comment in Footnote \ref{Hinv}).}
\be
{Y}\in \mathfrak{X}(P)\quad\text{and}\quad{\mathcal L}_{\mathfrak{h}^\#}{Y}=0\quad\Longleftrightarrow\quad{Y}\in \mathfrak{X}(P)^H\cong\Gamma\Big(\frac{TP}{H}\Big)\,.
\label{invcond}
\ee
These $H$-invariant vector fields are projectable on $P/H$, \textit{i.e.} $\Gamma(TP)^H\subset \Pi \Gamma(TP)$, and they generate flows of automorphisms of the principal bundle (\textit{i.e.} diffeomorphisms of $P$ mapping equivariantly fibres on fibres).
As can be checked from \eqref{CP/H} and \eqref{invcond},
the Atiyah algebra is a Lie algebra over ${C}^\infty(P/H)$ that can be seen as the centraliser of $\mathfrak{h}^\#$ inside the Lie algebra  $\mathfrak{X}(P)$ of vector fields on the total space. 

Two Lie subalgebras $\mathfrak{h_1}$ and $\mathfrak{h_2}$ of a Lie algebra $\mathfrak{g}$ form a \textbf{dual pair} in $\mathfrak{g}$ if $\mathfrak{h_1}$ and $\mathfrak{h_2}$ are mutual centralisers in $\mathfrak{g}$. 
The Lie algebra $\mathfrak{h}^\#$ of fundamental vector fields on $P$ and the Atiyah algebra $\mathfrak{X}(P)^H$ of invariant vector fields on $P$ form a dual pair in the Lie algebra $\mathfrak{X}(P)$ of vector fields on $P$.

Since $\mathfrak{h}$ transforms under the adjoint representation of $H$,
the associated vector bundle $P\times_{Ad_H}\mathfrak{h}$ over $P/H$ is called the \textbf{adjoint bundle} \cite[Appendix A.3]{Mackenzie}. It is a Lie algebra bundle (thus a Lie algebroid) over $P/H$ and as such is called the \textbf{adjoint algebroid}. The space $\Gamma(P\times_{Ad_H}\mathfrak{h})$ of sections of the adjoint bundle will be called the \textbf{adjoint sheaf} and will often be denoted ${C}^\infty(P)\otimes_{Ad_H}\mathfrak{h}$.
It is a Lie algebra over ${C}^\infty(P/H)$.
The sections of the adjoint bundle $\Gamma(P\times_{Ad_H}\mathfrak{h})$ can be thought of as $Ad_H$-equivariant $\mathfrak{h}$-valued functions on $P$, \textit{i.e.} 
\be
\sigma\in {C}^\infty(P)\otimes_{Ad_H}\mathfrak{h}\quad\Leftrightarrow\quad\sigma(ph)=Ad_{_{h^{-1}}}\sigma(p)\,,\quad \forall p\in P,\,\,\forall h\in H\,,
\ee
or, equivalently, $\sigma$ intertwines the right action with the adjoint representation
\be
\sigma\circ R_h =Ad_{_{h^{-1}}}\circ\sigma\quad\Leftrightarrow\quad Ad_h\circ\sigma\circ R_h =\sigma\,,\quad \forall h\in H\,.
\ee
Infinitesimally, this equivariance condition reads as $({\mathcal L}_{y^\#}+ad_y)\sigma=0$ for any $y\in \mathfrak{h}$.
More explicitly, if the section reads in a basis as $\sigma=\sigma^i\texttt{T}_i$ then its components transform
as: 
\be
{\mathcal L}_{\texttt{T}_j^\#}\sigma^{i}=-C^i_{jk}\sigma^{k}\,.
\label{strcts}
\ee
As can be checked from \eqref{strctss} and \eqref{strcts}, these sections are in one-to-one correspondence with $H$-invariant vertical vector fields on $P$ of the form 
\be
{Y}_\sigma = \sigma^{i}(x,y)\,\texttt{T}_i^\#\,.
\label{vertvectfprinc}
\ee
Consequently, the adjoint sheaf $\Gamma(P\times_{Ad_H}\mathfrak{h})$ is a ${C}^\infty(P/H)$-module with basis $\{\texttt{T}^\#_i\}$, and
one has the following isomorphisms of Lie algebras over ${C}^\infty(P/H)$: 
\be
\Gamma(P\times_{Ad_H}\mathfrak{h})\,\cong\,{C}^\infty(P)\otimes_{Ad_H}\mathfrak{h}\cong\Gamma(VP)^H\,.
\ee
This is the algebraic version of the canonical isomorphism \eqref{isofundact} of Lie algebroids over $P/H$.

At the level of tangent sheaves, the pushforward by $\pi$ is the right $H$-equivariant map $\pi_*:\Pi \Gamma(TP)\twoheadrightarrow  \Gamma(TP/H)$. Due to this property, one is interested in the $H$-invariant vector fields on $P$ which read in local coordinates as
\be
{Y}_\Pi = Y^{\mu}(x)\,\partial_{\mu}+\sigma^{i}(x,y)\,\texttt{T}_i^\#\,,
\label{Pidecompprinc}
\ee
where $\sigma$ is a section of the adjoint bundle.
Remember that any $H$-invariant vector field on $P$
is projectable on $P/H$, \textit{i.e.} $\Gamma(TP)^H\subset\Pi \Gamma(TP)$. This is geometrically clear but one can check it algebraically from the criterion \eqref{projectable} of projectability, since the Lie bracket of an $H$-invariant vector field on $P$
with a vertical vector field on $P$ is also vertical.
The adjoint sheaf $\Gamma(P\times_{Ad_H}\mathfrak{h})\cong\Gamma(VP)^H$ of
vertical invariant vector fields
is an ideal of the Atiyah algebra of invariant vector fields $\Gamma\big(\frac{TP}{H}\big)\cong\Gamma(TP)^H$.
Together with the tangent sheaf $ \Gamma(TP/H)$ of the base, these three algebras are Lie algebras over ${C}^\infty(P/H)$.
The tangent sheaf of $P/H$ is isomorphic to the corresponding quotient:
\be
 \Gamma\left(T\frac{P}{H}\right)\,\cong\,\Gamma(TP)^H\,\big/\,\Gamma(VP)^H\,\cong\,
\Gamma\left(\frac{TP}{H}\right)\Big/\Gamma(P\times_{Ad_H}\mathfrak{h})\,,
\ee
as is clear in coordinates \eqref{Pidecompprinc}. The translation of the canonical sequence \eqref{shortexactvectcan} in terms of the corresponding sheaves is the short exact sequence of Lie algebras
\be
0\to \Gamma(VP)^H\stackrel{i}{\hookrightarrow}\Gamma(TP)^H\stackrel{\pi_*}{\twoheadrightarrow} \Gamma\big(T\tfrac{P}{H}\big)\to 0\,,
\label{shortexactPH}
\ee
which can be expressed in words as follows: the Lie algebra of invariant vector fields is an extension of the Lie algebra of base vector fields by the Lie algebra of invariant vertical vector fields.

Let us summarise the various relations between the relevant sheaves for a principal $H$-bundle $P$:
\[
\begin{array}
	[c]{ccc}
	\text{Vertical} & & \text{Projectable}\\
	&&\\
	 \Gamma(VP)\cong{C}^\infty(P)\rtimes\mathfrak{h} & \subset & \Pi \Gamma(TP)
	\\
	&&\\
	\cup &  & \cup\\
	&&\\
	\Gamma(VP)^H\cong\Gamma\left(\frac{VP}{H}\right)\cong{C}^\infty(P)\otimes_{Ad_H}\mathfrak{h} &\quad\subset\quad & 
	\Gamma(TP)^H\cong\Gamma\left(\frac{TP}{H}\right)\\
	&&\\
	\text{Adjoint} & & \text{Atiyah}
\end{array}
\]

 \subsection{French connection (part I bis)\,: Charles Ehresmann}

A \textbf{principal connection} on the principal bundle $H\lefttorightarrow P \twoheadrightarrow P/H$ is an equivariant Ehresmann connection \cite{Ehresmann:1950}.\footnote{A widespread abuse of terminology refers to ``principal connections'' as ``Ehresmann connections'' for short (see \textit{e.g.} \cite[Definition 6.2.4]{Sharpe}). However, for the sake of conceptual clarity, we will not follow this common misuse here.} The equivariance implies that a principal connection is equivalent to a  linear splitting (see \textit{e.g.} \cite[Appendix A.4]{Mackenzie}))
\be
0\leftarrow \frac{VP}{H}\stackrel{\omega}{\twoheadleftarrow}\frac{TP}{H}\stackrel{\gamma}{\hookleftarrow}T\frac{P}{H}\leftarrow 0\,,
\label{shortexactvectAt2}
\ee
of the canonical sequence \eqref{shortexactvectcan} of vector bundles over $P/H$), where $\gamma$ is a section of the anchor $\pi_*$ of the Atiyah algebroid.

The horizontal lift of an (equivariant) Ehresmann connection is a field on $P$ of sections 
$\gamma|_p:T_{\pi(p)}\frac{P}{H}\hookrightarrow T_pP$ lifting (equivariantly) tangent vectors on the base manifold $P/H$ to tangent vectors 
on the total space $P$. It corresponds to a ${C}^\infty(P/H)$-linear map $\gamma:\Gamma\big(T\frac{P}{H}\big)\hookrightarrow \Gamma(TP)^H$ lifting vector fields on the base $P/H$ to invariant vector fields on the total space $P$.
For this reason, it corresponds in \eqref{shortexactvectAt2} to the section $\gamma:T\frac{P}{H}\hookrightarrow \frac{TP}{H}$ of the anchor of the Atiyah algebroid. 
In local coordinates, it reads as 
\be
\gamma: X^\mu\partial_\mu\mapsto X^{\mu}\,\nabla_\mu\,,\qquad\text{where}\quad \nabla_\mu\,:=\,\partial_{\mu}\,-\,\omega_\mu^i(x,y)\,\texttt{T}_i^\#\,,
\label{equivhorlift}
\ee
The equivariance property reads in components as
\be
{\mathcal L}_{\texttt{T}_j^\#}\omega_\mu^{i}\,=\,-\,C^i_{jk}\,\omega_\mu^{k}\,.
\label{strcts-Ad}
\ee
With a slight abuse of notation and terminology, these various maps are denoted with the same symbol $\gamma$ and may all be loosely called the \textbf{invariant horizontal lift}.

A principal connection can be defined equivalently \cite[Appendix A.4]{Mackenzie} as a linear splitting (\textit{i.e.} a splitting of vector bundles over $P/H$) of the Atiyah sequence \eqref{shortexactvectAt},
\be
0\leftarrow P\times_{Ad_H}\mathfrak{h}\stackrel{\omega^\mathfrak{h}}{\twoheadleftarrow}\frac{TP}{H}\stackrel{\gamma}{\hookleftarrow}T\frac{P}{H}\leftarrow 0\,,
\label{shortexactvectAt2bis}
\ee
where $\omega^\mathfrak{h}$ is a retraction of the fundamental action $\#$. This retraction $\omega^\mathfrak{h}:\frac{TP}{H}\hookrightarrow P\times_{Ad_H}\mathfrak{h}$ is a morphism of vector bundles over $P/H$, from the Atiyah bundle onto the adjoint bundle, but it can equivalently be seen as an $Ad_H$-equivariant $\mathfrak{h}$-valued differential one-form on $P$, in which case it is called the \textbf{principal connection one-form} $\omega^\mathfrak{h}\in \Omega^1(P)\otimes_{Ad_H} \mathfrak{h}$. 
The principal connection one-form can also be defined as a field of retractions $\omega_p^\mathfrak{h}:T_pP\hookrightarrow \mathfrak{h}$
such that $({\mathcal L}_{y^\#}+ad_y)\,\omega^\mathfrak{h}=0$ (for any $y\in \mathfrak{h}$)
and $\omega^\mathfrak{h}\circ \#=id_\mathfrak{h}$\,.

In order to see more explicitly what is a principal connection, one may look first at the pointwise version of the Atiyah sequence \eqref{shortexactvectAt}, \textit{i.e.} the short exact sequence of vector spaces
\be
0\to \mathfrak{h}\stackrel{\#_p}{\hookrightarrow}T_pP\stackrel{\pi_{*p}}{\twoheadrightarrow}T_{\pi(p)}\frac{P}{H}\to 0\,
\label{shortexactTP/H}
\ee
expressing that the tangent space at the base point $\pi(p)\in P/H$ is isomorphic to the 
quotient of the tangent space at $p\in P$ by the vertical subspace: 
\be
T_{\pi(p)}\frac{P}{H}\cong T_pP\big/V_pP\,,
\ee
since $V_pP\cong\mathfrak{h}$.
A principal connection is an $H$-equivariant field of splitting of the short exact sequences \eqref{shortexactTP/H} of vector spaces for all points $p\in P$
\be
0\leftarrow \mathfrak{h}\stackrel{\omega^\mathfrak{h}_p}{\twoheadleftarrow}T_pP\stackrel{\gamma|_p}{\hookleftarrow}T_{\pi(p)}\frac{P}{H}\leftarrow 0\,.
\label{splitTP/H}
\ee
In local coordinates, the principal connection one-form reads as
\be\label{omegamathfrakhlocal}
\omega^\mathfrak{h}: Y^{\mu}\partial_{\mu}+Y^{i}\,\texttt{T}_i^\#\mapsto \Big(Y^\mu\,\omega_\mu^i(x,y)+Y^i\Big)\,\texttt{T}_i\,.
\ee
Equivalently, one can write it as:
\be
\omega^\mathfrak{h}= \Big(\omega_\mu^i(x,y)\,dx^\mu+\texttt{T}^{\#*i}\Big)\otimes\texttt{T}_i\,,
\label{omegamathfrakh}
\ee
where $\{dx^\mu,\texttt{T}^{\#*i}\}$ denotes the basis of $T^*_pP$ dual to the basis $\{\frac{\partial}{\partial x^\mu},\texttt{T}_i^\#\}$of $T_pP$.
One should stress that the $Ad_H$-equivariance property \eqref{strcts-Ad} implies that the existence of a single horizontal section is enough to ensure that the principal connection is flat (\textit{i.e.} the horizontal distribution $HE$ is involutive) since all horizontal sections are obtained by right $H$-translation.

The sheaf version of the Atiyah sequence \eqref{shortexactvectAt} is the short exact sequence  of Lie algebras over ${C}^\infty(P/H)$,
\be
0\to {C}^\infty(P)\otimes_{Ad_H}\mathfrak{h}\stackrel{\#}{\hookrightarrow}\mathfrak{X}(P)^H\stackrel{\pi_*}{\twoheadrightarrow}\mathfrak{X}(P/H)\to 0\,,
\label{shortexactPH'}
\ee
which expresses that the Atiyah algebra $\Gamma\big(\frac{TP}{H}\big)\cong \mathfrak{X}(P)^H$ of
$H$-invariant vector fields on $P$ is an extension of the Lie algebra $\mathfrak{X}(P/H)$ of vector fields on $P/H$ by the Lie algebra $ \Gamma(P\times_{Ad_H}\mathfrak{h})\cong \Gamma(VP)^H$ of $H$-invariant vertical vector fields.
A principal connection provides a splitting of ${C}^\infty(P/H)$-modules
\be
0\leftarrow {C}^\infty(P)\otimes_{Ad_H}\mathfrak{h}
\stackrel{\omega^\mathfrak{h}}{\twoheadleftarrow}
\mathfrak{X}(P)^H
\stackrel{\gamma}{\hookleftarrow}\mathfrak{X}(P/H)\leftarrow 0\,,
\label{splitT(P/H)}
\ee
which is the sheaf version of \eqref{shortexactvectAt2}.
The image $\gamma\mathfrak{X}(P/H)$ will be called
the \textbf{invariant horizontal sheaf} on $P$. 
In general, the invariant horizontal sheaf $\gamma\mathfrak{X}(P/H)$ only has a structure of ${C}^\infty(P/H)$-module. It can be endowed with a Lie algebra structure via the Lie bracket of horizontal vector fields if and only if the horizontal distribution is involutive, \textit{i.e.} the connection is flat. In other words, a 
\textbf{flat horizontal lift} is a ${C}^\infty(P/H)$-linear Lie algebra morphism mapping vector fields on $P/H$ to invariant horizontal vector fields on $P$.

\vspace{5mm}
\begin{framed}
	\begin{center}
		\textbf{The many faces of principal connections}
	\end{center}
	
	\noindent
	Given a principal bundle $H\lefttorightarrow P \twoheadrightarrow P/H$, the following notions are equivalent:
	\begin{enumerate}
		\item a principal connection, 
		\item an $Ad_H$-equivariant Ehresmann connection on $P$,
		\item a linear splitting of the canonical (or, equivalently, of the Atiyah) sequence,
		\item an invariant horizontal lift, \textit{i.e.} a section $\gamma:T\frac{P}{H}\hookrightarrow \frac{TP}{H}$ of the anchor of the Atiyah algebroid,
		\item a principal connection one-form, \textit{i.e.} a retraction $\omega^\mathfrak{h}:\frac{TP}{H}\hookrightarrow P\times_{Ad_H}\mathfrak{h}$ of the fundamental action of $\mathfrak{h}$ on $P$.
	\end{enumerate}
	\vspace{3mm}
\end{framed}

An Ehresmann connection can be seen as a connection one-form on $P$ taking values in the vertical distribution, which is furthermore $Ad_H$-equivariant in the case of a principal connection.
Following the definition of Subsection \ref{curv2formsect}, the curvature two-form of an Ehresmann connection
is a horizontal differential two-form on $P$ taking values in the vertical distribution.
Composing with the canonical isomorphism $VP\stackrel{\sim}{\to}P\rtimes\mathfrak{h}$ implies that the curvature two-form of a principal connection can be seen as a horizontal differential two-form on $P$ taking values in the algebra $\mathfrak{h}$.

 \subsection{French connection (part II)\,: \'Elie Cartan}

A particularly important class of examples of principal bundles are Klein geometries. They are quite rich and deeply related to the differential geometry of Lie groups. They will be addressed on their own, at length, in the next section. However,
their infinitesimal counterparts, called Klein pairs, are very simple to introduce in terms of Lie algebras, so that we will already provide their definition in this subsection. 

The vast panorama of curved geometries envisioned by Cartan, that generalise both Klein geometries and curved Riemannian geometry, has a very concise definition in terms of the Atiyah bundle, which will be reviewed here. 

\subsubsection{An infinitesimal primer on Klein geometry}

A Lie algebra $\mathfrak{g}$ together with a Lie subalgebra $\mathfrak{h}\subseteq\mathfrak{g}$ is called \cite[Definition 4.3.16]{Sharpe} a \textbf{Klein pair} with \textbf{principal algebra} $\mathfrak{g}$ and \textbf{isotropy subalgebra} $\mathfrak{h}$. The adjoint representation of $\mathfrak{g}$ on itself can be restricted to the adjoint action of $\mathfrak{h}$ on $\mathfrak{g}$. This makes the principal algebra $\mathfrak{g}$ into an $\mathfrak{h}$-module, of which the isotropy subalgebra $\mathfrak{h}$ is a submodule. Accordingly, the quotient $\mathfrak{g}/\mathfrak{h}$ is also an $\mathfrak{h}$-module. As such, the quotient $\mathfrak{g}/\mathfrak{h}$ will be called 
\textbf{transvection module}.
Given a Lie algebra $\mathfrak{g}$, a Klein pair can also be defined as a short exact sequence of $\mathfrak{h}$-modules
\be
0\to \mathfrak{h}
	\hookrightarrow
\mathfrak{g}
	\twoheadrightarrow
\mathfrak{g}/\mathfrak{h}\to 0\,,\label{shortexactinfKlein}
\ee
where $\mathfrak{g}$ sits in the middle, all arrows are $\mathfrak{h}$-module morphisms and the $\mathfrak{h}$-module structure is inherited from the adjoint action of the submodule $\mathfrak{h}$ on itself or on $\mathfrak{g}$. 
This definition is indeed equivalent since the last requirement (\textit{i.e.} that the $\mathfrak{h}$-module structure arises from the adjoint action of $\mathfrak{g}$ on itself) implies that the subspace $\mathfrak{h}\subseteq\mathfrak{g}$ is actually a Lie subalgebra. In turn, this implies that the $\mathfrak{h}$-module morphism $\mathfrak{h}\hookrightarrow\mathfrak{g}$ is actually a Lie algebra morphism.
The short exact sequence \eqref{shortexactinfKlein} of $\mathfrak{h}$-modules will be called the \textbf{Klein sequence}.

The representation $r:\mathfrak{h}\to\mathfrak{gl}(\mathfrak{g}/\mathfrak{h})$ of the isotropy subalgebra $\mathfrak{h}$ on the transvection module $\mathfrak{g}/\mathfrak{h}$ is called the \textbf{isotropy representation}. It may not be faithful. In fact, an element $y\in\mathfrak{h}$ acts trivially on the transvection module $\mathfrak{g}/\mathfrak{h}$ (\textit{i.e.} $y\in\text{Ker}\,r$) if and only if $[y,\mathfrak{g}]\subseteq\mathfrak{h}$. The kernel of the isotropy representation $\text{Ker}\,r$ is an ideal of the isotropy subalgebra $\mathfrak{h}$ but it need not be an ideal of the principal algebra $\mathfrak{g}$. 
The maximal Lie ideal of $\mathfrak{g}$ which is contained in $\mathfrak{h}$ will be called the \textbf{ideal core} of the Klein pair $\mathfrak{h}\subseteq\mathfrak{g}$.  
The ideal core is a Lie subalgebra of the kernel of the isotropy representation.

\vspace{3mm}
\noindent{\small\textbf{Example (Lie algebra extension)\,:} If, in the Klein sequence \eqref{shortexactinfKlein}, all objects are Lie algebras and all arrows are Lie algebra morphisms, then the isotropy subalgebra $\mathfrak{h}$ is a Lie ideal of the principal algebra $\mathfrak{g}$. In other words, the ideal core of the Klein pair $\mathfrak{h}\subseteq\mathfrak{g}$ is the whole isotropy subalgebra $\mathfrak{h}$.
In such case, the transvection submodule $\mathfrak{p}:=\mathfrak{g}/\mathfrak{h}$ is indeed a Lie algebra. Accordingly, in such case $\mathfrak{h}$ and $\mathfrak{p}$ will sometimes be called, respectively, the \textbf{isotropy ideal} and the \textbf{transvection algebra}. Traditionally, in this case the (principal) algebra $\mathfrak{g}$ is called a \textbf{Lie algebra extension} of the (transvection) algebra $\mathfrak{p}$ by the (isotropy) ideal $\mathfrak{h}\subseteq\mathfrak{g}$.}

\vspace{3mm}
\noindent{\small\textbf{Example (Effective Klein pair)\,:} A Klein pair $\mathfrak{h}\subseteq\mathfrak{g}$ is called \textit{effective} \cite[Definition 4.3.16]{Sharpe} if there is no (non-trivial) ideal of the principal algebra $\mathfrak{g}$ inside the isotropy algebra $\mathfrak{h}$ or, equivalently, if the ideal core is trivial. 
If the Klein pair is effective, then the isotropy representation of $\mathfrak{h}$ on the transvection module $\mathfrak{g}/\mathfrak{h}$ is faithful.
}

\vspace{3mm}
If $\mathfrak{i}$ is any ideal of the principal algebra $\mathfrak{g}$ inside the isotropy subalgebra $\mathfrak{h}\subseteq\mathfrak{g}$, then $\mathfrak{h}/\mathfrak{i}\,\subset\,\mathfrak{g}/\mathfrak{i}$ is a Klein pair, which will be called the \textbf{quotient of the Klein pair} $\mathfrak{h}\subseteq\mathfrak{g}$ \textbf{by the ideal} $\mathfrak{i}$ \textbf{of the principal algebra} $\mathfrak{g}$ \textbf{inside the isotropy subalgebra} $\mathfrak{h}$. The transvection module $(\mathfrak{g}/\mathfrak{i})/(\mathfrak{h}/\mathfrak{i})$ of the quotient $\mathfrak{h}/\mathfrak{i}\,\subset\,\mathfrak{g}/\mathfrak{i}$ of the Klein pair, is isomorphic, as a vector space, to the transvection module $\mathfrak{g}/\mathfrak{h}$ of the original Klein pair. 
The quotient of an ineffective Klein pair $\mathfrak{h}\subseteq\mathfrak{g}$ by its ideal core $\mathfrak{i}$ is an effective Klein pair $\mathfrak{h}/\mathfrak{i}\,\subset\,\mathfrak{g}/\mathfrak{i}$. In this sense, there is no loss of generality in considering effective Klein pairs, as will be implicitly assumed from now on, whenever dealing with Klein or Cartan geometries.

 \subsubsection{Standard definition of Cartan connections}

Consider the following data: a principal bundle $H\lefttorightarrow P \twoheadrightarrow P/H$ and a Klein pair $\mathfrak{h}\subseteq\mathfrak{g}$ such that the dimension of the principal algebra $\mathfrak{g}$ is equal to the dimension of the total space $P$.

The embedding $\#:P\times\mathfrak{h}\hookrightarrow TP$ of vector bundles over $P$ is the anchor of the fundamental action algebroid $P\rtimes\mathfrak{h}$ corresponding to the free action of $H$ on $P$. 
An $Ad_H$-equivariant isomorphism 
\be\label{CartanisoHequiv}
\omega^\mathfrak{g}:TP\stackrel{\sim}{\to} P\times\mathfrak{g}
\ee
of vector bundles over $P$, which is a retraction of the anchor $\#:P\times\mathfrak{h}\hookrightarrow TP$, \textit{i.e.} which is such that $\omega^\mathfrak{g}\circ \#=id_{P\times \mathfrak{h}}$, 
is called a \textbf{Cartan connection} on the principal $H$-bundle $P$ \cite[Section IV.3]{Kob72}. 
A principal $H$-bundle $P$ endowed with a Cartan connection is called \textbf{a Cartan geometry modeled on the Klein pair} $\mathfrak{h}\subset \mathfrak{g}$ \cite[Definition 5.3.1]{Sharpe}.

There exists various equivalent definitions of Cartan connections, many of which will be reviewed here.\footnote{Cartan are so ubiquitous that several other equivalent characterisations of Cartan still exist, which will not be reviewed here (see \textit{e.g.} \cite{CrampinSaunders}, \cite{Attard:2019} and references therein).}
In order to disambiguate later the various equivalent definitions of a Cartan connection, an $Ad_H$-equivariant isomorphism \eqref{CartanisoHequiv}, with the above properties, will be called a \textbf{Cartan isomorphism}. 
The requirement of being a retraction of the anchor is equivalent to demanding that 
the restriction $\omega^\mathfrak{g}|_{{}_{VP}}$ of the Cartan isomorphism \eqref{CartanisoHequiv} identifies with the canonical isomorphism \eqref{fundiso'} of Lie algebroids between the vertical distribution $VP$ and the fundamental action algebroid $P\rtimes\mathfrak{h}$, \textit{i.e.}
\be
\omega^\mathfrak{g}|_{{}_{VP}}=\iota:VP\stackrel{\sim}{\to} P\times\mathfrak{h}
\ee 
In other words, a Cartan isomorphism is an isomorphism \eqref{CartanisoHequiv} of vector bundles over $P$ from the tangent bundle $TP$ to the Lie algebra bundle $P\rtimes\mathfrak{g}$, which is $Ad_H$-equivariant and whose restriction to the vertical distribution $VP\subset TP$ reduces to the canonical isomorphism \eqref{fundiso'}. One can thus think of a Cartan isomorphism as a (non-canonical) extension of the canonical isomorphism to the whole tangent bundle, retaining the property of equivariance.

Another perspective on the Cartan connection \eqref{CartanisoHequiv} is that it defines an $Ad_H$-equivariant field $\omega^\mathfrak{g}$ of isomorphisms $\omega^\mathfrak{g}|_p:T_pP\stackrel{\sim}{\to} \mathfrak{g}$ of vector spaces,
which are retractions of the anchor of the fundamental action algebroid, \textit{i.e.} $\omega^\mathfrak{g}|_p\circ \#_p=id_\mathfrak{h}$\,.
This will be called a \textbf{Cartan connection one-form} on the principal $H$-bundle $P$.
Equivalently, a Cartan connection one-form is a non-degenerate $Ad_H$-equivariant $\mathfrak{g}$-valued differential one-form on $P$, \textit{i.e.} $\omega^\mathfrak{g}\in \Omega^1(P)\otimes_{Ad_H} \mathfrak{g}$ whose vertical part is entirely determined by the canonical isomorphism.
Infinitesimally, the equivariance condition reads as $({\mathcal L}_{y^\#}+ad_y)\omega^\mathfrak{g}=0$ for any $y\in \mathfrak{h}$.

 \subsubsection{Relation between Cartan and Ehresmann connections}

A \textbf{split Klein pair} is a Klein pair $\mathfrak{h}\subset \mathfrak{g}$ such that the vector space $\mathfrak{g}$ is decomposed as the direct sum $\mathfrak{g}=\mathfrak{h}\oplus\mathfrak{p}$ of two vector subspaces: the isotropy subalgebra $\mathfrak{h}$ and the linear subspace $\mathfrak{p}\cong \mathfrak{g}/\mathfrak{h}$, which will be called the \textbf{transvection subspace} in this context.
The transvection subspace is a linear representative of the transvection module but it is \textit{not} an $\mathfrak{h}$-invariant subspace (\textit{i.e.} it is \textit{not} an $\mathfrak{h}$-submodule) in general. 
Equivalently, a Klein pair is split if the Klein sequence \eqref{shortexactinfKlein} splits linearly, \textit{i.e.} all arrows of the split sequence
\be\label{KLeinreduct}
0\leftarrow \mathfrak{h}
	\twoheadleftarrow
\mathfrak{g}
	\hookleftarrow
\mathfrak{p}\leftarrow 0\,.
\ee
are linear maps. Note that Klein pairs can always be split, \textit{i.e.} there always exists a linear splitting \eqref{KLeinreduct}.

Any Cartan geometry modeled on a split Klein pair defines an Ehresmann connection as follows: Let us restrict the codomain of the Cartan isomorphism \eqref{CartanisoHequiv} to the isotropy subalgebra $\mathfrak{h}$ of a split Klein pair. The corestriction $\omega^\mathfrak{h}:TP\twoheadrightarrow P\rtimes\mathfrak{h}$, composed with the fundamental isomorphism \eqref{fundiso}, is an Ehresmann connection $\omega:=\#_\bullet\circ\omega^\mathfrak{h}:TP\twoheadrightarrow VP$ on the total space $P$ of the principal $H$-bundle. Let us stress that, in general, this Ehresmann connection is \textit{not} equivariant and, thus, does \textit{not} define a principal connection (see the next subsection on the particular situation where it is).

In local coordinates, a Cartan connection one-form modeled on a split Klein pair  reads
\be
\omega^\mathfrak{g}: Y^{\mu}\partial_{\mu}+Y^{i}\,\texttt{T}_i^\#\,\mapsto\, Y^\mu\,\theta^a_\mu(x,y) \,\texttt{P}_a\,+\,
\Big(Y^\mu\,\omega_\mu^i(x,y)+Y^i\Big)\,\texttt{T}_i\,
\ee
where $\{\texttt{P}_a\}$ is a basis of the transvection subspace $\mathfrak{p}$ and $\{\texttt{T}_i\}$ is a basis of the isotropy subalgebra. 
The corresponding surjective  morphism $\omega^\mathfrak{h}:TP\twoheadrightarrow P\times h$ of vector bundles over $P$ reads as in
\eqref{omegamathfrakhlocal}.
Equivalently, one can decompose the Cartan connection one-form as the sum
\be
\omega^\mathfrak{g}\,=\,\omega^\mathfrak{p}+\omega^\mathfrak{h}\,,
\ee
where the $\mathfrak{h}$-valued one-form $\omega^\mathfrak{h}$ reads as in \eqref{omegamathfrakh}, while the $\mathfrak{p}$-valued one-form $\omega^\mathfrak{p}$ reads as
\be
\omega^\mathfrak{p}\,=\, \theta^a_\mu(x,y)\,dx^\mu \otimes\,\texttt{P}_a\,.
\ee

 \subsubsection{Relation between Cartan and  principal connections}

A \textbf{reductive Klein pair} is a Klein pair $\mathfrak{h}\subset \mathfrak{g}$ such that the adjoint $\mathfrak{h}$-module $\mathfrak{g}$ decomposes as the semidirect sum $\mathfrak{g}=\mathfrak{h}\inplus\mathfrak{p}$ of $\mathfrak{h}$-modules \cite[Definition 4.3.16]{Sharpe}, where $\mathfrak{p}\cong \mathfrak{g}/\mathfrak{h}$ will be called the \textbf{transvection submodule}. 
In other words, a split Klein pair is reductive if and only if $\mathfrak{p}$ is an $\mathfrak{h}$-invariant subspace (concretely, if and only if $[\mathfrak{h},\mathfrak{p}]\subseteq\mathfrak{p}$).
Equivalently, a Klein pair is reductive if the Klein sequence \eqref{shortexactinfKlein} splits as a sequence of $\mathfrak{h}$-modules, \textit{i.e.} all arrows of the short exact sequence \eqref{KLeinreduct} are $\mathfrak{h}$-module morphisms.
Let $\{\texttt{P}_a\}$ be a basis of the transvection submodule $\mathfrak{p}$ and $\{\texttt{T}_i\}$ be a basis of the isotropy subalgebra. 
In this basis of the principal algebra $\mathfrak{g}$, the reductivity reads in terms of the structure constants as $C^k_{ia}=0$.

A \textbf{reductive Lie algebra} is a finite-dimensional Lie algebra $\mathfrak{g}=\mathfrak{s}\oplus\mathfrak{z}$ which is the direct sum of a semisimple Lie algebra $\mathfrak{s}$ and an Abelian Lie algebra $\mathfrak{z}$ (the centre of $\mathfrak{g}$). The two distinct meanings of ``reductive'' are loosely linked by the following property: if the Klein pair $\mathfrak{h}\subset\mathfrak{g}$ is effective and if the isotropy subalgebra $\mathfrak{h}$ is a reductive Lie algebra, then the Klein pair $\mathfrak{h}\subset \mathfrak{g}$ is reductive (see \cite[Footnote 17]{Sharpe} and refs therein).
Another loose link is that reductive Lie algebras are \textbf{quadratic}, \textit{i.e.} they can be endowed with a non-degenerate adjoint-invariant bilinear form (see \textit{e.g.} \cite{Benito}). If the Lie algebra $\mathfrak{g}$ is quadratic (for instance, reductive) and if the restriction of the quadratic form to a Lie subalgebra $\mathfrak{h}\subset \mathfrak{g}$ remains non-degenerate, then the Klein pair $\mathfrak{h}\subset \mathfrak{g}$ is reductive.\footnote{This is true for instance, if the principal Lie algebra $\mathfrak{g}$ is a compact semisimple Lie algebra or if both $\mathfrak{h}$ and $\mathfrak{g}$ are semisimple.} This can be seen by considering as transvection submodule the orthogonal complement $\mathfrak{p}:=\mathfrak{h}^\perp$ of the isotropy subalgebra $\mathfrak{h}$ (with respect to the quadratic form).

The restriction $\omega^\mathfrak{h}$ of the codomain of a Cartan connection one-form $\omega^\mathfrak{g}$ to the isotropy subalgebra $\mathfrak{h}$ defines an Ehresmann connection on the principal $H$-bundle $P$. In particular, in the case of a reductive Klein pair it defines a principal connection one-form $\omega^\mathfrak{h}$ on the principal $H$-bundle $P$. The assumption on the Klein pair to be reductive is necessary in order that ${\mathcal L}_{{}_{\mathfrak{h}^\#}}\omega^\mathfrak{h}$ does \textit{not} receive contribution from $ad_{\mathfrak{h}}\,\omega^\mathfrak{p}$. Otherwise, the restriction $\omega^\mathfrak{h}$ is not $Ad_H$-equivariant.
The $Ad_H$-equivariance of the restriction $\omega^\mathfrak{p}$ of the codomain of a Cartan connection one-form $\omega^\mathfrak{g}$ to the transvection submodule $\mathfrak{h}$ of a reductive Klein pair reads in components as:
\be
{\mathcal L}_{\texttt{T}_i^\#}\theta_\mu^{a}\,=\,-\,C^a_{ib}\,\theta_\mu^{b}\,.
\label{strcts-Adp}
\ee
In this basis of $\mathfrak{g}$, the reductivity reads $C^k_{ia}=0$ which guarantees \eqref{strcts-Ad}.

 \subsubsection{Cartan connections via tractor algebroids}

The Lie algebra $\mathfrak{g}$ transforms under the adjoint representation of the Lie group $H$.
The associated vector bundle over $P/H$ is denoted $P\times_{Ad_H}\mathfrak{g}$ and called the \textbf{adjoint tractor bundle}.\footnote{The term ``tractor'' is a portmanteau word of ``\underline{Trac}y Thomas'' and ``twis\underline{tor}'', this type of bundle having been introduced in 1931 by Thomas as an alternative formulation of the Cartan conformal connection, then rediscovered in 1994 within the formalism of local twistors and generalised in \cite{BEG}.} The adjoint tractor bundle can be canonically endowed with a structure of Lie algebra bundle over $P/H$ with fibres isomorphic to $\mathfrak{g}$, \textit{i.e.} it is a Lie algebroid via the Lie bracket of $\mathfrak{g}$ and the trivial anchor, in which case it will be called the \textbf{adjoint tractor algebroid}. 
Accordingly, the space 
\be\label{PadHmathfrakg}
{C}^\infty(P)\otimes_{Ad_H} \mathfrak{g}:=
\Gamma(P\times_{Ad_H}\mathfrak{g})
\ee
of sections of the adjoint tractor algebroid $P\times_{Ad_H}\mathfrak{g}$ is a Lie algebra over ${C}^\infty(P/H)$ which will be called the \textbf{adjoint tractor algebra}.
It is isomorphic to the space of $Ad_H$-equivariant $\mathfrak{g}$-valued functions $\sigma^\mathfrak{g}$ on $P$.
Infinitesimally, this equivariance condition reads as $({\mathcal L}_{y^\#}+ad_y)\sigma^\mathfrak{g}=0$ for any $y\in \mathfrak{h}$.

The vector bundle $P\times_{Ad_H}\,\mathfrak{g}/\mathfrak{h}$ associated to the transvection module $\mathfrak{g}/\mathfrak{h}$ will be called the \textbf{transvection bundle}. A Cartan connection induces an isomorphism 
\be\label{isomorphismtransvbundle}
\Theta:T\frac{P}{H}\stackrel{\sim}{\to}P\times_{Ad_H}\,\mathfrak{g}/\mathfrak{h}
\ee
between the tangent bundle of the base manifold and the transvection bundle \cite[Theorem 5.3.15]{Sharpe}.
The Klein sequence \eqref{shortexactinfKlein} of $\mathfrak h$-modules is the model of the short exact sequence of associated vector bundles over $P/H$:
\be
0\to P\times_{Ad_H}\mathfrak{h}\hookrightarrow P\times_{Ad_H}\mathfrak{g}
\twoheadrightarrow P\times_{Ad_H}\,\mathfrak{g}/\mathfrak{h}\to 0\,.
\label{shortexactintract}
\ee
involving the adjoint bundle $P\times_{Ad_H}\mathfrak{h}$, the adjoint tractor bundle $P\times_{Ad_H}\mathfrak{g}$ and the transvection bundle $P\times_{Ad_H}\,\mathfrak{g}/\mathfrak{h}$.
For any principal $H$-bundle $P$ and any Klein pair $\mathfrak{h}\subset\mathfrak{g}$,
the short exact sequence \eqref{shortexactintract} of morphisms of vector bundles over $P/H$ is canonical and will be called the \textbf{tractor sequence}.

The Cartan connection one-form can be seen as a ${C}^\infty(P)$-linear map,
\begin{eqnarray}
	\omega^\mathfrak{g}&:& \Gamma(TP)\twoheadrightarrow {C}^\infty(P)\otimes\mathfrak{g} \label{CPmorphism}\\
	&:&Y^{\mu}(x,y)\,\partial_{\mu}+Y^{i}(x,y)\,\texttt{T}_i^\#\nonumber\\
	&&\qquad\mapsto Y^{\mu}(x,y)\,\theta^a_{\mu}\,\texttt{P}_a \,\,+\,\,\Big(Y^\mu(x,y)\,\omega_\mu^i(x,y)\,+Y^{i}(x,y)\,\Big)\texttt{T}_i\,,\nonumber
\end{eqnarray}
sending vector fields on $P$ to $\mathfrak{g}$-valued functions on $P$. Its $Ad_H$-equivariance
property implies that it sends $H$-invariant vector fields on $P$ to $Ad_H$-equivariant $\mathfrak{g}$-valued functions on $P$.
Moreover, as a corollary the Cartan connection even provides an isomorphism of ${C}^\infty(P/H)$-modules 
between the  Atiyah algebra and the adjoint tractor algebra: 
\be\label{isomAtitract}
\omega^\mathfrak{g}:\mathfrak{X}(P)^H\stackrel{\sim}{\rightarrow}{C}^\infty(P)\otimes_{Ad_H} \mathfrak{g}\,.
\ee
Accordingly, this isomorphism is also true at the geometric level of vector bundles over $P/H$. Indeed, an equivalent definition \cite[Section 5]{Crampin} of a Cartan isomorphism \eqref{CartanisoHequiv} is as an isomorphism of vector bundles over $P/H$
\be\label{Crampindef}
\omega^\mathfrak{g}:\frac{TP}{H}\stackrel{\sim}{\rightarrow}P\times_{Ad_H}\mathfrak{g}
\ee
from the Atiyah bundle to the adjoint tractor bundle that reduces to the canonical isomorphism \eqref{fundisofundact} upon restriction to the vertical sub-bundle $\frac{VP}{H}\subset\frac{TP}{H}$.\footnote{Remember that, canonically, the Atiyah bundle $\frac{TP}{H}$ is a transitive Lie algebroid while the adjoint tractor bundle $P\times_{Ad_H}\mathfrak{g}$ is a Lie algebra bundle over $P/H$, \textit{i.e.} the anchor of the former (latter) is surjective (vanishing). Therefore, they are not isomorphic as Lie algebroids. However, the vector bundle isomorphism between them allows to endow each of them with two distinct anchors (see \cite{Crampin} for more comments).}
This point of view further highlights the fact that the Klein pair is the underlying model of the Cartan geometry defined by a Cartan connection. In fact, the tractor sequence \eqref{shortexactintract} is isomorphic, to the Atiyah sequence \eqref{shortexactvectAt} of the principal $H$-bundle $P$ via the Cartan isomorphism \eqref{Crampindef}.
This can be summarised in the following isomorphism of short exact sequences of vector bundles over $P/H$:
\be\label{salsiciettatrac}
\begin{array}
	[c]{ccccccccc}
	0&\to& \frac{VP}{H}&\hookrightarrow&\frac{TP}{H}&\twoheadrightarrow&T\frac{P}{H}&\to& 0\\
	&&&&&&&&\\
	&&\iota\downarrow\sim&&\omega^\mathfrak{g}\downarrow\sim&&\Theta\downarrow\sim&&\\
	&&&&&&&&\\
	0&\to& P\times_{Ad_H}\mathfrak{h}&\hookrightarrow&P\times_{Ad_H}\mathfrak{g}&\twoheadrightarrow&P\times_{Ad_H}\,\mathfrak{g}/\mathfrak{h}&\to& 0
\end{array}
\,.
\ee
This further enlightens the relation between Cartan and principal connections: when the Klein pair is reductive, the splitting \eqref{KLeinreduct} of $\mathfrak{h}$-modules corresponds to the splitting
\be
0\leftarrow P\times_{Ad_H}\mathfrak{h}
	\twoheadleftarrow
P\times_{Ad_H}\mathfrak{g}
	\hookleftarrow
P\times_{Ad_H}\mathfrak{p}\leftarrow 0\,.
\ee
of vector bundles over $P/H$. Due to \eqref{Crampindef}, this short exact sequence is, in turn, isomorphic 
to the short exact sequence \eqref{shortexactvectAt2} of a principal connection. In such case, this indirectly leads to an isomorphism between the base tangent bundle $T\frac{P}{H}$ and the transvection bundle $P\times_{Ad_H}\mathfrak{p}$ associated to the transvection submodule:
\be\label{salsiciettatracred}
\begin{array}
	[c]{ccccccccc}
	0&\leftarrow& \frac{VP}{H}&\hookleftarrow&\frac{TP}{H}&\stackrel{\gamma}{\twoheadleftarrow}&T\frac{P}{H}&\leftarrow& 0\\
	&&&&&&&&\\
	&&\iota\downarrow\sim&\omega^\mathfrak{h}\swarrow&\omega^\mathfrak{g}\downarrow\sim&\omega^\mathfrak{p}\searrow&\Theta\downarrow\sim&&\\
	&&&&&&&&\\
	0&\leftarrow& P\times_{Ad_H}\mathfrak{h}&\hookleftarrow&P\times_{Ad_H}\mathfrak{g}&\twoheadleftarrow&P\times_{Ad_H}\,\mathfrak{p}&\leftarrow& 0
\end{array}
\,.
\ee

\vspace{2mm}
\noindent\textbf{Remark:} Despite some formal similarities between principal and Cartan connection one-forms, there are also some important differences. On the one hand, they are both $Ad_H$\,-equivariant one-forms $\omega$ on $P$, \textit{i.e.} $\omega$ intertwines the right action with the adjoint representation
\be\label{equivarianceomega}
\omega\circ (R_h)_* =Ad_{_{h^{-1}}}\circ\omega\quad\Leftrightarrow\quad Ad_h\circ\omega\circ (R_h)_* =\omega\,,\quad \forall h\in H\,,
\ee
and their restriction to the vertical sub-bundle is the inverse of the canonical isomorphism, \textit{i.e.} $\omega|_{{}_{VP}}=\iota$. On the other hand,
they respectively take values in the isotropy algebra $\mathfrak h$ or in the principal algebra $\mathfrak{g}$ and define a field of retractions or isomorphisms of vector spaces, respectively. These features are summarised in Table \ref{PrincipvsCartan}. 
\begin{table}
	\begin{center}
		\begin{tabular}{
				|c|c|c|}
			\hline
			Connection & Morphism of vector bundles & Vertical restriction \\ 
			\hline
			&&\\
			Principal & $\omega^\mathfrak{h}:\frac{TP}{H}\twoheadrightarrow P\times_{Ad_H}\mathfrak{h}$ &\\
			&& $\omega^\mathfrak{h}|_{\frac{VP}{H}}=\omega^\mathfrak{g}|_{\frac{VP}{H}}=\iota\,:\,\frac{VP}{H}\stackrel{\sim}{\rightarrow}P\times_{Ad_H}\mathfrak{h}$\\
			Cartan & $\omega^\mathfrak{g}:\frac{TP}{H}\stackrel{\sim}{\rightarrow}P\times_{Ad_H}\mathfrak{g}$ &\\
			&&\\
			\hline
		\end{tabular}
	\end{center}
	\caption{Principal vs Cartan connection}
	\label{PrincipvsCartan}
\end{table}

 \subsubsection{Cartan connections as intertwiners}

The constant $\mathfrak{g}$-valued functions on $P$ can be identified with the vector space $\mathfrak{g}$.
The codomain $\Gamma(P\times\mathfrak{g})$ of the ${C}^\infty(P)$-linear map \eqref{CPmorphism} can be restricted to the space of constant $\mathfrak{g}$-valued functions on $P$. The preimages of constant $\mathfrak{g}$-valued functions on $P$ by the Cartan connection one-form \eqref{CPmorphism} are nowhere (or everywhere) vanishing vector fields. They are sometimes called \textbf{constant vector fields} on $P$. For instance, the fundamental vector fields identify with the vertical constant vector fields.
One should stress that, in general, constant vector fields are \textit{not} invariant (nor projectable) vector fields. In fact,  invariant (respectively, constant) vector fields are the preimages of  $\mathfrak{g}$-valued functions on $P$ that are adjoint-equivariant (respectively, constant).

The injective linear map  
\be
D:\mathfrak{g}\hookrightarrow  \Gamma(TP)
\label{universalcovder}
\ee
that sends any element of the principal algebra, interpreted as a constant $\mathfrak{g}$-valued function on $P$, to its corresponding constant vector field on $P$, will be called the \textbf{fundamental derivative} defined by the Cartan connection $\omega^\mathfrak{g}$.
The restriction of the fundamental derivative to the isotropy subalgebra $\mathfrak{h}$
is nothing but the fundamental action \eqref{fund}, \textit{i.e.} $D|_\mathfrak{h}=\#:\mathfrak{h}\hookrightarrow  \Gamma(VP)$. 
The fundamental derivative \eqref{universalcovder} reads in components
\be\label{Cartanfunder}
\texttt{P}_a\mapsto \hat{D}_a=e_a^\mu(x,y)\nabla_\mu\,,\qquad \texttt{T}_i\mapsto \texttt{T}_i^\#
\ee
where $\theta^a_\mu e_b^\mu=\delta^a_b$ and with $\nabla_\mu$ defined in \eqref{equivhorlift}.
The equivariance property \eqref{equivarianceomega} of the Cartan connection implies that the fundamental derivative \eqref{universalcovder} intertwines the adjoint representation of $H$ on $\mathfrak{g}$ with the right action of $H$ on $P$, \textit{i.e.}
\be\label{equivariancefunder}
(R_h)_* \circ D = D\circ Ad_{_{h^{-1}}}\quad\Leftrightarrow\quad (R_h)_*\circ D\circ Ad_h =D\,,\quad \forall h\in H\,.
\ee
The infinitesimal counterpart of the equivariance implies that a fundamental derivative can be defined as an intertwiner (\textit{i.e.} an $\mathfrak{h}$-module morphism) whose restriction to the isotropy algebra is the fundamental action.
More precisely, the fundamental derivative intertwines the following $\mathfrak{h}$-actions: the adjoint action $ad$ on $\mathfrak{g}$ and the fundamental action ${\mathcal L}\circ \#$ on constant vector fields, \textit{i.e.} 
\be
D\circ ad=({\mathcal L}\circ \#)\circ D\,.
\ee
In components, this is equivalent to the property 
\be
[\texttt{T}_i^\#,\hat{D}_a]=C_{ia}^b\hat{D}_b+C_{ia}^j\texttt{T}_j^\#\,.
\ee
Actually, there is a one-to-one correspondence between Cartan connection one-forms and fundamental derivatives: they are dual to each other (as emphasised in Table \ref{dualityCD}). 

\begin{table}
	\begin{center}
		\begin{tabular}{
				|c|c|c|}
			\hline
			Property & Cartan connection & Fundamental derivative \\ 
			\hline\hline
			Field of & $\omega|_p:T_pP\stackrel{\sim}{\to}\mathfrak{g}$ & $D|_p:\mathfrak{g}\stackrel{\sim}{\to} T_pP$ \\
			linear isomorphisms &&\\\hline
			Equivariance & ${\mathcal L}_{y^\#}\omega=ad_y\omega$ & ${\mathcal L}_{y^\#}(Dz)=D(ad_yz)$ \\
			&$\forall y\in\mathfrak{h}$&$\forall z\in\mathfrak{g}$\\\hline
			Relation with  & $\omega\circ\#=id_\mathfrak{h}$ & $D|_\mathfrak{h}=\#$ \\
			fundamental action &&\\\hline
		\end{tabular}
	\end{center}
	\caption{\label{dualityCD}Duality between Cartan connection and fundamental derivative}
\end{table}

A fundamental derivative \eqref{universalcovder} can be extended to a  
${C}^\infty(P)$-linear map sending $\mathfrak{g}$-valued functions on $P$ to vector fields on $P$. Its intertwining property implies that it sends $Ad_H$-equivariant $\mathfrak{g}$-valued functions on $P$ to $H$-invariant vector fields on $P$.
In this sense, the fundamental derivative
\be
D=(\omega^\mathfrak{g})^{-1}:{C}^\infty(P)\otimes_{Ad_H} \mathfrak{g}\stackrel{\sim}{\rightarrow} \mathfrak{X}(P)^H
\label{universalcovderiso}
\ee
is an isomorphism between the adjoint tractor algebra and the Atiyah algebra, which is the inverse 
of the Cartan isomorphism \eqref{isomAtitract}. 
In more geometric terms, the inverse of a Cartan isomorphism \eqref{CartanisoHequiv} is an $H$-equivariant isomorphism of vector bundles over $P$
\be\label{Crampindef*}
D=(\omega^\mathfrak{g})^{-1}:P\times\mathfrak{g}\stackrel{\sim}{\rightarrow}TP
\ee
from the vector bundle $P\times\mathfrak{g}$ to the tangent bundle $TP$ that extends the fundamental isomorphism \eqref{fundactanchor}
from the vector sub-bundle $P\times\mathfrak{h}$ to the vertical distribution $VP$. Such an isomorphism \eqref{Crampindef*} will be called a \textbf{Cartan parallelism}.
Equivalently, one can go to the quotient by the $H$-action. In such case, a Cartan parallelism can be defined as an isomorphism of vector bundles over $P/H$
\be\label{Crampindef**}
D=(\omega^\mathfrak{g})^{-1}:P\times_{Ad_H}\mathfrak{g}\stackrel{\sim}{\rightarrow}\frac{TP}{H}
\ee
from the adjoint tractor bundle to the Atiyah bundle, extending the fundamental isomorphism \eqref{isofundact} 
from the adjoint bundle $P\times_{Ad_H}\mathfrak{h}$ to the vertical sub-bundle $\frac{VP}{H}$ of the Atiyah bundle. It is indeed
the inverse of a Cartan isomorphism \eqref{Crampindef}.
All these comments can be summarised in the following isomorphism of short exact sequences of vector bundles over $P/H$ from the tractor sequence to the canonical sequence:
\[
\begin{array}
	[c]{ccccccccc}
	0&\to& P\times_{Ad_H}\mathfrak{h}&\hookrightarrow&P\times_{Ad_H}\mathfrak{g}&\twoheadrightarrow&P\times_{Ad_H}\,\mathfrak{g}/\mathfrak{h}&\to& 0\\
	&&&&&&&&\\
	&&\#\downarrow\sim&&D\downarrow\sim&&\Theta^{-1}\downarrow\sim&&\\
	&&&&&&&&\\
	0&\to& \frac{VP}{H}&\stackrel{i}{\hookrightarrow}&\frac{TP}{H}&\stackrel{\pi_*}{\twoheadrightarrow}&T\frac{P}{H}&\to& 0
\end{array}
\]

A fundamental derivative $D:\mathfrak{g}\hookrightarrow \Gamma(TP)$ can also be defined as an intertwiner, \textit{i.e.} an $\mathfrak{h}$-module morphism from the principal algebra $\mathfrak{g}$ to the Lie algebra of nowhere (or everywhere) vanishing vector fields on $P$ whose restriction to the isotropy algebra $\mathfrak{h}$ reduces to the fundamental action. 
When the Klein pair is reductive, the restriction of the fundamental derivative to the transvection submodule remains an $\mathfrak{h}$-module morphism, 
\be
D|_\mathfrak{p}\,:\,\mathfrak{p}\hookrightarrow \Gamma(TP)\,:\,\texttt{P}_a\to \hat{D}_a\,.
\ee
In components, one has
\be
[\texttt{T}_i^\#,\hat{D}_a]=C_{ia}^b\hat{D}_b 
\ee
since $C_{ia}^j=0$ in the reductive case.

\vspace{5mm}
\begin{framed}
	\begin{center}
		\textbf{The many faces of Cartan connections}
	\end{center}
	
	\noindent
	Consider a Klein pair $\mathfrak{h}\subseteq\mathfrak{g}$ and a principal bundle $H\lefttorightarrow P \twoheadrightarrow P/H$ whose total space $P$ is parallelisable and of dimension equal to the one of the principal algebra $\mathfrak{g}$. Given this data, the following notions are equivalent:
	\begin{enumerate}
		\item a Cartan geometry defined on the principal $H$-bundle $P$ and modeled on the Klein pair $\mathfrak{h}\subseteq\mathfrak{g}$, 
		\item a Cartan isomorphism $\omega^\mathfrak{g}:\frac{TP}{H}\stackrel{\sim}{\rightarrow}P\times_{Ad_H}\mathfrak{g}$, \textit{i.e.} an isomorphism of vector bundles over $P/H$ from the Atiyah bundle to the adjoint tractor bundle that extends the canonical isomorphism $\iota:\frac{VP}{H}\stackrel{\sim}{\rightarrow}P\times_{Ad_H}\mathfrak{h}$,
		\item a Cartan connection one-form $\omega^\mathfrak{g}\in \Omega^1(P)\otimes_{Ad_H} \mathfrak{g}$, \textit{i.e.} a non-degenerate $Ad_H$-equivariant $\mathfrak{g}$-valued differential one-form on $P$ whose vertical part is entirely determined by the canonical isomorphism $\iota:VP\stackrel{\sim}{\rightarrow}P\times\mathfrak{h}$,
		\item a fundamental derivative $D:\mathfrak{g}\hookrightarrow  \Gamma(TP)$, \textit{i.e.} an injective intertwiner of $\mathfrak{h}$-modules, from the principal algebra $\mathfrak{g}$ into a Lie algebra of nowhere (or everywhere) vanishing vector fields on $P$, that extends the fundamental action: $\#:\mathfrak{h}\hookrightarrow  \Gamma(TP)$,
		\item a Cartan parallelism $D:P\times_{Ad_H}\mathfrak{g}\stackrel{\sim}{\rightarrow}\frac{TP}{H}$, \textit{i.e.} an isomorphism of vector bundles over $P/H$ from the adjoint tractor bundle to the Atiyah bundle that extends the fundamental isomorphism $\#:P\times_{Ad_H}\mathfrak{h}\stackrel{\sim}{\rightarrow}\frac{VP}{H}$.
	\end{enumerate}
	\vspace{3mm}\end{framed}

 \subsubsection{Curvature of Cartan connection one-forms}

Due to close similarity between connection one-forms for principal and Cartan connections, there is a natural way to adapt the notion of curvature from the principal connection and unify both definitions. The \textbf{curvature two-form of a principal (or Cartan) connection one-form} 
$\omega$ is the $Ad_H$-equivariant horizontal differential two-form on $P$ taking values in the isotropy algebra $\mathfrak{h}$ (or the principal algebra $\mathfrak{g}$) defined as
\be\label{curv2form'}
\Omega=d\omega+\frac12\,[\omega,\omega]\,,
\ee 
where the Lie bracket on the right-hand-side is the one of the isotropy algebra $\mathfrak{h}$ (or the principal algebra $\mathfrak{g}$).
This two-form is horizontal, in the sense that it gives zero if any of its two argument belongs to the vertical distribution $VP$ (see \textit{e.g.} \cite[Corollary 5.3.10]{Sharpe} for the case of Cartan connections). The horizontality can be shown by making use of formula \eqref{domegaformula}, \textit{i.e.}
\be\label{Omegaformula}
\Omega({X}_1,{X}_2)={\mathcal L}_{{X}_1}\,\big(\,\omega({X}_2)\,\big)\,-\,{\mathcal L}_{{X}_2}\,\big(\,\omega({X}_1)\,\big)\,-\,\omega\big(\,[{X}_1,{X}_2]\,\big)+\big[\,\omega({{X}_1})\,,\,\omega({{X}_2})\,\big]\,,
\ee
and by computing the action of \eqref{curv2form'} on a pair of vector fields ${X}_1$ and ${X}_2$ on $P$ where one of them is a fundamental vector field, \textit{e.g.} ${X}_1=x_1^\#$ with $\omega({X}_1)=x_1\in\mathfrak{h}$:
\ba
\Omega({X}_1,{X}_2)
&=&{\mathcal L}_{x_1^\#}\,\big(\,\omega({X}_2)\,\big)\,-\,\underbrace{{\mathcal L}_{{X}_2}\,x_1}_{=0}\,-\,\omega\big(\,\underbrace{[x_1^\#,{X}_2]}_{={\mathcal L}_{x_1^\#}{X}_2}\,\big)+\underbrace{\big[\,x_1\,,\,\omega({{X}_2})\,\big]}_{=(ad_{x_1}\,\omega)({X}_2)}\nonumber\\
&=&{\mathcal L}_{x_1^\#}\,\big(\,\omega({X}_2)\,\big)\,-\,\omega\big(\,{\mathcal L}_{x_1^\#}{X}_2\,\big)\,-\,\big(\,{\mathcal L}_{x_1^\#}\omega\,\big)({X}_2)\,=\,0\,.\nonumber
\ea
where the $Ad_H$-equivariance of the connection one-form was used to get the last line.

The linear map 
\be
\Delta\,:\,\mathfrak{g}\wedge\mathfrak{g}\to\mathfrak{X}(P)
\,:\,y_1\wedge y_2\mapsto D\big(\,[y_1,y_2]_{\mathfrak{g}}\,\big)-[\,D(y_1)\,,\,D(y_2)\,]_{\mathfrak{X}(P)}
\ee
measures the failure of the fundamental derivative $D$ to be a Lie algebra action on $P$ and will be called the \textbf{curvature of the fundamental derivative}.
It is a vector-field-valued two-form on the principal algebra, which is horizontal in the sense that it vanishes if any of its argument belongs to the isotropy algebra $\mathfrak{h}$, \textit{i.e.} $\Delta\in \wedge^2(\mathfrak{g}/\mathfrak{h})^*\,\otimes\, \Gamma(TP)$.
The property of $ad_\mathfrak{h}$-equivariance of the fundamental derivative is strictly equivalent to the horizontality of its curvature: 
\be
{\mathcal L}_{y^\#}(Dz)=D(ad_yz)\quad\Leftrightarrow\quad [\,y^\#\,,\,Dz\,]_{ \mathfrak{X}(P)}=D\big(\,[y,z]_{\mathfrak{g}}\,\big)\,,\quad
\forall y\in\mathfrak{h}\,,\,\,\forall z\in\mathfrak{g}\,.
\ee
Finally, the composition $K=\omega^\mathfrak{g}\circ\Delta$ of the curvature of the fundamental derivative and the Cartan connection one-form defines a map
\be
K_\bullet\,:\,P\to \wedge^2(\mathfrak{g}/\mathfrak{h})^*\otimes\mathfrak{g}
\,:\,p\mapsto K_p 
\ee
called the \textbf{curvature function of the Cartan geometry} where \cite[Definition 5.3.22 \& Exercise 5.3.27]{Sharpe}
\ba
K_p&:& \wedge^2\mathfrak{g}\to\mathfrak{g}\\
&:&y_1\wedge y_2\mapsto\Omega_p(Dy_1,Dy_2) \,=\,[y_1,y_2]_{\mathfrak{g}}\,-\,\omega^\mathfrak{g}_p\Big([\,D(y_1)\,,\,D(y_2)\,]_{\mathfrak{X}(P)}\Big)\,.\nonumber
\ea

The \textbf{torsion two-form of a Cartan connection one-form} is the projection $\Omega^{\mathfrak{g}/\mathfrak{h}}$ of the curvature two-form to the transvection module $\mathfrak{g}/\mathfrak{h}$. 
A Cartan connection one-form $\omega^\mathfrak{g}$ is said \textbf{torsionless} if the torsion two-form vanishes, \textit{i.e.} if its curvature two-form $\Omega^\mathfrak{g}$ only takes values in the isotropy subalgebra $\mathfrak{h}$.
This condition is equivalent to the condition 
\be
[\,\omega^\mathfrak{g}({{X}_1})\,,\,\omega^\mathfrak{g}({{X}_2})\,]_\mathfrak{g}-\omega^\mathfrak{g}\big(\,[{X}_1,{X}_2]_{ \mathfrak{X}(P)}\,\big)\in {C}^\infty(P)\otimes_{Ad_H}\mathfrak{h}
\ee
for any pair of constant vector fields $X_1,X_2$ on $P$, as follows from \eqref{Omegaformula} and the fact that their images $\omega^\mathfrak{g}({{X}})$ are constant.
Similarly to the above argument, this is also equivalent to the condition  
\be
D\big(\,[y_1,y_2]_{\mathfrak{g}}\,\big)-[\,D(y_1)\,,\,D(y_2)\,]_{ \mathfrak{X}(P)}\in\Gamma(VP)
\ee
for any pair of Lie algebra elements $y\in\mathfrak{g}$. This shows that a Cartan geometry is torsionless if and only if the curvature $\Delta$ of the fundamental derivative $D$ is vertical-valued, \textit{i.e.} the fundamental derivative fails to be a Lie algebra action only up to vertical vector fields.

 \subsubsection{Flat Cartan geometries as principal homogeneous spaces}

A connection one-form $\omega
$ is said \textbf{flat} when its curvature two-form $\Omega
$ vanishes. A Cartan geometry will be said \textbf{flat} if its Cartan connection one-form is flat.
The vanishing of the curvature two-form $\Omega^\mathfrak{g}$ of a Cartan connection one-form $\omega^\mathfrak{g}$ is equivalent to the condition $[\,\omega^\mathfrak{g}({{X}_1})\,,\,\omega^\mathfrak{g}({{X}_2})\,]_\mathfrak{g}=\omega^\mathfrak{g}\big(\,[{X}_1,{X}_2]_{ \mathfrak{X}(P)}\,\big)$ for any pair of constant vector fields ${X}$ on $P$.
Equivalently, in terms of the fundamental derivative $D$ the flatness condition states that $D\big(\,[y_1,y_2]_{\mathfrak{g}}\,\big)=[\,D(y_1)\,,\,D(y_2)\,]_{ \mathfrak{X}(P)}$ for any pair of Lie algebra elements $y\in\mathfrak{g}$. In conclusion, a Cartan connection one-form $\omega^\mathfrak{g}$ is flat if and only if its fundamental derivative $D$ is a (regular) action of $\mathfrak{g}$ on $P$ (extending the fundamental action $\#$ of $\mathfrak{h}$ on $P$), which will be called the \textbf{fundamental action of the principal algebra}. In fact, locally a flat Cartan geometry defines a Maurer-Cartan geometry, \textit{i.e.} the principal bundle $P$ can be endowed with a structure of principal homogeneous space for a Lie group $G$ corresponding to the principal algebra $\mathfrak{g}$.\footnote{Maurer-Cartan geometries will be discussed at length in the next section.} More precisely, along the lines of \cite{Bernshtein} a regular action of $\mathfrak{g}$ on a principal $H$-bundle $P$ extending the fundamental action of $\mathfrak{h}$ will be called a \textbf{principal homogeneous} $\mathfrak{g}$\textbf{-space}.\footnote{This notion is particularly useful when the Lie algebra $\mathfrak{g}$ is infinite-dimensional, in which case there may not exist any corresponding Lie group $G$.} The fundamental derivative of a flat Cartan connection modeled on a Klein pair $\mathfrak{h}\subset\mathfrak{g}$ endows the total space $P$ with the structure of a principal homogeneous $\mathfrak{g}$-space.

\vspace{5mm}
\begin{framed}
	\begin{center}
		\textbf{The many faces of flat Cartan connections}
	\end{center}
	
	\noindent
	Given a principal bundle $H\lefttorightarrow P \twoheadrightarrow P/H$, modeled on a Klein pair $\mathfrak{h}\subseteq\mathfrak{g}$ such that the dimension of the principal algebra $\mathfrak{g}$ is equal to the dimension of the total space $P$, the following notions are equivalent:
	\begin{enumerate}
		\item a flat Cartan geometry defined on the principal $H$-bundle $P$ and modeled on the Klein pair $\mathfrak{h}\subseteq\mathfrak{g}$, 
		\item a flat Cartan connection one-form $\omega^\mathfrak{g}\in \Omega^1(P)\otimes_{Ad_H} \mathfrak{g}$, \textit{i.e.} a Cartan connection one-form with vanishing curvature two-form,
		\item a flat fundamental derivative $D:\mathfrak{g}\hookrightarrow  \Gamma(TP)$, \textit{i.e.} a regular action of the principal algebra $\mathfrak{g}$ on $P$ that extends the fundamental action: $\#:\mathfrak{h}\hookrightarrow  \Gamma(TP)$.
	\end{enumerate}
	\vspace{3mm}\end{framed}

 \subsubsection{Moving frames and Cartan gauge connections}

An element $p\in P$ of the total space of a principal $H$-bundle $\pi:P\twoheadrightarrow P/H$ is sometimes called an $H$-\textbf{frame at} $\pi(p)$.\footnote{This definition corresponds to Cartan's view of ``frames'' as any collection of geometric shapes on which a group acts regularly (\textit{i.e.} freely and transitively) \cite{Cartan}.}
Accordingly, a section $s:P/H\hookrightarrow P$ of a principal bundle is sometimes called a \textbf{moving frame on the base} or 
a \textbf{(complete) gauge of the principal bundle}. It will also be referred to as a moving frame of the principal bundle. 

An automorphism of the principal $H$-bundle $P$ is an $H$-equivariant diffeomorphism, \textit{i.e.} a diffeomorphism $f:P\stackrel{\sim}{\to}P$ commuting with the right $H$-action, \textit{i.e.} $f(ph)=f(p)h$, for all $H$-frames $p\in P$ and all elements $h\in H$ of the structure group.
Any automorphism $f:P\stackrel{\sim}{\to}P$ over $b:M\stackrel{\sim}{\to}M$
will modify the gauge $s$ into another gauge $s'=f\circ s\,\circ\, b^{-1}$. For this reason, the vertical automorphisms (\textit{i.e.} the automorphisms over the identity such that  $b=id_M\,\Leftrightarrow\,\pi\circ f=\pi$) are usually called \textbf{gauge transformations}.
They are generated by the vertical invariant vector fields over $P$. The latter acquire in this way the interpretation of \textbf{infinitesimal gauge transformations}. The canonical isomorphism $\iota$ of Lie algebras from $\Gamma(VP)^H$ onto ${C}^\infty(P)\otimes_{Ad_H}\mathfrak{h}$, 
models the infinitesimal gauge transformations of a principal $H$-bundle $P$ as $H$-covariant fields on $P/H$ taking values in the isotropy algebra of a Klein geometry. 
This explains why the adjoint sheaf ${C}^\infty(P)\otimes_{Ad_H}\mathfrak{h}\cong\Gamma(VP)^H$ is sometimes called the \textbf{Lie algebra of gauge transformations}. More generally, the invariant vector fields over $P$ correspond to the infinitesimal automorphisms of the principal bundle $H$-bundle $P$. 

\vspace{2mm}
\noindent\textbf{Remark:}
In physicist language, fundamental vector fields ${Y}_y = y^{i}\,\texttt{T}_i^\#$
correspond to infinitesimal ``global'' (or ``rigid'') symmetries while
invariant vertical vector fields \eqref{vertvectfprinc}, where the components $\sigma^i$ are such that \eqref{strcts} holds, correspond to infinitesimal ``local'' (or ``gauge'') symmetries.
\vspace{2mm}

The pullback, $s^*\omega$, of the principal (or Cartan) connection one-form 
$\omega$, along a gauge $s:P/H\hookrightarrow P$, is a differential one-form on the base space $P/H$ valued in $\mathfrak{h}$ (or $\mathfrak{g}$) called the principal (or Cartan) \textbf{gauge connection} with respect to the given gauge.\footnote{See \textit{e.g.} \cite[Definition 5.1.8 \& Proposition 5.2.5]{Sharpe} for the case of a Cartan gauge connection.} They are the usual objects physicists work with.
The pullback $s^*\omega$ of the corresponding curvature two-form $\Omega$ along the gauge $s$ is a differential two-form on $P/H$ valued in $\mathfrak{h}$ (or $\mathfrak{g}$) that
will be called the \textbf{gauge curvature} \cite[Definition 5.1.8 \& Proposition 5.2.5]{Sharpe}.
A gauge connection $s^*\omega$ is said to be flat if and only if its gauge curvature $s^*\Omega$ vanishes everywhere on $P/H$. This fact is a necessary and sufficient condition for the principal (or Cartan) connection $\omega$ on $P$ to be flat (due to the equivariance).

In the physics literature, one usually considers the action of gauge transformations on gauge connections as follows: given a fixed connection $\omega$, a gauge transformation maps a section $s$ into a section $s'$ and, correspondingly, transforms the gauge connection $s^*\omega$ into the gauge connection $s^{\prime*}\omega$. The usual gauge transformation rules of (principal or Cartan) connection one-forms can be obtained in this way. Another perspective is to consider the action of infinitesimal gauge transformations on $\omega$. More generally, one may consider the Lie derivative of $\omega$ along any (\textit{i.e.} not necessarily vertical) vector field on $P$. 
Cartan's magic formula ${\mathcal L}_{{X}}=i_{{X}}\circ d+d\circ i_{{X}}$ and the derivation property $i_{{X}}[\omega,\omega]_{\mathfrak g}=2\,[i_{{X}}\omega,\omega]_{\mathfrak g}$ imply
\be
{\mathcal L}_{{X}}\omega=i_{{X}}\Omega+\underbrace{d\chi-[\chi,\omega]}_{
	=(d+ad_\omega)\chi}\,,
\ee
where $\chi=\omega({{X}})$\,.
For vertical vector fields,
\be
{X}=\chi^\#\in \Gamma(VP)={C}^\infty(P)\otimes\mathfrak{h}^\#\,,\quad\text{where}\quad\chi\in{C}^\infty(P)\rtimes\mathfrak{h}\,,
\ee
this transformation rule simplifies
\be
{\mathcal L}_{{}_{\chi^\#}}\omega=d\chi-[\chi,\omega]
=(d+ad_\omega)\chi\,,
\ee
because the curvature is horizontal. This implies that the transformation rule of a gauge connection under an infinitesimal gauge transformation identifies with the Lie derivative of the gauge connection under the corresponding vertical invariant vector field.

\pagebreak

\section{Model geometries of Cartan connections}\label{modelgeo}

The subject of this section is the particular example of flat Cartan geometries known as Klein geometries.
Cartan succeeded to unify Klein's and Riemann's visions of geometry in a single approach. His perspective on differential geometries (\textit{e.g.} Riemannian, Weyl, etc) was to view them as (curved) ``deformations'' of (flat) ``model'' geometries offered by the group-theoretical view on geometry due to Klein. This explains the importance of Klein geometries for Cartan geometries.

 \subsection{Maurer-Cartan geometry}

Let $G$ be a Lie group.

 \subsubsection{Left and right multiplications} 

The \textbf{left multiplication} is the homomorphism\footnote{In other words, $L_{g_1g_2}=L_{g_1}\circ L_{g_2}$ ($\forall g_1,g_2\in G$).} of groups
\be\label{leftmultiplication}
L_\bullet\,:\,G\to {Diff}(G)\,:\,g\mapsto L_g\,.
\ee
It provides a canonical left action of the Lie group $G$ on itself, where the diffeomorphism $L_g:g'\mapsto gg'$ stands for the \textbf{left translation} by $g\in G$. The pushforward $(L_g)_*:T_{g'}G\stackrel{\sim}{\to} T_{gg'}G$ of the left translation provides an isomorphism of vector spaces between all tangent spaces. 
Left-invariant vector fields on $G$ will be called \textbf{fundamental vector fields of the Lie group} $G$.
Endowed with the Lie bracket of vector fields, the space of fundamental vector fields defines the Lie algebra $\mathfrak{g}$ of the Lie group $G$: 
\be
\mathfrak{g}\,=\,\mathfrak{X}_{\text{linv}}(G)\,=\, \Gamma\left(\frac{TG}{G}\right)\,.
\ee
However, as a vector space the Lie algebra is traditionnally identified with the tangent space at the origin via the left multiplication. The latter identification will always be implicitly assumed from now on, though $T_eG\cong \mathfrak{g}$ is \textit{a priori} only an isomorphism of vector spaces.

The \textbf{right multiplication} is the antihomomorphism\footnote{In other words, $R_{g_1g_2}=R_{g_2}\circ R_{g_1}$ ($\forall g_1,g_2\in G$).} of groups
\be\label{rightmultiplication}
R_\bullet\,:\,G\to {Diff}(G)\,:\,g\mapsto R_g\,.
\ee 
It provides a canonical right action of the Lie group $G$ on itself, where the diffeomorphism $R_g:g'\mapsto g'g$ stands for the \textbf{right translation} by $g\in G$. The \textbf{fundamental action of the Lie algebra} $\mathfrak{g}$ \textbf{on the Lie group} $G$
is the canonical isomorphism $\#\,:\,\mathfrak{g}\stackrel{\sim}{\to}\mathfrak{X}_{\text{linv}}(G)$ between the Lie algebra $\mathfrak{g}$ and the Lie algebra of left-invariant vector fields on $G$. More concretely, the fundamental action of the Lie algebra can be expressed in terms of the pullback of the right action of the Lie group as the map
\be
\#\,:\,\mathfrak{g}\stackrel{\sim}{\to}\mathfrak{X}_{\text{linv}}(G)\,:\,y\mapsto y^\#:=\left.\frac{d}{dt}(R_{\exp ty})^*\,\right|_{t=0}
\label{fundamentalG}
\ee
where $y^\#$ is defined by its effect on the structure algebra ${C}^\infty(G)$. Geometrically, this can be described as follows: Firstly, one may associate to any Lie algebra element $y\in\mathfrak{g}$ a one-parameter subgroup of $G$ (in the neighborhood of the identity) \cite[Theorem 20.1 and Proposition 20.5]{Lee2} with typical element denoted $\exp ty\in G$ for some parameter $t\in\mathbb R$. Secondly, the action of the latter subgroup by right multiplication defines a flow on $G$, \textit{i.e.} a one-parameter group of diffeomorphisms of $G$ with typical element $R_{\exp ty}\in Diff(G)$ or, equivalently, a one-parameter group of automorphisms of the structure algebra ${C}^\infty(G)$ with typical element $(R_{\exp ty})^*\in{Aut}(\,{C}^\infty(G)\,)$. Thirdly, the vector field generating this flow is a fundamental vector field given by the expression in \eqref{fundamentalG}.
The fundamental action of the Lie algebra $\mathfrak{g}$ on the Lie group $G$ is regular and defines a structure of regular action Lie algebroid $G\rtimes\mathfrak{g}$ over $G$ with anchor
\be\label{anchordiesel}
\#_\bullet\,:\,G\rtimes\mathfrak{g}\stackrel{\sim}{\to} TG\,:\,(g,y)\mapsto y^\#|_g\,,
\ee
which will be called the \textbf{fundamental action algebroid on the Lie group $G$}.
A canonical isomorphism from $\mathfrak{g}$ to all tangent spaces to $G$ is provided by the field of linear isomorphisms 
\be
\#_g\,:\,\mathfrak{g}\stackrel{\sim}{\to} T_gG\,:\,y\mapsto y^\#|_g
\label{diesel}
\ee

The anchor \eqref{anchordiesel} can be expressed in terms of the pushforward of the left multiplication \eqref{leftmultiplication}.
Indeed, the fundamental action can be equivalently defined in terms of the pushforward of the left multiplication. In fact, $R_{g'}(g)=gg'=L_g(g')$ implies that for $g'=e+t\,y+{\mathcal O}(t^2)$ infinitesimally close to the identity, one finds $(L_g)_*:T_e G\stackrel{\sim}{\to} T_gG:\,y\mapsto y^\#|_g$. Comparing with \eqref{diesel}, one deduces that: $\#_\bullet=(L_\bullet)_*$.

Obviously, all previous constructions have a ``mirror'' analogue where the role of ``left'' and ``right'' are exchanged. For instance, the following action of the Lie algebra $\mathfrak{g}$ on $G$
\be
\sharp\,:\,\mathfrak{g}\stackrel{\sim}{\to} \mathfrak{X}_{\text{rinv}}(G)\,:\,y\mapsto y^{\sharp}:=\left.\frac{d}{dt}(L_{\exp ty})^*\,\right|_{t=0}
\label{mfundamentalG}
\ee
will be called the \textbf{principal action of the Lie algebra} $\mathfrak{g}$ \textbf{on the Lie group} $G$, where $\mathfrak{X}_{\text{rinv}}(G)$ denotes the Lie algebra of right-invariant vector fields that will be called \textbf{principal vector fields} on $G$ because they are the vector fields corresponding to the left action of $G$ on itself.
Note that 
\be
\forall x^\#\in\mathfrak{X}_{\text{linv}}(G)\,,\,\forall y^{\sharp}\in \mathfrak{X}_{\text{rinv}}(G)\,:
\qquad{\mathcal L}_{x^\#}y^{\sharp}=[x^\#,y^{\sharp}]=-{\mathcal L}_{y^{\sharp}}{x^\#}=0\,.
\ee
As one can see, the fundamental and principal $\mathfrak{g}$-actions commute with each other, consistently with the fact that the left and right $G$-multiplication commute.
Note also that $\#_g=(L_g)_*$ and $\sharp_g=(R_g)_*$.
The Lie algebras $\mathfrak{g}^\#$ and $\mathfrak{g}^{^{\sharp}}$ of right and left fundamental vector fields on the Lie group $G$ form a dual pair in the Lie algebra $\mathfrak{X}(G)$ of vector fields on $G$. In other words, the generators of the left/right translations (\textit{i.e.} respectively of the principal/fundamental $\mathfrak{g}$-action) are the right/left-invariant (\textit{i.e.} the principal/fundamental) vector fields on $G$. This is summarised in Table \ref{MirrorLiegroup}.

\vspace{1mm}
\noindent{\small \textbf{Notation:} For spaces of fields on $G$, the superscript ``H'' will stand for $H$-invariance under right multiplication $R$ or, equivalently, $\mathfrak{h}$-invariance under the fundamental action $\#$\,,	while the subscript ``rinv'' (respectively ``linv'') will stand for $G$-invariance under right (respectively, left) multiplication or, equivalently, $\mathfrak{g}$-invariance under the (extended) fundamental action $\#$ (respectively, the principal action $\sharp$).}
\vspace{1mm}

\begin{table}
	\begin{center}
		\begin{tabular}{
				|c|c|c|}
			\hline
			Geometry & Fundamental & Principal \\ 
			\hline\hline
			&&\\
			$G$-actions & Right multiplication $R_\bullet$ & Left multiplication $L_\bullet$ \\
			&&\\\hline
			&&\\
			Pushforwards & Pushfor. right mult. & Pushfor. left mult.\\
			& $\sharp_\bullet=(R_\bullet)_*$ & $\#_\bullet=(L_\bullet)_*$\\
			&&\\\hline
			&&\\
			$\mathfrak{g}$-actions& Infinit. right mult. & Infinit. left mult. \\
			&$\bullet^{\#}=\left.\frac{d}{dt}(R_{\exp t\bullet})^*\,\right|_{t=0}$ & $\bullet^{\sharp}=\left.\frac{d}{dt}(L_{\exp t\bullet})^*\,\right|_{t=0}$ \\
			&&\\\hline
			&&\\
			Generators & Left-inv. vect. fields & Right-inv. vect. fields \\
			& $\mathfrak{X}_{\text{linv}}(G)=\mathfrak{g}^\#$ & $\mathfrak{X}_{\text{rinv}}(G)=\mathfrak{g}^\sharp$
			\\
			&&\\\hline
		\end{tabular}
	\end{center}
	\caption{\label{MirrorLiegroup}Summary of mirror geometric structures on a Lie group}
\end{table}

 \subsubsection{Maurer-Cartan one-forms} 

The inverse of the bijective anchor \eqref{anchordiesel} of the fundamental action algebroid is a canonical isomorphism of Lie algebroids over $G$
\be\label{MCisomorphism}
\omega_{MC}^\mathfrak{g}\,:\,TG\stackrel{\sim}{\to} G\rtimes \mathfrak{g}
\ee
from the tangent bundle $TG$ to the fundamental action algebroid $G\rtimes\mathfrak{g}$, which will be called the \textbf{Maurer-Cartan isomorphism}. In other words, $\omega_{MC}^\mathfrak{g}=\#^{-1}$. In order to obtain an explicit formula, let us recall that $\#_\bullet=(L_\bullet)_*$, hence $\omega_{MC}^\mathfrak{g}|_\bullet=(L_{\bullet})^{-1}_*$. Usually, this last formula is written differently. The left multiplication \eqref{leftmultiplication} is a group homomorphism, hence $id_G=L_e=L_{g^{-1}g}=L_{g^{-1}}\circ L_{g}$ ($\forall g\in G$). Moreover, the pushforward is a covariant functor, hence $id_{TG}=(L_e)_*=(L_{g^{-1}g})_*=(L_{g^{-1}}\circ L_{g})^*=(L_{g^{-1}})_*\circ (L_{g})_*$. Therefore, $(L_g)^{-1}_*=(L_{g^{-1}})_*$\,.
To conclude, one obtains the usual formula (see \textit{e.g.} \cite[Definition 3.1.3]{Sharpe})
\be\label{explicitformulaMC}
\omega_{MC}^\mathfrak{g}|_g=(L_{g^{-1}})_*\,.
\ee

The Maurer-Cartan isomorphism \eqref{MCisomorphism} defines a canonical left-invariant field of linear isomorphisms from the tangent spaces of a Lie group $G$ to $\mathfrak{g}$, which is called the \textbf{Maurer-Cartan one-form}. More concretely, the Maurer-Cartan one-form is the left-invariant\footnote{The pullback $(L_g)^*:T^*_{gg'}G\stackrel{\sim}{\to} T^*_{g'}G$ of the left multiplication by the element $g\in G$ shows that
	the left-invariance property reads $(L_g)^*\omega_{MC}^\mathfrak{g}|_{gg'}=\omega_{MC}^\mathfrak{g}|_{g'}$.} and right $Ad_G$-equivariant $\mathfrak{g}$-valued differential one-form on $G$
defined by the field of isomorphisms 
\be
\omega_{MC}^\mathfrak{g}|_g:T_gG\stackrel{\sim}{\to} T_eG:{X}|_g\mapsto (L_{g^{-1}})_*({X}|_g)
\ee
of vector spaces. 
Infinitesimally, the left-invariance and right-equivariance read, respectively, as
\be
\forall y\in \mathfrak{g}\,:\quad {\mathcal L}_{y^\sharp}\omega_{MC}^\mathfrak{g}=0 \,,
\quad {\mathcal L}_{y^\#}\omega_{MC}^\mathfrak{g}=ad_y\omega_{MC}^\mathfrak{g}\,.
\ee

\vspace{2mm}
{\small\noindent\textbf{Remark 1:} Note that it is traditional to define the Maurer-Cartan one-form $\omega_{MC}^\mathfrak{g}$ as above but there is also of course a mirror analogue of the latter, where left and right are exchanged. 
	More explicitly, there is also a right invariant and left $Ad_G$-equivariant $\mathfrak{g}$-valued differential one-form $\tilde{\omega}_{MC}^\mathfrak{g}$ on $G$
	defined by the field of isomorphisms $\tilde{\omega}_{MC}^\mathfrak{g}|_g:{X}|_g\mapsto (R_{g^{-1}})_*({X}|_g)$
	of vector spaces. 
	The relation between these two 1-forms is $\tilde{\omega}_{MC}^\mathfrak{g}|_g=(Ad_g)_*{\omega}_{MC}^\mathfrak{g}|_g$\,.}

\vspace{2mm}
{\small\noindent\textbf{Remark 2:} As a side remark, note that the tangent bundle $TG$ of any Lie group $G$ can be endowed with a structure of Lie group via the pushforward $\mu_*:TG\times TG\to TG$ of the product $\mu:G\times G\to G$. In this case, the tangent bundle $TG$ is called the \textbf{tangent Lie group}. It is isomorphic to the semidirect product $G\ltimes\mathfrak{g}$ of the Lie group $G$ with the additive group $\mathfrak{g}$, via the adjoint action $Ad:G\hookrightarrow GL(\mathfrak{g})$ of $G$ on $\mathfrak{g}$.\footnote{A remarkable example is the tangent Lie group of the rotation group in three dimensions which is isomorphic to the Euclidean Lie group in three dimensions, $T\,SO(3)\cong ISO(3)$. Let us stress that this example is exceptional in that the collusion of dimensions between the rotation and translation groups does \textit{not} work in the higher-dimensional versions, $T\,SO(n)\ncong ISO(n)$ for $n>3$.} In this sense, the Maurer-Cartan one-form can also be seen as an isomorphism $\omega_{MC}^\mathfrak{g}:TG\stackrel{\sim}{\to} G\ltimes \mathfrak{g}$ of Lie groups, but this point of view will not be followed since our focus is on Lie algebroid structures.}
\vspace{1mm}

A vector field ${X}$ is left-invariant if and only if its image $\omega_{MC}^\mathfrak{g}({X})$ by the Maurer-Cartan one-form is a ($\mathfrak{g}$-valued) constant function over $G$. That is to say, for a Maurer-Cartan geometry constant vector fields identify with left-invariant vector fields \cite[Lemma 3.2.2]{Sharpe}. In fact, the Maurer-Cartan one-form can also be seen as a Lie algebra isomorphism $\omega_{MC}^{\mathfrak{g}}:\mathfrak{X}_{\text{linv}}(G)\stackrel{\sim}{\to}\mathfrak{g}$ which is the inverse of the fundamental action \eqref{fundamentalG}: $\omega_{MC}^\mathfrak{g}\circ \#=id_\mathfrak{g}$.

\subsubsection{Matrix Lie groups} 

A particularly important case (the one usually considered in physics litterature) is when $G$ admits a matrix realisation (\textit{e.g.} classical Lie groups), \textit{i.e.} a finite-dimensional faithful linear representation of Lie group 
\be\label{matrixrealisation}
\texttt{g}\,:\,G\hookrightarrow GL(N)\,:\,g\mapsto \textsl{g}\,,
\ee
where $\textsl{g}$ is an invertible $N\times N$ matrix. The group morphism \eqref{matrixrealisation} will be called the \textbf{defining representation of the matrix Lie group} $G$ (see \cite{Hall} for a thorough introduction to the theory of matrix Lie groups).  
The pushforward 
\be
d\texttt{g}\,:\,TG\to GL(N)\times \mathfrak{gl}(N)\,:\,(g,y)\mapsto (\textsl{g},\textsl{y})
\ee
of the defining representation \eqref{matrixrealisation} of the matrix Lie group $G$ induces a finite-dimensional faithful representation of the Lie algebra $\mathfrak{g}$
\be
(d\texttt{g})_e\,:\,T_eG\hookrightarrow \mathfrak{gl}(N)\,:\,y\mapsto \textsl{y},
\ee
where $\textsl{y}$ is an $N\times N$ matrix.

Via such a representation, the pushforward $(L_g)_*$ of a left translation by $g$ 
is merely a left matrix multiplication by $\textsl{g}$.
Consequently, the Maurer-Cartan one-form $\omega_{MC}^{\mathfrak{gl}(N)}$ on $GL(N)$ pullbacked on $G$ via the defining representation \eqref{matrixrealisation}
is a matrix-valued differential one-form $\texttt{g}^*\omega_{MC}^{\mathfrak{gl}(N)}$ on $G$ usually written, with a slight abuse of notation, as
\be
\omega^\mathfrak{g}\,=\,\texttt{g}^{-1}d\texttt{g}\,.
\ee
The right-hand-side provides a very concrete interpretation of the formula \eqref{explicitformulaMC} in the case of matrix Lie groups (see \textit{e.g.} \cite[Example 3.1.7]{Sharpe}).

\subsubsection{Exponential map} 

To any Lie algebra element $y\in\mathfrak{g}$, one may associate a one-parameter subgroup of elements $\exp(\,t\,y)\in G$, where $t$ is a real parameter. Geometrically, one considers the principal vector field $y^\sharp\in\mathfrak{X}_{\text{linv}}(G)$ generating the flow
\begin{equation}
	\exp(t\,y^\sharp)=R_{\exp(t\,y)}\in Diff(G).
\end{equation} 
The map 
\be\label{expgsmall}
\exp\,:\,\mathfrak{g}\stackrel{\sim}{\to}G\,:\,y\mapsto \exp(y)\,,
\ee
sends a Lie algebra element $y\in\mathfrak{g}$ to the image of the identity element $e\in G$ by the flow $\exp(\,t\,y^\sharp)$ evaluated at $t=1$, 
\be
\exp(y)\,:=\,\exp{y}^\sharp(e)\in G\,.
\ee
The (local) diffeomorphism \eqref{expgsmall} is called the \textbf{exponential map from the Lie algebra} $\mathfrak{g}$ \textbf{onto the Lie group} $G$ \cite[Section I.4]{Kob63}. The exponential map allows to make use of the coordinates of the Lie algebra $\mathfrak{g}$ in a basis as (local) coordinates on the Lie group $G$ (strictly speaking, only on the image of the exponential map).

A remarkable property of the exponential map is that it intertwines the adjoint representation $ad:\mathfrak{g}\to\mathfrak{gl}(\mathfrak{g})$ of the Lie algebra with the adjoint representation $Ad:G\to Diff(G)$ of the Lie group,
\be
Ad \circ \exp\,=\, \exp\circ\, ad\,.
\ee
More explicitly, for matrix Lie groups this reads
\be
e^y\,g\,e^{-y}=e^{ad_y}(g)\,,
\ee
for any $y\in\mathfrak{g}$ and $g\in G$\,.
Another useful property of the exponential map of matrix Lie groups is the formula\footnote{See \textit{e.g.} \cite[Chapter 2]{Hall} for a proof.}
\be\label{Duham}
e^{y+\varepsilon z}\,=\,e^y\Big(\,1\,+\,\varepsilon\,\Omega_{{}_{-ad_y}} z\,+\,{\mathcal O}(\varepsilon^2)\,\Big)\,,
\ee
where
\be\label{Omega}
\Omega_z\,:=\,\frac{\exp z-1}{z}\,=\,1\,+\,\frac12\,z\,+\,\frac16\,z^2\,+\,{\mathcal O}(z^3)
\ee
is an analytic function of one variable.

The \textbf{differential of the exponential map from the Lie algebra onto the Lie group} is the pushforward of \eqref{expgsmall}
\be\label{dexpgsmall}
\exp_*\,:\,T\mathfrak{g}\stackrel{\sim}{\to}TG\,.
\ee
The explicit expression of the map $\exp_*|_y:T_y\mathfrak{g}\stackrel{\sim}{\to}T_{\exp y}G$ for $y\in\mathfrak{g}$
can be deduced from \eqref{Duham} and reads:
\be\label{Duhamel}
d(e^y)\,=\,e^y\,\,\frac{1-e^{-ad_y}}{ad_y}\,dy\,.
\ee
This is sometimes known as the Duhamel formula.

\subsubsection{Coordinates on matrix Lie groups} 

From the Duhamel formula \eqref{Duhamel}, a very explicit expression for the Maurer-Cartan one-form can be obtained in terms of the structure constants.
Let us introduce a basis $\{T_i\}$ for the Lie algebra $\mathfrak{g}$, with $C^i_{jk}$ the corresponding structure constants: $[T_j,T_k]=C^i_{jk}T_i$. One can use the coordinates of an element $y=y^i T_i$ of the Lie algebra $\mathfrak{g}$ in order to label the element $g=\exp(y)$ of the Lie group $G$. The Maurer-Cartan one-form
\be
\texttt{g}^{-1}d\texttt{g}\,=\,W^i{}_j\,dy^j\otimes\textsl{T}_i\,,
\ee
is a matrix-valued one-form on $G$, whose entries $W^i{}_j$ are defined by
\be
W\,=\,(e^{M}-I)M^{-1}\,,\quad\text{with}\quad M^i{}_j\,=\,C^i_{jk}\,y^k\,,
\ee
and are power series in the coordinates $y^i$ of the element $\texttt{g}=\exp\textsl{y}$ (since the matrix $M$ is linear in the coordinates).
The first few terms are given by
\be
W^i{}_j,=\,\delta^i_j\,+\,\frac12\,C^i_{jk}\,y^k\,+\,\frac16\,C^i_{km}C^k_{jn}\,y^my^n \,+\,{\mathcal O}\big(\,|y|^3\,\big)\,.
\ee

For matrix Lie groups, the independent components $\textsl{g}^a{}_b$ ($a,b=1,2,\ldots, N$) of the corresponding matrix $\textsl{g}\in GL(N)$ provide an alternative to the coordinates $y^i$ for the element $g\in G$ (with $\textsl{e}^a{}_b=\delta^a_b$ for the identity). In such case, the vector fields $\frac{\partial}{\partial \textsl{g}^a{}_b}$ (respectively, the differential one-forms $d\textsl{g}^a{}_b$) define the coordinate bases of the (co)tangent spaces to $G$. Therefore, the tangent vectors $\texttt{T}_a{}^b:=\frac{\partial}{\partial \textsl{g}^a{}_b}$ at the identity provide a basis for the classical Lie algebra $\mathfrak{g}$.
Moreover, the fundamental action then reads 
\be
\#\,:\,\textsl{y}=\textsl{y}^a{}_b\frac{\partial}{\partial \textsl{g}^a{}_b}\mapsto \textsl{y}_\textsl{g}^\#=\textsl{g}^a{}_d\textsl{y}^d{}_b\,\frac{\partial}{\partial \textsl{g}^a{}_b}
\ee
and the Maurer-Cartan one-form 
\be
\omega^\mathfrak{g}=(\textsl{g}^{-1})^c{}_a\,\, d\textsl{g}^a{}_b\otimes \frac{\partial}{\partial \textsl{g}^c{}_b}\,.
\ee

\noindent{\small\textbf{Example (Semidirect sum with an Abelian ideal)\,:} Consider a semidirect product $G=G_0\ltimes G_1$ of a Lie subgroup $G_0$ with an Abelian normal subgroup $G_1$. The corresponding Lie algebra is a semidirect sum $\mathfrak{g}=\mathfrak{g}_0\inplus \mathfrak{g}_1$ of a Lie subalgebra $\mathfrak{g}_0$ with an Abelian ideal $\mathfrak{g}_1$. This defines a reductive Klein pair $\mathfrak{h}:=\mathfrak{g}_0\subset\mathfrak{g}$ whose transvection module $\mathfrak{p}:=\mathfrak{g}_1\cong\mathfrak{g}/\mathfrak{g}_0$ is an Abelian ideal, which will be called the \textbf{translation ideal}.
	When the adjoint action of the isotropy subalgebra $\mathfrak{h}=\mathfrak{g}_0$ on the translation ideal $\mathfrak{p}=\mathfrak{g}_1$ is faithful, then the isotropy subalgebra can be considered as a Lie algebra of affine transformations of the translation ideal. Moreover, at the level of the corresponding Klein geometry $H=G_0\subset G$, the additive translation group $G_1$ acts regularly on the affine space $V=G/G_0$ on which the isotropy group $G_0$ acts faithfully by assumption.}

\vspace{3mm}
\noindent{\small\textbf{Notation:}
To emphasise all these facts, one will denote the corresponding Lie algebras as follows (inspired by physicist common usage). First, the translation ideal will be denoted in the same way as the affine space $V:= \mathfrak{g}_1\cong \mathfrak{g}/\mathfrak{g}_0$ on which the corresponding additive group acts regularly. Second, the isotropy subalgebra will be denoted as $\mathfrak{h}(V):=\mathfrak{g}_0\subset\mathfrak{gl}(V)$ to emphasise that it is a Lie subalegbra of the general linear algebra of endomorphisms of $V$. Third, the principal Lie algebra will be denoted $\mathfrak{ih}(V):=\mathfrak{g}\subset\mathfrak{igl}(V)$ to emphasise that it is a Lie subalegbra of the affine algebra. In physics, it is standard to say that generic elements of $\mathfrak{ih}(V)$ are ``inhomogeneous'' transformations of $V$, to contrast them from elements of $\mathfrak{h}(V)$ which are said ``homogeneous''.}

\subsubsection{Affine Maurer-Cartan geometry} 

Consider the affine group 
\be
IGL(n)=GL(n)\ltimes{\mathbb R}^n 
\ee
with faithful matrix representation 
as the subgroup of $GL(n+1)$ preserving the affine plane with first coordinate kept constant (and nonvanishing)\,:
\ba\label{IGLn}
&\texttt{g}\in GL(n+1)\quad\text{and}\quad \texttt{g}^T\left(
\begin{array}{c}
	1 \\
	0 \\
\end{array}
\right)=\left(
\begin{array}{c}
	1 \\
	0 \\
\end{array}
\right)\quad\Longleftrightarrow&\nonumber
\\
&\texttt{g}=\left(%
\begin{array}{cc}
	1 & 0 \\
	\vec{x} & \texttt{h} \\
\end{array}%
\right)=\left(%
\begin{array}{cc}
	1 & 0 \\
	\vec{x} & 1 \\
\end{array}%
\right)\left(%
\begin{array}{cc}
	1 & 0 \\
	0 & \texttt{h} \\
\end{array}%
\right)\in IGL(n)\subset GL(n+1)\,,&\nonumber
\ea
where $\texttt{h}\in GL(n)$ is an invertible $n\times n$ matrix of entries $e^\mu_a:=(\,\texttt{h})^\mu{}_a$, and $\vec{x}\in{\mathbb R}^n$ is a column of length $n$ with entries $x^\mu$ ($\mu=1,\ldots,n$).
The inverse reads
\be
\texttt{g}^{-1}=\left(%
\begin{array}{cc}
	1 & 0 \\
	-\texttt{h}^{-1}\vec{x} & \texttt{h}^{-1} \\
\end{array}%
\right)=
\left(%
\begin{array}{cc}
	1 & 0 \\
	0 & \texttt{h}^{-1} \\
\end{array}%
\right)
\left(%
\begin{array}{cc}
	1 & 0 \\
	-\vec{x} & 1 \\
\end{array}%
\right)
\in GL(n+1)\,.
\ee
The coordinates on $IGL(n)$ will be taken as the entries $x^\mu$ and $e^\mu_a$ of, respectively, the column $\vec{x}\in{\mathbb R}^n$ and of the square matrix $\texttt{h}\in GL(n)$.
The entries of $\texttt{h}^{-1}$ will be denoted $(\texttt{h}^{-1})^a{}_\mu=\theta_\mu^a$.
The Maurer-Cartan one-form of $IGL(n)$ reads in this representation:
\be
\omega^{\mathfrak{igl}(n)}=\texttt{g}^{-1}d\texttt{g}=\left(%
\begin{array}{cc}
	0 & 0 \\
	\texttt{h}^{-1}d\vec{x} & \texttt{h}^{-1}d\,\texttt{h} \\
\end{array}%
\right)=\left(%
\begin{array}{cc}
	0 & 0 \\
	\omega^{{\mathbb R}^n} & \omega^{\mathfrak{gl}(n)} \\
\end{array}%
\right)\in \mathfrak{gl}(n+1)\,.
\ee
In components, the restriction of the Maurer-Cartan one-form to the codomains of the translation ideal and to the general linear subalgebra are, respectively, 
\be
\omega^{{\mathbb R}^n}=\theta^a_\mu\,dx^\mu\otimes \texttt{P}_a\quad\text{and}\quad 
\omega^{\mathfrak{gl}(n)}=\theta^a_\mu\,de_b^\mu\otimes \texttt{T}_a{}^b\,, 
\ee
where $\{\texttt{P}_a\}$ and $\{\texttt{T}_a{}^b\}$ stand for some bases of the Lie algebras ${\mathbb R}^n$ and $\mathfrak{gl}(n)$.

Consider an element $\overline{g}\in IGL(n)$ represented by the matrix 
\be
\overline{\texttt{g}}=\left(%
\begin{array}{cc}
	1 & 0 \\
	\vec{y} & \overline{\texttt{h}} \\
\end{array}%
\right)
\ee
with entries $y^\mu$ for the column $\vec{y}$ and $\overline{\,h}^\mu{}_\nu$ for the invertible $n\times n$ matrix $\overline{\texttt{h}}$,
and an element $\widetilde{g}\in IGL(n)$ represented by the matrix 
\be
\widetilde{\texttt{g}}=\left(%
\begin{array}{cc}
	1 & 0 \\
	\vec{z} & \widetilde{\,\texttt{h}} \\
\end{array}%
\right)
\ee
with entries $z^a$ for the column $\vec{z}$ and $\widetilde{\,h}^b{}_a$ for the invertible $n\times n$ matrix $\widetilde{\,\texttt{h}}$.
On the one hand, the left translation $L_{\overline g}$ acts on $\texttt{g}$ in \eqref{IGLn} as follows: $g^\prime=L_{\overline g}\,\,g={\overline g}\,g$ represented by the matrix 
$\texttt{g}^\prime=\overline{\texttt{g}}\,{\texttt{g}}$
with entries 
\be
x^{\prime\mu}=\overline{h}^\mu{}_\nu\, x^\nu+y^\mu \,,\qquad e^{\prime \mu}_a=\overline{h}^\mu{}_\nu\, e^{\nu}_a\,,
\ee
whose infinitesimal version for group elements near the identity $\overline{\texttt{g}}=I_{n\times n}+\delta\overline{\texttt{g}}$ leads to the principal action of $\mathfrak{igl}(n)$ on $IGL(n)$
\be
\delta x^{\mu}=(\delta\overline{h})^\mu{}_\nu\, x^\nu+\delta y^\mu \,,\qquad \delta e^{\mu}_a=(\delta\overline{h})^\mu{}_\nu\, e^{\nu}_a\,.
\ee
Accordingly, a basis of the Lie algebra $\mathfrak{X}_{\text{rinv}}\big(IGL(n)\big)=\mathfrak{igl}(n)^{^\sharp}$ of right-invariant vector fields on $IGL(n)$
is provided by 
\be\label{sharpgenerators}
\texttt{P}^\sharp_\mu=\frac{\partial}{\partial x^\mu}\,,\qquad
(\texttt{T}{}_\mu{}^\nu)^\sharp=\,x^\nu\,\frac{\partial}{\partial x^\mu}+e^\nu_a\frac{\partial}{\partial e_a^\mu}\,.
\ee
On the other hand, the right translation $R_{\tilde g}$ acts on $\texttt{g}$ in \eqref{IGLn} as follows: $g^\prime=R_{\widetilde g}\,\,g=g\,{\widetilde g}$ represented by the matrix 
$\texttt{g}^\prime={\texttt{g}}\,\widetilde{\texttt{g}}$
with entries 
\be
x^{\prime\mu}=x^\mu+e^\mu_a\,z^a \,,\qquad e^{\prime \mu}_a=e^\mu_b\widetilde{\,h}^b{}_a\,,
\ee
whose infinitesimal version for group elements near the identity $\overline{\texttt{g}}=I_{n\times n}+\delta\overline{\texttt{g}}$ leads to the fundamental action of $\mathfrak{igl}(n)$ on $IGL(n)$
\be
\delta x^{\mu}=e^\mu_a\,\delta z^a \,,\qquad \delta e^{\prime \mu}_a=e^\mu_b (\delta\widetilde{\,h})^b{}_a\,.
\ee
Accordingly,
a basis of the Lie algebra $\mathfrak{X}_{\text{linv}}\big(IGL(n)\big)=\mathfrak{igl}(n)^\#$ of left-invariant vector fields on $IGL(n)$
is provided by 
\be\label{fundvectfieldsIGLn}
\texttt{P}^\#_a=e_a^\mu\frac{\partial}{\partial x^\mu}\,,\qquad 
(\texttt{T}{}_a{}^b)^\#=\,e^\mu_a\frac{\partial}{\partial e_b^\mu}\,.
\ee

 \subsection{Klein geometry}

The paradigmatic examples of principal bundles are Klein geometries.

A closed Lie subgroup $H\subseteq G$ of a connected Lie group $G$, such that the left coset space $G/H$ is connected,
is called a \textbf{Klein geometry with isotropy group} $H$, \textbf{principal group} $G$ \textbf{and homogeneous space} $G/H$ \cite[Definition 4.3.2]{Sharpe}. The degenerate case $H=G$ was called \textbf{Maurer-Cartan geometry} in the previous section. 
The subgroup-subalgebra theorem in the theory of Lie groups states that, for any Lie group $G$, there is a bijection between connected Lie
subgroups $H\subset G$ and Lie subalgebras $\mathfrak{h}\subseteq\mathfrak{g}$ \cite[Theorem 5.20]{Hall}. In this way, given a principal group $G$ there is a one-to-one correspondence between Klein geometries (with connected isotropy subgroups) and Klein pairs.
A Klein geometry defines a short exact sequence of connected manifolds
\be
e\to H\stackrel{i}{\hookrightarrow}G\stackrel{\pi}{\twoheadrightarrow}G/H\to e\,,\label{shortexactKlein}
\ee
The first two arrows on the left in \eqref{shortexactKlein} are Lie group morphisms and the last two arrows are connected manifold morphisms.

The left multiplication \eqref{leftmultiplication} induces an action of the principal group $G$ on the homogeneous space $G/H$
\be\label{leftmultiplication'}
L_\bullet\,:\,G\to {Diff}(G/H)\,:\,g\mapsto L_g\,,
\ee
where the action of $g$ on left cosets is $L_g:g'H\mapsto gg'H$.
This action \eqref{leftmultiplication'} of the principal group $G$ on the homogeneous space $G/H$ may not be effective. In fact, the kernel of \eqref{leftmultiplication'} is the maximal normal subgroup of the principal group $G$ inside the isotropy group $H$, which is called the \textbf{kernel of the Klein geometry} (or \textbf{normal core of the subgroup}) $H\subseteq G$ \cite[Definition 4.3.2]{Sharpe}. 

\vspace{3mm}
\noindent{\small\textbf{Example (Lie group extension)\,:} If all arrows in \eqref{shortexactKlein} are Lie group morphisms, then the homogeneous space $G/H$ must be, by assumption, endowed with a structure of Lie group. In fact, if the kernel $H$ of the surjective morphism $\pi:G\,\twoheadrightarrow\,G/H$ is a normal subgroup of $G$ (\textit{i.e.} $H\trianglelefteqslant G$), then the quotient $G/H$ is endowed with a natural group structure. In such case, the Lie group $G$ is called a \textbf{Lie group extension} of the (quotient) group $G/H$ by the (normal sub)group $H$. The kernel of such a Klein geometry  $H\subseteq G$ is equal to the isotropy subgroup. Equivalently, for a Lie group extension $G$ of $G/H$ by a Lie subgroup $H$, the normal core of $H\subseteq G$ is the subgroup $H$ itself.
}

\vspace{3mm}
\noindent{\small\textbf{Example (Effective Klein geometry)\,:} A Klein geometry $H\subseteq G$ is called \textit{effective} \cite[Definition 4.3.2]{Sharpe} iff there is no (non-trivial) normal subgroup of the principal group $G$ inside the isotropy subgroup $H$, iff the normal core is trivial, iff the representation of the isotropy subgroup $H$ on the homogeneous space $G/H$ is effective. The latter necessary and sufficient condition justifies the term ``effective Klein geometry''.\footnote{This equivalence is not obvious at first sight, but follows as a corollary of elementary results about group actions: Firstly, the kernel of a group action is the intersection of all the stabiliser subgroups (for each point). Secondly, for the action of $G$ on $G/H$ the stabiliser of the left coset $gH$ is the subgroup $gHg^{-1}\subseteq G$.
	Finally, one concludes by observing that the intersection of the subgroups $gHg^{-1}$ (for all $g\in G$) can be seen as the maximal normal subgroup of $G$ inside $H$.}  
The quotient of an ineffective Klein geometry $H\subset G$ by its normal core $N=\text{Ker}\, L_\bullet$ is an effective Klein geometry $H/N\,\subset\,G/N$ \cite[Definition 4.3.3]{Sharpe}. In this sense, there is no loss of generality in considering effective Klein geometries, which will be implicitly assumed from now on.}

 \subsubsection{Maurer-Cartan vs Klein vs Cartan geometries}

The restriction of the right multiplication of $G$ to the free right action of $H$ on $G$ defines a structure of principal $H$-bundle over the homogeneous space $G/H$ whose total space is the principal group $G$ \cite[Example I.5.1]{Kob63}, \textit{i.e.} 
\be
H\lefttorightarrow G\twoheadrightarrow G/H\,.
\ee 
The algebraic version of the $H$-bundle fibration $\pi:G\twoheadrightarrow G/H$ is the embedding $\pi^*:{C}^\infty(G/H)\hookrightarrow {C}^\infty(G)$ where one may identify, as in \eqref{CP/H}, the structure algebra ${C}^\infty(G/H)$ of the homogeneous space $G/H$ as the commutative subalgebra ${C}^\infty(G)^H=\pi^*{C}^\infty(G/H)$ of the structure algebra ${C}^\infty(G)$ of the principal group, spanned by $H$-invariant functions on $G$.

The fundamental vector fields of this principal $H$-bundle span the image $\mathfrak{h}^\#$ of the fundamental $\mathfrak{h}$-action on $G$, thus they are the \textit{vertical}, left $G$-invariant, right $Ad_G$-equivariant, vector fields on $G$. The Lie algebra $\mathfrak{h}^\#$ is the intersection of the vertical sheaf $ \Gamma(VG)$ with the Lie algebra $\mathfrak{X}_{\text{linv}}(G)=\mathfrak{g}^\#$ of left $G$-invariant vector fields, \textit{i.e.} $\mathfrak{h}^\#= \Gamma(VG)\cap\mathfrak{X}_{\text{linv}}(G)$. 
It is important to distinguish it from the adjoint sheaf
$\Gamma(VP)^H\cong \Gamma(P\times_{Ad_H}\mathfrak{h})$ of vertical right $H$-invariant vector fields on $P$, which is
the intersection of the vertical sheaf $ \Gamma(VG)$ with the Atiyah algebra $\mathfrak{X}(G)^H$ of right $H$-invariant vector fields on $P$, \textit{i.e.} $\Gamma(VP)^H= \Gamma(VG)\cap\mathfrak{X}(G)^H$.

\begin{table}
	\begin{center}
			\begin{tabular}{
					|c||c|c|}
				\hline
				Geometry & Klein & Cartan \\ 
				\hline\hline
				Total space & Principal homogeneous space & Principal $H$-bundle \\
				$P$ & diffeomorphic to $G$ & over $P/H$ \\\hline
				Group action & Free and transitive & Free right \\
				on $P$ & $G$-action & $H$-action \\\hline
				$\mathfrak{g}$-valued one-form & Maurer-Cartan one-form & Cartan connection \\
				on total space $P$ & (flat) & (curved) \\\hline
				$\mathfrak{g}$-valued one-form & Darboux & Cartan gauge \\
				on quotient manifold $P/H$ & derivative & connection \\\hline
			\end{tabular}
	\end{center}
	\caption{\label{KvsC}Flat vs Curved Cartan geometries}
\end{table}

Klein geometries are flat examples (and models) of Cartan geometries, as summarised in the table \ref{KvsC}.
A Klein pair $\mathfrak{h}\subseteq\mathfrak{g}$ is somehow the infinitesimal version of a Klein geometry.
In fact, the canonical sequence \eqref{shortexactvectcan} of a Klein geometry is precisely modeled on the Klein sequence \eqref{shortexactinfKlein}. More precisely, there is a canonical isomorphism (of short exact sequences of morphisms of vector bundles over $G/H$) from the canonical sequence \eqref{shortexactvectcan} of a Klein geometry to the tractor sequence \eqref{shortexactintract} via the Maurer-Cartan one-form $\omega_{MC}^\mathfrak{g}$:
\be\label{salsicietta'}
\begin{array}
	[c]{ccccccccc}%
	0&\to& \frac{VG}{H}&\stackrel{i}{\hookrightarrow}&\frac{TG}{H}&\stackrel{\pi_*}{\twoheadrightarrow}&T\frac{G}{H}&\to& 0\\
	&&&&&&&&\\
	&&\omega_{MC}^\mathfrak{h}\downarrow\sim&&\omega_{MC}^\mathfrak{g}\downarrow\sim&&\Theta_{MC}\downarrow\sim&&\\
	&&&&&&&&\\
	0&\to& G\times_{Ad_H}\mathfrak{h}&\hookrightarrow&G\times_{Ad_H}\mathfrak{g}&\twoheadrightarrow&G\times_{Ad_H}\,\mathfrak{g}/\mathfrak{h}&\to& 0
\end{array}
\,.
\ee
For instance, 
the Maurer-Cartan one-form $\omega_{MC}^\mathfrak{g}$ is a particular case of Cartan connection one-form \eqref{Crampindef} and provides an isomorphism between the Atiyah bundle $\frac{TG}{H}$ of a Klein geometry $H\subseteq G$ and the \textbf{homogeneous adjoint tractor bundle} $G\times_{Ad_H}\mathfrak{g}$. Moreover, the map $\Theta_{MC}$ is a particular instance of \eqref{isomorphismtransvbundle} and provides an isomorphism between the tangent bundle  
$T\frac{G}{H}$ of a homogeneous space $G/H$ and the transvection bundle $G\times_{Ad_H}\mathfrak{g}/\mathfrak{h}$, which is called in this context the \textbf{homogeneous vector bundle associated to the transvection module}.

Let us summarise the various relations between the relevant sheaves for the principal bundle $H\lefttorightarrow G\twoheadrightarrow G/H$ defined by the Klein geometry $H\subset G$\,:
\vspace{3mm}
\begin{footnotesize}
	\[%
	\begin{array}
		[c]{ccccc}%
		\Gamma_{\text{linv}}(TG)=\mathfrak{g}^\# && \quad\subset&& \Gamma(TG)\cong C^\infty(G)\rtimes_\#\mathfrak{g}\\
		&&&&\\
		&&&& \cup\\
		&&&&\\
		\cup&& \Gamma(VG)
		\cong C^\infty(G)\rtimes_\#\mathfrak{h} & \subset & \Pi\Gamma(TG)
		\\
		&&&&\\
		&&\cup &  & \cup\\
		&&&&\\
		\Gamma_{\text{linv}}(VG)=\mathfrak{h}^\#&\quad\subset\quad&\Gamma(VG)^H
		\cong C^\infty(G)\otimes_{Ad_H}\mathfrak{h}^\#
		&\quad\subset\quad & 
		\Gamma(TG)^H
		\cong C^\infty(G)\otimes_{Ad_H}\mathfrak{g}^\#
		\\
		&&
		&& 
		\cong C^\infty(G/H)\rtimes_\sharp\mathfrak{g}\\
		&&&&\\
		&&
		&  & \cup\\
		&&&&\\
		&&
		&& 
		\Gamma_{\text{rinv}}(TG)=\mathfrak{g}^\sharp
	\end{array}
	\]
\end{footnotesize}

 \subsubsection{Reductive Klein geometries}\label{MCKC}

For a reductive Klein pair, the splitting \eqref{KLeinreduct} of the Klein sequence implies the splitting of the tractor sequence in the second line of \eqref{salsicietta'}, which in turn implies the splitting of the canonical sequence in the first line. Therefore, in this case the Maurer-Cartan one-form $\omega_{MC}^\mathfrak{g}$ on the Lie group $G$ induces a principal connection one-form $\omega^\mathfrak{h}$ on the principal $H$-bundle $G$. Beware that the latter principal connection one-form $\omega^\mathfrak{h}$ may not be flat, although  the Maurer-Cartan one-form $\omega_{MC}^\mathfrak{g}$ is obviously flat.

With a slight abuse of notation, for all concrete examples of Klein geometries the corresponding groups should, from now on, be implicitly understood as the connected component thereof. 

\vspace{3mm}
\noindent{\small\textbf{Example (Klein's view on non-Euclidean geometries)\,:} Consider the orthogonal group as isotropy subgroup $O(n)$ of a Klein geometry. Taking respectively as principal groups either the orthogonal group $O(n+1)$, the Euclidean group $IO(n)$ or the Lorentz group $O(n,1)$ leads to Klein's view on the following non-Euclidean geometries: spherical geometry on $\mathbb{S}^n=O(n+1)/O(n)$, Euclidean geometry on $\mathbb{R}^n=IO(n)/O(n)$ and hyperbolic geometry on $\mathbb{H}^n=O(n,1)/O(n)$. These Klein geometries are reductive and the corresponding principal connections are curved in the cases of the spherical and hyperbolic geometries, and flat in the case of Euclidean geometry (since the transvection submodule $\mathbb{R}^n\subset \mathfrak{io}(n)$ is an Abelian ideal).}

\begin{table}
	\begin{center}
		\begin{tabular}{
				|c|c|c|c|}
			\hline
			Riemannian & Isotropy & Principal & Homogeneous \\ 
			Klein geometry & subgroup $H$ & group $G$ & space $G/H$ \\ 
			\hline\hline
			Spherical & $O(n)\qquad\subset$ & $O(n+1)$ & Hypersphere $\mathbb{S}^n$ \\\hline
			Euclidean & $O(n)\qquad\subset$ & $IO(n)$  & Euclidean space $\mathbb{R}^n$ \\\hline
			Hyperbolic& $O(n)\qquad\subset$ & $O(n,1)$ & Hyperbolic space $\mathbb{H}^n$ \\\hline
		\end{tabular}
	\end{center}
	\caption{The three models of Riemannian geometry}
\end{table}

\vspace{3mm}
\noindent{\small\textbf{Counter-example (conformal Klein geometries)\,:} The paradigmatic example of non-reductive geometry is conformal geometry.
In this context, the linear similarity 
group of the Euclidean space ${\mathbb R}^n$ is often denoted
\be
CO(n)\,:=\,{\mathbb R}\times O(n)\,\subset\, GL(n)\,,
\ee
where ${\mathbb R}$ stands for the dilation subgroup of invertible matrices proportional to the identity matrix. The similarity group (including the translations) will be denoted accordingly,
\be
ICO(n)\,:=\,CO(n)\ltimes {\mathbb R}^n\cong {\mathbb R}\ltimes IO(n)\,\subset\, IGL(n)\,,
\ee
and is sometimes called \textbf{Weyl group} in this context

On the one hand, the \textbf{Weyl model} is the Klein geometry with the similarity group $ICO(n)$ as principal group and 
the linear similarity group $CO(n)$ as isotropy subgroup. The homogeneous space of the Weyl model is the Euclidean space ${\mathbb R}^n$ (seen as a conformal space here). The similarity group $ICO(n)$ acts as the group of globally defined conformal transformations of the conformal plane ${\mathbb R}^n$ (generated by translations, rotations and dilations). 

On the other hand, the \textbf{M\" obius model} is the Klein geometry with the \textbf{conformal group} $O(n+1,1)$ as principal group, also called \textbf{M\" obius group} in this case, and the \textbf{conformal isotropy subgroup} generated by rotations, dilations and special conformal transformations (also called conformal boosts), as isotropy subgroup.  Remember that the conformal boosts are not globally well-defined on ${\mathbb R}^n$ because they send one point to infinity and vice versa. In fact, the homogeneous space of the M\" obius model is the hypersphere $\mathbb{S}^n\cong{\mathbb R}^n\bigcup\infty$, the conformal compactification of the Euclidean space ${\mathbb R}^n$ via the addition of the point at infinity. The M\" obius group $O(n+1,1)$ acts as the group of global conformal transformations of $\mathbb{S}^n$ (generated by translations, rotations, dilations and special conformal transformations). Its conformal isotropy subgroup is isomorphic to the full similarity group $ICO(n)$ since special conformal transformations are conjugated to translations via the inversion mapping the origin to infinity.}

\begin{table}
	\begin{center}
		\begin{tabular}{
				|c|c|c|c|}
			\hline
			Conformal & Isotropy & Principal & Homogeneous \\ 
			Klein geom. & subgroup $H$ & group $G$ & space $G/H$ \\ 
			\hline\hline
			Weyl & Linear sim. & Weyl group  & Conformal \\
			model & $CO(n)$ & $ICO(n)$ & plane $\mathbb{R}^n$ \\\hline
			M\" obius & Conf. isotrop. & M\" obius group & Conformal sphere \\
			model & $\cong ICO(n)$ & $O(n+1,1)$ & $\mathbb{S}^n\cong \mathbb{R}^n\bigcup\infty$ \\\hline
		\end{tabular}
	\end{center}
	\caption{The two models of conformal geometry}
\end{table}

\begin{table}
	\begin{center}
		\footnotesize
		\begin{tabular}{
				|c|c|c|c|c|}
			\hline
			Conformal  & Symmetry & With a fixed point  & Globally defined & Whole \\
			Klein geom. & transfos & (isotropy subgroup) & (principal group) & model space \\
			\hline\hline
			Weyl & Conformal isom. & Linear  & Similarities & Plane \\
			model & of the plane & similarities & = Isom. + Dilat. &   \\\hline
			M\" obius & Conformal isom. & Linear sim. & All conformal  & Sphere \\
			model & of the sphere & + Conf. boosts & isometries & = Plane + pt at $\infty$ \\\hline
		\end{tabular}
	\end{center}
	\caption{Globally defined symmetry transformations in the two models of conformal geometry}
\end{table}

 \subsubsection{Principal action}

The left multiplication plays a very important role for Klein geometry since it induces a transitive $G$-action on the left coset space $G/H$ that will be called the \textbf{principal action} of $G$ on $G/H$.
More explicitly, this action  \eqref{leftmultiplication'} is defined in terms of the left multiplication of left cosets $L_g:g'H\mapsto gg'H$.
The base space $G/H$ of the principal $H$-bundle $\pi:G\twoheadrightarrow G/H$ is in fact a homogeneous $G$-space under this transitive action of the principal group $G$ with the subgroup $H$ as isotropy subgroup stabilising the identity coset $eH$.

Aside from global (\textit{i.e.} topological) considerations, an important conceptual distinction between a homogeneous $G$-space $M$ (respectively, a principal homogeneous $G$-space $P$) and a coset space $G/H$ (respectively, a Lie group $G$) is that, for the former, no point is privileged while, for the latter, the identity coset $eH$ (respectively, the neutral element $e$) is distinguished. In other words, the identification implicitly assumed in the text between the former and the latter holds only up to a non canonical choice of fixed point (respectively, origin). 

\vspace{3mm}
\noindent{\small\textbf{Example (affine vs vector space)\,:}
	For instance, an affine space $A$ modeled on a vector space $V$ is a principal homogeneous space with the additive group $V$ as principal group. Conversely, a vector space $V$ is an affine space $A$ together with a choice of origin.}

\subsubsection{Isotropy group squared as principal group}\label{H^2}

Consider the subclass of Klein geometries where the principal group is the direct product of two copies of the isotropy group, $G=H\times H$, and where the isotropy group is embedded inside the principal group as the diagonal subgroup $i(H)\subset H\times H$ via the inclusion 
\be
i\,:\,H\hookrightarrow H\times H\,:\,h\mapsto (h,h)\,.
\ee
This class of examples is of interest because Einstein gravity with negative cosmological constant (or, respectively, higher-spin gravity with finite spectrum of spins from $2$ to $N$) in three dimensions is modeled on such Klein geometries, where the group $H$ is the special linear group $SL(2,\mathbb{R})$ (or, respectively $SL(N,\mathbb{R})$, see \textit{e.g.} \cite{Campoleoni:2010zq} or the last entry in \cite{Reviews} which is a comprehensive review on the subject).
Moreover, these gravity theories are described on-shell by flat Cartan geometries. In other words, their solutions are flat Cartan connections, hence their geometry locally looks like the Klein geometry discussed here.

The fundamental action of the diagonal subgroup $H$ on $H\times H$ is the direct product of the right multiplications, \textit{i.e.} the group antihomomorphism
\be
R_\bullet \times R_\bullet\,:\, H\hookrightarrow {Diff}(H\times H)\,:\,h\mapsto R_{h}\times R_{h}\,,
\ee
where the corresponding vertical translation is the diffeomorphism 
\be
R_{h}\times R_{h}:(\bar h_L,\bar h_R)\mapsto (\bar h_Lh\,,\bar h_Rh)\,.
\ee
The homogenous space $\frac{H\times H}{H}$ is made of the $H$-orbits, \textit{i.e.} the equivalence classes $[\bar h_L,\bar h_R]$ for the equivalence relation 
\be
(\bar h_Lh\,,\bar h_Rh)\,\sim\,(\bar h_L,\bar h_R)\,,
\ee
for which we will select representatives of the form $(\bar h,e)$, \textit{i.e.} $[\bar h_L,\bar h_R]=[\bar h,e]$ with $\bar h:=\bar h_L\bar h_R^{-1}$. As one can see, as a manifold the homogeneous space $\frac{H\times H}{H}$ is diffeomorphic to $H$ since one can identify base points with elements $\bar h\in H$. The principal action of the whole group $H\times H$ on the homogeneous space $\frac{H\times H}{H}$
comes from the direct product of the left multiplications, \textit{i.e.} the group homomorphism
\be\label{principactHxH}
L_\bullet \times L_\bullet\,:\,H\times H\to {Diff}\left(\frac{H\times H}{H}\right)\,:\,(h_L,h_R)\mapsto L_{h_L}\times L_{h_R}\,,
\ee
where the corresponding left translations induce the following diffeomorphism of the homogeneous space
\be
L_{h_L}\times L_{h_R}\,:\,\big[\,\bar h,e\,\big]\mapsto \big[\,h_L^{}\bar h\,,\, h_R \,\big]=\big[\,h_L^{}\bar h\, h_R^{-1} \,,\,e\,\big]\,,
\ee
Therefore, in terms of the manifold $H$ diffeomorphic to the homogeneous space $\frac{H\times H}{H}$ the principal action identifies with 
\be
L_\bullet \times R^{-1}_\bullet\,:\,H\times H\to {Diff}\left(H\right)\,:\,(h_L,h_R)\mapsto L_{h_L}\times R^{-1}_{h_R}\,,
\ee
where the transformation law of the representatives $\bar h\in H$ of base points is given by
\be
L_{h_L}\times R^{-1}_{h_R}\,:\,\bar h\mapsto h_L^{}\bar h\, h_R^{-1}\,,
\ee
The restriction of the principal action to the diagonal subgroup $i(H)\subset H\times H$ is nothing but the adjoint action of $H$ on itself since $L_{h}\times R^{-1}_{h}=Ad_h$. 

The fixed points $\big[\bar h,e\big]\in\frac{H\times H}{H}$ of the principal action \eqref{principactHxH} are defined by the condition: $h_L^{}\bar h\, h_R^{-1}=\bar h\,\Leftrightarrow\,h_L^{}\bar h= \bar h\, h_R\,\Leftrightarrow\, Ad_{\bar h}h_R=h_L$. As it should, the stabiliser of the identity coset $[e,e]\in\frac{H\times H}{H}$ is the diagonal subgroup: $h_L=h_R$.
All points of the homogeneous space $\frac{H\times H}{H}$ are kept fixed if and only if $h_R=h_L=h$ belongs to center $Z(H)\subset H$ of the diagonal subgroup. Consequently, the action of the quotient group $$\frac{H\times H}{Z(H)}$$ induced by the principal action of $H\times H$ on the homogeneous space $\frac{H\times H}{H}$ is effective. Remember that if the isotropy algebra $\mathfrak{h}$ is semisimple (\textit{i.e.} it has no Abelian ideal) then the center of the Lie group $H$ is a discrete subgroup  (since the center of any semisimple Lie algebra $\mathfrak{h}$ is trivial: $\mathfrak{z}(\mathfrak{h})=\{0\}$).

The tangent spaces to the principal group $H\times H$ are isomorphic to the direct sum of two copies of the isotropy subalgebra: $T_{(h_L,\,h_R)}\,(H\times H)\cong \mathfrak{h}\oplus\mathfrak{h}$.
The isotropy subalgebra $\mathfrak{h}$ is embedded into the direct sum of two copies of the isotropy subalgebra $\mathfrak{h}\oplus\mathfrak{h}$ as
\be
i_*\,:\,\mathfrak{h}\hookrightarrow \mathfrak{h}\oplus\mathfrak{h}\,:\,y\mapsto y\oplus 0\,+\,0\oplus y\,.
\ee 
The image will be denoted as $\mathfrak{g}_+:= i_*\mathfrak{h}$. An obvious linear complement is the transvection submodule 
$\mathfrak{g}_-\cong(\mathfrak{h}\oplus\mathfrak{h})\,/\,\mathfrak{h}_+$ spanned by the elements of the form $y\oplus 0\,-\,0\oplus y$.
In other words, 
\be\label{gplusminus}
\mathfrak{g}_\pm\,=\,\{\,y\oplus 0\,\pm\,0\oplus y\,\in\,\mathfrak{h}\oplus\mathfrak{h}\,\mid\,y\in\mathfrak{h}\,\}\,.
\ee

\vspace{3mm}
\noindent\textbf{Remark:} In the case of an inclusion $H\subset G$, on may consider the construction above for both Lie groups $H$ and $G$, separately leading to the following embedding of principal bundles summarised in the following diagram
\be
\begin{array}
	[c]{ccccc}%
	G &  \hookrightarrow & G\times G & \twoheadrightarrow & \frac{G\times G}{G}\\
	& & & & \\ 
	\uparrow  & & \uparrow & & \uparrow \\ 
	& & & & \\ 
	H & \hookrightarrow & H\times H & \twoheadrightarrow & \frac{H\times H}{H}
\end{array}
\ee
where the first two vertical arrows are embeddings of groups while the last one is an embedding of manifold. The first two embeddings are trivial, while the last embedding is clear if one uses representatives of the form $(\bar h,e)$ for $\frac{H\times H}{H}$ and $(\bar g,e)$ for $\frac{G\times G}{G}$.
\vspace{3mm}

A \textbf{symmetric Lie algebra} $\mathfrak{g}$ is a ${\mathbb Z}_2$-graded Lie algebra\footnote{Note that the terminology can be source of confusion here: ${\mathbb Z}_2$-graded Lie algebras (\textit{i.e.} Lie algebras endowed with a ${\mathbb Z}_2$-grading) are to be distinguished from Lie superalgebras (\textit{i.e.} ${\mathbb Z}_2$-graded vector spaces endowed with a graded bracket which is ${\mathbb Z}_2$-graded antisymmetric).} decomposing as the semidirect sum $\mathfrak{g}=\mathfrak{g}_+\inplus\mathfrak{g}_-$ of $\mathfrak{g}_+$-modules. More explicitly, one has the following relations on the Lie bracket
\be
[\mathfrak{g}_+,\mathfrak{g}_+]\subset\mathfrak{g}_+\,,\quad [\mathfrak{g}_+,\mathfrak{g}_-]\subset\mathfrak{g}_-\,,\quad [\mathfrak{g}_-,\mathfrak{g}_-]\subset\mathfrak{g}_+\,.
\ee
Equivalently, a symmetric Lie algebra can be defined as a Lie algebra equipped with an involutive automorphism $\tau$. The subspaces $\mathfrak{g}_\pm$ are the eigenspaces of $\tau$ with eigenvalue $\pm1$. 
A symmetric Lie algebra $\mathfrak{g}=\mathfrak{g}_+\inplus\mathfrak{g}_-$ defines a reductive Klein pair $\mathfrak{g}_+\subset\mathfrak{g}$ (where the isotropy subalgebra $\mathfrak{g}_+$ is spanned by the even elements while the transvection submodule $\mathfrak{g}_-$ is spanned by the odd elements) which will be called a \textbf{symmetric Klein pair}.

\vspace{3mm}
\noindent{\small\textbf{Example (Non-Euclidean geometries)\,:} The three paradigmatic examples of non-Euclidean geometries, \textit{i.e.} spherical geometry on $\mathbb{S}^n=O(n+1)/O(n)$, Euclidean geometry on $\mathbb{R}^n=IO(n)/O(n)$ and hyperbolic geometry on $\mathbb{H}^n=O(n,1)/O(n)$, are reductive Klein geometries based on three symmetric Klein pairs where the (even) isotropy algebra is $\mathfrak{so}(n)$. The involutive automorphisms $\tau$ flips the signs of the transvection generators.}

\vspace{3mm}
\noindent{\small\textbf{Example (Isotropy algebra twice)\,:}
	The Lie algebra $\mathfrak{g}=\mathfrak{h}\oplus\mathfrak{h}$ which is the direct sum of two copies of the Lie algebra $\mathfrak{h}$
	provides an example of symmetric Lie algebra,
	\be
	\mathfrak{g}=\mathfrak{h}\oplus\mathfrak{h}\cong\mathfrak{g}_+\inplus\mathfrak{g}_-
	\ee
	via the isotropy subalgebra $\mathfrak{g}_+\cong\mathfrak{h}$ and transvection submodule $\mathfrak{g}_-$ defined in \eqref{gplusminus}. Therefore, the flat Cartan connections which are solutions of Einstein gravity and its higher-spin generalisation in three spacetime dimensions provide examples (up to important global subtleties) of reductive Klein geometries based on a symmetric Klein pair.}

 \subsubsection{Moving frames}

A point $g\in G$ of the principal $H$-bundle $\pi:G\twoheadrightarrow G/H$ is called a \textbf{frame on the homogeneous space} $G/H$,
while a section $s:G/H\hookrightarrow G$  is called a \textbf{moving frame on the homogeneous space} $G/H$.

\vspace{3mm}
\noindent{\small\textbf{Example (Tangent frame bundle over the affine space)\,:} The reductive Klein geometry $GL(n)\subset IGL(n)$ defines a principal $GL(n)$-bundle over ${\mathbb R}^n$, where ${\mathbb R}^n$ is seen as an affine space.
	Given a choice of origin, the general linear group $GL(n)$ is the isotropy group of the origin in ${\mathbb R}^n$, where ${\mathbb R}^n$ is now seen as a vector space.
	The principal $GL(n)$-bundle $IGL(n)$ over ${\mathbb R}^n$, \textit{i.e.} \be\label{affgeombundle}
	GL(n)\lefttorightarrow IGL(n) \twoheadrightarrow {\mathbb R}^n\,,
	\ee
	can be thought of as the tangent frame bundle $F\,{\mathbb R}^n$ over the affine space ${\mathbb R}^n$. The points $p$ of $F\,{\mathbb R}^n$ are tangent frames at points $m$ of ${\mathbb R}^n$. The affine group $IGL(n)$ acts regularly on $F\,{\mathbb R}^n$, therefore $IGL(n)\cong F\,{\mathbb R}^n$ as manifolds, and points of $IGL(n)$ can indeed be identified with frames on the affine space ${\mathbb R}^n$. In particular, the general linear group $GL(n)$ acts freely on the space $F_m{\mathbb R}^n$ of tangent frames at a point $m$.}
\vspace{3mm}

A moving frame defines a splitting of the short exact sequence \eqref{shortexactKlein} of connected manifolds associated to the Klein geometry,
\be
e\leftarrow H\stackrel{r}{\twoheadleftarrow}G\stackrel{s}{\hookleftarrow}G/H\leftarrow e\,.\label{splitshortexactKlein}
\ee
where $r:G\twoheadrightarrow H$ is an $H$-equivariant retraction of the embedding $i:H\hookrightarrow G$.

\vspace{5mm}
\begin{framed}
	\begin{center}
		\textbf{The many faces of moving frames on homogeneous spaces}
	\end{center}
	
	\noindent
	For a Klein geometry $H\subset G$, the following notions are equivalent:
	\begin{enumerate}
		\item a gauge of the principal bundle $H\lefttorightarrow G\twoheadrightarrow G/H$,
		
		\item a moving frame on the homogeneous space $G/H$,
		
		\item a section $s:G/H\hookrightarrow G$ of the fibration $\pi:G\twoheadrightarrow G/H$,
		
		\item an $H$-equivariant retraction $r:G\twoheadrightarrow H$ of the embedding $i:H\hookrightarrow G$,
		
		\item a splitting \eqref{splitshortexactKlein} of the short exact sequence \eqref{shortexactKlein} of manifolds.
	\end{enumerate}
	\vspace{3mm}
\end{framed}
\vspace{3mm}

The pullback of the Maurer-Cartan one-form $s^*\omega_{MC}^\mathfrak{g}$ along a moving frame $s:G/H\hookrightarrow G$ is a flat Cartan gauge connection $A^\mathfrak{g}:=s^*\omega_{MC}^\mathfrak{g}$ on $G/H$ called \textbf{the Darboux (or left logarithmic) derivative of the primitive} $s\,$ \cite[Definition 3.5.1]{Sharpe}. The equation $dA^\mathfrak{g}+\frac12[A^\mathfrak{g},A^\mathfrak{g}]=0$ that expresses the vanishing of the curvature two-form ($s^*\Omega_{MC}^\mathfrak{g}=0$) of the Darboux derivative is called the \textbf{structural equation of the homogeneous space} $G/H$.
For any $\mathfrak{g}$-valued one-form $A^\mathfrak{g}$ on a manifold $M$ satisfying to the structural equation, there exists (at least locally) a primitive $s:M\hookrightarrow G$ of which the one-form is the Darboux derivative (\textit{i.e.} $A^\mathfrak{g}:=s^*\omega_{MC}^\mathfrak{g}$) \cite[Section 3.6]{Sharpe}. In this sense, the structural equation provides a differential criterion for a manifold $M$ to be locally diffeomorphic to the homogeneous space of a Klein geometry $H\subset G$. This is part of Cartan's method of moving frames. Note that the primitive is determined uniquely up to the global left translation by an element of the principal group, \textit{i.e.} $s_1^*\omega_{MC}^\mathfrak{g}=s_2^*\omega_{MC}^\mathfrak{g}$ if and only if $s_1=L_g\circ s_2$ for some $g\in G$  \cite[Theorem 3.5.2]{Sharpe}.

\vspace{2mm}
\noindent\textbf{Remark:} When $G$ is a matrix Lie group, the pullback of the defining representation by the section may be written as the map 
\be
s^*\texttt{g}\,:\,G/H\hookrightarrow GL(n)\,:\,x\mapsto \textsl{g}(x)\,.
\ee
In such case, the Darboux derivative composed with the linear representation would be the following matrix-valued one-form on $G/H$: $s^*(\texttt{g}^{-1}d\texttt{g})=\textsl{g}^{-1}(x)\,\partial_\mu\textsl{g}(x)\,dx^\mu$.

\vspace{3mm}
\noindent\textbf{Example (Affine Klein geometry)\,:} The reductive Klein geometry $GL(n)\subset IGL(n)$
defines a principal bundle 
\eqref{affgeombundle}
with fibration
\be
\pi\,:\,IGL(n)\twoheadrightarrow {\mathbb R}^n\,:\,\left(%
\begin{array}{cc}
	1 & 0 \\
	\vec{x} & \texttt{h} \\
\end{array}%
\right)\mapsto \,\vec{x}
\ee
where the decomposition \eqref{IGLn} has been used. 
The general linear group $GL(n)$ is the isotropy group of the origin in ${\mathbb R}^n$.

The fundamental action of the Lie algebra $\mathfrak{igl}(n)$ on the Lie group $IGL(n)$ is determined by the fundamental vector fields of $IGL(n)$, written in \eqref{fundvectfieldsIGLn}.
The fundamental vector fields $\texttt{T}^\#{}_a{}^b=\,e^\mu_a\frac{\partial}{\partial e_b^\mu}$ of the principal $GL(n)$-bundle \eqref{affgeombundle} span the Lie algebra $\mathfrak{gl}(n)^\#$ of vertical left-invariant vector fields on $IGL(n)$. Thus the algebra ${C}^\infty({\mathbb R}^n)\cong{C}^\infty\big(\,IGL(n)\,\big)^{GL(n)}$ is realised as the space of functions on $IGL(n)$ that are independent of the vertical coordinates $e^\mu_a\,$. 
The horizontal constant vector fields $\texttt{P}^\#_a=e_a^\mu\frac{\partial}{\partial x^\mu}$ of the reductive flat Cartan geometry on the principal bundle \eqref{affgeombundle} are the left-invariant vector fields on $IGL(n)$ spanning the Abelian Lie subalgebra of translations.

The principal action of $\mathfrak{igl}(n)$ on the affine space ${\mathbb R}^n$ is determined by the vector fields on the base ${\mathbb R}^n$:
$\pi_*\texttt{P}^\sharp_\mu=\frac{\partial}{\partial x^\mu}$ and $\pi_*(\texttt{T}_\mu{}^\nu)^\sharp=\,x^\nu\frac{\partial}{\partial x^\mu}$, where we used \eqref{sharpgenerators}.
The right-invariant vector fields on $IGL(n)$ are of the form:
\be
{X}\in\mathfrak{X}\big(IGL(n)\big)^{GL(n)}\cong \Gamma\left(\frac{T\big(\,IGL(n)\,\big)}{GL(n)}\right)\cong \Gamma\left(IGL(n)\times_{Ad_{\,GL(n)}} \mathfrak{igl}(n)\right)
\ee
\ba
\Longleftrightarrow\quad{X}&=&Z^\mu(x)\,\texttt{P}^\sharp_\mu+Y^\mu{}_\nu(x)\,(\texttt{T}{}_\mu{}^\nu)^\sharp\\
&=&X^\mu(x)\,\frac{\partial}{\partial x^\mu}+Y^\mu{}_a(x)\,\frac{\partial}{\partial e_a^\mu}\label{coefslin}\\
&=&X^a(x)\,\texttt{P}^\#_a+Y^a{}_b(x)\,(\texttt{T}{}_a{}^b)^\#\label{coefsnonline}
\ea
where $X^\mu(x):=Z^\mu(x)+x^\nu \,Y^\mu{}_\nu(x)$ due to \eqref{sharpgenerators}. Note that the coefficients $Y^\mu{}_a(x):=Y^\mu{}_\nu(x)\,e^\nu_a$ in \eqref{coefslin} depend linearly on the vertical coordinates. However, the coefficients in \eqref{coefsnonline} depend non-linearly on the vertical coordinates: $X^a(x):=\theta_\mu^a\,X^\mu(x)$
and $Y^a{}_b(x):=\theta_\mu^a\,e^\nu_b\,Y^\mu{}_\nu(x)$
so that they transform in the adjoint representation of $GL(n)$ on $\mathfrak{igl}(n)$ as they should.
The Maurer-Cartan one-form $\omega^{\mathfrak{igl}(n)}=\theta^a_\mu(dx^\mu\otimes \texttt{P}_a+de_b^\mu\otimes \texttt{T}_a{}^b)$
indeed maps ${X}$ given by the second line \eqref{coefslin} to the vector in the third line \eqref{coefsnonline} with $\#$ removed:
\be
\omega^{\mathfrak{igl}(n)}\,:\,X^\mu(x)\,\frac{\partial}{\partial x^\mu}+Y^\mu{}_a(x)\,\frac{\partial}{\partial e_a^\mu}\mapsto
X^a(x)\,\texttt{P}_a+Y^a{}_b(x)\,\texttt{T}{}_a{}^b\,.
\ee
The corresponding horizontal lift maps a vector field $X^\mu(x)\,\frac{\partial}{\partial x^\mu}$ on the base ${\mathbb R}^n$ to a vector field on $IGL(n)$ with the same component expression (though its meaning is of course distinct) because 
\be
X^a(x)\,\texttt{P}^\#_a=X^\mu(x)\,\theta_\mu^a\,e_a^\nu\frac{\partial}{\partial x^\nu}=X^\mu(x)\,\frac{\partial}{\partial x^\mu}\,.
\ee

The principal $GL(n)$-bundle $IGL(n)$ can be thought of as the tangent frame bundle of the affine space ${\mathbb R}^n$, \textit{i.e.}
$IGL(n)\cong F\,{\mathbb R}^n$. A moving frame of this principal $GL(n)$-bundle tantamounts to a map ${\mathbb R}^n\to GL(n):\vec{x}\mapsto \texttt{h}(\vec{x})$ whose components $e_a^\mu(x)$ are seen to provide a field of bases $\texttt{e}_a:=e_a^\mu(x)\frac{\partial}{\partial x^\mu}$ for the tangent spaces to ${\mathbb R}^n$. 
In components, the corresponding Darboux derivative restricted to the codomains of the translation ideal and to the general linear subalgebra are, respectively, 
\be
A^{{\mathbb R}^n}=\theta^a_\mu(x)\,dx^\mu\otimes \texttt{P}_a\,,\quad A^{\mathfrak{gl}(n)}=\theta^a_\nu(x)\,\partial_\mu e_b^\nu(x)\,dx^\mu\otimes \texttt{T}_a{}^b\,.
\ee
The holonomic basis is the particular moving frame $\vec{x}\mapsto e$ whose image is the neutral element of $GL(n)$, \textit{i.e.} $e_\mu^a(x)=\delta_\mu^a$, in which case the only non-vanishing part of the Darboux derivative
$A^{\mathfrak{igl}(n)}=\,dx^\mu\otimes \texttt{P}_\mu$ is the canonical solder form on $T\,{\mathbb R}^n$.

The corestriction $\omega^{\mathfrak{gl}(n)}=\theta^a_\mu\,de_b^\mu\otimes \texttt{T}_a{}^b$
of the Maurer-Cartan one-form to the isotropy subalgebra defines a flat principal connection on the principal bundle 
\eqref{affgeombundle} whose horizontal sections are constant moving frames, of which the above holonomic basis is the simplest case.

\pagebreak

 \section{Koszul connections on vector bundles}\label{vectb}

In this section, one will generically have a fibration $\pi:\mathbb{V}\twoheadrightarrow M$ defining a vector bundle over $M$ where each fibre is isomorphic to the same given vector space $V$. 
The dimension of the typical fibre $V$ is called the \textbf{rank} of the vector bundle $\mathbb{V}$.\footnote{See \textit{e.g.} \cite[Chapter 10]{Lee2} for general definitions and facts about vector bundles.}

 \subsection{Frame bundle}

At the point $m\in M$, the identification of the fibre with the vector space $V$ is performed by a \textbf{linear frame} of $\mathbb{V}_m$, which is an isomorphism 
\be
\texttt{e}|_m\,:\,V\stackrel{\sim}{\to} \mathbb{V}_m\,.
\ee
In the present case, the vector space $V$ should be thought as given and fixed. It may even be thought as endowed with an ordered basis. If $V$ is of finite dimension $n$ (the rank of $\mathbb{V}$), then this is equivalent to replace it with the vector space ${\mathbb R}^n$ (since this one has a canonical basis).
In this view, a frame is equivalent to an ordered basis $\{\texttt{e}_a|_m\}$ of the fibre $\mathbb{V}_m$. The inverse $(\texttt{e}|_m)^{-1}:\mathbb{V}_m\stackrel{\sim}{\to} V$
of a linear frame is called a \textbf{linear coframe}.

The set $F_m\mathbb{V}$ of linear frames at a point is a principal homogeneous space of the general linear group $GL(V)$.
In fact, a principal bundle for the general linear group is necessarily a \textbf{linear frame bundle} $F\mathbb{V}=\bigcup_m F_m\mathbb{V}$, which can be seen as the principal bundle 
\be
GL(V)\lefttorightarrow F\mathbb{V}\twoheadrightarrow M\,.
\ee
More precisely, the linear frame bundle functor $F:\mathbb{V}\mapsto F\mathbb{V}$ mapping any vector bundle to its linear frame bundle is an equivalence between the category of vector bundles of rank $n$ and the category of principal bundles with the general linear group $GL(n)$ as structure group \cite[Theorem 14.2.7]{Dieck}. 
A section of the linear\footnote{Notice that the adjective ``linear'' is often dropped when the context (vector bundles) makes it clear.}  frame bundle $F\mathbb{V}$ is called a \textbf{moving frame of the vector bundle} $\mathbb{V}$. The corresponding space of sections will be denoted ${\mathcal F}\mathbb{V}(M)$ and called the \textbf{frame sheaf}. A global moving frame exists if and only if the linear frame bundle is trivial (since it is a principal bundle), which is true if and only if the vector bundle itself is trivial (since a global moving frame provides a trivialisation), see \textit{e.g.} \cite[Corollary 10.20]{Lee2}.

Let $\{\texttt{e}_a\}$ denote a moving frame of $\mathbb{V}$.
Then the components $y^a$ of vectors ($y=y^a\texttt{e}_a$) provide natural coordinates along each fibre.
Let $x^\mu$ denote coordinates on the base $M$. Thence $(x^\mu,y^a)$ provide coordinates on the vector bundle $\mathbb{V}$.  Obviously, the range of the base index $\mu$ equals the dimension of the base $M$ while the range of the fibre index $a$ equals the rank of the vector bundle. 
A vector bundle automorphism is a diffeomorphism of the total space $\mathbb{V}$ that maps, linearly, fibres to fibres. In general, it induces a diffeomorphism of the base $M$. In coordinates, it corresponds to the transformation rule 
\be
(x^\mu,y^a)\,\to\,\Big(x^{\prime\mu}(x)\,,\,y^{\prime a}=\Lambda^a{}_b(x)\,y^b\,\Big)\,.
\ee

 \subsection{French connection (part III)\,: Jean-Louis Koszul}

Given the importance of vector bundles in the theory of fundamental interactions, we will devote some time to present the various equivalent definitions of linear connections. In fact, what physicist refer to as ``covariant derivatives'' very often stand for what mathematicians would call ``linear connections'' on associated vector bundles.\footnote{The precise relation between the loose term ``covariant'' and the mathematical term ``associated'' will be discussed in Section \ref{associatedbundles}).}

 \subsubsection{Linear connection \textit{\`a la} Charles Ehresmann}

A \textbf{linear connection} on a vector bundle $\mathbb{V}$ is an Ehresmann connection on the fibre bundle $\mathbb{V}$ 
whose horizontal distribution is linear with respect to each fibre ($\cong V$) \cite[Paragraph 17.16]{Dieudonne1970} . In local coordinates, this simply means that the components in \eqref{lincon} depend \textit{linearly} on the fibre coordinates:
$\omega_\mu^a(x,y)=\omega_\mu{}^a{}_b(x)\,y^b$. An Ehresmann connection on a vector bundle $\mathbb{V}$ which is \textit{not} a linear connection is often called a \textbf{nonlinear connection on the vector bundle} $\mathbb{V}$. 

 \subsubsection{Linear connection \textit{\`a la} Jean-Louis Koszul}

The celebrated definition of a linear connection on a vector bundle as a Koszul connection \cite{Koszul} is by now the standard one (it can be found in most textbooks) because it is much simpler to implement than the above definiton as an Ehresmann connection. A \textbf{Koszul connection} is a ${C}^\infty(M)$-linear map 
\be
\nabla_\bullet\,:\,\mathfrak{X}(M)\to \mathfrak{gl}\big(\Gamma({\mathbb{V}})\big)\,:\,{X}\mapsto \nabla_{{X}}
\ee
where 
\be
\nabla_{{X}}\,:\,\Gamma(\mathbb{V})\to \Gamma(\mathbb{V})\,:\,\sigma\mapsto\nabla_{{X}}\sigma
\ee
is a \textbf{covariant derivative} on the left ${C}^\infty(M)$-module $\Gamma(\mathbb{V})$ \textbf{along the vector field} $X\in\mathfrak{X}(M)$, in the sense that it obeys to the following Leibniz rule:
\be
\nabla_{{X}} (f\cdot\sigma)\,=\,{{X}}[f]\cdot \sigma\,+\,f\cdot\nabla_{{X}}\,\sigma\,,\quad\forall f\in{C}^\infty(M)\,,\quad\forall \sigma\in\Gamma(\mathbb{V})\,.
\ee
These covariant derivatives along vector fields form a Lie algebra via the commutator (see \textit{e.g.} \cite[Subsection 3.6.2]{Bekaert:2023jvl}), which is called the \textbf{Atiyah algebra of the vector bundle $\mathbb{V}$} and which will be denoted $\mathfrak{CD}^1\,(\mathbb{V})$. They are sections of a vector bundle called the \textbf{Atiyah bundle of the vector bundle $\mathbb{V}$} which will be denoted $CD^1\,\mathbb{V}$. In other words, one has $\Gamma\big(CD^1\mathbb{V}\big)=\mathfrak{CD}^1(\mathbb{V})$.

Equivalently, a Koszul connection on a vector bundle $\mathbb{V}$ can be defined as a \textbf{covariant differential}, that is to say a linear map 
\be\label{Koszulconnection}
\nabla\,:\,\Gamma(\mathbb{V})\to \Gamma(\,T^*M\underset{M}{\otimes} \mathbb{V}\,)\,:\,\sigma\mapsto\nabla\sigma
\ee
obeying to the following Leibniz rule:
\be
\nabla (f\cdot\sigma)\,=\,df\otimes \sigma\,+\,f\cdot\nabla\sigma\,,\quad\forall f\in{C}^\infty(M)\,,\quad\forall \sigma\in\Gamma(\mathbb{V})\,,
\ee
where $\nabla\sigma\in \Gamma(\,T^*M\underset{M}{\otimes} \mathbb{V}\,)$ is called the \textbf{covariant differential of the section} $\sigma$ of the vector bundle $\mathbb{V}$. In the simplest case where the vector bundle is the trivial line bundle, the covariant differential is simply the differential of functions (\textit{i.e.} $\nabla=d$ for $\mathbb{V}=M\times\mathbb R$).
The covariant differential $\nabla\sigma\in \Gamma(\,T^*M\underset{M}{\otimes} \mathbb{V}\,)$ of the section $\sigma\in \Gamma(\mathbb{V})$ can equivalently be seen as the following left ${C}^\infty(M)$-module morphism 
\be
\nabla_\bullet\sigma\,:\,\Gamma(TM)\to \Gamma(\mathbb{V})\,:\,{X}\mapsto\nabla_{{X}}\sigma\,.
\ee
The action of the covariant derivative  $\nabla_\mu:=\nabla_{\partial_\mu}$ along a coordinate basis vector field $\partial_\mu$ on a section $\sigma=\sigma^a\texttt{e}_a$ reads in components as 
\be
(\nabla_\mu\sigma)^a(x)=\partial_\mu\sigma^a(x)+\omega_\mu{}^a{}_b(x)\,\sigma^b(x)\,,
\ee
as can be checked by applying the Leibniz rule and by defining $\nabla_\mu\texttt{e}_b= \texttt{e}_a\,\omega_\mu{}^a{}_b(x)$.

Note that the same suggestive notation $\nabla_{\mu}$ has been used intentionally for the horizontal lift of a coordinate basis vector field $\partial_\mu$ on $M$ (with respect to a linear connection on $\mathbb{V}$) and for the covariant derivative along the same vector field, in order to emphasise that they are in one-to-one correspondence and can therefore be essentially identified.
The precise relation between the Ehresmann and Koszul definitions of a connection on a vector bundle 
relies on the isomorphism between the fibres of the bundle $\mathbb{V}$ and of its vertical distribution $V\mathbb{V}$: 
\be\label{verticalidentific}
\mathbb{V}_m\cong\text{V}\cong V_p\mathbb{V}\,,
\ee
where $m=\pi(p)$.
This isomorphism justifies the loose identification between the basis elements $\frac{\partial}{\partial y^a}$ of $V_p\mathbb{V}$ with the basis elements $\texttt{e}_a$ of $\mathbb{V}_m$ which we will perform in the sequel.

\vspace{5mm}
\begin{framed}
	\begin{center}
		\textbf{The many faces of linear connections}
	\end{center}
	
	\noindent
	Given a vector bundle $\mathbb{V}$ over a base manifold $M$, the following notions are equivalent:
	\begin{enumerate}
		\item a linear connection on the vector bundle $\mathbb{V}$,
		
		\item an Ehresmann connection on $\mathbb{V}$ which is linear with respect to each fibre,
		
		\item a Koszul connection on the vector bundle $\mathbb{V}$,
					
		\item a principal connection on the linear frame bundle $F\mathbb{V}$.
	\end{enumerate}
	\vspace{3mm}
\end{framed}

\pagebreak

 \section{Induced connections on associated bundles}\label{associatedbundles}

The importance of principal and associated bundles stems from the reconstruction theorem asserting that any fibre bundle $E$ over $M$ with typical fibre $F$ and structure group $H$ is the associated bundle $E=P\times_H F$
of a principal $H$-bundle of total space $P=E\times_H F$, through the left action of the group $H$ on the typical fibre $F$
(see \textit{e.g.} \cite[Subsection 24.4.1]{Dubrovin}).
In practice, we will mostly focus on the subclass of vector bundles associated to principal bundles.

Two particular applications of the notion of associated bundles will be of interest:

\vspace{2mm}
\noindent$\bullet$ Firstly, reduction of principal bundles: they are intimately related to Klein geometries through the notion of associated bundles.

\vspace{2mm}
\noindent$\bullet$ Secondly, faithful linear representations: they provide a one-to-one correspondence between principal and vector bundles, where the associated principal bundle to a vector bundle is a reduction of the linear frame bundle. 

 \subsection{Associated vector bundles}

Consider a principal $H$-bundle with total space $P$ over the base manifold $M=P/H$,
\be
H\lefttorightarrow P \stackrel{\pi}{\twoheadrightarrow} M=P/H\,.
\ee
Consider a linear representation 
\be
r\,:\,H\to GL({V})\,:\,h\mapsto r_h
\ee
of the Lie group $H$ on the vector space $V$. 

The \textbf{vector bundle} $P\times_{r_H}V$ \textbf{associated to the principal bundle} $P$ \textbf{through the linear representation} $r$ \textbf{of the structure group} $H$ \textbf{on the vector space} $V$ (or $r_H$-associated bundle, for short) is defined as the vector bundle over $M$ obtained as the quotient of the Cartesian product $P\times V$ by the equivalence relation: 
\be\label{equivrelassoc}
(p,v)\sim \left(\,p\,h,r(h^{-1})\,v\,\right)\quad\Longleftrightarrow\quad (p\,h,\,w)\sim \big(p,r(h)\,w\big)\,,
\ee
where $p\in P$ is an $H$-frame, $v,w\in V$ are vectors and $h\in H$ is a group element.
The points of the associated vector bundle are equivalence classes $[p\,,v]\in P\times_{r_H}V$. Therefore, the fibre above a point $m\in M$ of the base is the collection of equivalence classes $[p\,,v]$ with $\pi(p)=m$\,. Although the choice of representative is not canonical, each fibre is isomorphic to the original vector space, \textit{i.e.} $(P\times_{r_H}V)_m\cong V\,$. 

When the linear representation of $H$ on $V$ is faithful (\textit{i.e.} the morphism $r:H\hookrightarrow GL({V})$ of Lie groups is injective), the associated vector bundle  $P\times_{r_H}V$ has slightly more structure than a generic vector bundle $\mathbb{V}$ over $M=P/H$ with typical fibre $V$ because the structure group of $P\times_{r_H}V$ is reduced to the subgroup $H\subseteq GL(V)$.
Conversely, given a vector bundle $\mathbb{V}$ with structure group $H$
one can construct\footnote{See \textit{e.g.} \cite[Paragraphs 8.1 and 9.1]{Steenrod} for more general constructions.} a principal $H$-bundle from a vector bundle $\mathbb{V}$ with structure group $H$ and fibres $\mathbb{V}_m\cong\text{V}$, in which case the former bundle is called the \textbf{principal bundle} $P$ \textbf{associated to the vector bundle} $\mathbb{V}$ \textbf{with structure group} $H$ \textbf{through the faithful linear representation} $r$ \textbf{of the Lie group} $H$ \textbf{on the fibre} $V$. In such case, the vector bundle $\mathbb{V}$ can be seen retrospectively as the associated vector bundle $P\times_{r_H}V$. 
This reconstruction is detailed now in the degenerate case of the fundamental representation of the general linear group.

\vspace{3mm}
\noindent{\small\textbf{Example (Linear frame bundle of a vector bundle)\,:}
	A principal $GL(V)$-bundle $P$ whose structure group is the general linear group $GL(V)$ is necessarily the linear frame bundle $P=F\mathbb{V}$ of the associated vector bundle $\mathbb{V}=P\times_{GL(V)}V$ through the fundamental representation of the general linear group $GL(V)$ on the vector space $V$. In fact, there is a bijective correspondence (up to isomorphisms) between vector bundles over $M$ with typical fibre $V$ and principal $GL(V)$-bundles over $M$, where the bijection is the functor $F:\mathbb{V}\mapsto F\mathbb{V}$ \cite[Theorem 14.2.7]{Dieck}.\footnote{In this sense, the theory of principal bundles encompass the theory of vector bundles. This explains why geometers love so much principal bundles, although vector bundles are somewhat more easy to handle.}
	Given an ordered basis $\{\texttt{e}_a|_m\}$ of the fibre $\mathbb{V}_m$ (\textit{i.e.} a linear frame $\texttt{e}|_m\in F_m\mathbb{V}$), a vector $\texttt{v}|_m\in\mathbb{V}_m$ of this fibre decomposes as $\texttt{v}|_m=v^a\,\texttt{e}_a|_m$ and its components represent the vector $v\in V$, where $V$ is assumed to be endowed with an ordered basis $\{\texttt{e}_a\}$.
	The general linear group is represented by invertible matrices $\textsl{h}\in GL(V)$ acting on the basis as $\texttt{e}'_a=\texttt{e}_b\,\textsl{h}^b{}_a$ and on the components as $v'{}^a=(\textsl{h}^{-1})^a{}_b\,v^b$. Retrospectively, the collection of ordered bases $\{\texttt{e}_a|_m\}$ above each point $m\in M$ of the base defines a moving frame $\texttt{e}:M\hookrightarrow F\mathbb{V}$ of the linear frame bundle $F\mathbb{V}$ which allows to reconstruct exactly the vector bundle associated to $F\mathbb{V}$ through the fundamental representation of $V$ with the original vector bundle $\mathbb{V}$.
	To any principal connection on the linear frame bundle $F\mathbb{V}$ is associated an induced linear connection on the vector bundle $\mathbb{V}$. Conversely, any linear connection on a vector bundle $\mathbb{V}$ is induced from a unique principal connection on the linear frame bundle $F\mathbb{V}$ \cite[Paragraph 19.11]{Michor}.
}

\subsection{Reductions}

Let $H\subseteq G$ be a Klein geometry. A submanifold $Q\subseteq P$ is called a \textbf{reduction} (or \textbf{partial gauge}) of the principal $G$-bundle $P$ to the principal $H$-bundle $Q$ if the free action of $G$ on $P$ restricts to a free action of $H$ on $Q$ and defines a principal $H$-bundle $Q$ over the same base, \textit{i.e.} $\frac{P}G\cong\frac{Q}H$ \cite[Definition 4.2.12]{Sharpe}. This may be depicted as follows:
\be\label{reduxprincbundl}
\begin{array}
	[c]{ccccc}%
	G &  \lefttorightarrow & P & &\\
	& & & \searrow & \\ 
	\uparrow & & \uparrow & & M\cong \frac{P}{G}\cong\frac{Q}{H} \\ 
	& & & \nearrow & \\ 
	H & \lefttorightarrow & Q & & 
\end{array}
\,.
\ee
The left vertical arrow is a Lie group embedding while the right vertical arrow is an embedding of principal bundle over $M$. The two diagonal arrows are the respective fibrations of the two principal bundles on their common base $M$.

\vspace{3mm}
\noindent{\small\textbf{Example (Reduction of the linear frame bundle)\,:} Consider a matrix group, \textit{i.e.} a subgroup $H\subset GL(V)$ of the general linear group of the vector space $V$, \textit{i.e.} a faithful representation $i:H\hookrightarrow GL({V})$.
Consider the vector bundle $\mathbb{V}:=P\times_{i_H}V$ associated to a principal $H$-bundle $P$ through the faithful representation defined by the embedding of Lie groups. 
Conversely, the principal $H$-bundle $P$ can be seen as a reduction of the principal $GL(V)$-bundle $F\mathbb{V}$ of linear frames of the associated bundle. The embedding is 
	\be
	\textsl{i}\,:\,P\hookrightarrow F\mathbb{V}\,:\, p\mapsto \texttt{e}(p)
	\ee
	where $\texttt{e}(p)$ is a linear frame at $m=\pi(p)$ defined by
	\be
	\label{noncaninvsisjhomsh}
	\texttt{e}(p)\,:\,V\to \mathbb{V}_m\,:\,v\mapsto [p\,,v]\,.
	\ee
}

 \subsubsection{Covariance \textit{vs} equivariance}

The notions of covariance and equivariance turn out to be so intimately related that they are, in some sense, equivalent. Nevertheless, this relation is subtle, thus we consider that it is important to start by distinguishing them neatly at the beginning. This may seem pedantic but it can be useful for conceptual clarity.

\paragraph{Covariance:}

Let $P\times_{r_H}V$ be the vector bundle associated to the principal $H$-bundle $P$ through the representation $r$ of the structure group $H$ on the vector space $V$. The sections $\sigma:M\hookrightarrow P\times_{r_H}V$ of the $r_H$-associated vector bundle will be called $r_H$-\textbf{covariant fields on} $P/H$. The vector space $\Gamma(P\times_{r_H}\text{V})$ of such sections will be called the \textbf{associated sheaf}.

The map 
\be\label{reconstructionmap}
\texttt{r}\,:\,P\underset{M}{\times}\big(P\times_{r_H}V\big)\,\to\, V\,:\,\big(\,p\,,[p\,,v]\,\big)\mapsto v
\ee
will be called the \textbf{reconstruction map}. 
Given an $H$-frame $p\in P$ at the point $m=\pi(p)\in P/H$ of the base, the map 
\be
\texttt{r}|_p\,:\,(P\times_{r_H}V)_m\to V\,:\,[p\,,v]\mapsto v
\ee
reconstructs the corresponding vector $v\in V$. The reconstruction map is $r_H$-equivariant in the sense that $\texttt{r}|_{ph}=r(h^{-1})\,\texttt{r}|_{p}$\,, \textit{i.e.}
\be
\texttt{r}|_{ph}:[p\,,v]\mapsto r(h^{-1}) v\,,
\ee
since $[p,v]= \big[\,p\,h,r(h^{-1})\,v\,\big]$
by the definition \eqref{equivrelassoc}.

In the case of the linear frame bundle (\textit{i.e.} the principal $GL(V)$-bundle $F\mathbb{V}$ of the vector bundle $\mathbb{V}\cong F\mathbb{V}\times_{i_H}V$), it is clear that given a linear frame $\texttt{e}|_m:V\stackrel{\sim}{\to} \mathbb{V}_m$ one can reconstruct the vector of $V$ corresponding to any element of the fibre $\mathbb{V}_m$ via the linear coframe $(\texttt{e}|_m)^{-1}:\mathbb{V}_m\stackrel{\sim}{\to} V$.

 \paragraph{Equivariance:}

A map $f:P\to V$ such that 
\be
f(ph)\,=\,r(h^{-1})\,f(p)\,, \qquad\forall p\in P,\quad\forall h\in H\,,
\ee 
is called an $r_H$-\textbf{equivariant map from the principal bundle} $P$ \textbf{to the vector space} $V$. 
The subspace of ${C}^\infty(P)\otimes V$ spanned by the maps $f$ from $P$ to $V$ which are $r_H$-{equivariant}, will be called the \textbf{equivariant sheaf on} $P$ and will be denoted as ${C}^\infty(P)\otimes_{r_H} V$. 
An equivalent way to define the equivariant sheaf is to consider the representation of the Lie group $H$ on the vector space ${C}^\infty(P)\otimes V$ defined as
\be
\texttt{r}\,:\,H\to GL\big({C}^\infty(P)\otimes V\big)\,:\,h\to \texttt{r}_h
\ee
where
\be
(\texttt{r}_h f)(p)\,:=\,r(h)\,f(ph)\,, \qquad\forall p\in P,\quad\forall h\in H\,.
\ee
In this way, an $r_H$-equivariant map from the principal bundle $P$ to the vector space $V$ can be seen as an $\texttt{r}_H$-invariant element of the representation space ${C}^\infty(P)\otimes V$. Accordingly, the equivariant sheaf ${C}^\infty(P)\otimes_{r_H} V$ can be seen as the subspace of $\texttt{r}_H$-invariant elements in ${C}^\infty(P)\otimes V$.

Infinitesimally, the equivariance condition of $f\in {C}^\infty(P)\otimes_{r_H} V$ reads as $({\mathcal L}_{y^\#}+r_y)f=0$ for all $y\in\mathfrak{h}$, where we denoted the Lie algebra representation
$r:\mathfrak{h}\to\mathfrak{gl}(V)$ with the same symbol as the Lie group representation (although, strictly speaking, it should read $r_{*e}$ for the pushforward at the identity). In other words,
the equivariant sheaf ${C}^\infty(P)\otimes_{r_H} V$ is the subspace of ${C}^\infty(P)\otimes V$ which is annihilated by the $\mathfrak{h}$-action $({\mathcal L}\circ\#)\otimes id_V+id_{{C}^\infty(P)}\otimes r$.
More explicitly, 
if the Lie algebra representation is 
\be\label{Liealg}
r\,:\,\mathfrak{h}\to \mathfrak{gl}(V)\,:\,\texttt{T}_i\mapsto \textsc{T}_i=(\textsc{T}_i)^a{}_b \,\texttt{e}_a\otimes \texttt{e}^{*b}\,,
\ee
then the equivariance condition reads in components
\be\label{equivcondition}
{\mathcal L}_{\texttt{T}_i^\#}f^a=-(\textsc{T}_i)^a{}_bf^b\,.
\ee

The space of constant $r_H$-equivariant maps $f:P\to V$ is an $H$-invariant subspace of ${C}^\infty(P)\otimes_{r_H} V$ isomorphic to the submodule $V^H\subseteq V$ of $H$-invariant elements. Indeed, let $v:=f(p)\in V$ be the value of the constant map $f$, then the $r_H$-equivariance implies $v=r(h^{-1})v$ for all $h\in H$, hence $v\in V^H$. 

 \subsubsection{Covariance $\Leftrightarrow$ Equivariance}

A crucial observation is the existence of a one-to-one correspondence between $r_H$-covariant fields $\sigma$ on $P/H$ and $r_H$-equivariant maps $f$ from $P$ to $V$. More precisely, there is a canonical isomorphism of ${C}^\infty(P/H)$-modules between the associated sheaf and the equivariant sheaf
\be
\label{isjhomsh}
i\,:\,\Gamma(P\times_{r_H}\text{V})\stackrel{\sim}{\to}{C}^\infty(P)\otimes_{r_H} V\,:\,\sigma\mapsto f
\ee
where $\sigma:P/H\hookrightarrow P\times_{r_H}V$ is an $r_H$-covariant field on $P/H$ (\textit{i.e.} a section of the $r_H$-associated vector bundle) whose image is an $r_H$-equivariant map $f:P\to V$ from the principal bundle $P$ to the vector space $V$ defined as follows:
\be
f\,:=\,\texttt{r}\circ\big[id_P\times (\sigma\circ \pi)\big]
\ee
where $\texttt{r}$ is the reconstruction map \eqref{reconstructionmap} and $\pi:P\twoheadrightarrow P/H$ is the fibration of the principal $H$-bundle. Notice that the map
\be
id_P\times (\sigma\circ \pi)\,:\,P\to P\underset{M}{\times}(P\times_{r_H}V)
\,:\,p\mapsto\Big(\,p,\sigma\big(\pi(p)\big)\,\Big)
\ee
sends the total space $P$ of the principal $H$-bundle $P$ to the fibrewise product between $P$ and $P\times_{r_H}V$.
The image $f=i(\sigma)$ is an $r_H$-equivariant map (because the reconstrunction map is). 

The inverse of \eqref{isjhomsh} is the isomorphism
\be
\label{invsisjhomsh}
i^{-1}\,:\,{C}^\infty(P)\otimes_{r_H} V\stackrel{\sim}{\to}\Gamma(P\times_{r_H}\text{V})\,:\,f\mapsto\sigma
\ee
sending an $r_H$-equivariant map $f$ from the principal bundle $P$ to the vector space $V$ onto an $r_H$-covariant field 
on $P/H$ defined by the relation:
\be\label{sigmasectdef}
\sigma\circ\pi\,=\,\pi_{r_H}\circ\big(id_P\times f\big)\,,
\ee
where
\be
\pi_{r_H}\,:\,P\times V\twoheadrightarrow P\times_{r_H}V\,:\,(p\,,v)\mapsto [p\,,v]
\ee
is the projection onto $H$-orbits.
The equation \eqref{sigmasectdef} defines $\sigma$ uniquely because the arbitrariness in the choice of $p$ along each fibre will not affect the endvalue.
In fact, the correspondence \eqref{invsisjhomsh} is much easier to describe  in practice than \eqref{isjhomsh}. Concretely, it reads:
\be
\sigma(m)\,:=\,\big[\,p\,,\,f(p)\,\big]\,,\qquad \text{with}\quad m=\pi(p)\,,
\ee
which is well-defined since $\big[\,ph\,,\,f(ph)\,\big]=\big[\,ph\,,\,R(h^{-1})f(p)\,\big]=\big[\,p\,,\,f(p)\,\big]$ by definition of the equivalence relation \eqref{equivrelassoc}.

 \subsubsection{Covariance \textit{as} Equivariance}

The isomorphism \eqref{isjhomsh} allows to show that any linear representation $r:H\hookrightarrow GL({V}):h\mapsto r_h$ of the Lie group $H$ on the vector space $V$ defines an \textbf{induced representation} 
\begin{equation}\label{assrepr}
	\textsl{r}\,:\,\frac{TP}H\,\,\to\,\, CD^1\,\mathbb{V}
\end{equation}
\textbf{of the Atiyah algebroid of the} $H$-\textbf{principal bundle} $P$ \textbf{on the associated vector bundle} $\mathbb{V}=P\times_{r_H}V$. In other words, there is an induced morphism of Lie algebroids from the Atiyah algebroid of the principal bundle to the Atiyah algebroid of the associated vector bundle. It induces a morphism from the Atiyah sequence \eqref{shortexactvectAt} of the principal $H$-bundle $P$ and the Atiyah sequence  of the $r_H$-associated bundle $\mathbb V$:
\be\label{salsiciettab}
\begin{array}
	[c]{ccccccccc}%
	0&\to& \frac{VP}{H}&\hookrightarrow&\frac{TP}{H}&\twoheadrightarrow&T\frac{P}{H}&\to& 0\\
	&&&&&&&&\\
	&&\textsl{r}|_{\mathfrak{h}^\#}\downarrow&&\textsl{r}\downarrow&& id\downarrow\sim&&\\
	&&&&&&&&\\
	0&\to& \mathfrak{gl}(\mathbb{V})&\hookrightarrow&CD^1\,\mathbb{V}&\twoheadrightarrow&T\frac{P}{H}&\to& 0
\end{array}
\ee
where $\mathfrak{gl}(\mathbb{V})$ is the Lie algebra bundle of fibre $\mathfrak{gl}(\mathbb{V}_m)$ above $m\in M$ and the vertical arrows are Lie algebroid morphisms.
In fact, the representation 
\be
\textsl{r}\,:\,\frac{VP}H\to \mathfrak{gl}(\mathbb{V})\,:\,\texttt{T}_i^\#\mapsto -\textsc{T}_i
\ee
of the adjoint bundle on the $r_H$-associated vector bundle is quite natural, since one expects to have $\textsl{r}\circ\#:P\times_{Ad_H}\mathfrak{h}\to \mathfrak{gl}(\mathbb{V}):\texttt{T}_i\mapsto \textsc{T}_i$ consistently with \eqref{equivcondition}.

The associated representation \eqref{assrepr} reads, in terms of spaces of sections, as
\begin{equation}\label{assreprsheaf}
	\textsl{r}:\mathfrak{X}(P)^H\to \mathfrak{CD}^1(\mathbb{V})\,.
\end{equation}
Concretely, this action of the Atiyah algebra $\mathfrak{X}(P)^H$ on the associated sheaf $\Gamma(\mathbb{V})$ 
relies on the two isomorphisms $\Gamma(\frac{TP}H)\cong\mathfrak{X}(P)^H$
and $\Gamma(\mathbb{V}) \cong {C}^\infty(P)\,\otimes_{r_H} V$. The first isomorphism provides an action of the Lie algebra $\Gamma(\frac{TP}H)$ on the principal $H$-bundle $P$ commuting with the fundamental action, via the Lie derivative of functions on $P$ along invariant vector fields. Together with the second isomorphism, the latter action induces an action of the Atiyah algebra $\Gamma(\frac{TP}H)$ on the associated sheaf $\Gamma(\mathbb{V})$. 
Indeed, let ${X}\in\mathfrak{X}(P)^H$ be an $H$-invariant vector field on $P$. Then ${X}\otimes id_V$ has a well-defined action on $r_H$-equivariant maps $f\in{C}^\infty(P)\otimes_{r_H} V$ in the sense that: if $f$ is $r_H$-equivariant, then $f':=({X}\otimes id_V)[f]$ also is. In other words, if $f=i(\sigma)$ where $i$ stands for the isomorphism \eqref{isjhomsh}, then there exists an equivariant map $\sigma'$ such that $f'=i(\sigma')$. We identify the corresponding linear map 
\be
\textsl{r}({X})=i^{-1}\circ({X}\otimes id_V)\circ i:\sigma\mapsto\sigma'
\ee
as defining the representation \eqref{assreprsheaf}.
More precisely, one should finally check that the endomorphim $\textsl{r}({X})$ is a first-order differential operator on the associated vector bundle whose principal symbol is a tangent vector field on $M$ (times the identity).
To be explicit, an $H$-invariant vector field on $P$ reads ${X}=X^\mu(x)\,\partial_\mu+Y^{i}(x,y)\,\texttt{T}_i^\#$ where $Y=Y^{i}(x,y)\,\texttt{T}_i$ is an adjoint-equivariant map from $P$ to $\mathfrak h$. The image of this vector field ${X}$ by the representation of
the Atiyah algebroid $\frac{TP}H$ on the associated vector bundle $\mathbb{V}$ is simply the first-order differential operator $\textsl{r}({X})=X^\mu(x)\,\partial_\mu+Y^{i}(x,y)\,\textsc{T}_i$ on $\mathbb{V}$.

\vspace{3mm}
\noindent{\small\textbf{Example (Adjoint representation)\,:} Let $H\subseteq G$ be a Klein geometry and $P$ a principal $H$-bundle.
	A particular example of induced representation of the Atiyah algebroid $\frac{TP}H$ arises from the adjoint representation $Ad:H\hookrightarrow GL(\mathfrak{g})$ which induces the canonical representation on the adjoint tractor bundle $P\times_{Ad_H}\mathfrak{g}$:
	\be\label{Adrep}
	\textsl{Ad}\,:\,\frac{TP}H\,\to\, CD^1\,(\,P\times_{Ad_H}\mathfrak{g}\,)\,:\,X^\mu\partial_\mu+Y^{i}(x,y)\,\texttt{T}_i^\#\mapsto X^\mu\partial_\mu+Y^{i}(x,y)\,ad_{{\texttt{T}_i}}\,,
	\ee
	which will be called the \textbf{adjoint representation of the Atiyah algebroid of a principal bundle on its adjoint tractor bundle}.
}

\vspace{3mm}
\noindent{\small\textbf{Example (Linear frame bundle)\,:} Consider a vector bundle $\pi:\mathbb{V}\twoheadrightarrow M$ over $M$ where each fibre is isomorphic to the vector space $V$. The linear frame bundle $F\mathbb{V}$ is a principal $GL(V)$-bundle over $M\cong\frac{F\mathbb{V}}{GL(V)}$. Obviously, the identity map $id_{GL({V})}$ is the defining representation on $V$ of the general linear group $GL(V)$. Its induced representation provides a canonical isomorphism between the Atiyah algebroid $\frac{T\,F\mathbb{V}}{GL(V)}$ of the linear frame bundle and the Atiyah algebroid $CD^1\,\mathbb{V}$ on the vector bundle $\mathbb{V}\cong F\mathbb{V}\times_{GL(V)}V$.\footnote{This isomorphism explains the use of the same terminology ``Atiyah algebroid'' for both principal and vector bundles.} Therefore, the morphism \eqref{salsiciettab} between their Atiyah sequences is an isomorphism of short exact sequences of Lie algebroids
	\be
	\begin{array}
		[c]{ccccccccc}%
		0&\to& \frac{V\,F\mathbb{V}}{GL({V})}&\hookrightarrow&\frac{T\,F\mathbb{V}}{GL({V})}&\twoheadrightarrow&TM&\to& 0\\
		&&&&&&&&\\
		&&\downarrow\sim&&\downarrow\sim&& \downarrow\sim&&\\
		&&&&&&&&\\
		0&\to& \mathfrak{gl}(\mathbb{V})&\hookrightarrow&CD^1\,\mathbb{V}&\twoheadrightarrow&TM&\to& 0
	\end{array}
	\ee
}

 \subsubsection{Covariance \textit{of} derivatives}

The induced representation of the Atiyah algebroid of the principal $H$-bundle $P$ on the associated vector bundle $\mathbb{V}=P\times_{r_H}V$ allows to obtain from any principal connection $\omega^\mathfrak{h}$ on $P$ a linear connection $\nabla^{\mathbb{V}}$ on the $r_H$-associated vector bundle $\mathbb{V}$, which will be called the \textbf{associated connection}. Indeed, it is clear from \eqref{salsiciettab} that a splitting \eqref{shortexactvectAt2} of the Atiyah sequence of the principal bundle on the first line (\textit{i.e.} a principal connection) implies a splitting  of the Atiyah sequence of the associated vector bundle on the second line (\textit{i.e.} a linear connection).
Concretely, the horizontal lift 
\be
\nabla^{\mathbb{V}}_{\bullet}:=\textsl{r}\circ \gamma\,\,:\,\,TM\,\,\hookrightarrow\,\, CD^1\mathbb{V}\,:\,X
\mapsto 
\nabla^{\mathbb{V}}_{X}
\ee
of the $r_H$-associated vector bundle is the composition of the horizontal lift $\gamma:TM\hookrightarrow\frac{TP}H$ with the
induced representation $\textsl{r}:\frac{TP}H\to  CD^1\mathbb{V}$.
More explicitly, the $H$-invariant horizontal lift of the vector field ${X}=X^\mu(x)\,\partial_\mu$ on $P/H$   is an invariant vector field $\gamma({{X}})=X^\mu(x)\nabla^P_\mu$ on $P$, where $\nabla^P_\mu\,:=\,\partial_{\mu}\,-\,\omega_\mu^i(x,y)\,\texttt{T}_i^\#$. Moreover, $\gamma({{X}})\otimes id_V$ has a well-defined action on functions $f\in{C}^\infty(P)\otimes_{r_H} V$. If $f$ is $r_H$-equivariant, then $\gamma({{X}})[f]$ also is. In other words, if $f=i(\sigma)$ where $i$ stands for the isomorphism \eqref{isjhomsh}, then $\gamma({{X}})[f]=i(\sigma')$. We identify the corresponding linear map $\sigma\mapsto\sigma'$ as the covariant derivative along ${X}$.
Thus, the latter is defined by $\gamma({{X}})\,[\,i(\sigma)\,]=i(\nabla^{\mathbb{V}}_{{X}}\sigma)$. To summarise,
\begin{eqnarray}
	\nabla^{\mathbb{V}}_\bullet:=i^{-1}\circ (\gamma
	\otimes id_V)\circ\, i&:&\mathfrak{X}(M)\hookrightarrow \mathfrak{CD}^1(\mathbb{V})\\
	&:&{X}
	\mapsto 
	\nabla^{\mathbb{V}}_{{X}}=i^{-1}\circ \big(\gamma({X})
	\otimes id_V\big)\circ\, i\nonumber
\end{eqnarray}
defines a Koszul connection on $M$. 

One may observe that the elements of the submodule $V^H\subseteq V$ of $H$-invariant elements define 
$r_H$-covariant fields $\sigma\in\Gamma(\mathbb{V})$ on $M$ which are covariantly constant, $\nabla^{\mathbb{V}}\sigma=0$, since the corresponding $r_H$-equivariant maps $f\in {C}^\infty(P)\otimes_{r_H} V$ are constant functions on $P$.

 \subsubsection{Covariance in components}

Let $v^a$ denote the components of a vector $v\in V$ in an ordered basis $\{\texttt{e}_a\}$ of the vector space $V$. 
The matrix Lie group $H$ is represented faithfully through the defining representation $r:H\hookrightarrow GL(V):h\mapsto\textsl{h}$ by invertible matrices $\textsl{h}=\textsl{h}^b{}_a\,\texttt{e}_b\otimes \texttt{e}^{*a}$ acting as 
\be
\texttt{e}'_a=\texttt{e}_b\,\textsl{h}^b{}_a\,,\quad\text{and}\quad v'{}^a=(\textsl{h}^{-1})^a{}_b\,v^b\,.
\ee
The independent components $\textsl{h}^a{}_b$ are taken as suggestive coordinates on the matrix Lie group $H$. 

Let $(x^\mu,y^i)$ be some coordinates on the principal $H$-bundle $P$ with $\{\texttt{T}_i\}$ as basis for $\mathfrak{h}$ and $y^i$ as coordinates on the fibres. In this coordinate system, a moving frame reads 
\be
s\,:\,M\hookrightarrow P\,:\,x^\mu\mapsto y^i(x)\,.
\ee
Once a (local) moving frame is given, one can take instead $(x^\mu,\textsl{h}^a{}_b)$ as (local) coordinates on the principal $H$-bundle $P$ (with the moving frame taken as the identity section).
In practice, the moving frame on the principal $H$-bundle $P$ also defines a moving frame of the linear frame bundle $F\mathbb{V}$ of the associated vector bundle $\mathbb{V}:=P\times_{r_H}V$. Indeed, the principal $H$-bundle can be seen as a sub-bundle thereof, $P\subseteq F\mathbb{V}$\,. The moving linear frame reads in local coordinates as
\be
\texttt{e}\,:\,M\hookrightarrow F\mathbb{V}\,:\,x^\mu\mapsto\texttt{e}_a(x)\,.
\ee
Given such a moving linear frame, an $H$-frame of local coordinates $(x^\mu,\textsl{h}^a{}_b)$ can be identified with the linear frame $\texttt{e}'_a(x,h)=\texttt{e}_b(x)\,\textsl{h}^b{}_a(x)$ where the group element $h\in H$ relates the linear frame $\texttt{e}(x)$ and the linear frame $\texttt{e}'_a(x,h)$.
Moreover, local coordinates on the vector bundle $\mathbb{V}$ are provided by $(x^\mu,v^a)$ where $x^\mu$ are still coordinates on the base manifold $M$ while $v^a$ are Cartesian coordinates on the vector space $V$. In this way, $r_H$-covariant fields on $M$ read as
\be\label{rHcovf}
\sigma\,:\,M\hookrightarrow \mathbb{V}\,:\,x\mapsto\texttt{v}(x)=v^a(x)\,\texttt{e}_a(x)\,.
\ee
The corresponding $r_H$-equivariant map $f=i(\sigma)$ from $P$ to $V$ is 
\be
f\,:\,P\hookrightarrow V\,:\,(x,h)\mapsto\texttt{f}(x,h)=v'{}^a(x,h)\texttt{e}_a\,,
\ee
where 
\be
v'{}^a(x,h)=\big(\textsl{h}^{-1}\big)^a{}_b(x)\,v^b(x)\,,\quad\text{for}\quad\texttt{e}'_a(x,h)=\texttt{e}_b(x)\,\textsl{h}^b{}_a(x)\,.
\ee
This ensures that $\texttt{v}(x)$ in \eqref{rHcovf} is well-defined, \textit{i.e.} $v'{}^a(x,h)\texttt{e}'_a(x,h)=v^a(x)\,\texttt{e}_a(x)$.

Let $C^i_{jk}$ be the structure constants in the basis $\{\texttt{T}_i\}$ of the Lie algebra $\mathfrak{h}$. 
Vertical $H$-invariant vector fields are interpreted as infinitesimal gauge transformations and they read in coordinates as $\hat{Z}=Z^{i}(x,y)\,\texttt{T}_i^\#$ where $Z=Z^{i}(x,y)\,\texttt{T}_i$ is a section of the adjoint bundle $P\times_{r_H}\mathfrak h$, \textit{i.e.} an $Ad_H$-equivariant map from $P$ to $\mathfrak h$. Explicitly, its infinitesimal transformation under the adjoint action of another vertical $H$-invariant vector field ${Y}=Y^{i}(x,y)\,\texttt{T}_i^\#$ reads
\be
{\mathcal L}_{{Y}}Z^{i}=-\,C^i_{jk}Y^jZ^{k}\,.
\ee
The equivariance condition together with
\be
{\mathcal L}_{\texttt{T}_j^\#}\texttt{T}_k^\#=C^i_{jk}\texttt{T}_i^\#
\ee
implies the $H$-invariance $[\texttt{T}_i^\#,\hat{Z}]={\mathcal L}_{\texttt{T}_i^\#}\hat{Z}=0$.
The Lie algebra $\mathfrak{h}$ is represented on $V$ through \eqref{Liealg}, thus
the $r_H$-equivariance of the map $f:P\to V$ reads infinitesimally as
\be
{\mathcal L}_{{Y}}f^a\,=\,-\,Y^i\,(\textsc{T}_i)^a{}_b\,f^b\,.
\ee
More generally, consider an $H$-invariant vector field on $P$ reading in coordinates as $\hat{Z}=X^\mu(x)\,\partial_\mu+Y^{i}(x,y)\,\texttt{T}_i^\#$\,. It acts on the $r_H$-equivariant  map $f$ as
\be
{\mathcal L}_{\hat{Z}}f^a\,=\,X^\mu(x)\,\partial_\mu f^a\,-\,Y^i\,(\textsc{T}_i)^a{}_b\,f^b\,.
\ee

Let  $\omega^{\mathfrak{h}}=\big(\,\omega_\mu^i(x,h)\,dx^\mu+\texttt{T}^{\#*i}\,\big)\otimes\texttt{T}_i$ be the expression in components of a principal connection one-form on the principal $H$-bundle $P$. The $H$-invariant horizontal lift of a vector field
${X}=X^\mu(x)\,\partial_\mu$ on the base $M$ reads in components as 
\be
\nabla^P_{{X}}\,=\,X^\mu(x)\,\big(\partial_{\mu}\,-\,\omega_\mu^i(x,h)\,\texttt{T}_i^\#\big)\,.
\ee
Such an $H$-invariant vector field of $\mathfrak{X}(P)^H$ acts on the $r_H$-equivariant maps from $P$ to $V$ as the matrix-valued differential operator
\be
\big(\nabla^{\mathbb V}_{{X}}\big)^a{}_b\,=\,X^\mu(x)\,\Big(\,\delta^a{}_b\,\partial_{\mu}\,+\,\omega_\mu^i(x,h)\,(\textsc{T}_i)^a{}_b\,\Big)\,.
\ee
The corresponding covariant derivative along ${X}$ acts on $r_H$-covariant fields $\sigma$ on $M$ as
\be
\nabla_{{X}}v^a(x)\,=\,X^\mu(x)\,\Big(\,\,\partial_{\mu}v^a(x)\,+\,\omega_\mu^i\big(x,h(x)\big)\,(\textsc{T}_i)^a{}_bv^b(x)\,\Big)\,.
\ee
Therefore the components of the Koszul connection associated to the gauge principal connection one-form $s^*\omega^{\mathfrak{h}}$
are given by $\omega_\mu{}^a{}_b(x)=\omega_\mu^i\big(x,h(x)\big) (\textsc{T}_i)^a{}_b$. 
Notice that, although a gauge is necessary to write down explicitly the components of the associated Koszul connection in some coordinates, the underlying geometric object does not require such a choice of gauge. In this sense, a gauge choice has the same status as a choice of coordinates for manifolds (or, as a choice of basis for vector spaces)\,: it is just an auxiliary tool for making explicit computations.

Inverting the (active vs passive) points of view, a finite gauge transformation can be seen as a change of section $s\to s'$ in which case it acts on the $r_H$-covariant field on $M$ as $v'{}^a(x)=\textsl{h}^a{}_b(x)\,v^b(x)\,.$
Accordingly, the vertical $H$-invariant vector fields can also be interpreted as an infinitesimal change of section, in which case it acts on the $r_H$-covariant field on $M$ as 
\be
{\mathcal L}_{{Y}}v^a(x)\,=\,{\mathcal Y}^i(x)\,(\textsc{T}_i)^a{}_b\,v^b(x)\,,
\ee
where ${\mathcal Y}^i(x):=Y^i\big(x,h(x)\big)$ is then interpreted as the infinitesimal gauge parameter.
By construction, the Lie derivative ${\mathcal L}_{{Y}}$ along a vertical $H$-invariant vector field ${Y}$ and the covariant derivative $\nabla^{\mathbb V}_{{X}}$ along a base vector field ${X}$ commute (and similarly for a finite gauge transformation). In the physics literature, this is basically the defining property of a ``covariant'' derivative (in the sense that it transforms covariantly under gauge transformations). Notice that this property provides a simple way to deduce the transformation law of the components of the Koszul connection, \textit{i.e.} ${\mathcal L}_{{Y}}\,\omega_\mu{}^a{}_b=\partial_\mu{\mathcal Y}^a{}_b+\omega_\mu{}^a{}_c\,{\mathcal Y}^c{}_b$, in agreement with the transformation law of the gauge principal connection, ${\mathcal L}_{{Y}}\,\omega_\mu^i=\partial_\mu{\mathcal Y}^i+C^i_{jk}\,\omega_\mu^j\,{\mathcal Y}^k$.

\subsection{Tractor bundles}\label{tractbdles}

Let $\pi:P\twoheadrightarrow P/H$ be a principal $H$-bundle $P$.
Consider a Klein geometry $H\subseteq G$ and a representation $r:G\to GL(V)$ of the principal group $G$ on the vector space $V$. 

The vector bundle $P\times_{r_H}V$ over $P/H$ associated to the restriction of the representation $r$ to the isotropy subgroup $H$ is called the \textbf{tractor bundle associated to principal} $H$-\textbf{bundle} $P$ \textbf{via the the representation} $r$ \textbf{of the principal group} $G$ \textbf{on the vector space} $V$. 

\vspace{3mm}
\noindent{\small\textbf{Example (Adjoint tractor bundle)\,:}
	For a principal $H$-bundle $P$ and a Klein geometry $H\subseteq G$, the adjoint representation $Ad:G\to GL(\mathfrak{g})$ of the Lie group $G$ on its Lie algebra $\mathfrak g$ leads to the adjoint tractor bundle $P\times_{Ad_H}\mathfrak{g}$\,.}

\vspace{3mm}
The representation
$r:\mathfrak{g}\to\mathfrak{gl}(V)$ of the principal algebra on a $\mathfrak{g}$-module $V$ defines an \textbf{induced representation}
\be\label{rGtractbundle}
\texttt{r}\,:\,P\times_{Ad_H}\mathfrak{g}\,\to\, CD^1{\mathbb V}\,,
\ee
\textbf{of the adjoint tractor algebroid on the tractor bundle} $\mathbb{V}:=P\times_{r_H}V$ \textbf{associated to the} $\mathfrak{g}$-\textbf{module} $V$. 
Therefore, given a $\mathfrak{g}$-module $V$ there are two canonical induced representations on the tractor bundle: on the one hand, there is the induced representation $\textsl{r}:\frac{TP}H\to CD^1\,\mathbb{V}$ of the Atiyah algebroid \eqref{assrepr}, and on the other hand, there is the induced representation $\texttt{r}:P\times_{Ad_H}\mathfrak{g}\to CD^1\mathbb{V}$ of the adjoint tractor algebroid \eqref{rGtractbundle}. Their respective restrictions, $\textsl{r}|_{\frac{VP}{H}}\,:\,\frac{VP}H\to \mathfrak{gl}(\mathbb{V})$
and $\texttt{r}|_{P\times_{r_H}\mathfrak{h}}\,:\,P\times_{r_H}\mathfrak{h}\to \mathfrak{gl}(\mathbb{V})$ to the adjoint algebroid $\frac{VP}{H}\cong P\times_{Ad_H}\mathfrak{h}$ coincide. More precisely, the fundamental action $\#:P\times_{r_H}\mathfrak{h}\stackrel{\sim}{\to}\frac{VP}{H}$ is a canonical isomorphism of Lie algebra bundles over $P/H$ relating these two restricted representations: 
\be
\texttt{r}|_{P\times_{r_H}\mathfrak{h}}\,=\,\textsl{r}|_{\frac{VP}{H}}\,\circ\,\#\,.
\ee 
This can be summarised in the Lie algebroid morphism from the tractor sequence \eqref{shortexactintract} of the principal $H$-bundle $P$ to the Atiyah sequence of the tractor bundle $\mathbb V$:
\be\label{salsiciettabc}
\begin{array}
	[c]{ccccccccc}%
	0&\to& P\times_{Ad_H}\mathfrak{h}&\hookrightarrow&P\times_{Ad_H}\mathfrak{g}&\twoheadrightarrow&P\times_{Ad_H}\,\mathfrak{g}/\mathfrak{h}&\to& 0\\
	&&&&&&&&\\
	&&\textsl{r}\circ\#\downarrow&&\texttt{r}\downarrow&& \downarrow&&\\
	&&&&&&&&\\
	0&\to& \mathfrak{gl}(\mathbb{V})&\hookrightarrow&CD^1\mathbb{V}&\twoheadrightarrow&T\frac{P}{H}&\to& 0
\end{array}
\,.
\ee
For instance, the composition $\texttt{Ad}=\textsl{Ad}\,\circ \#:P\times_{Ad_H}\mathfrak{h}\,\to\,\mathfrak{gl}(P\times_{Ad_H}\mathfrak{h})$ of the fundamental action with the adjoint representation \eqref{Adrep} provides the \textbf{canonical self-representation of the adjoint algebroid} $P\times_{Ad_H}\mathfrak{h}$.

Consider a Cartan geometry modeled on the Klein pair $\mathfrak{h}\subset\mathfrak{g}$, \textit{i.e.} the principal $H$-bundle $P$ is endowed with a Cartan connection $\omega^{\mathfrak g}:\frac{TP}{H}\stackrel{\sim}{\to}P\times_{r_H}\mathfrak{g}$\,.
The composition 
\be\label{funderD}
\texttt{D}:=\textsl{r}\,\circ D:P\times_{Ad_H}\mathfrak{g}\,\to\, CD^1{\mathbb V}
\ee
of the Cartan intertwiner $D:P\times_{Ad_H}\mathfrak{g}\stackrel{\sim}{\to}\frac{TP}H$ with the induced representation $r:\frac{TP}H\to CD^1{\mathbb V}$ of the Atiyah algebroid $\frac{TP}H$ on the $r_H$-associated bundle ${\mathbb V}$ is a Lie algebroid connection of the adjoint tractor bundle $P\times_{Ad_H}\mathfrak{g}$ on the tractor bundle ${\mathbb V}=P\times_{r_H}V$, which will be called the \textbf{fundamental derivative on the tractor bundle} $P\times_{r_H}V$\,.
The underlying idea is the following composition of (vector bundle morphisms) of the short exact sequences:
\be\label{salsiciettafund}
\begin{array}
	[c]{ccccccccc}%
	0&\to& P\times_{Ad_H}\mathfrak{h}&\hookrightarrow&P\times_{Ad_H}\mathfrak{g}&\twoheadrightarrow&P\times_{Ad_H}\,\mathfrak{g}/\mathfrak{h}&\to& 0\\
	&&&&&&&&\\
	&&\#\downarrow\sim&&D\downarrow\sim&&\Theta^{-1}\downarrow\sim&&\\
	&&&&&&&&\\
	0&\to& \frac{VP}{H}&\stackrel{i}{\hookrightarrow}&\frac{TP}{H}&\stackrel{\pi_*}{\twoheadrightarrow}&T\frac{P}{H}&\to& 0\\
	&&&&&&&&\\
	&&\textsl{r}|_{\mathfrak{h}^\#}\downarrow&&\textsl{r}\downarrow&& id\downarrow\sim&&\\
	&&&&&&&&\\
	0&\to& \mathfrak{gl}(\mathbb{V})&\hookrightarrow&CD^1\mathbb{V}&\twoheadrightarrow&T\frac{P}{H}&\to& 0
\end{array}
\,.
\ee
All arrows are morphisms of vector bundles over $P/H$ (actually, almost all arrows are even morphisms of Lie algebroids over $P/H$ except the fundamental derivative $D$ which is Lie algebroid morphism if and only if the Cartan connection is flat).
The induced representations \eqref{assrepr} and \eqref{rGtractbundle} of, respectively, the Atiyah algebroid and the adjoint tractor algebroid, are not directly related in general, in fact the latter algebroids are not isomorphic (their anchor and Lie bracket usually differ). 
They only coincide in the case of a Klein geometry.

With the help of a Cartan geometry, a tractor bundle comes equipped with two Lie algebroid connection of the adjoint tractor algebroid $P\times_{Ad_H}\mathfrak{g}$ on the tractor bundle ${\mathbb V}=P\times_{r_H}V$: on the one hand, there is the induced representation \eqref{rGtractbundle}, seen as a flat Lie algebroid connection $\texttt{r}:P\times_{Ad_H}\mathfrak{g}\to CD^1\mathbb{V}$, and on the other hand, there is the fundamental derivative \eqref{funderD}, which is a (possibly curved) Lie algebroid connection
$\texttt{D}:P\times_{Ad_H}\mathfrak{g}\to CD^1{\mathbb V}$. Their restrictions to the adjoint bundle $P\times_{Ad_H}\mathfrak{h}$ coincide. This fact can be used in order to define a Koszul connection on the tractor bundle.
Using the isomorphism $T\frac{P}{H}\cong \frac{TP}H\,/\,\frac{VP}H$, one can see that the difference between the fundamental derivative \eqref{funderD} and the induced representation \eqref{rGtractbundle} defines a Koszul connection $\nabla:T\frac{P}{H}\to CD^1\mathbb V$ on the tractor bundle via $\nabla:=(\texttt{D}-\texttt{r})\circ \Theta$. This map is well-defined because $\texttt{D}|_{P\times_{Ad_H}\mathfrak{h}}=\textsl{r}\circ\#=\texttt{r}|_{P\times_{Ad_H}\mathfrak{h}}$\,. This Koszul connection is called the \textbf{tractor connection associated to the representation} $r$ \textbf{of the principal algebra}. 
Let the representation $r:\mathfrak{g}\to\mathfrak{gl}(V)$ of the principal algebra $\mathfrak{g}$ of basis $\{\texttt{P}_a,\texttt{T}_i\}$ (adapted to the Klein pair $\mathfrak{h}\subset\mathfrak{g}$ which is not necessarily reductive) on a $\mathfrak{g}$-module $V$ with basis $\{\texttt{e}_\alpha\}$ read
\be
r\,:\qquad\texttt{P}_a\mapsto \textsc{P}_a=(\textsc{P}_a)^\alpha{}_\beta\,\texttt{e}_\alpha\otimes\texttt{e}^{*\beta}\,,\qquad \texttt{T}_i\mapsto \textsc{T}_i=(\textsc{T}_i)^\alpha{}_\beta\,\texttt{e}_\alpha\otimes\texttt{e}^{*\beta}\,,
\ee
while the Cartan intertwiner on the principal bundle is the map \eqref{Cartanfunder}
\be
D\,:\qquad\texttt{P}_a\mapsto \hat{D}_a=e_a^\mu(x,y)\,\big(\partial_{\mu}\,-\,\omega_\mu^i(x,y)\,\texttt{T}_i^\#\big)\,,\qquad \texttt{T}_i\mapsto \texttt{T}_i^\#\,.
\ee
Therefore, the fundamental derivative on the tractor bundle takes the form
\be
\texttt{D}\,:\qquad\texttt{P}_a\mapsto \hat{D}_a=e_a^\mu(x,y)\,\big(\partial_{\mu}\,-\,\omega_\mu^i(x,y)\,\textsc{T}_i\big)\,,\qquad \texttt{T}_i\mapsto \textsc{T}_i\,.
\ee
Consequently, the tractor connection is the Koszul connection
\be
\nabla_\mu=\partial_{\mu}\,-\,\theta^a_\mu(x)\,\textsc{P}_a-\,\omega_\mu^i\big(x,y(x)\big)\,\textsc{T}_i\,,
\ee
where $y(x)$ is a moving frame of the principal $H$-bundle $P$.

When the Klein pair $\mathfrak{h}\subset \mathfrak{g}$ is reductive, the Cartan connection $\omega^{\mathfrak{g}}$ induces a vector bundle isomorphism $\Theta:T\frac{P}{H}\stackrel{\sim}{\to}P\times_{Ad_H}\mathfrak{p}$ between the base tangent bundle and the vector bundle associated to the transvection submodule $\mathfrak{p}\cong\mathfrak{g}/\mathfrak{h}$. Remember that Cartan geometries for a reductive Klein pair are endowed with a principal connection $\omega^{\mathfrak{h}}$ on the $H$-bundle $P$ which will define a Koszul connection on any associated vector bundle. 
The importance of tractor bundles is that this class of associated vector bundles is endowed with a Koszul connection inherited from the Cartan connection, \textit{even} when the Klein pair is \textit{not} reductive.

\pagebreak

\section{Relations between principal bundles}\label{relprincbundles}

A Lie group morphism between two Lie groups may induce a morphism of principal bundles over the same base with these two Lie groups as respective fibers. The goal of this chapter is to consider several instances of such constructions and their interplay with Klein and Cartan geometries.

\subsection{Klein geometry bundles}

A \textbf{Klein geometry bundle} is a fibre bundle $E$ over $M$ such that the typical fibres are homogeneous spaces of the structure group $G$, \textit{i.e.} $E_m\cong G/H$ for some subgroup $H\subseteq G$.
For instance, given a principal $G$-bundle $P$, the associated bundle $P\times_G G/H\,\cong\, P/H$ is a Klein geometry bundle over $P/G$ \cite[Proposition I.5.5]{Kob63}.

The composition of a Lie group morphism $\nu:H\to G$ with the right multiplication $L_\bullet: G\to {Diff}(G)$ is a right action $L_\bullet\circ\nu:H\to {Diff}(G)$ of the Lie group $H$ on the manifold $G$. Therefore, any Lie group morphism $\nu:H\to G$ induces a right action of the Lie group $H$ on any principal $G$-bundle.
For any principal bundle $G\lefttorightarrow P \twoheadrightarrow P/G$, an injective morphism $i:H\hookrightarrow G$ of Lie groups induces a free $H$-action on the total space $P$ and, accordingly, defines a principal bundle $H\lefttorightarrow P \twoheadrightarrow P/H$.
The respective fibrations will be denoted $\pi_G:P\twoheadrightarrow P/G$ and $\pi_H:P\twoheadrightarrow P/H$ in order to distinguish them carefully.

\subsection{Reduction of principal bundles}

\subsubsection{Reductions of principal bundles as sections of Klein geometry bundles}\label{reductionsprincipal}

Let $H\subseteq G$ be a Klein geometry. 

For any principal $H$-bundle $Q$ over $M\cong Q/H$, one may consider the associated principal $G$-bundle $P=Q\times_H G$. The principal $H$-bundle $Q$ is then a reduction of the principal $G$-bundle $P$. 
Conversely, finding a reduction of the $G$-bundle $P$ is equivalent to finding a principal $H$-bundle $Q$ whose associated bundle $Q\times_H G$ is isomorphic to $P$ as a principal $G$-bundle.

More geometrically, a submanifold $i:Q\hookrightarrow P$ is a reduction of the $G$-bundle $P$ if the free action of $G$ on $P$ restricts to a free action of $H$ on $Q$ and defines a principal $H$-bundle $Q$ over the same base $\frac{P}G\cong\frac{Q}H$\,. 
In other words, a reduction of the principal $G$-bundle $P$ to the principal $H$-bundle $Q$ can be defined as an embedding $i:Q\hookrightarrow P$ of total spaces with the property that it is an $H$-equivariant morphism of fibre bundles over the same base $M\cong\frac{P}G\cong\frac{Q}H$\,. The $H$-equivariance of the embedding $i:Q\hookrightarrow P$ implies that it projects on the quotient by $H$, \textit{i.e.} it defines an embedding $\sigma:Q/H\hookrightarrow P/H$ of manifolds. This injection is actually a section of the Klein geometry bundle $P\times_G G/H\,\cong\, P/H$ over $P/G\cong Q/H$.

Conversely, for any principal $G$-bundle $P$ over $P/G$, a section $\sigma:P/G\hookrightarrow P/H$ of the Klein geometry bundle $P/H$ over $P/G$ defines a reduction $Q:=\sigma^*P$ via pullback along this distinguished section (see \textit{e.g.}  \cite{Catren:2014vza}). In this way, the principal $H$-bundle $Q$ is seen as a submanifold of $P$. Given a Klein geometry $H\subset G$, there is a one-to-one correspondence between reductions $i:Q\hookrightarrow P$ of a $G$-bundle $P$ to an $H$-bundle $Q$ and sections $\sigma:P/G\hookrightarrow P/H$ of the Klein geometry bundle \cite[Proposition I.5.6]{Kob63}.
The reduction process can be summarised by means of the following diagram, adapted from \cite{Catren:2014vza}\,:
\begin{equation}
	\xymatrix@R=.8in @C=.55in{
		Q=\sigma^* P\, \ar@{^{(}->}[r]^{i\,=\,\pi_H^*(\sigma)}\ar[d]_{\sigma^*(\pi_H)} & P\cong Q\times_H G \ar[d]^{\pi_H} \ar[ld]^{\pi_G}
		\\
		M=\frac{Q}{H}=\frac{P}{G}   \ar@/^1.2pc/[r]^{\sigma}  &  \frac{P}{H}\cong P\times_{G} \frac{G}{H}  \ar[l]                                     &
	}
\end{equation}
This can also be summarised in the following diagram (with a different orientation) that completes \eqref{reduxprincbundl}: 
\be\label{reductionprincbundl}
\begin{array}
	[c]{ccccc}%
	G &  \lefttorightarrow & P & \stackrel{\pi_H}{\twoheadrightarrow} & E\cong P\times_G G/H\cong \frac{P}{H}\\
	& & & & \\ 
	\uparrow & &i \uparrow & \stackrel{\pi_G}{\searrow} & \downuparrows\sigma \\ 
	& & & & \\ 
	H & \lefttorightarrow & Q & \stackrel{\pi_H}{\twoheadrightarrow} & M\cong \frac{P}{G}\cong\frac{Q}{H}
\end{array}
\ee
where, with a slight abuse of the terminology, we denoted by $\pi_H$ the projection to the $H$-orbit space for both $P$ and $Q$.  Note that for any principal $G$-bundle with total space $P$, a Klein geometry $H\subseteq G$ defines a principal $H$-bundle with the same total space  \cite[Proposition I.5.5]{Kob63}. This fact can be summarised as the implication
\begin{equation}
	G\lefttorightarrow P \stackrel{\pi_G}{\twoheadrightarrow} P/G\qquad\stackrel{H\subseteq G}{\Longrightarrow}\qquad	H\lefttorightarrow P \stackrel{\pi_H}{\twoheadrightarrow} P/H\,.
\end{equation}

\vspace{5mm}
\begin{framed}
	\begin{center}
		\textbf{Partial gauges as reductions}
	\end{center}
	
	\noindent
	Given a Klein geometry $H\subseteq G$ and a principal $G$-bundle $P$ over $G/H$, the following notions are equivalent:
	\begin{enumerate}
		\item  a reduction of the principal $G$-bundle $P$ to the principal $H$-bundle $Q$,
		\item  a submanifold $Q\subseteq P$ such that the free action of $G$ on $P$ restricts to a free action of $H$ on $Q$ and defines a principal $H$-bundle $Q$ over the same base $\frac{P}G\cong\frac{Q}H$\,,
		\item  a principal $H$-bundle $Q$ whose associated bundle $Q\times_H G$ is isomorphic to $P$ as a principal $G$-bundle,
		\item  a partial gauge of the principal bundle $G$-bundle $P$,
		\item a section of the Klein geometry bundle $\frac{P}{H}$ over the base $\frac{P}{G}\cong\frac{Q}{H}$.
	\end{enumerate}
	\vspace{3mm}
\end{framed}

Infinitesimally, the second definition (as a submanifold where the free action of the principal group restricts to the isotropy subgroup) requires that the restrictions of the fundamental vector fields of the subalgebra $\mathfrak{h}\subseteq\mathfrak{g}$ to the reduced sub-bundle $Q$ are tangent to this submanifold:
\be
\mathfrak{h}^\#|_{i(Q)}\subset i_*\Gamma(TQ)\,,\quad\text{and}\quad\text{dim}(P)-\text{dim}(\mathfrak{g})\,=\,\text{dim}(Q)-\text{dim}(\mathfrak{h})\,,
\ee
where $\#:\mathfrak{g}\hookrightarrow  \Gamma(VP)$ is the fundamental action of the $G$-bundle $P$.
There are two classes of degenerate cases: the class of Maurer-Cartan geometries, $H=G$, where the reduction is trivial (the embedding is the identity, $i=id_{M}$) and the class with trivial isotropy group, $H=\{e\}$, where the reduction is a moving frame on $P/G$ (the embedding is a complete gauge, $i:P/G\hookrightarrow P$).

 \subsubsection{Reducibility of connections}

Let the principal $H$-bundle $Q$ be a reduction of the principal $G$-bundle $P$ via the embedding $i:Q\hookrightarrow P$. 

Consider $i_*TQ\subset TP$ the image of the tangent bundle of the reduced bundle by the pushforward $i_*:TQ\hookrightarrow TP$ of the embedding.
Consider also the restriction $H_{i(Q)}P\subset HP$ of the horizontal distribution to the reduced bundle, \textit{i.e.} the pullback of the horizontal distribution $HP$ along the embedding $i:Q\hookrightarrow P$. There are two extreme cases which will be discussed here: when the restriction $H_{i(Q)}P\subset HP$ of the horizontal distribution is either \textit{everywhere} tangent to the reduced bundle, \textit{i.e.} $H_{i(Q)}P\subset i_*TQ$, or \textit{nowhere} tangent to the reduced bundle, \textit{i.e.} $H_{i(Q)}P\,\cap\,i_*TQ=\zeta(M)$ where $\zeta$ stands for the zero section. 

 \paragraph{Horizontal distribution nowhere tangent to the reduced bundle} 

A \textbf{principal connection} on the $G$-bundle $P$ over $M$ is \textbf{reducible to a Cartan connection} on the reduced $H$-bundle $Q$ over $M$ if and only if $\text{dim}\,Q\,=\,\text{dim}\,\mathfrak{g}$ and if the pushforward of the tangent bundle $TQ$ of the submanifold by the embedding $i:Q\hookrightarrow P$ is \textit{nowhere} horizontal, \textit{i.e.} 
\be
H_{i(Q)}P\,\,\cap\,\,i_*TQ\,=\,\zeta(M)\,.
\ee
Since $i_*TQ$ nowhere belongs to the kernel of the connection one-form $\omega^\mathfrak{g}:TP\to \mathfrak{g}$ on the $G$-bundle $P$, the pullback $i^*\omega^\mathfrak{g}$ of this connection one-form along the embedding $i:Q\hookrightarrow P$ defines a Cartan connection on the reduced bundle $Q$ modeled on the Klein pair $\mathfrak{h}\subseteq \mathfrak{g}$. Conversely, any Cartan connection on a principal $H$-bundle $Q$
defines such a principal connection on the associated principal $G$-bundle $P=Q\times_H G$,
by equivariance (see \cite[Proposition A.3.1]{Sharpe} for a proof of the one-to-one correspondence). 

The section $\sigma:P/G\hookrightarrow P/H$ of the Klein geometry bundle $P\times_G G/H\cong P/H$ associated to the reduction of a principal connection on the $G$-bundle $P$ to a Cartan connection on the reduced $H$-bundle $Q$ 
is sometimes called a \textbf{compensator field}. More explicitly, in the reductive case a basis of the principal algebra $\mathfrak{g}$ reads $\{\texttt{T}_I\}=\{\texttt{P}_a,\texttt{T}_i\}$ where $\texttt{P}_a$ is a basis of the transvection submodule and $\texttt{T}_i$ is a basis of the isotropy algebra $\mathfrak{h}$.
Therefore,
\ba
\omega^\mathfrak{g}&=& \Big(\omega_\mu^I(x,Y)\,dx^\mu+\texttt{T}^{\#*I}\Big)\otimes\texttt{T}_I\\
\Rightarrow i^*\omega^\mathfrak{g}&=&\theta^a_\mu(x,y)\,dx^\mu \otimes\,\texttt{P}_a\,+\,\Big(\omega_\mu^i\big(\,x,Y(x,y)\,\big)\,dx^\mu+\texttt{T}^{\#*i}\Big)\otimes\texttt{T}_i\nonumber
\ea
where $\theta^a_\mu(x,y)\,dx^\mu:=\omega_\mu^a\big(\,x,Y(x,y)\,\big)\,dx^\mu+i^*\texttt{P}^{\#*a}$. For the sake of notional simplicity, we have not distinguished $\texttt{T}^{\#*i}$ and its pullback. One can see explicitly that the principal connection is reducible to a Cartan connection iff
$\theta^a_\mu$ is everywhere non-degenerate.

In terms of Lie algebroids, the reduction $i:Q\hookrightarrow P$ of principal bundles induces an isomorphism $Q\times_{Ad_H}\mathfrak{g}\cong P\times_{Ad_G}\mathfrak{g}$ of Lie algebra bundles over $M$. In other words, the adjoint tractor bundle of the principal $H$-bundle $Q$ is isomorphic to the adjoint bundle of the principal $G$-bundle $P$. The pushforward of the reduction $i:Q\hookrightarrow P$ of principal bundles induces an embedding $i_*:\frac{TQ}{H}\hookrightarrow \frac{TP}{G}$ of vector bundles over $M$ from the Atiyah bundle of the principal $H$-bundle $Q$ to the Atiyah bundle of the principal $G$-bundle $P$. Finally, the principal connection on the $G$-bundle $P$ is the surjection $\omega^\mathfrak{g}:\frac{TP}{G}\twoheadrightarrow P\times_{Ad_G}\mathfrak{g}$. The property that this principal connection on the $G$-bundle $P$ is reducible to a Cartan connection on the reduced $H$-bundle $Q$, ensures that the following commutative diagram
\be
\begin{array}
	[c]{ccc}%
	\frac{TP}{G}&\stackrel{\omega^\mathfrak{g}}{\twoheadrightarrow}&P\times_{Ad_G}\mathfrak{g}\\
	&&\\
	i_*\uparrow\quad&&\downarrow \sim\\
	&&\\
	\frac{TQ}{H}&\stackrel{i^*\omega^\mathfrak{g}}{\to} & Q\times_{Ad_H}\mathfrak{g}
\end{array}
\ee
defines a vector bundle isomorphism
$i^*\omega^\mathfrak{g}$
from the Atiyah bundle $\frac{TQ}{H}$ of the $H$-bundle $Q$ to the adjoint tractor bundle $Q\times_{Ad_H}\mathfrak{g}$.

 \paragraph{Horizontal distribution everywhere tangent to the reduced bundle} 

The following statements are equivalent:
\begin{itemize}
	\item[$\bullet$] A \textbf{principal connection} on the $G$-bundle $P$ over $M$ is \textbf{reducible to a principal connection} on the reduced $H$-bundle $Q$ defined by the embedding $i:Q\hookrightarrow P$ or, equivalently by the section $\sigma:P/G\hookrightarrow P/H$ of the Klein geometry bundle $P\times_G G/H\cong P/H$.
	\item[$\bullet$] The restriction $H_{i(Q)}P$ of the horizontal distribution $HP$ to the reduced bundle is \textit{everywhere} tangent to the reduced bundle, \textit{i.e.} is a vector sub-bundle of the pushforward of the tangent bundle $TQ$ by the embedding:
	\be
	H_{i(Q)}P\subset i_*TQ\quad\Longleftrightarrow\quad\sigma_*TM\subset(\,\pi_H)_*H_{i(Q)}P\,,
	\ee
	where the embedding $i:Q\hookrightarrow P$ corresponds to the section $\sigma:M\hookrightarrow P/H$ of the Klein geometry bundle.  
	\item[$\bullet$] The compensator field $\sigma$ is horizontal with respect to the Ehresmann connection on the associated bundle $P\times_G G/H\cong P/H$ induced from the principal connection on the $G$-bundle $P$.
	\item[$\bullet$] The pullback along the embedding $i:Q\hookrightarrow P$ of the connection one-form $\omega^\mathfrak{g}:TP\hookrightarrow \mathfrak{g}$ on the $G$-bundle $P$ defines an $\mathfrak{h}$-valued one-form $\omega^\mathfrak{h}:=i^*\omega^\mathfrak{g}$ on the reduced $H$-bundle $Q$, via the commutativity of the diagram of vector bundles
	\be
	\begin{array}
		[c]{ccc}%
		\frac{TP}{G}&\stackrel{\omega^\mathfrak{g}}{\twoheadrightarrow}&P\times_{Ad_G}\mathfrak{g}\\
		&&\\
		i_*\uparrow\quad&&\uparrow \iota\\
		&&\\
		\frac{TQ}{H}&\stackrel{\omega^\mathfrak{h}}{\twoheadrightarrow} & Q\times_{Ad_H}\mathfrak{h}
	\end{array}
	\ee
	where the vertical arrows are injective. The reduction $i:Q\hookrightarrow P$ of principal bundles induces an embedding $\iota:Q\times_{Ad_H}\mathfrak{h}\hookrightarrow P\times_{Ad_H}\mathfrak{g}$ of Lie algebra bundles over $M$, such that the adjoint tractor bundle of the principal $H$-bundle $Q$ is a vector sub-bundle of the adjoint bundle of the principal $G$-bundle $P$.
\end{itemize} 

Let $\mathfrak{h}\subseteq\mathfrak{g}$ be a reductive Klein pair and $\omega^\mathfrak{g}$ a principal connection one-form on the $G$-bundle $P$. In such case, even if this principal connection is not reducible to a principal connection on the reduced $H$-bundle $Q$, one can nevertheless extract such a principal connection. In fact, the restriction to the codomain $\mathfrak{h}$ of the pullback $i^*\omega^\mathfrak{g}$
of the one-form $\omega^\mathfrak{g}$ along the embedding $i:Q\hookrightarrow P$ always define a principal connection one-form $\omega^\mathfrak{h}$ on the reduced bundle $Q$ which will be called the \textbf{reduced principal connection} on the reduced $H$-bundle $Q$.

 \subsubsection{Reduction of Klein geometries}

The Klein geometry $H_2 \subseteq G_2$ is reducible to the Klein geometry $H_1 \subseteq G_1$ if the principal and isotropy groups $H_1$ and $G_1$ are respective subgroups of the corresponding groups $H_2$ and $G_2$,
\be
\label{redKlgeo1}
\begin{array}
	[c]{ccc}%
	H_2 & \subset & G_2\\
	\cup & & \cup\\
	H_1 & \subset & G_1
\end{array}
\,,
\ee
and if the corresponding homogeneous spaces are diffeomorphic, 
\be
\label{redKlgeo2}
\frac{G_2}{H_2}\cong\frac{G_1}{H_1}\,\,.
\ee
The last condition guarantees that it defines a reduction of the $H_2$-bundle $G_2$ to the $H_1$-bundle $G_1$. 
As is manifest from the conditions \eqref{redKlgeo1}-\eqref{redKlgeo2}, there is a duality for reducible Klein geometries  (under some technical assumptions about the embeddings)\,:
the Klein geometry $H_2 \subseteq G_2$ is reducible to the Klein geometry $H_1 \subseteq G_1$ iff
the Klein geometry $G_1 \subseteq G_2$ is reducible to the Klein geometry $H_1 \subseteq H_2$. 

The above conditions \eqref{redKlgeo1}-\eqref{redKlgeo2} can be summarised in a single diagram, analogous to the diagram \eqref{reduxprincbundl},
\be\label{reduxprincbundlklein'}
\begin{array}
	[c]{ccccc}%
	H_2 &  \subset & G_2 & &\\
	& & & \searrow & \\ 
	\cup & & \cup & & \frac{G_2}{H_2}\cong\frac{G_1}{H_1} \\ 
	& & & \nearrow & \\ 
	H_1 & \subset & G_1 & & 
\end{array}
\,.
\ee
This diagram can also be completed to include the corresponding section of the Klein geometry bundle $G_2/H_1$ over the homogeneous space
$G_2/H_2\cong G_1/H_1$ with typical fibre the homogeneous space $H_2/H_1$.
The analogue of \eqref{reductionprincbundl} is the diagram
\be\label{reductionprincbundlklein}
\begin{array}
	[c]{ccccc}%
	H_2 &  \lefttorightarrow & G_2 & \stackrel{\pi_{H_1}}{\twoheadrightarrow} & \frac{G_2}{H_1}\\
	& & & & \\ 
	\uparrow & & \uparrow i & \stackrel{\pi_{G_2}}{\searrow} & \downuparrows\sigma \\ 
	& & & & \\ 
	H_1 & \lefttorightarrow & G_1 & \stackrel{\pi_{H_1}}{\twoheadrightarrow} & M\cong \frac{G_2}{H_2}\cong\frac{G_1}{H_1}
\end{array}
\ee

\vspace{3mm}
\noindent{\small\textbf{Example (Grassmannians)\,:} The Grassmannian $Gr(m,n)$ is the manifold of all $m$-dimensional linear subspaces of an $n$-dimensional vector space. This flag variety is the homogeneous space of the parabolic Klein geometry $B(m,n-m)\subset SL(n)$, where the isotropy subgroup $B(m,n-m)$ is the subgroup of block upper-triangular matrices with unit determinant, for blocks of size $m\times m$ and $(n-m)\times(n-m)$. This (parabolic) Klein geometry is reducible to the (\textit{non} parabolic) Klein geometry $SO(m)\times SO(n–m)\subset SO(n)$.
	In other words,
	\be
	\label{redKlGrass}
	\begin{array}
		[c]{ccccc}%
		B(m,n-m) &  \subset & SL(n) & &\\
		& & & \searrow & \\ 
		\cup & & \cup & & Gr(m,n)\cong\frac{SL(n)}{B(m,n-m)}\cong\frac{SO(n)}{SO(m)\times SO(n–m)} \\ 
		& & & \nearrow & \\ 
		SO(m)\times SO(n–m) & \subset & SO(n) & & 
	\end{array}
	\,.
	\ee
}

\vspace{3mm}
\noindent{\small\textbf{Example (Reduction of Klein affine geometry to Euclidean geometry)\,:}
	The Klein affine geometry $GL(V)\subset IGL(V)$ can be reduced to the Euclidean geometry $O(V) \subset IO(V)$ since, in both cases, the homogeneous space is $V\cong IGL(V)/GL(V)\cong IO(V)/O(V)$, either seen as an affine space in the first case or as an Euclidean space in the second case.
	\be\label{reduxprincbundlkleinEucl}
	\begin{array}
		[c]{ccccc}%
		GL(V) &  \subset & IGL(V) & &\\
		& & & \searrow & \\ 
		\cup & & \cup & & V \\ 
		& & & \nearrow & \\ 
		O(V) & \subset & IO(V)  & & 
	\end{array}
	\,.
	\ee
	The reduction \eqref{reduxprincbundlkleinEucl} of Klein geometries can be seen as a reduction of the corresponding principal bundles, summarised in the commutative diagram corresponding to \eqref{reductionprincbundlklein}:
	\be\label{reductAfftoEucl}
	\begin{array}
		[c]{ccccc}%
		GL(V) &  \lefttorightarrow & IGL(V) & \stackrel{\pi_{O(V)}}{\twoheadrightarrow} & \frac{IGL(V)}{O(V)}\\
		& & & & \\ 
		\uparrow & & \uparrow i & \stackrel{\pi_{GL(V)}}{\searrow} & \downuparrows \sigma\\ 
		& & & & \\ 
		O(V) & \lefttorightarrow & IO(V) & \stackrel{\pi_{O(V)}}{\twoheadrightarrow} & V
	\end{array}
	\ee
	The fibre of the Klein geometry bundle $IGL(V)/O(V)$ over $V$ is the homogeneous space $GL(V)/O(V)$ which can be interpreted as the space of Euclidean metrics on $V$. In fact, the reduction $i:IO(V)\hookrightarrow IGL(V)$ is equivalent to a choice of Euclidean metric on $V$, \textit{i.e.} a section $\sigma:V\hookrightarrow IGL(V)/O(V)$ of the Klein geometry bundle. 
	Dually, the reduction \eqref{reduxprincbundlkleinEucl} of Klein geometries also defines the reduction of the Klein geometry $IO(V)\subset IGL(V)$ to the Klein geometry $O(V)\subset GL(V)$ which share the space $GL(V)/O(V)\cong IGL(V)/IO(V)$ of Euclidean metrics as homogeneous space.
}

\vspace{3mm}
\noindent{\small\textbf{Example (Reduction of Weyl model to Euclidean geometry)\,:}
	The Weyl model $CO(V)\subset ICO(V)$ can be reduced to the Euclidean geometry $O(V) \subset IO(V)$ since, in both cases, the homogeneous space is $V\cong ICO(V)/CO(V)\cong IO(V)/O(V)$, either seen as a conformal space in the first case or as an Euclidean space in the second case.
	\be\label{reduxprincbundlkleinconf}
	\begin{array}
		[c]{ccccc}%
		CO(V) &  \subset & ICO(V) & &\\
		& & & \searrow & \\ 
		\cup & & \cup & & V \\ 
		& & & \nearrow & \\ 
		O(V) & \lefttorightarrow & IO(V)  & & 
	\end{array}
	\,.
	\ee
	The corresponding reduction of principal bundles is
	\be
	\begin{array}
		[c]{ccccc}%
		CO(V) &  \lefttorightarrow & ICO(V) & \stackrel{\pi_{O(V)}}{\twoheadrightarrow} & \frac{ICO(V)}{O(V)}\cong V\times\mathbb R\\
		& & & & \\ 
		\uparrow & & \uparrow i& \stackrel{\pi_{CO(V)}}{\searrow} & \downuparrows \sigma\\ 
		& & & & \\ 
		O(V) & \lefttorightarrow & IO(V) & \stackrel{\pi_{O(V)}}{\twoheadrightarrow} & V
	\end{array}
	\ee
	The fibre of the Klein geometry bundle $ICO(V)/O(V)\cong V\times\mathbb R$ is the homogeneous space $CO(V)/O(V)\cong \mathbb R$ which can be interpreted as the space of scales on $V$. In fact, the reduction $i:IO(V)\hookrightarrow ICO(V)$ is equivalent to a choice of scale on $V$, \textit{i.e.} a section $\sigma:V\hookrightarrow ICO(V)/O(V)$ of the Klein geometry bundle. 
	The dual view of \eqref{reduxprincbundlkleinconf} is as a reduction of the Klein geometry $IO(V)\subset ICO(V)$ to the Klein geometry $O(V)\subset CO(V)$ who share the space $\mathbb{R}\cong ICO(V)/IO(V)\cong CO(V)/O(V)$ of scales as homogeneous space.}

 \subsubsection{Reduction of Klein pairs}

A \textbf{reduction of the Klein pair} $\mathfrak{h}_2 \subseteq \mathfrak{g}_2$ to the Klein pair $\mathfrak{h}_1 \subseteq \mathfrak{g}_1$ is an embedding $i:\mathfrak{g}_1 \hookrightarrow \mathfrak{g}_2$ of principal algebras such that its restriction to the isotropy subalgebras is also an embedding 
$i|_{\mathfrak{h}_1}:\mathfrak{h}_1 \hookrightarrow \mathfrak{h}_2$\,,
\be\label{diagreduxKleinpair}
\begin{array}
	[c]{ccc}%
	\mathfrak{h}_2 & \subset & \mathfrak{g}_2\\
	\cup & & \cup\\
	\mathfrak{h}_1 & \subset & \mathfrak{g}_1
\end{array}
\,,
\ee
and such that the transvection modules are isomorphic vector spaces, 
\be\label{diagreduxKleinpair2}
\frac{\mathfrak{g}_1}{\mathfrak{h}_1}\cong\frac{\mathfrak{g}_2}{\mathfrak{h}_2}\,\,.
\ee
As is manifest from the diagram \eqref{diagreduxKleinpair}, 
the Klein pair  $\mathfrak{h}_2 \subseteq \mathfrak{g}_2$ is reducible to the Klein pair $\mathfrak{h}_1 \subseteq \mathfrak{g}_1$ iff
the Klein pair  $\mathfrak{g}_1 \subseteq \mathfrak{g}_2$ is reducible to the Klein pair $\mathfrak{h}_1 \subseteq \mathfrak{h}_2$. 
The second reduction will be called the \textbf{dual reduction of Klein pairs}.
The reduction $\mathfrak{h}_1 \subseteq \mathfrak{g}_1$ of a reductive Klein pair $\mathfrak{h}_2 \subseteq \mathfrak{g}_2$ is also a reductive Klein pair, with the same transvection submodule, \textit{i.e.} $\mathfrak{p}_1\cong\mathfrak{p}_2$ due to \eqref{diagreduxKleinpair2}. However, the dual reduction is not necessarily reductive.
\vspace{2mm}

Let $\mathfrak{h}_1 \subseteq \mathfrak{g}_1$ be a reduction of the Klein pair $\mathfrak{h}_2 \subseteq \mathfrak{g}_2$ and  let the principal $H_1$-bundle $P_1$ be a reduction $P_1\subset P_2$ of the principal $H_2$-bundle $P_2$. 
A \textbf{Cartan geometry on the principal bundle $H_2\lefttorightarrow P_2\twoheadrightarrow M$, modeled on the Klein pair $\mathfrak{h}_2 \subseteq \mathfrak{g}_2$}, is \textbf{reducible to a Cartan geometry on the reduced bundle $H_1\lefttorightarrow P_1\twoheadrightarrow M$, modeled on the reduced Klein pair} $\mathfrak{h}_1 \subseteq \mathfrak{g}_1$,
if and only if the pullback along the embedding $i:P_1\hookrightarrow P_2$ defines a $\mathfrak{g}_1$-valued one-form $\omega^\mathfrak{g_1}:=i^*\omega^\mathfrak{g_2}$ on the reduced bundle $P_1$.
This condition is equivalent to the property that the restriction of Cartan fundamental vector fields
of the principal subalgebra $\mathfrak{g_1}\subset\mathfrak{g}_2$ to the reduced bundle $i:P_1\hookrightarrow P_2$ are tangent to this submanifold:
\be\label{reducedCartan}
D_\mathfrak{g_1}\Big|_{i(P_1)}\subset i_* \Gamma(TP_1)\,.
\ee
Notice that the fact that the principal $H_1$-bundle $P_1$ is the reduction of the principal $H_2$-bundle $P_2$ already guarantees that \eqref{reducedCartan} holds for the isotropy subalgebra $\mathfrak{h}_1\subseteq \mathfrak{h}_2$ (because $\mathfrak{h}_1^\#|_{i(P_1)}\subset i_* \Gamma(TP_1)$ holds by assumption).

Let $\mathfrak{h}_1 \subseteq \mathfrak{h}_2$ be the reduction  of a reductive Klein pair $\mathfrak{g}_1 \subseteq \mathfrak{g}_2$. Let $\omega^{\mathfrak{g}_2}$ be a Cartan connection one-form on the principal $H_2$-bundle $P_2$. Let $i:P_1\hookrightarrow P_2$ be a reduction of  the principal $H_2$-bundle $P_2$ to the principal $H_1$-bundle $P_1$ such that $\text{dim}\,P_1=\text{dim}\,\mathfrak{g}_1$. Consider the pullback $i^*\omega^{\mathfrak{g}_2}$ of the Cartan connection one-form $\omega^{\mathfrak{g}_2}$ along the embedding $i:P_1\hookrightarrow P_2$.
To discuss the inverse point of view, let us consider the restriction $\omega^{\mathfrak{g}_2/\mathfrak{g}_1}$ of the one-form $\omega^{\mathfrak{g}_2}$ to the transvection submodule $\mathfrak{g}_2/\mathfrak{g}_1$ of the reducible Klein pair $\mathfrak{g}_1 \subseteq \mathfrak{g}_2$. Its kernel $\text{Ker}\,\omega^{\mathfrak{g}_2/\mathfrak{g}_1}$ is an $H_1$-equivariant distribution on $P_1$. 
The isomorphism 
\be
i_*TP_1\,\cong\,\frac{T_{i(P_1)}P_2}{\text{Ker}\,\omega^{\mathfrak{g}_2/\mathfrak{g}_1}}
\ee
of vector budles over $P_1/H_1$ allows to define a fundamental derivative from the restriction of the fundamental derivative $D|_{\mathfrak{g}_1}:\mathfrak{g}_1\hookrightarrow  \Gamma(TP_2)$ defined by $\omega^{\mathfrak{g}_2}$. It defines a Cartan connection one-form $\omega^{\mathfrak{g}_1}$ on the reduced bundle $P_1$ which will be called the \textbf{reduced Cartan connection} on the reduced $H_1$-bundle $P_1$.

 \subsection{Projection of principal bundles}

Consider a Lie group extension $G$ of the Lie group $H$ by the closed Lie group $I$, \textit{i.e.} a short exact sequences of Lie group morphisms
\be
e\to I\stackrel{i}{\hookrightarrow}G\stackrel{\pi}{\twoheadrightarrow}H\to e\,,
\ee
encoding that $H\cong G/I$, that is to say the Lie group $H$ is isomorphic to the quotient of the Lie group $G$ by its normal subgroup  $I\trianglelefteqslant G$.
In other words, we are in the situation of Subsection \ref{reductionsprincipal} in the particular case where the Klein geometry is such that the isotropy group $I$ is a normal subgroup of the principal group $G$ and the homogeneous space  $H\cong G/I$ is thus a Lie group.
Accordingly, any principal $G$-bundle $P$ over $P/G$ is also a principal $I$-bundle over $Q:=P/I$ and the latter is itself a principal $H$-bundle $Q$ over $M:=\frac{P}G\cong\frac{Q}H$. 
This can be summarised in the following diagram where the two vertical arrows are fibrations of principal $I$-bundles while the two diagonal arrows are fibrations of fibre bundles over the same base $M$:
\be
\begin{array}
	[c]{ccccccc}%
	I&\trianglelefteqslant & G &  \lefttorightarrow & P & &\\
	& &  & & & \searrow & \\ 
	& &\pi\downarrow & & \downarrow \pi& & M\cong \frac{P}{G}\cong\frac{Q}{H} \\ 
	& & & & & \nearrow & \\ 
	& & H\cong \frac{G}{I} & \lefttorightarrow & Q\cong \frac{P}{I} & & 
\end{array}
\,.
\ee
The projection $\pi:P\twoheadrightarrow Q$ is an $H$-equivariant morphism of fibre bundles over the same base $M$. Note that the free action of $G$ on $P$ projects to the free action of $H$ on $Q$\,.
Accordingly, the principal $H$-bundle $Q$ will be called a \textbf{projection of the principal} $G$-\textbf{bundle} $P$.

Let $ \Gamma(VP)$ denote the sheaf of vertical vector fields of the fibre bundle $P$ over $Q$.
Infinitesimally,
\be
\mathfrak{i}^\#\subset \Gamma(VP)
\quad\text{and}\quad\text{dim}(P)-\text{dim}(\mathfrak{g})\,=\,\text{dim}(Q)-\text{dim}(\mathfrak{h})\,,
\ee
where $\mathfrak{i}$ denotes the Lie ideal of $\mathfrak{g}$ corresponding to the normal Lie subgroup $I\trianglelefteqslant G$ and
$\mathfrak{i}^\#$ is the Lie algebra of fundamental vector fields of the principal bundle $I\lefttorightarrow P \twoheadrightarrow Q$.
A principal (respectively, Cartan) connection on the principal $G$-bundle $P$ over $M$ always defines a principal (respectively, Cartan) connection on the principal $I$-bundle $P$ over $Q$. This is somewhat obvious since one simply restrict the right action to the normal subgroup $I\trianglelefteqslant G$.

\vspace{3mm}
\noindent\textbf{Example (Projection of Klein geometry)\,:}
The {projection of the Klein geometry} $H_2 \subseteq G_2$ onto the Klein geometry $H_1 \subseteq G_1$ is a surjective morphism $G_2 \twoheadrightarrow G_1$ of Lie groups such that 
\begin{enumerate}
	\item its restriction to the subgroups is also a surjective morphism $H_2 \twoheadrightarrow H_1$ of Lie groups,
	\be
	\begin{array}
		[c]{ccc}%
		H_2 & \hookrightarrow & G_2\\
		\downarrow & & \downarrow\\
		H_1 & \hookrightarrow & G_1
	\end{array}
	\,,
	\ee
	\item the kernels of the two surjective Lie group morphisms $G_2 \twoheadrightarrow G_1$ and $H_2 \twoheadrightarrow H_1$ coincide to the same normal subgroups, \textit{i.e.} $G_2$ (respectively, $H_2$) is a Lie group extension of $G_1$ (respectively, $H_1$) by the normal subgroup $I$, and
	\item the following homogeneous spaces are diffeomorphic, 
	\be
	\frac{G_1}{H_1}\cong\frac{G_2}{H_2}\,\,.
	\ee
	
\end{enumerate}
There is a fibration $G_2\twoheadrightarrow G_1$ of the principal $H_2$-bundle $G_2$ over the principal $H_1$-bundle $G_1$\,, which is a morphism of fibre bundles over the same $M\cong\frac{G_2}{H_2}\cong\frac{G_1}{H_1}$\,.
Moreover, the embedding $H_2\hookrightarrow G_2$ is a morphism of principal $I$-bundles, from the principal $I$-bundle $H_2$ over $H_1$ into the principal $I$-bundle $G_2$ over $G_1$.
All these facts can be summarised in the following commutative diagram where all arrows are Lie group morphisms, except the two diagonal ones on the right which are merely fibrations.
\be
\begin{array}
	[c]{ccccc}%
	& I & & & \\
	\qquad\swarrow & & \searrow\qquad & & \\
	& & & & \\
	H_2 &  \hookrightarrow & G_2 & &\\
	& & & \searrow & \\ 
	\downarrow & & \downarrow & & 
	\frac{G_2}{H_2}\cong \frac{G_1}{H_1} \\ 
	& & & \nearrow & \\ 
	H_1\cong \frac{H_2}{I} & \hookrightarrow & G_1\cong \frac{G_2}{I} & & \\
	& & & & 
\end{array}
\label{projKleingeom}
\ee
Note that following successively, either the horizontal or vertical arrows, one has short exact sequences of Lie groups.

The infinitesimal analogue is a \textbf{projection of the Klein pair} $\mathfrak{h}_2 \subseteq \mathfrak{g}_2$ onto the Klein pair $\mathfrak{h}_1 \subseteq \mathfrak{g}_1$ is a surjective morphism $\mathfrak{g}_2 \twoheadrightarrow \mathfrak{g}_1$ of Lie algebras such that its restriction to the subalgebras is also a surjective morphism $\mathfrak{h}_2 \twoheadrightarrow \mathfrak{h}_1$ of Lie algebras,
and such that the kernel $\mathfrak{i}\subseteq\mathfrak{h}_2\subseteq\mathfrak{g}_2$, of the surjections $\mathfrak{g}_2 \twoheadrightarrow \mathfrak{g}_1$ and
$\mathfrak{h}_2 \twoheadrightarrow \mathfrak{h}_1$, defines two extensions by the Lie algebra $\mathfrak{i}$: 
\be\label{projKleinpair}
\begin{array}
	[c]{cccc}%
	& \mathfrak{i} &   \\
	\qquad\swarrow & & \searrow\qquad  \\
	& &  \\
	\mathfrak{h}_2 & \hookrightarrow & \mathfrak{g}_2 \\
	\downarrow & & \downarrow \\
	\mathfrak{h}_1 & \hookrightarrow & \mathfrak{g}_1 
\end{array}
%
\ee
The images of these two vertical surjections are, respectively, isomorphic to the quotient spaces $\mathfrak{h}_1\cong \mathfrak{h}_2/\mathfrak{i}$ and $\mathfrak{g}_1\cong \mathfrak{g}_2/\mathfrak{i}$.
Therefore, the transvection modules are isomorphic vector spaces 
\be
\frac{\mathfrak{g}_1}{\mathfrak{h}_1}\cong\frac{\mathfrak{g}_2}{\mathfrak{h}_2}\,\,.
\ee

 \subsection{Klein geometries and split extensions}

 \subsubsection{Split extensions and semidirect sums}

A Lie algebra extension $\mathfrak{g}$ of $\mathfrak{h}$ by $\mathfrak{i}$ is called a
\begin{itemize}
	\item \textbf{central extension} if the isotropy ideal $\mathfrak{i}$ belongs to the centre of the principal algebra $\mathfrak{g}$.
	\item \textbf{split extension} if $\mathfrak{h}$ is a Lie subalgebra (which will be called the transvection subalgebra) via a splitting of the Klein sequence of $\mathfrak h$-modules, 
	\be
	0\to \mathfrak{i}
		\hookrightarrow
	\mathfrak{g}
		\twoheadrightarrow
	\mathfrak{h}\to 0
	\ee
	such that the section (\textit{i.e.} the first arrow from right to left) in the splitting 
	\be\label{KLeinreductt}
	0\leftarrow\mathfrak{i}
		\twoheadleftarrow
	\mathfrak{g}
		\hookleftarrow
	\mathfrak{h}\leftarrow 0\,.
	\ee
	is a Lie algebra morphism.
	\item \textbf{trivial extension} if it is a split extension such that the transvection subalgebra $\mathfrak{h}$ and the isotropy ideal $\mathfrak{i}$ commute with each other.
\end{itemize}
Note that, in a split extension, the retraction (\textit{i.e.} the second arrow from right to left) in the splitting \eqref{KLeinreductt} is \textit{not} a Lie algebra morphism in general, but only a linear map. In fact, all arrows in the split sequence \eqref{KLeinreductt} are Lie algebra morphisms if and only if the extension is trivial (since both $\mathfrak{i}$ and $\mathfrak{h}$ are Lie ideals in such case).

\vspace{5mm}
\begin{framed}
	\begin{center}
		\textbf{The many faces of direct sums of Lie algebras}
	\end{center}
	
	\noindent
	Given two Lie algebras $\mathfrak{i}$ and $\mathfrak{h}$, the following notions are equivalent with each other:
	\begin{enumerate}
		\item a direct sum $\mathfrak{g}=\mathfrak{i}\oplus\mathfrak{h}$ of Lie algebras.
		\item a trivial extension $\mathfrak{g}$ of $\mathfrak{h}$ by $\mathfrak{i}$.
		\item a reductive Klein pair $\mathfrak{h}\subseteq \mathfrak{g}$ such that the transvection module $\mathfrak{i}=\mathfrak{g}/\mathfrak{h}$ is a Lie ideal of the principal algebra $\mathfrak{g}$.
		\item a Lie algebra extension $\mathfrak{g}\supseteq \mathfrak{h}$ of the isotropy subalgebra $\mathfrak{h}$ by a transvection algebra $\mathfrak{i}=\mathfrak{g}/\mathfrak{h}$ which is a Lie ideal $\mathfrak{i}\subseteq\mathfrak{g}$ of the principal algebra.
		\item a splitting of a Klein sequence of Lie algebra morphisms such that all arrows in the splitting are also Lie algebra morphisms.
	\end{enumerate}
	\vspace{3mm}\end{framed}

Note that the Lie algebras $\mathfrak{i}$ and $\mathfrak{h}$ play a symmetric role in a direct sum and, in fact, one can exchange them everywhere in the above box. As an important generalisation of this notion, one should mention:

\vspace{5mm}
\begin{framed}
	\begin{center}
		\textbf{The many faces of semidirect sums of Lie algebras}
	\end{center}
	
	\noindent
	Given two Lie algebras $\mathfrak{i}$ and $\mathfrak{h}$, the following notions are equivalent with each other:
	\begin{enumerate}
		\item a semidirect sum $\mathfrak{g}=\mathfrak{h}\inplus\mathfrak{i}$ of Lie algebras
		\item a decomposition of a Lie algebra $\mathfrak{g}$ as the linear sum of a Lie subalgebra $\mathfrak{h}\subseteq\mathfrak{g}$ and a Lie ideal $\mathfrak{i}\subseteq\mathfrak{g}$
		\item a split extension $\mathfrak{g}$ of $\mathfrak{h}$ by $\mathfrak{i}$.
		\item a reductive Klein pair $\mathfrak{h}\subseteq \mathfrak{g}$ such that the transvection submodule $\mathfrak{i}=\mathfrak{g}/\mathfrak{h}$ is a Lie subalgebra of the principal algebra $\mathfrak{g}$.
		\item a Lie algebra extension $\mathfrak{g}\supseteq \mathfrak{h}$ of the isotropy subalgebra $\mathfrak{h}$ by a transvection algebra $\mathfrak{i}=\mathfrak{g}/\mathfrak{h}$ which is a Lie subalgebra $\mathfrak{i}\subseteq\mathfrak{g}$ of the principal algebra.
		\item a splitting of the Klein sequence of Lie algebra morphisms such that the splitting section is a Lie algebra morphism.
	\end{enumerate}
	\vspace{5mm}\end{framed}

Similarly, the following definitions of a semidirect product of Lie groups are equivalent with each other:

\vspace{5mm}
\begin{framed}
	\begin{center}
		\textbf{The many faces of semidirect products of Lie groups}
	\end{center}
	
	\noindent
	Given two Lie groups $I$ and $H$, the following notions are equivalent with each other:
	\begin{enumerate}
		\item a semidirect product $G=H\ltimes I$
		\item a split extension $G$ of $H$ by $I$.
		\item a reductive Klein geometry $H\subseteq G$ such that the homogeneous space $I=G/H$ is a closed normal Lie subgroup of $G$. The isotropy subgroup $H$ is then isomorphic to the quotient group $G/I$.
		\item a Klein geometry $I\trianglelefteqslant G$ such that the homogeneous space $G/I$ is a quotient group isomorphic to a closed Lie subgroup $H$ of $G$.
		\item a moving frame $s:G/I\hookrightarrow G$ of the principal $I$-bundle $\pi:G\twoheadrightarrow G/I$ which is a Lie group morphism embedding the Lie group $H=G/I$ into the principal group $G$.
		\item a splitting of the Klein sequence such that the splitting embedding itself defines a Klein geometry.
	\end{enumerate}
	\vspace{3mm}
\end{framed}

Given a semidirect product $G=H\ltimes I$ and a principal $G$-bundle $P$ over $M:=P/G$, a moving frame of the principal $I$-bundle $P$ over $Q:=P/I$ is a reduction of the principal $G$-bundle $P$ to the principal $H$-bundle $Q$, which is in turn equivalent to a section of the associated Klein geometry bundle $E:=P/H$ over $M$.
This can be seen in the following diagram: 
\be
\begin{array}
	[c]{ccccc}%
	I & & & & \\
	& & & & \\ 
	\downuparrows & & & & \\ 
	& & & & \\ 
	G &  \lefttorightarrow & P & \twoheadrightarrow & E\cong \frac{P}{H}\\
	& & & & \\ 
	\downuparrows  & & \downuparrows & & \downuparrows \\ 
	& & & & \\ 
	H\cong \frac{G}{I} & \lefttorightarrow & Q\cong \frac{P}{I} & \twoheadrightarrow & M\cong \frac{P}{G}\cong\frac{Q}{H}
\end{array}
\ee
The vertical arrows on the left are the split short exact sequences defining the semidirect product. 
The two horizontal arrows on the right are fibrations of principal $H$-bundles.
One should stress that a \textit{complete} gauge $i:Q\hookrightarrow P$ of the principal $I$-bundle $P$ over $Q$ is just a \textit{partial} gauge of the principal $G$-bundle $P$ over $M$.
Let $\omega^\mathfrak{g}$ be a principal connection one-form on $P$. 
The Klein pair $\mathfrak{h}\subseteq \mathfrak{g}$ is reductive thus the corestriction $\omega^{\mathfrak{h}}$ of the gauge principal connection one-form $i_*\omega^\mathfrak{g}$ to the codomain $\mathfrak{h}$ is the reduced principal connection one-form on $Q$.

A reduction of a Klein pair $\mathfrak{g}_1 \subseteq \mathfrak{g}_2$ to a Klein pair $\mathfrak{h}_1 \subseteq \mathfrak{h}_2$
providing a splitting of the projection \eqref{projKleinpair} of Klein pairs, in the sense that $\mathfrak{h}_2=\mathfrak{h}_1\inplus\mathfrak{i}$ and $\mathfrak{g}_2=\mathfrak{g}_1\inplus\mathfrak{i}$,
will be called a \textbf{splitting of Klein pairs},
\be\label{reduxsplitKleinpair}
\begin{array}
	[c]{ccc}%
	\mathfrak{i}& &\\
	\downuparrows & & \\
	\mathfrak{h}_2 & \hookrightarrow & \mathfrak{g}_2\\
	\downuparrows & & \downuparrows\\
	\mathfrak{h}_1 & \hookrightarrow & \mathfrak{g}_1
\end{array}
\,.
\ee
In particular, the Klein pairs $\mathfrak{h}_1\subseteq\mathfrak{h}_2$ and $\mathfrak{g}_1\subseteq\mathfrak{g}_2$ are reductive.
Moreover, the definition implies that the isotropy and principal algebras of the Klein pair $\mathfrak{h}_2\subseteq\mathfrak{g}_2$
decompose into the semidirect sums $\mathfrak{h}_2=\mathfrak{h}_1\inplus\mathfrak{i}$ and $\mathfrak{g}_2=\mathfrak{g}_1\inplus\mathfrak{i}$ of Lie algebras and of $\mathfrak{h}_1$-modules.

Let $\mathfrak{h}_1 \subseteq \mathfrak{h}_2$ be a splitting of the Klein pair $\mathfrak{g}_1\subseteq \mathfrak{g}_2$ as defined above.
Let $i:P_1\hookrightarrow P_2$ be a gauge of the principal $I$-bundle $P_2$ over $P_1$ defining a reduction of the $H_2$-bundle $P_2$ to the $H_1$-bundle over $P_1$. Let $\omega^{\mathfrak{g}_2}$ be a Cartan connection one-form on $P_2$ modeled on the Klein pair
$\mathfrak{h}_2\subseteq\mathfrak{g}_2$.
The corestriction $\omega^{\mathfrak{g}_1}$ of the gauge Cartan connection one-form $i_*\omega^{\mathfrak{g}_2}$ 
to the codomain $\mathfrak{g}_1$ is a Cartan connection on $P_1$ modeled on the Klein pair
$\mathfrak{h}_1\subseteq\mathfrak{g}_1$
which will be called the \textbf{projected Cartan connection} on the reduced bundle.
Notice that, in general, the projection of a flat Cartan connection is \textit{not} flat. Indeed, because the restriction $\Omega^{\mathfrak{g}_1}$ of the gauge curvature two-form $i^*\Omega^{\mathfrak{g}_2}$ to the codomain $\mathfrak{g}_1$ may include contributions coming from the Lie bracket $[\omega^{\mathfrak{g}_1},\omega^{\mathfrak{i}}]$ where $\omega^{\mathfrak{i}}$ is the restriction of the connection one-form
$\omega^{\mathfrak{g}_2}$ to the codomain $\mathfrak{i}$.

\section*{Acknowledgments}

I thank Kevin Morand for various discussions on Cartan and Ehresmann connections, as well as Yannick Herfray for discussions on Weyl connections and conformal geometry in general. I am particularly grateful to Thomas Basile for his patient reading and useful comments on some early version of these notes.

\pagebreak

\section*{\Huge Extra material}

\addcontentsline{toc}{section}{Extra material}

\appendix

\stopcontents[main]

\startcontents[appendices]

\printcontents[appendices]{}{1}

\clearpage

\section{Induced representations, parabolic geometries and higher-spin algebras}\label{inducedreps}

\subsection{Homogeneous vector bundles }\label{homvectbundles}

Let $\pi:G\twoheadrightarrow G/H$ be the principal $H$-bundle defined by the Klein geometry $H\subseteq G$.
Consider a linear representation $r\,:\,H\to GL({V})\,:\,h\mapsto r_h$
of the isotropy group $H$ on the vector space $V$. 
The associated vector bundle $\mathbb{V}=G\times_{r_H}V$ is called a \textbf{homogeneous vector bundle}. 

\vspace{2mm}\textbf{Examples:}
\begin{itemize}
	\item[$\bullet$] \underline{Trivial bundle:} If the linear representation of the isotropy subgroup $H$ on the vector space $V$ extends to a linear representation $r:G\to GL({V}):g\mapsto r_g$ of the principal group $G$, then the corresponding homogeneous vector bundle is trivial
	\be
	G\times_{r_H}V\cong\frac{G}{H}\times V
	\ee
	where the above isomorphism is the map 
	\be
	G\times_{r_H}V\stackrel{\sim}{\to}\frac{G}{H}\times V:[g,v]\mapsto(gH,gv)\,,
	\ee 
	which is well-defined since $[gh,h^{-1}v]\mapsto\big(\,(gh)H\,,\,(gh)h^{-1}v\,\big)=(gH,gv)$\,. Actually, any homogeneous vector bundle $G\times_{r_H}V$
	is trivial as a homogeneous vector bundle if and only if $r_H$ is
	the restriction to $H$ of a representation $r:G\to GL({V})$, see \textit{e.g.} \cite[Observation II.A.i]{Green:2014}.
	\item[$\bullet$] \underline{Ati}y\underline{ah bundle:} For $H$ acting on the principal algebra $\mathfrak{g}$ via the adjoint representation,  the corresponding homogeneous vector bundle is the adjoint tractor bundle isomorphic to the Atiyah bundle	of the principal $H$-bundle $G$,
	\be
	G\times_{Ad_H}\mathfrak{g}\cong	\frac{TG}{H}\,.
	\ee
	The adjoint action can be restricted to the isotropy subalgebra and to the transvection module.
	\item[$\bullet$] \underline{Ad}j\underline{oint bundle:} For $H$ acting on the isotropy subalgebra $\mathfrak{h}$ via the adjoint representation, the corresponding homogeneous vector bundle is the adjoint bundle of the principal $H$-bundle $G$ isomorphic to the vertical sub-bundle of the Atiyah bundle,
	\be
	G\times_{Ad_H}\mathfrak{h}\cong	\frac{VG}{H}\,.
	\ee 
	\item[$\bullet$] \underline{Tan}g\underline{ent bundle:} For $H$ acting on the transvection module $\mathfrak{g}/\mathfrak{h}$ via the adjoint representation, the corresponding homogeneous vector bundle is the tangent bundle 	of the homogeneous space $G/H$,
	\be
	T\frac{G}{H}=G\times_{Ad_H}\mathfrak{g}/\mathfrak{h}\,.
	\ee
	This works as well for its dual and corresponding tensor products.
\end{itemize}
\vspace{2mm}

\subsubsection{Induced representations for Lie groups}

The principal group $G$ acts on the $r_H$-associated vector bundle $\mathbb{V}$ from the left via the left multiplication on $G$, via $g(p,v):=(gp,v)$.
The $r_H$-equivariant sheaf ${C}^\infty(G)\otimes_{r_H} V$ on the principal group is called the \textbf{induced representation space} because, via the left multiplication, it carries a linear representation of $G$ called the \textbf{representation of the principal group} $G$ \textbf{induced from the representation} $r$ \textbf{of the isotropy subgroup} $H$. In this case, the induced representation space is denoted 
\be
\text{Ind}{}^G_H(r)\,:=\,{C}^\infty(G)\,\otimes_{r_H} V\,.
\ee 
Sometimes, it will also be denoted $\text{Ind}{}^G_H(V)$ when considering the relation between induced representations for Lie groups and for Lie algebras.

\vspace{3mm}
\noindent{\small\textbf{Example (Free module)\,:} If the linear representation of $H$ on $V$ extends to a linear representation $r$ of the principal group $G$, then the corresponding homogeneous vector bundle is trivial $G\times_{r|_{{}_H}}V\cong\frac{G}{H}\times V$ and, therefore, the induced representation space is the tensor product of the structure sheaf ${C}^\infty(G/H)$ with the vector space $V$:
	\be
	\text{Ind}{}^G_H(r|_{{}_H})\,\cong\,{C}^\infty(G/H)\,\otimes\,V\,.
	\ee 
}

\vspace{3mm}

Infinitesimally, the induced representation space $\text{Ind}{}^G_H(r)$ is a $\mathfrak{g}$-module defined by the principal action $\sharp$ of $\mathfrak{g}$ on $G$ and by the representation $r$ of $\mathfrak h$ on $V$. The corresponding representation will be called the \textbf{representation of the principal algebra} $\mathfrak g$ \textbf{induced from the representation} $r$ \textbf{of the isotropy subalgebra} $\mathfrak h$.
The associated sheaf $\Gamma({\mathbb V})$ of a homogeneous vector bundle $\mathbb{V}=G\times_{r_H}\text{V}$ will be called the \textbf{induced sheaf}.
The isomorphism \eqref{isjhomsh} implies that the induced representation space $\text{Ind}{}^G_H(r)$ and the induced sheaf $\Gamma(\mathbb{V})$ carry equivalent representations, which motivates our choice of terminology. The action of $G$ on $r_H$-equivariant maps $f\in\text{Ind}{}^G_H(r)$ 
is as follows: $(g\cdot f)(g'):=r(g)\big(\,f(g^{-1}g')\,\big)$.
Note that, given a Klein geometry $H\subseteq G$, there is a one-to-one correspondence between linear representations of the isotropy subalgebra $\mathfrak{h}$ on the vector space $V$ and covariant actions of the principal algebra $\mathfrak{g}$ on the homogeneous vector bundle $\mathbb{V}=G\times_{r_H}\text{V}$.\footnote{See \textit{e.g.} \cite[Example 2.6]{Kosmann-Schwarzbach} for a proof.} Or, more concisely, there is a one-to-one correspondence between $\mathfrak{h}$-modules $V$ and induced $\mathfrak{g}$-modules $\text{Ind}{}^G_H(r)\cong\Gamma(G\times_{r_H}\text{V})$.

\subsubsection{Induced representations for Lie algebras}

Consider a linear representation 
$r\,:\,\mathfrak{h}\to \mathfrak{gl}(V)\,:\,y\mapsto r_y$
of the isotropy subalgebra $\mathfrak{h}$ on the vector space $V$. Let $V^*$ denote the dual $\mathfrak{h}$-module carrying the contragredient representation $r^*$.

The relative tensor product  of the universal enveloping algebra ${\mathcal U}(\mathfrak{g})$ of the principal algebra with the dual $\mathfrak{h}$-module $V^*$ quotiented by the universal enveloping algebra ${\mathcal U}(\mathfrak{h})$ of the isotropy subalgebra will be denoted  
\be\label{reltensproduct}
\text{ind}^{\mathfrak{g}}_{\mathfrak{h}}(V)\,:=\,{\mathcal U}(\mathfrak{g})\otimes_{\,{\mathcal U}(\mathfrak{h})}V^*
\ee
and called the \textbf{module of the principal algebra} $\mathfrak{g}$ \textbf{induced from the module} $V$
\textbf{of the isotropy subalgebra} $\mathfrak{h}$.
The choice of performing the relative tensor product with the dual module is motivated by the observation that the representation $\text{Ind}{}^G_H(V)$ of the principal group $G$, induced from the representation $r$ of the isotropy subgroup $H$ on $V$, is related via duality (see next subsection) to the module $\text{ind}^{\mathfrak{g}}_{\mathfrak{h}}(V)$ of the principal algebra $\mathfrak{g}$, induced from the corresponding representation $r$ of the isotropy subalgebra $\mathfrak{h}$ on $V$.

Consider a principal algebra which is the semidirect sum of Lie algebras $\mathfrak{g}=\mathfrak{h}\inplus\mathfrak{p}$ of the transvection ideal $\mathfrak{p}$ and the isotropy subalgebra $\mathfrak{h}$.  
The Poincar\'e-Birkhoff-Witt theorem implies the isomorphism of $\mathfrak{g}$-modules between the module of $\mathfrak{g}=\mathfrak{h}\inplus\mathfrak{p}$ induced from the $\mathfrak{h}$-module $V$ and
the tensor product of the universal enveloping algebra of the transvection ideal $\mathfrak{p}$ with the dual $\mathfrak{h}$-module $V$,
\be\label{isoPBWindmod}
\text{ind}^{\mathfrak{g}}_{\mathfrak{h}}(V)\,\cong\,{\mathcal U}(\mathfrak{p})\otimes V^*\,.
\ee
More generally, the isomorphism \eqref{isoPBWindmod} holds at the level of vector spaces as long as the principal algebra decomposes as a sum $\mathfrak{g}=\mathfrak{h}\oplus\mathfrak{p}$ of vector spaces where each summand is a subalgebra (but $\mathfrak{p}$ may not be a submodule under the adjoint action of $\mathfrak{h}$).

\subsubsection{Relation between the induced representations for Lie algebras and Lie groups}

The linear dual of the $\mathfrak{g}$-module induced from the representation $r$ on the $\mathfrak{h}$-module $V$ is related to the representation of the principal group $G$ induced from the representation $r$ of the isotropy subgroup $H$ via the infinite-jet bundle construction.

The notion of a jet of order $k$ is very general and is defined, for any fibre bundle $\pi:E\twoheadrightarrow M$, as an equivalence class of sections $\sigma:M\hookrightarrow E$ that ``touch'' till order $k$ at a point (in the sense that all their partial derivatives coincide till order $k$ at this point). We will follow the notation of \cite{Saunders}, so the $k$-jet bundle of $\pi$ will be denoted as $J^k\pi$ and its fibration on the base manifold $M$ is written $j^k\pi:J^k\pi\twoheadrightarrow M$ (with $J^0\pi=E$ and $\pi_0=\pi$). The other fibrations on the lower-jet bundles are written $\pi_{k\to l}:J^k\pi\twoheadrightarrow J^l\pi$ (with $l<k$). In a local trivialisation, the coordinates of a point in $J^k\pi$ read as ($x^\mu$, $y^i$, $y_\mu^i$, \ldots, $y_{\mu_1\ldots\mu_k}^i$) where the coordinates $y_{\mu_1\ldots\mu_k}^i$ denote the possible values of the $k$th derivatives of a section at that point.

A representation $r:H\to GL(V):h\mapsto r_h$ of the isotropy group $H$ on $V$ defines a representation $r:\mathfrak{h}\to \mathfrak{gl}({V}):y\mapsto r_y$ of the isotropy subalgebra on the same vector space. 
In turn, this defines the contragredient representation $r^*:\mathfrak{h}\to \mathfrak{gl}({V}^*):y\mapsto r^*_y$ of the isotropy subalgebra on the dual vector space $V^*$.

Let $\pi:\mathbb{V}\twoheadrightarrow G/H$ be the fibration of the homogeneous vector bundle $\mathbb{V}=G\times_{r_H}V$\,.
The linear dual of the $\mathfrak{g}$-module induced from the representation $r$ on the $\mathfrak{h}$-module $V$ is isomorphic to the jet space $J^\infty_{eH}\pi$ of $\infty$-jets of $r_H$-covariant fields, at the identity left coset\footnote{For more details, see \textit{e.g.} the paper \cite{Eastwood:1987ki} or the lecture notes \cite[Lecture 3]{Eastwood:1996}.}
\be\label{reltensprod}
J^\infty_{eH}\pi\,\cong\,\Big(\,\text{ind}^{\mathfrak{g}}_{\mathfrak{h}}(V)\,\Big)^*\,.
\ee
Consequently, the linear dual of the homogeneous vector bundle $G\times_{r_H}\text{ind}^{\mathfrak{g}}_{\mathfrak{h}}(V)$ is the $\infty$-jet bundle $J^\infty \pi$ of the homogeneous vector bundle $\mathbb{V}=G\times_{r_H}V$.

\subsubsection{Higher-spin algebras}

The quotient algebra of the universal enveloping algebra ${\mathcal U}(\mathfrak{g})$ of the principal algebra by the annihilator $\text{Ann}\,\Gamma(\mathbb{V})$ of the induced sheaf, will be called the \textbf{off-shell higher-spin algebra associated to the homogeneous vector bundle} $G\times_{r_H}\text{V}$ \textbf{for the Klein pair} $\mathfrak{h}\subset\mathfrak{g}$ and will be denoted as 
\be
\texttt{hs}(\mathfrak{g}/\mathfrak{h};\text{V})\,:=\,{\mathcal U}(\mathfrak{g})/\text{Ann}\,\Gamma(G\times_{r_H}\text{V})\,.
\ee
By construction, the corresponding representation of the off-shell higher-spin algebra on the induced sheaf is faithful.

\vspace{3mm}
\noindent{\small\textbf{Example (Maurer-Cartan geometry)\,:} If the isotropy subgroup is trivial, $H=\{e\}$, then the homogeneous vector bundle $G\times V$ is trivial and the induced representation space is the tensor product $\text{Ind}{}^G_{\{e\}}(V)={C}^\infty(G)\otimes V$ of the structure sheaf ${C}^\infty(G)$ with the vector space $V$. 
	The representation of the universal enveloping algebra ${\mathcal U}(\mathfrak{g})$ of the Lie algebra $\mathfrak{g}$ on the structure algebra ${C}^\infty(G)$ of the Lie group $G$ is faithful, therefore the off-shell higher-spin algebra of a Maurer-Cartan geometry coincides with  the universal enveloping algebra: $\texttt{hs}(\mathfrak{g}/\{0\};\text{V})={\mathcal U}(\mathfrak{g})$.
}

\vspace{3mm}
\noindent{\small\textbf{Example (Trivial representation)\,:} In the particular case of the one-dimensional trivial representation
	of the isotropy subgroup $H$, the homogeneous vector bundle $\mathbb{V}=\frac{G}{H}\times\mathbb{R}$ is the trivial line bundle over the homogeneous space $G/H$ and
	the induced representation space $\text{Ind}{}^G_H(1)\cong{C}^\infty(G/H)$ is isomorphic to the structure sheaf. The off-shell higher-spin algebra associated to this homogeneous line bundle will be called the \textbf{higher-spin structure algebra} \textbf{for the Klein pair} $\mathfrak{h}\subset\mathfrak{g}$ and denoted as $\texttt{hs}(\mathfrak{g}/\mathfrak{h})$. 
}

\vspace{3mm}

The annihilator $\text{Ann}\,W\subset{\mathcal U}(\mathfrak{g})$ of an irreducible $\mathfrak{g}$-module $W$ is called a \textbf{primitive ideal} of the universal enveloping algebra. If the irreducible $\mathfrak{g}$-module $W$ is obtained as a quotient of an induced representation space $\text{Ind}^G_H(r)$ then, the quotient algebra ${\mathcal U}(\mathfrak{g})/\text{Ann}\,W$ of the universal enveloping algebra ${\mathcal U}(\mathfrak{g})$ by the primitive ideal $\text{Ann}\,W$, will be called the \textbf{on-shell higher-spin algebra associated to the homogeneous vector bundle} $\mathbb{V}$ \textbf{for the Klein pair} $\mathfrak{h}\subset\mathfrak{g}$\,. By construction, the induced representation of the universal algebra ${\mathcal U}(\mathfrak{g})$ of the principal algebra on the quotient of the equivariant module $\text{Ind}^G_H(r)$ is faithful.
If the induced representation space $\text{Ind}^G_H(r)$ is irreducible, then the corresponding higher-spin algebra is both an off-shell and an on-shell higher-spin algebra, in which case it will be called a \textbf{on/off-shell higher-spin algebra}.

For all classical Lie algebras $\mathfrak{g}$, there is an irreducible $\mathfrak{g}$-module $J_{\mathfrak{g}}$ called the \textbf{minimal representation} (see \cite{Joseph:1974} for a rigorous definition) which is ``minimal'' in the sense that it arises as a quotient of an induced representation space with minimal Bernstein and Gelfand-Kirillov dimensions.\footnote{See \textit{e.g.} \cite{Vogan} for the technical definition of ``Bernstein dimension'' and ``Gelfand-Kirillov dimension'' of Harish-Chandra (infinite-dimensional) modules.} Intuitively, the dimension of the corresponding transvection module $\mathfrak{g}/\mathfrak{h}$ must be minimal (\textit{i.e.} the dimension on the coset space $G/H$ on which this field lives) and the dimension of the $\mathfrak{h}$-module $V$ must also be minimal (\textit{i.e.} the vector space in which this field takes values). Its annihilator ${\mathcal J}({\mathfrak{g}}):=$\,Ann\,$J_{\mathfrak{g}}\,$, is a primitive ideal called the \textbf{Joseph ideal} \cite{Joseph:1976}. The quotient algebra ${\mathcal U}(\mathfrak{g})/{\mathcal J}({\mathfrak{g}})$ is an on-shell higher-spin algebra, which will be called the \textbf{higher-spin algebra of the classical Lie algebra} $\mathfrak{g}$ with the natural notation $\texttt{hs}(\mathfrak{g})$
\cite{Joung:2014qya}.

\subsection{Parabolic Klein geometries}

Let $\mathfrak{g}$ be a finite-dimensional semisimple Lie algebra and $\mathfrak{g}_0\subset\mathfrak{g}$ be a Cartan subalgebra.\footnote{See e.g. \cite[Subsection 3.2.1]{Cap} for a very concise introduction to the material reviewed in the following paragraph.}

A $\mathfrak{g}_0$-diagonal module is called a \textbf{weight space}, the elements of which are called \textbf{weight vectors}. A \textbf{weight module} is a $\mathfrak{g}$-module equal to the (automatically, direct) sum of its weight spaces.
The eigenvalues of the Cartan subalgebra are encoded in an element $\lambda\in\mathfrak{g}^*_0$ called \textbf{weight}.
A non-vanishing weight for the adjoint representation $ad:\mathfrak{g}\to\mathfrak{gl}(\mathfrak{g})$ is called a \textbf{root}.
A semisimple Lie algebra admits a \textbf{root space decomposition} $\mathfrak{g}=\mathfrak{g}_-\oplus\,\mathfrak{g}_0\,\oplus\mathfrak{g}_+$ as a vector space, where $\mathfrak{g}_\pm$ are maximal nilpotent Lie subalgebras associated with the positive/negative roots. They will be called \textbf{raising/lowering nilpotent subalgebras} $\mathfrak{g}_\pm$ of the finite-dimensional semisimple Lie algebra $\mathfrak{g}$. 
The maximal sovable Lie algebras $\mathfrak{b}_\pm=\mathfrak{g}_0\inplus\mathfrak{g}_\pm$ will be called \textbf{raising/lowering Borel subalgebras} of $\mathfrak{g}$ for this root system.
A $\mathfrak{g}$-module ${\mathcal W}_\pm:={\mathcal U}(\mathfrak{g}_{{}_\mp})v_\pm$ generated by a non-vanishing weight vector $v_\pm$ annihilated by the raising/lowering Borel subalgebra, \textit{i.e.} $\mathfrak{g}_\pm v_\pm=0$, is called a \textbf{highest/lowest weight module}. 

\paragraph{Basic results in the representation theory of semisimple Lie algebras:}
Highest/lowest weight modules are always indecomposable, but not necessarily irreducible (when they are infinite-dimensional). However, a finite-dimensional module is always fully reducible and it is highest/lowest-weight if and only if it is irreducible. Therefore, every finite-dimensional module decomposes into a (finite) sum of (finite-dimensional) highest/lowest-weight modules.

\subsubsection{Parabolic Klein geometries}\label{parabolicKleingeom}

A proper Lie subalgebra $\mathfrak{h}$ containing a Borel subalgebra (\textit{i.e.} ${\mathfrak b}_\pm\subseteq{\mathfrak h}\subset{\mathfrak g}$) is called a \textbf{parabolic subalgebra} (with similar terminology for Lie groups) \cite[Definition 3.2.1]{Cap}.\footnote{Strictly speaking, this definition applies to complex Lie algebras. There are some subtleties for real Lie algebras. In such case, the proper definition is to reduce the problem to the complex case: a Lie subalgebra $\mathfrak{h}\subset\mathfrak{g}$ of a \textit{real} Lie algebra $\mathfrak{g}$ is said parabolic if the complexification $\mathfrak{h}_{\mathbb C}\subset\mathfrak{g}_{\mathbb C}$ of the isotropy subalgebra is a parabolic subalgebra of the complexification $\mathfrak{g}_{\mathbb C}$ of the principal algebra.} In down-to-earth terms, a raising/lowering Borel subalgebra is generated by a Cartan subalgebra together with all its raising/lowering elements while for a parabolic subalgebra one may add extra lowering/raising operators.

A Klein geometry $H\subseteq G$ with a principal group $G$ which is semisimple and with an isotropy subgroup $H$ which is parabolic, will be called a \textbf{parabolic Klein geometry}. The homogeneous space $G/H$ of a parabolic Klein geometry is called a \textbf{generalised flag manifold} \cite[Subsection 3.2.6]{Cap}.

\vspace{3mm}
\noindent{\small\textbf{Example (Flag manifolds)\,:} A \textbf{flag} in a vector space $V$ of finite dimension $n\in\mathbb N$ is a collection $\{V_i\}$ of proper linear subspaces, such that $V_i\subset V_j$ for $i<j$, where $i\in\{0,1,\cdots,k\}$ (with $1\leqslant k\leqslant n$) such that $V_0=\{0\}$ and $V_k=V$.\footnote{The definition of $\mathbb N$-filtered vector space is a generalisation of the definition of flag to the infinite-dimensional case.} The dimensions of the subspaces is called the \textbf{signature of the flag}. A \text{flag variety (or flag manifold) of given signature} is the space of all flags with this fixed signature. It is a homogeneous space with principal group $G=SL(n)$, the semisimple group of matrices with unit determinant and with isotropy subgroup $H\subset G$, the parabolic subgroup of unit determinant matrices which are block upper-triangular, where the dimensions of the blocks are the differences of the dimensions of subspaces.}

\vspace{3mm}
\noindent{\small\textbf{Example (Grassmannians)\,:} A Grassmannian $Gr(m,n)$ is a flag variety with $k=2$ and signature $0<m<n$, since it is the manifold of all $m$-dimensional linear subspaces $V_1$ of an $n$-dimensional vector space $V_2$. It is the homogeneous space of the parabolic Klein geometry $B(m,n-m)\subset SL(n)$ where the isotropy subgroup $H=B(m,n-m)$ is the subgroup of block upper-triangular matrices of blocks $m\times m$ and $(n-m)\times(n-m)$ with unit determinant. The corresponding Klein pair is the principal algebra $\mathfrak{g}=\mathfrak{sl}(n)$ spanned by traceless $n\times n$ matrices and the isotropy subalgebra $\mathfrak{h}=\mathfrak{b}(m,n-m)$ spanned by traceless upper-triangular matrices for blocks $m\times m$ and $(n-m)\times(n-m)$. The isotropy subalgebra decomposes as the semidirect sum $\mathfrak{h}=\mathfrak{h}_0\inplus\mathfrak{i}$ of the subalgebra $\mathfrak{h}_0\cong\big(\,\mathfrak{gl}(m)\oplus\mathfrak{gl}(n-m)\,\big)/\mathfrak{u}(1)$ spanned by traceless block-diagonal matrices and the Lie ideal $\mathfrak{i}$ spanned by linear maps from $\mathbb{R}^{n-m}$ to $\mathbb{R}^m$.\footnote{See \textit{e.g.} \cite[Example 3.2.8]{Cap} for more details.}}
	
	\vspace{3mm}
	\noindent{\small\textbf{Example (Klein geometry of the real projective space)\,:} The Klein geometry with the projective group $G=PGL(n+1)$ as semisimple principal group and the linear projective group $H$ (generated by general linear and special projective transformations) as parabolic isotropy subgroup is the model of projective geometry  \cite[Subsection 1.1.3]{Cap}. Its homogeneous space $\mathbb{R}\mathbb{P}^n$ is the projective space, \textit{i.e.} the space of rays of the ambient space ${\mathbb R}^{n+1}$, it is diffeomorphic to the quotient $\mathbb{S}^n/\mathbb{Z}_2$.\footnote{The projective space $\mathbb{R}\mathbb{P}^n$ can also be understood recursively as the affine space $\mathbb{R}^n$ to which one adds a projective hyperplane at infinity (diffeomorphic to $\mathbb{R}\mathbb{P}^{n-1}$).}}
	
	\vspace{3mm}
	\noindent{\small\textbf{Example (M\" obius model)\,:} The Klein geometry with the conformal group $G=O(n+1,1)$ as semisimple principal group and the isotropy subgroup $H\subset G$, with $H\cong ICO(n)$ generated by rotations, dilations and special conformal transformations as parabolic\footnote{Note that the parabolic subgroup $ICO(n)\subset O(n+1,1)$ is not a Borel subgroup because it contains the \textit{semisimple} group $O(n)$ or rotations.} subgroup is the M\" obius model \cite[Subsection 1.6.2]{Cap}. Its homogeneous space is the space of rays of the null cone inside the ambient Minkowski spacetime ${\mathbb R}^{n+1,1}$. Being  diffeomorphic to the hypersphere $\mathbb{S}^n$, this homogeneous space is often called the \textbf{celestial (or conformal) sphere}.}
	
	\vspace{3mm}
	\noindent{\small\textbf{Counter-examples (Hyperbolic and spherical geometries)\,:} A semisimple algebra necessarily contains a Cartan subalgebra together with all its raising \textit{and} lowering elements. For this reason, a parabolic subalgebra containing a semisimple proper subalgebra \textit{cannot} be a Borel subalgebra. Similarly, a parabolic subalgebra \textit{cannot} be semisimple. Therefore, spherical geometry on $\mathbb{S}^n=O(n+1)/O(n)$ and hyperbolic geometry on $\mathbb{H}^n=O(n,1)/O(n)$ are \textit{not} parabolic geometries since their isotropy algebras are semisimple.}
	
	\vspace{3mm}
	
	A filtration of a $\mathfrak{g}$-module $V$, \textit{i.e.}
	\be\label{compositionseries}
	0=V_0\subset V_1\subset\cdots \subset V_{\ell-1}\subset V_{\ell}=V\,,
	\ee
	where each inclusion $V_i\subset V_{i+1}$ is strict and such that each $V_i$ is a maximal $\mathfrak{g}$-submodule of $V_{i+1}$, is called a \textbf{composition series} of a $\mathfrak{g}$-module \cite[Definition 2.5.1]{Beachy} and $\ell$ is called the length of $V$ \cite[Definition 2.5.3]{Beachy}.
	The quotient modules $W_{i+1}:=V_{i+1}/V_i$ are simple and called the \textbf{composition factors} of $V$,
	\textit{i.e.} one has the short exact sequences of $\mathfrak{g}$-module morphisms:
	\be\label{compositionfactors}
	0\to V_i\hookrightarrow V_{i+1}\twoheadrightarrow W_{i+1}\to 0\,,
	\ee
	often summarised in the notation $V_{i+1}=W_{i+1}\inplus V_i$.
	A standard notation encoding both the filtration \eqref{compositionseries} and the composition factors \eqref{compositionfactors} is the iteration of the previous notation:
	\be
	V=W_{\ell}\inplus W_{\ell-1}\inplus \cdots\inplus W_1
	\ee
	The Jordan–H\"older theorem asserts that the number of occurrences of each simple module as a composition factor does not depend on the choice of composition series (up to isomorphisms) \cite[Theorem 2.5.2]{Beachy}.
	
	\vspace{3mm}
	\noindent{\small\textbf{Example (Null line)\,:} Consider a null line $N$ inside the Minkowski spacetime ${\mathbb R}^{n+1,1}$. Its stabiliser is the parabolic subgroup $ICO(n) \subset O(n+1,1)$ \cite[Subsection 1.6.3]{Cap}.\footnote{Remember the remark in Subsection \ref{MCKC}: with a slight abuse of notation, for all concrete examples of Klein geometries the corresponding groups should be implicitly understood as the connected component thereof.}
		The hyperplane $N^\perp$ orthogonal to $N$ is null since it contains the line $N$ itself. There is a composition series
		\be
		0\subset N\subset N^\perp \subset {\mathbb R}^{n+1,1}\,,
		\ee
		of invariant spaces of $ICO(n)$ (or, infinitesimally, of $\mathfrak{ico}(n)$-modules).
	}
	
	\vspace{3mm}
	The \textbf{Langlands decomposition} of the parabolic subgroup $H$ is its decomposition $H=MAN$ as a product of three subgroups where $M$ is reductive, $A$ is Abelian and $N$ is nilpotent. An induced representation for which the representation $r$ of the parabolic subgroup $H=MAN$ is a product $\mu\otimes\alpha\otimes 1$ of a finite-dimensional representation $\mu$ of $M$, a one-dimensional representation of $\alpha$ and the trivial representation of $N$ carries a representation of $G$ called an \textbf{elementary representation} of $G$. The importance of these representation comes from the theorem of Langlands establishing that every irreducible (admissible) representation of a real connected semisimple
	Lie group $G$ (with finite center) is equivalent to a subrepresentation of an elementary representation of $G$ (see \textit{e.g.} \cite[Section 4.3.2]{Dobrev}).
	The induced representation $\mathfrak{g}\to\mathfrak{gl}\big(\Gamma(G\times_{r_H}V)\big)$ of the principal algebra $\mathfrak{g}$ on the induced sheaf $\Gamma(G\times_{r_H}V)$ is equivalent to the Lie algebra representation arising from the elementary representation $G\to GL\big(\text{Ind}{}^G_H(r)\big)$ of the principal group on the induced representation space. The theorem of Langlands above thus suggests that every on-shell higher-spin algebra based on a semisimple principal algebra arises as a quotient of an off-shell higher-spin algebra.
	
	\vspace{3mm}
	\noindent{\small\textbf{Example (M\" obius model)\,:} 
		Consider the M\" obius model of conformal geometry with parabolic Klein pair $\mathfrak{h}\subset\mathfrak{g}$, where the conformal isotropy subalgebra $\mathfrak{h}\cong\mathfrak{ico}(n)$ forms a parabolic subalgebra of the conformal algebra $\mathfrak{g}=\mathfrak{o}(n+1,1)$. The Iwasawa decomposition (\textit{i.e.} the decomposition of a real semisimple Lie algebra into its maximal
		compact subalgebra, an abelian and a nilpotent subalgebra  \cite[Proposition 2.3.5]{Cap}) for the conformal algebra is its decomposition into its maximal compact subalgebra $\mathfrak{k}=\mathfrak{o}(n+1)$, the abelian subalgebra $\mathfrak{a}=\mathfrak{o}(1,1)$ generated by the dilation operator, and the nilpotent subalgebra $\mathfrak{n}=\mathbb{R}^{n*}$ generated by the special conformal transformations. The centraliser of the one-dimensional subalgebra $\mathfrak{a}=\mathfrak{o}(1,1)$ generated by the dilation operator is the subalgebra $\mathfrak{m}\subset\mathfrak{k}$ generated by the rotations: $\mathfrak{m}=\mathfrak{o}(n)$.
		Their direct sum $\mathfrak{m}\oplus\mathfrak{a}=\mathfrak{co}(n)$ is generated by the linear similarities. Together with the nilpotent subalgebra, they form the parabolic subalgebra $\mathfrak{h}=\mathfrak{m}\oplus(\mathfrak{a}\inplus\mathfrak{n})$ of the conformal algebra.\footnote{Of course, one can introduce the corresponding Iwasawa decomposition at the group level \cite[Theorem 2.3.5]{Cap}: $G=KAN$ with $K$, $A$ and $N$ the subgroups obtained by exponentiation of $\mathfrak{k}$, $\mathfrak{a}$ and $\mathfrak{n}$ respectively. One can further introduce the centraliser $M$ of $A$ in $K$, and consider the parabolic subgroup $H=MAN$.} The corresponding elementary representations are labeled by $\lambda=(\Delta,\vec s)$, where $\Delta$ is the ``scaling dimension'' (\textit{i.e.} the eigenvalue of the dilation operator generating $\mathfrak{a}$) and $\vec s =(s_1,s_2,\cdots, s_r)$ is the ``spin'' (\textit{i.e.} the half/integer labels of a finite-dimensional irreducible representation of the rotation subalgebra $\mathfrak{m}$ of rank $r$). The corresponding covariant fields on the celestial sphere $\mathbb{S}^n$ are called \textbf{conformal primary fields}.
	}
	
	\subsection{Generalised Verma modules}\label{GeneralisedVermamodules}
	
	A Klein pair $\mathfrak{h}\subseteq\mathfrak{g}$ with a semisimple principal algebra $\mathfrak g$ and a parabolic isotropy subalgebra $\mathfrak h$ will be called a \textbf{parabolic Klein pair}. 
	The $\mathfrak{g}$-module induced from a finite-dimensional $\mathfrak{h}$-module $V^*$ is a filtered $\mathfrak{g}$-module 
	\be
	\text{ind}^{\mathfrak{g}}_{\mathfrak{h}}(V^*)={\mathcal U}(\mathfrak{g})\otimes_{\,{\mathcal U}(\mathfrak{h})}V\,,
	\ee
	which will be called the \textbf{generalised Verma module for the parabolic Klein pair} $\mathfrak{h}\subset\mathfrak{g}$, \textbf{induced from the finite-dimensional} $\mathfrak{h}$-\textbf{module} $V$.\footnote{Note the different terminology about the induction with respect to induced modules between $V$ and $V^*$. Since $V$ is finite-dimensional, it is isomorphic to its dual $V^*$ as a vector space (though not as a module) so one will sometimes be sloppy about the distinction (especially when $V$ is one-dimensional).} (See \textit{e.g.} \cite[Subsection 3.2.13]{Cap} for a very concise introduction to generalised Verma modules.)
	
	They are highest/lowest-weight modules (of which the highest/lowest weight $\lambda$ is the highest/lowest weight of the module $V$), decomposing as a direct sum of finite-dimensional weight spaces. 
	The $G$-equivariant differential operators on the homogeneous space $G/H$ of a parabolic Klein geometry $H\subseteq G$ are in one-to-one correspondence with morphisms between generalised Verma modules of its Klein pair \cite[Theorem 1.4.10]{Cap}. When the morphisms are injective, those operators can be identified with the image of the finite-dimensional $\mathfrak{h}$-module inside the other generalised Verma module.
	
	\vspace{3mm}
	\noindent{\small\textbf{Example (M\" obius model)\,:} For the parabolic Klein pair $\mathfrak{h}\subset\mathfrak{g}$ where the conformal isotropy subalgebra $\mathfrak{h}\cong\mathfrak{ico}(n)$ is a parabolic subalgebra of the conformal algebra $\mathfrak{g}=\mathfrak{o}(n+1,1)$, the transvection subalgebra $\mathfrak{p}=\mathbb{R}^n$ is generated by the translations. The weight $\lambda=(\Delta,\vec s)$ of a finite-dimensional module $V_{(\Delta,\vec s)}$ of the conformal isotropy subalgebra $\mathfrak{h}$ is made of the scaling dimension $\Delta$ and the spin labels $\vec s$. Therefore, the corresponding generalised Verma module reads
		\be
		{\mathcal V}_{(\Delta,\vec s)}=\text{ind}^{\mathfrak{o}(n+1,1)}_{\mathfrak{ico}(n)}\big(\,V^*_{(\Delta,\vec s)}\big)\cong\bigodot(\mathbb{R}^n)\otimes V_{(\Delta,\vec s)}\,,
		\ee
		which is the dual of the jet space of conformal primary fields on the celestial sphere $\mathbb{S}^n$.}
	\vspace{3mm}
	
	For the sake of simplicity and because of its own interest, let us write explicitly the case corresponding to a Borel subalgebra (most of the discussion extends to the more general case of a parabolic subalgebra). The quotient ${\mathcal U}(\mathfrak{g})/{\mathcal U}(\mathfrak{g})\mathfrak{b}_\pm$ of a parabolic Klein pair $\mathfrak{b}_\pm\subseteq\mathfrak{g}$ with raising/lowering Borel isotropy subalgebra is isomorphic to the universal enveloping algebra ${\mathcal U}(\mathfrak{g}_\mp)$ of the lowering/raising nilpotent subalgebra $\mathfrak{g}_\pm$. 
	
	\subsubsection{Verma modules}
	
	Consider an irreducible finite-dimensional $\mathfrak{b}_\pm$-module of the raising/lowering Borel subalgebra $\mathfrak{b}_\pm\subseteq\mathfrak{g}$ of a semisimple Lie algebra $\mathfrak{g}$. This irreducible $\mathfrak{b}_\pm$-module is necessarily a one-dimensional weight space annihilated by the raising/lowering nilpotent subalgebra $\mathfrak{g}_\pm$.\footnote{In fact, a corollary of Lie's theorem is that any finite-dimensional irreducible representation of a solvable Lie algebra is one-dimensional \cite[Theorem 3']{Serre:2001}.} If $\lambda\in\mathfrak{g}_0$ is its weight, then this $\mathfrak{b}_\pm$-module will be denoted as $V^\pm_\lambda$.
	For the parabolic Klein pair $\mathfrak{b}_\pm\subset\mathfrak{g}$, the generalised Verma module ${\mathcal U}(\mathfrak{g})\otimes_{{\mathcal U}(\mathfrak{b}_\pm)}\,V^\pm_\lambda$ induced from the $\mathfrak{b}_\pm$-module $V^\pm_\lambda$ is called the \textbf{Verma module of highest/lowest-weight} $\lambda$ and it will be denoted ${\mathcal V}^\pm_\lambda$. The importance of Verma modules comes from the fact that any highest/lowest-weight irreducible module of a semisimple Lie algebra is a  quotient of a Verma module \cite[Section 20.3]{Humphreys}. The isomorphism \eqref{isoPBWindmod} implies for the Verma module that
	\be
	{\mathcal V}^\pm_\lambda=\text{ind}^{\mathfrak{g}}_{\mathfrak{b}_\pm}(V^{\pm*}_\lambda)\cong{\mathcal U}(\mathfrak{g}_\mp)\otimes V^\pm_\lambda\,.
	\ee
	
	\begin{table}
		\begin{center}
			\footnotesize
			\begin{tabular}{
					|c|c|c|c|}
				\hline
				Principal & Isotropy & Representation of & Representation of \\
				group & subgroup & Lie group $G$ & Lie algebra $\mathfrak{g}$\\ 
				\hline\hline
				Generic & Generic & Induced representation & Induced module \\
				$G$ & $H$ & $\text{Ind}{}^G_H(V)={C}^\infty(G)\,\otimes_{r_H}V\cong\Gamma(\pi)$ & $\text{ind}^{\mathfrak{g}}_{\mathfrak{h}}(V)={\mathcal U}(\mathfrak{g})\otimes_{\,{\mathcal U}(\mathfrak{h})}V^*\cong (J_{eH}\pi)^*$ \\ 
				\hline
				Semisimple & Parabolic & Elementary representation & Generalised Verma module \\
				& $H=MAN$ & $\text{Ind}{}^G_{MAN}(V_\lambda)$ &   $\text{ind}^{\mathfrak{g}}_{\mathfrak{h}}(V_\lambda)\cong{\mathcal U}(\mathfrak{p})V^*_\lambda$ \\
				\hline
				Semisimple & Borel & Elementary representation  & Verma module \\
				& $B_\pm$ & $\text{Ind}{}^G_{B_\pm}(V^{\pm}_\Delta)$ & $\text{ind}^{\mathfrak{g}}_{\mathfrak{b}_\pm}(V^{\pm}_\Delta)\cong{\mathcal U}(\mathfrak{g}_\mp)V^{\pm*}_\Delta$  \\ 
				\hline
			\end{tabular}
		\end{center}
		\caption{Induced representations}
	\end{table}
	
	For generic weights $\lambda\in\mathfrak{g}^*_0$, the annihilators Ann\,${\mathcal V}^\pm_\lambda$ are primitive ideals (but the Verma module may be reducible for integral weights). Such associative ideals are the \textit{minimal} primitive ideals of ${\mathcal U}(\mathfrak{g})$ \cite[Section 10.6]{Humphreys2}. Therefore, an off-shell higher-spin algebra $\texttt{hs}[\mathfrak{g}/\mathfrak{b}_\pm;{\mathcal V}^\pm_\lambda]$ for a parabolic Klein pair $\mathfrak{b}_\pm\subset\mathfrak{g}$ whose isotropy subalgebra $\mathfrak{b}_\pm$ is a raising/lowering Borel subalgebra, and a Verma module ${\mathcal V}^\pm_\lambda$ will be called a \textbf{maximal off-shell higher-spin algebra}.
	A theorem of Duflo asserts that any primitive ideal of a semisimple Lie algebra is the annihilator of (the quotient of) a Verma module \cite{Duflo}. 
	This suggests that any on-shell higher-spin algebra based on a semisimple Lie algebra $\mathfrak{g}$ is the quotient of a maximal off-shell higher-spin algebra $\texttt{hs}[\mathfrak{g}/\mathfrak{b}_\pm;{\mathcal V}^\pm_\lambda]$ for a parabolic Klein pair $\mathfrak{b}_\pm\subset\mathfrak{g}$.
	The table \ref{hsalgs} summarises the main abstract higher-spin algebras that provide a vast generalisation of the ones encountered in higher-spin gravity, such as the celebrated one based on the Klein pair $\mathfrak{ico}(n-1,1)\subset\mathfrak{so}(n,2)$ suited for the simplest example of higher-spin holography in any dimension.
	
	\begin{table}
		\begin{center}
			\small
			\begin{tabular}{
					|c|c|c|}
				\hline
				Higher-spin algebra & Faithful module & Ideal of ${\mathcal U}(\mathfrak{g})$ \\
				\hline\hline
				off-shell & induced representation & annihilator of\\
				$\texttt{hs}(\mathfrak{g}/\mathfrak{h};\text{V})$ & $\text{Ind}{}^G_H(r)\cong\Gamma(G\times_{r_H}\text{V})$ & induced representation\\
				\hline
				on-shell & quotient of & primitive \\
				& induced representation & ideal \\
				\hline
				structure & structure algebra & annihilator of \\
				$\texttt{hs}(\mathfrak{g}/\mathfrak{h})$ & ${C}^\infty(G/H)$ & structure algebra\\
				\hline
				classical & minimal & Joseph ideal\\
				$\texttt{hs}(\mathfrak{g})$ & representation & $J_{\mathfrak{g}}$\\
				\hline
				maximal & Verma & minimal\\ 
				off-shell & module & primitive ideal\\ 
				\hline
			\end{tabular}
		\end{center}
		\caption{\label{hsalgs}Various abstract higher-spin algebras}
	\end{table}
	
	\subsubsection{Klein geometry of the real projective line}
	
	To illustrate a wide array of notions that have been introduced, one may consider the simplest non-trivial example of parabolic Klein geometry: one-dimensional projective geometry\footnote{An inspiring pedagogical introduction to one-dimensional projective geometry can be found in \cite{Ovsienko}.} for the Klein model of the real projective line (or, equivalently, one-dimensional conformal geometry for the M\" obius model of the celestial circle).
	
	\paragraph{Real projective line} The projectivisation ${\mathbb P}{\mathbb R}^2$ of the plane, \textit{i.e.} the set of lines through the origin, defines the real projective line ${\mathbb R}{\mathbb P}^1$. The \textbf{linear coordinates} on ${\mathbb P}{\mathbb R}^2={\mathbb R}{\mathbb P}^1$ are the equivalence classes $(x,y)\sim \lambda (x,y)$, with $\lambda\neq 0$ of Cartesian coordinates of points of ${\mathbb R}^2-\{(0,0)\}$. They are denoted $[x:y]$. The real line is embedded into the projective line via the injection 
	\be
	i\,:\,{\mathbb R}\hookrightarrow{\mathbb R}{\mathbb P}^1\,:\,t\mapsto [t:1]\,,
	\ee
	where $t$ is the Cartesian coordinate on the real line. The missing point has linear coordinates $[1:0]$, it is called the \textbf{point at infinity} and is denoted $\infty$. The coordinate $t$ is called the \textbf{affine coordinate} on the chart ${\mathbb R}{\mathbb P}^1-\infty\cong\mathbb R$. The arithmetic on ${\mathbb R}$ can be partly extended to ${\mathbb R}{\mathbb P}^1$ via the rules: $\frac{1}{0}=\infty$, $\frac{1}{\infty}=0$, $t\cdot\infty=\infty$ for $t\neq 0$ and $t+\infty=\infty$ for $t\neq\infty$. Notice that $-\infty$ is identified with $\infty$ and the real projective line has indeed the topology of a circle. Nevertheless, this intuitive arithmetic is not complete (and cannot be completed in a consistent way)\,: \textit{e.g.} $\frac{0}{0}$, $\frac{\infty}{\infty}$, $0\cdot\infty$, $\infty-\infty$ remain indeterminate (as we all learn in kindergarten calculus).
	
	\paragraph{Projective linear group} The faithful representation of the general linear group $GL(2,{\mathbb R})$ on the plane ${\mathbb R}^2$ by $2\times 2$ invertible matrices induces the faithful representation on the real projective line ${\mathbb R}{\mathbb P}^1$ of the projective linear group $PGL(2,{\mathbb R})$ of $2\times 2$ invertible matrices defined up to proportionality. The projective general linear group $PGL(2,{\mathbb R})$ is isomorphic to the connected component of the Lorentz group $SO(2,1)$ (which can be interpreted as the M\" obius group of the circle) and it has the topology of a circle times a plane. The double covering $Spin(2,1)$ of the (connected component) of the Lorentz group $SO(2,1)$ is isomorphic to the special linear group $SL(2,{\mathbb R})$ (in agreement with the fact that a $2\times 2$ invertible matrix is of course proportional to a $2\times 2$ matrix of determinant one).
	
	\paragraph{Fractional linear transformations} The linear transformations of the linear coordinates induce fractional linear transformations
	\be
	t\mapsto\frac{at+b}{ct+d}\,,\qquad ac-bd=1\,,
	\ee
	of the affine coordinate. The isotropy group 
	of the origin $t=0$ is determined by the condition $b=0$ and is the Abelian group ${\mathbb R}\ltimes{\mathbb R}$ of similarities of the plane, where the first factor is the Abelian subgroup $A={\mathbb R}$ of dilations which acts on the nilpotent subgroup $N={\mathbb R}$ of special conformal transformations. 
	
	\paragraph{Parabolic Klein geometry} The real projective line ${\mathbb R}{\mathbb P}^1$ is the homogeneous space of the parabolic Klein geometry ${\mathbb R}\ltimes{\mathbb R}\subset PGL(2,{\mathbb R})$. The principal action of $\mathfrak{sl}(2,{\mathbb R})$ on ${\mathbb R}{\mathbb P}^1$ is via the following three vector fields:
	\be
	\hat{P}:=\frac{d}{dt}\,,\qquad\hat{D}:=t\frac{d}{dt}\,,\qquad\hat{K}:=t^2\frac{d}{dt}\,.
	\ee 
	with $\hat{P}$ generating the translations, $\hat{D}$ generating the dilations and $\hat{K}$ generating the special conformal transformations. The parabolic Klein pair of one-dimensional projective geometry is ${\mathbb R}\inplus{\mathbb R}\subset \mathfrak{sl}(2,{\mathbb R})$, where the first summand is the Cartan subalgebra $\mathfrak{g}_0={\mathbb R}$ spanned by the dilation generator $\hat{D}$ and the second summand is the lowering nilpotent subalgebra $\mathfrak{g}_{-}={\mathbb R}$ spanned by the special conformal transformation generator $\hat{K}$. 
	
	\paragraph{Densities} The densities of weight $\Delta$ on a one-dimensional manifold $M$ are the sections $\phi_\Delta$ of the line bundle 
	\be
	F_\Delta M\,:=\,\otimes^\Delta\, T^*M\,,\qquad \Delta\in\mathbb{R}\,.
	\ee
	They can be thought as some generalisations of symmetric covariant tensor fields on $M$ where the ``rank'' $\Delta$ is allowed to be non-integer.
	In terms of a local coordinate $t$ on $M$, they read explicitly as 
	\be
	\phi_\Delta\,=\,\phi_\Delta(t)\,(dt)^\Delta\,.
	\ee
	The space ${\mathcal F}_\Delta(M)\,:=\Gamma(F_\Delta M)$ of densities on $M$ will be called the \textbf{density sheaf of weight} $\Delta$. In particular, the density sheaf of weight zero is the structure sheaf, ${\mathcal F}_0(M)={C}^\infty(M)$ since $F_0M=M\times\mathbb{R}$ is the trivial line bundle. The density sheaves of negative integer weights, $\Delta=-r$ with $r\in\mathbb{N}$, are isomorphic to the sheaves of symmetric contravariant tensors, ${\mathcal F}_{-r}(M)=\odot^r \Gamma(TM)$ since $F_{-r}M=\odot^r TM\,$. Two density sheaves of distinct weights are isomorphic as vector spaces but not as $ \Gamma(TM)$-modules. When $M$ is compact (\textit{i.e.} diffeomorphic to the circle), the integral on $M$
	of the product of a density of weight $\Delta$ and a density of weight $1-\Delta$ defines a non-degenerate pairing 
	\be
	\int_M \phi_\Delta\phi_{1-\Delta}
	\ee
	between ${\mathcal F}_\Delta(M)$ and ${\mathcal F}_{1-\Delta}(M)$.
	The space of densities ${\mathcal F}(M)$ of all possible weights can be thought as an $\mathbb{R}$-graded algebra in the sense that 
	${\mathcal F}_{\Delta_1}(M){\mathcal F}_{\Delta_2}(M)\subseteq{\mathcal F}_{\Delta_1+\Delta_2}(M)$.
	The transformation law of densities under infinitesimal diffeomorphisms can be seen as a canonical representation \begin{equation}
		L^\Delta:\mathfrak{X}(M)\hookrightarrow\mathfrak{D}^1\big(\,{\mathcal F}_\Delta(M)\,\big)
	\end{equation} of the Lie algebra $\mathfrak{X}(M)$ of vector fields on $M$ on the density sheaf ${\mathcal F}_\Delta(M)$, where $\mathfrak{D}^1\big(\,{\mathcal F}_\Delta(M)\,\big)$ denotes the Lie algebra of first-order differential operators acting on densities of weight $\Delta$. This representation reads in terms of the affine coordinate as follows 
	\be
	L^\Delta_{{X}}\,=\,X(t)\,\frac{d}{dt}\,+\,\Delta\,\frac{dX}{dt}(t)\,,
	\ee
	where ${X}=X(t)\frac{d}{dt}$ is a vector field on $M$. In particular, for trivial weight $\Delta=0$ one recovers the representation ${\mathcal L}=L^{\Delta=0}$ of the Lie algebra $\mathfrak{X}(M)$ on the structure algebra ${C}^\infty(M)$ as the usual Lie derivatives ${\mathcal L}_{{X}}=X(t)\,\frac{d}{dt}$\,.
	
	\paragraph{Elementary representations} At the algebra level, the irreducible representations of the lowering Borel subalgebra $\mathfrak{b}_{-}={\mathbb R}\inplus{\mathbb R}$ are one-dimensional representations $r_\Delta$ labeled by the lowest weight $\Delta$ which is equal to the eigenvalue of the dilation generator $\hat{D}$. The corresponding $\mathfrak{b}_{-}$-modules are weight spaces $V_\Delta$ annihilated by $\mathfrak{g}_{-}$\,, \textit{i.e.} by the action of the generator $\hat{K}$ of special conformal transformations. 
	At the group level, the irreducible representations of the lowering Borel subgroup $B_-={\mathbb R}\ltimes{\mathbb R}$ are one-dimensional representations carried by the spaces $V_\Delta$ where the nilpotent subgroup of special conformal transformations acts trivially. 
	
	\paragraph{Associated vector bundles} Let $P$ be a principal bundle over $M$ of structure group $H={\mathbb R}\ltimes{\mathbb R}$. 
	The vector bundle $P\times_{r_\Delta} V_\Delta$ associated to the principal bundle $P$ over $M$ through the linear representation
	$r_\Delta$ is isomorphic to the line bundle $F_\Delta M$. Therefore, the associated sheaf $\Gamma\big(P\times_{r_\Delta} V_\Delta\,\big)$ of $r_\Delta$-covariant fields on $M$ is isomorphic to the sheaf ${\mathcal F}_\Delta(M)$ of densities on $M$ of weight $\Delta$.
	In other words, $r_\Delta$-covariant fields on $M$ identify with weight-$\Delta$ densities.
	
	\paragraph{Homogenous vector bundles} In the particular case of the parabolic Klein geometry for one-dimensional projective geometry, the homogeneous vector bundle $PGL(2,{\mathbb R})\times_{r_\Delta} V_\Delta$, associated to the principal bundle $PGL(2,{\mathbb R})$ over ${\mathbb R}{\mathbb P}^1$ through the linear representation
	$r_\Delta$, is isomorphic to the line bundle $F_\Delta{\mathbb R}{\mathbb P}^1$.
	The induced representation space 
	\be
	\text{Ind}^{PGL(2,{\mathbb R}}_{{\mathbb R}\ltimes{\mathbb R}}(V_\Delta)={\mathcal C}\big(PGL(2,{\mathbb R})\,\big)\otimes_{r_\Delta} V_\Delta\cong\Gamma\big(PGL(2,{\mathbb R})\times_{r_\Delta} V_\Delta\,\big)\cong{\mathcal F}_\Delta({\mathbb R}{\mathbb P}^1)
	\ee
	carries the elementary representations of the projective linear group $PGL(2,{\mathbb R})$. They are isomorphic to the sheaves ${\mathcal F}_\Delta({\mathbb R}{\mathbb P}^1)$ of densities on ${\mathbb R}{\mathbb P}^1$ of weight $\Delta$.
	The corresponding induced representation $L^\Delta:\mathfrak{sl}(2,{\mathbb R})\mapsto\mathfrak{D}^1({\mathbb R}{\mathbb P}^1)$ of the linear projective algebra on the real projective line is determined by the three differential operators of order one reading, in affine coordinates, as:
	\be
	L^\Delta_{\hat{P}}=\frac{d}{dt}\,, \quad L^\Delta_{\hat{D}}=t\frac{d}{dt}+\Delta\,,\quad 
	L^\Delta_{\hat{K}}=t^2\frac{d}{dt}+2\Delta t\,.
	\ee
	
	\paragraph{Primary field}
	A density $\phi_\Delta$ on the real projective line ${\mathbb R}{\mathbb P}^1$ of weight $\Delta$ will be called a primary field of weight $\Delta$ when seen as an element of the induced sheaf $\Gamma\big(PGL(2,{\mathbb R})\times_{r_\Delta} V_\Delta\,\big)$.
	The $\infty$-prolongation $j^\infty\phi_\Delta$ of a primary field will be called a \textbf{primary field with its tower of descendants}. They are sections of the $\infty$-jet bundle $J^\infty\pi_\Delta$ of the homogeneous vector bundle 
	\be
	\pi_\Delta\,:\, PGL(2,{\mathbb R})\times_{r_\Delta}V_\Delta\twoheadrightarrow{\mathbb R}{\mathbb P}^1\,.
	\ee
	
	\paragraph{Verma modules} For the Klein pair ${\mathbb R}\inplus{\mathbb R}\subset \mathfrak{sl}(2,{\mathbb R})$, the Verma module ${\mathcal V}_\Delta$ of lowest weight $\Delta$ induced from the weight space $V_\Delta$ is the relative tensor product 
	\be\label{Vermmod}
	{\mathcal V}_\Delta=\text{ind}^{\mathfrak{pgl}(2,{\mathbb R}}_{{\mathbb R}\inplus{\mathbb R}}(V_\Delta)={\mathcal U}\big(\mathfrak{sl}(2,{\mathbb R})\,\big)\otimes_{{\mathcal U}({\mathbb R}\inplus{\mathbb R})}V_\Delta\cong {\mathcal U}({\mathbb R})\otimes V_\Delta\,,
	\ee 
	where the weight space $V_\Delta$ and its dual are denoted by the same symbol for the sake of simplicity (this space being one-dimensional this abuse of notation is particularly harmless).
	As a module of the Cartan subalgebra $\mathfrak{g}_0={\mathbb R}$, it decomposes as a direct sum of one-dimensional weight spaces 
	\be
	{\mathcal U}({\mathbb R})\otimes V_\Delta\,\cong\, \bigoplus_{p=0}^\infty V_{\Delta+p}\,.
	\ee
	The universal enveloping algebra ${\mathcal U}({\mathbb R})$ of the Abelian Lie algebra of translations is isomorphic to the Abelian algebra ${\mathbb R}[\frac{d}{dt}]$ of polynomials in the translation generator, \textit{i.e.} it is spanned by \textit{finite} linear combinations of monomials $\left(\frac{d}{dt}\right)^p$. In other words, $\{\left(\frac{d}{dt}\right)^p\}_{p\in{\mathbb N}}$ is a basis of ${\mathbb R}[\frac{d}{dt}]$. 
	According to the isomorphism \eqref{Vermmod}, a basis of the Verma module ${\mathcal V}_\Delta$ is $\{\left(\frac{d}{d\varepsilon}\right)^p\Big|_\Delta\}_{p\in{\mathbb N}}$ where $\left(\frac{d}{dt}\right)^p\Big|_\Delta$ denotes the linear form on $\mathcal{F}_\Delta({\mathbb R}{\mathbb P}^1)$ sending a density $\phi_\Delta$ on the projective real line to its $p$th derivative at the origin $\frac{d^p\phi_\Delta}{dt^p}(0)$. 
	
	\paragraph{Dual of Verma modules} Let $\phi^{(p)}_\Delta$ be understood as coordinates on the $\infty$-jet space $J_0\pi_\Delta$ of the homogeneous vector bundle $PGL(2,{\mathbb R})\times_{r_\Delta}V_\Delta$, at the origin of ${\mathbb R}{\mathbb P}^1$. 
	Indeed, the coordinates of the $\infty$-jet at the origin of a primary field $\phi_\Delta$ reads $\phi^{(p)}_\Delta=\frac{d^p\phi_\Delta}{dt^p}(0)$\,. The $\infty$-jet space $J_0\pi_\Delta$ at the origin is isomorphic to the linear dual ${\mathcal V}^*_\Delta$ of the Verma module ${\mathcal V}_\Delta$.
	The linear dual of the universal enveloping algebra ${\mathcal U}({\mathbb R})^*$ of the Lie algebra of translations is isomorphic to the Abelian Lie algebra ${\mathbb R}\llbracket \varepsilon\rrbracket $ of formal power series in one coordinate $\phi_\Delta(\varepsilon)=\sum_{p=0}^\infty\frac1{p!}\phi^{(p)}_\Delta\varepsilon^p$.
	The isomorphism
	$J_0\pi_\Delta\,\cong\,{\mathcal V}_\Delta^*\,\cong\,\big(\,{\mathcal U}({\mathbb R})\otimes V_\Delta\,\big)^*$
	incarnates into the fact that the $\infty$-jet at the origin of a primary field $\phi_\Delta(t)$ can be represented as a formal power series 
	\be
	\phi_\Delta(\varepsilon)=\Bigg(\exp\Big(\varepsilon\frac{d}{dt}\Big)\phi_\Delta\Bigg)(0)\,.
	\ee
	
	\paragraph{Feigin algebra} The Verma module ${\mathcal V}_\Delta$ is an eigenspace of the quadratic Casimir operator $\hat{C}_2$ of $\mathfrak{sl}(2,{\mathbb R})$ of eigenvalue equal to $\Delta(1-\Delta)$.
	In fact, the annihilator Ann$\,{\mathcal V}_\Delta$ of the Verma module of $\mathfrak{sl}(2,{\mathbb R})$ with lowest weight $\Delta$, inside the universal enveloping algebra ${\mathcal U}\big(\mathfrak{sl}(2,{\mathbb R})\big)$ is the associative ideal $I_\Delta$ generated by $C_2+\Delta(\Delta-1)$.
	The commutator algebra of the quotient of the universal enveloping algebra ${\mathcal U}\big(\mathfrak{sl}(2,{\mathbb R})\big)$ by the associative ideal $I_\Delta$ will be called the Feigin algebra \cite{Feigin1988} and denoted $\mathfrak{gl}[\lambda]$ where $\Delta=\frac{1-\lambda}2$.
	
	\paragraph{Higher-spin algebra} For $\lambda\neq \mathbb N$, the Verma module ${\mathcal V}_\Delta$ is irreducible,  therefore the on/off-shell higher-spin algebra  $\texttt{hs}[\mathfrak{sl}(2,{\mathbb R})/({\mathbb R}\inplus{\mathbb R})\,;\,{\mathcal V}_\Delta]$ associated to the line bundle $F_\Delta{\mathbb R}{\mathbb P}^1$ for the Klein pair corresponding to the one-dimensional projective geometry is isomorphic to the Feigin algebra $\mathfrak{gl}[\lambda]$. The latter decomposes into the direct sum $\mathfrak{gl}[\lambda]={\mathbb R}\oplus \mathfrak{hs}[\lambda]$, where $\mathfrak{hs}[\lambda]:=\mathfrak{gl}[\lambda]\,/\,{\mathbb R}$ is a simple Lie algebra. 
	
	\paragraph{Higher-spin algebra}
	For $\lambda=N$ with $N\in\mathbb N$, the Verma module ${\mathcal V}_{\frac{1-N}2}$ is reducible with submodule ${\mathcal V}_{\frac{1+N}2}$. The quotient is the finite-dimensional irreducible $\mathfrak{sl}(2,{\mathbb R})$-module ${\mathcal D}_N$ of dimension $2N+1$. Its annihilator Ann$\,{\mathcal D}_N$ 
	inside the universal enveloping algebra ${\mathcal U}\big(\mathfrak{sl}(2,{\mathbb R})\big)$ is an associative ideal $J_N\supset I_{\frac{1-N}2}$, and the on-shell higher-spin algebra $\texttt{hs}[\mathfrak{sl}(2,{\mathbb R})/({\mathbb R}\inplus{\mathbb R})\,;\,{\mathcal D}_N]$ associated to the line bundle $F_{\frac{1-N}2}{\mathbb R}{\mathbb P}^1$ for the Klein pair corresponding to the one-dimensional projective geometry is isomorphic to the finite-dimensional general linear algebra $\mathfrak{gl}(N)$.
	
	\paragraph{Primitive ideals}
	The previous discussion described the complete list of primitive ideals of the complex semisimple algebra $\mathfrak{sl}(2,{\mathbb C})$ \cite[Section 4.3]{Mazorchuk}. Consequently, this presentation is also the complete classification of higher-spin algebras based on this Lie algebra. The algebra $\mathfrak{sl}(2,{\mathbb R})$ is the smallest semisimple Lie algebra and it appears to be the only semisimple one for which the complete list of primitive ideals is known.

\section{Tangent structures}\label{Gstructures}

The paradigmatic example of vector bundle is the tangent bundle $\tau_M:TM\twoheadrightarrow M$, see \textit{e.g.} \cite[Proposition 10.4]{Lee2}. It is associated to the \textbf{tangent frame bundle} $FM:=F\,TM$ \cite[Examples I.5.2-3]{Kob63} through the fundamental representation of the general linear group $GL(V)$ on each tangent space $T_mM$, where $V$ is any vector space with dimension equal to the one of the manifold $M$.

\subsection{Tangent (co)frame bundle}

A \textbf{tangent frame} at $m\in M$ is an element of the fibre $F_mM$, \textit{i.e.} it is a linear frame of $T_mM$ but it can be defined more abstractly as a 1-jet $j^1_0f$ at the origin of a diffeomorphism $f:V\to M$ from (a neighborhood of the origin in) the vector space $V$ to the manifold $M$. Indeed, such a 1-jet is equivalent to a 0-jet (\textit{i.e.} the point $f(\,0)=m\in M$) together with an element of $T^*_0V\otimes T_mM$ (\textit{i.e.} the derivative of $f$ at the origin). Since $T^*_0V\cong V^*$, the latter is in turn equivalent to a linear frame $\texttt{e}|_m:V\stackrel{\sim}{\to} T_mM$. To any coordinate system is associated a (local) moving tangent frame: the coordinate basis. 
Therefore, in coordinates a tangent frame reads as a collection $\texttt{e}_a={e}_a^\mu\partial_\mu$ of tangent vectors providing a basis of the tangent space or, equivalently, as the set ${e}_a^\mu=\frac{\partial x^\mu}{\partial \varepsilon^a}$ of partial derivatives at that point of a diffeomorphism $f:V\to M:\varepsilon^a\mapsto x^\mu$. 

A \textbf{tangent coframe} at $m\in M$ is a linear coframe $(\texttt{e}|_m)^{-1}:T_mM\stackrel{\sim}{\to} V$ of a tangent frame $\texttt{e}|_m$ at $m$. In coordinates, it can be seen as a collection $\texttt{e}^{*a}=\theta^a_\mu dx^\mu$ of tangent covectors providing a basis of the cotangent space.

\subsection{G-structures}

Let $G\subset GL(V)$ be a subgroup of the general linear group. A \textbf{$G$-structure} on $M$ is a principal $G$-bundle over $M$ obtained as a reduction $P\subset FM$ of the tangent frame bundle by restriction of $GL(V)$ to the subgroup $G$. As a diagram a $G$-structure $P$ is depicted as:
\be
\begin{array}
	[c]{ccccc}%
	GL(V) &  \lefttorightarrow & FM & \twoheadrightarrow & \frac{FM}{G}\\
	& & & & \\ 
	\uparrow & & \uparrow i & & \downuparrows s\\ 
	& & & & \\ 
	G & \lefttorightarrow & P & \twoheadrightarrow & M
\end{array}
\ee
where the ascending vertical arrows are embeddings corresponding respectively to the Klein geometry $G\subset GL(V)$, to the reduction $i:P\hookrightarrow FM$ of principal bundles and to the section $s:M\hookrightarrow FM\,/\,G$ of the Klein geometry bundle with the coset space $GL(V)/G$ as fibre.

Let $f:M_1\stackrel{\sim}{\to}M_2$ be a diffeomorphism between two manifolds. It induces a diffeomorphism $f_*:FM_1\stackrel{\sim}{\to}FM_2$ between the corresponding frame bundles. An \textbf{isomorphism of $G$-structures} $\varphi:P_1\stackrel{\sim}{\to}P_2$ between a $G$-structure $P_1$ on $M_1$ and a $G$-structure $P_2$ on $M_2$ is a diffeomorphism $f:M_1\stackrel{\sim}{\to}M_2$ of which the induced diffeomorphism $f_*:FM_1\stackrel{\sim}{\to}FM_2$ between the frame bundles is such that $f_*\big(i_1(P_1)\big)=i_2(P_2)$. In such case, the isomorphism reads $\varphi:=i_2^{-1}\circ f_*\circ i_1$\,.

A \textbf{flat $G$-structure} is a $G$-structure $P\subset FM$ on a manifold $M$ of dimension $n$ that admits a global section $e:M\hookrightarrow P$ consisting of $n$ commuting vector fields. A $G$-structure is \textbf{integrable (or locally flat)} if it is locally (\textit{i.e.} on sufficiently small neighborhoods) isomorphic to a flat $G$-structure. 

 \subsubsection{Symplectic geometry as integrable symplectic structure} 

Let $V$ be a vector space of even dimension.
An $Sp(V)$-structure identifies with a bundle of \textbf{symplectic frames} and is sometimes called an \textbf{almost-symplectic structure} on a manifold $M$. The latter notion is equivalent to a manifold $M$ endowed with a non-degenerate differential two-form $\Omega$ since the symplectic two-form on $V$ preserved by $Sp(V)$ induces a symplectic two-form on each tangent space $T_mM$ via any symplectic frame $\texttt{e}|_m:V\stackrel{\sim}{\to} T_mM$. 
The almost-symplectic structure is integrable if and only if the non-degenerate differential two-form $\Omega$ is closed. In other words, a symplectic manifold $M$ is nothing but an integrable symplectic structure on $M$.

 \subsubsection{Riemannian geometry as orthonormal frame bundle} 

An $O(V)$-structure identifies with a \textbf{bundle of orthonormal frames} $OFM\subset FM$ on $M$. Moreover, the latter notion is equivalent to a Riemannian manifold (\textit{i.e.} a manifold $M$ endowed with a positive-definite metric) since the scalar product on $V$ preserved by $O(V)$ induces a scalar product on each tangent space $T_mM$ via any orthonormal frame $\texttt{e}|_m:V\stackrel{\sim}{\to} T_mM$. 

The \textbf{polar decomposition} of a square invertible real matrix $A\in GL(V)$ is the unique decomposition of the form $A=|A|\,R$, where $|A|=(AA^T)^{\frac12}$ is a positive-definite matrix and $R=|A|^{-1}A\in O(V)$ is an orthogonal matrix.
Therefore, the homogeneous space $GL(V)\,/\,O(V)$ identifies with the space of all positive-definite scalar products on $V$, thus a metric can be identified with a section $g:M\hookrightarrow FM/O(V)$ of the Klein geometry bundle $FM\,/\,O(V)$ over $M$ which will be called the \textbf{metric bundle}. This can be summarised in the commutative diagram 
\be\label{Riemannorthoframe}
\begin{array}
	[c]{ccccc}%
	GL(V) &  \lefttorightarrow & FM & \twoheadrightarrow & \frac{FM}{O(V)}\\
	& & & & \\ 
	\uparrow & & \uparrow i & & \downuparrows g\\ 
	& & & & \\ 
	O(V) & \lefttorightarrow & OFM & \twoheadrightarrow & M
\end{array}
\ee
where $i:OFM\hookrightarrow FM$ is the reduction of principal bundles.

An isometry between two Riemannian manifolds $M$ and $M'$ with respective metrics $g:M\hookrightarrow FM/O(V)$ and $g':M'\hookrightarrow FM'/O(V')$ can be defined as a  map $f:M\to M'$ such that the pullback of the metric $g'$ on $M'$ along $f$ is equal to the metric $g$ on $M$, \textit{i.e.} $f^*g'=g$.
The orthonormal frame bundles of the homogeneous spaces of the three standard non-Euclidean geometries (spherical geometry on $\mathbb{S}^n=O(n+1)/O(n)$, Euclidean geometry on $\mathbb{R}^n=IO(n)/O(n)$ and hyperbolic geometry on $\mathbb{H}^n=O(n,1)/O(n)$\,) are diffeomorphic to their corresponding principal groups (\textit{i.e.} $OF\mathbb{S}^n\cong O(n+1)$, $OF\mathbb{R}^n\cong IO(n)$ and $OF\mathbb{H}^n=O(n,1)$\,) which identify with their isometry groups.
For Euclidean geometry, this reduction of Klein geometries can be summarised in the commutative diagram 
\be\label{reductAfftoEuclframe}
\begin{array}
	[c]{ccccc}%
	GL(V) &  \subset & IGL(V)\cong FV & \twoheadrightarrow & \frac{FV}{O(V)}\\
	& & & & \\ 
	\cup & & \cup & & \downuparrows \\ 
	& & & & \\ 
	O(V) & \subset & IO(V)\cong OFV & \twoheadrightarrow & V
\end{array}
\ee
where the upper row is the Klein affine geometry $GL(V)\subset IGL(V)$ while the second row is the Euclidean geometry $O(V) \subset IO(V)$. Retrospectively, the definition of Riemannian geometry as an $O(V)$-structure is modelled on the reduction \eqref{reductAfftoEucl} of Klein affine geometry to Euclidean geometry.   
At the level of infinitesimal symmetries, the reduction \eqref{reductAfftoEucl} of Klein geometries corresponds to following reduction of Klein pairs
\be\label{diagreduxKleinpairconf1}
\begin{array}
	[c]{ccc}%
	\mathfrak{gl}(V) & \subset & \mathfrak{igl}(V)\\
	\cup & & \cup\\
	\mathfrak{so}(V) & \subset & \mathfrak{iso}(V)
\end{array}
\ee
with the following isomorphism of transvection modules: 
\be
\mathfrak{igl}(V)/\mathfrak{gl}(V)\cong V\cong\mathfrak{iso}(V)/\mathfrak{so}(V)\,,
\ee
where $V$ stands for the Lie subalgebra of translations. The dual version of the latter is:
\be
\mathfrak{igl}(V)/\mathfrak{iso}(V)\cong \odot^2V\cong\mathfrak{gl}(V)/\mathfrak{so}(V)\,,
\ee
where $\odot^2V/{\mathbb R}$ stands for the space of symmetric matrices.

 \subsubsection{Conformal geometry as conformal frame bundle} 

A \textbf{conformal frame} at a point $m\in M$ is a ray of orthonormal frames at $m$. Let $CO(V):={\mathbb R}\times O(V)$ denote the group of linear similarities of the Euclidean space $V$. A $CO(V)$-structure is a \textbf{bundle of conformal frames} $CFM\subset FM$ on $M$, where $CFM$ denotes the bundle over $M$ of all conformal frames on $M$. A \textbf{conformal metric} is a conformal class of metrics, \textit{i.e.} a ray of positive-definite metrics. A manifold $M$ endowed with a conformal metric is called a \textbf{conformal manifold}. A $CO(V)$-structure is equivalent to a conformal manifold.

The homogeneous space $GL(V)\,/\,CO(V)$ identifies with the space of all positive-definite scalar products on the Euclidean space $V$ modulo a scale (since the general linear group $GL(V)$ acts transitively on the space of all inner products while $CO(V)$ stabilises the collection of all inner products proportional to the standard one). 
A conformal class $[g]$ of metrics, $g_{\mu\nu}(x)\sim \Omega^2(x) g_{\mu\nu}(x)$ with $\Omega\in{C}^\infty(M)$ a nowhere vanishing function (\textit{i.e.} $\Omega(x)\neq0$, $\forall x$), can be identified with a section of the Klein geometry bundle $FM\,/\,CO(V)$ which will be called the \textbf{conformal metric bundle}. 
This reduction of principal bundles can be summarised in the commutative diagram 
\be\label{confgeomreduction}
\begin{array}
	[c]{ccccc}%
	GL(V) &  \lefttorightarrow & FM & \twoheadrightarrow & \frac{FM}{CO(V)}\\
	& & & & \\ 
	\uparrow & & \uparrow i & & \downuparrows [g]\\ 
	& & & & \\ 
	CO(V) & \lefttorightarrow & CFM & \twoheadrightarrow & M
\end{array}
\ee

A \textbf{conformal isometry} between two conformal manifolds $M$ and $M'$ with respective conformal metrics $[g]:M\hookrightarrow FM/CO(V)$ and $[g']:M'\hookrightarrow FM'/CO(V)$ can be defined as a  map $f:M\to M'$ such that the pullback of the conformal metric $[g']$ on $M'$ along $f$ is equal to the conformal metric $[g]$ on $M$, \textit{i.e.} $[f^*g']=[g]$.

The conformal frame bundle of the Euclidean space $V$ which is the homogeneous space of the Weyl model $CO(V)\subset ICO(V)$ is diffeomorphic to its principal group (the Weyl group), $CF\,V\cong ICO(V)$, which identifies with the group of global conformal isometries of $V$.
This reduction of Klein geometries can be summarised in the commutative diagram 
\be\label{reductKleinconf}
\begin{array}
	[c]{ccccc}%
	GL(V) &  \subset & IGL(V)\cong FV & \twoheadrightarrow & \frac{FV}{CO(V)}\\
	& & & & \\ 
	\cup & & \cup & & \downuparrows \\ 
	& & & & \\ 
	CO(V) & \subset & ICO(V)\cong CFV & \twoheadrightarrow & V
\end{array}
\ee
where the upper row is the Klein affine geometry $GL(V)\subset IGL(V)$ while the second row is the Weyl model $CO(V) \subset ICO(V)$ of conformal geometry. Retrospectively, the definition of conformal geometry as a $CO(V)$-structure \eqref{confgeomreduction} is modeled on the reduction \eqref{reductKleinconf} of Klein affine geometry to the Weyl model.
At infinitesimal level, the reduction \eqref{reductKleinconf} of Klein geometries corresponds to following reduction of Klein pairs
\be\label{diagreduxKleinpairconf1'}
\begin{array}
	[c]{ccc}%
	\mathfrak{gl}(V) & \subset & \mathfrak{igl}(V)\\
	\cup & & \cup\\
	\mathfrak{co}(V) & \subset & \mathfrak{ico}(V)
\end{array}
\ee
with the following isomorphism of transvection modules: 
\be
\mathfrak{igl}(V)/\mathfrak{gl}(V)\cong V\cong\mathfrak{ico}(V)/\mathfrak{co}(V)\,,
\ee
where $V$ stands for the Lie subalgebra of translations. The dual version of the latter is:
\be
\mathfrak{igl}(V)/\mathfrak{ico}(V)\cong \odot^2V\,/\,{\mathbb R}\cong\mathfrak{gl}(V)/\mathfrak{co}(V)\,,
\ee
where $\odot^2V/{\mathbb R}$ stands for the space of symmetric matrices modulo a scale.

 \subsubsection{Oriented manifolds as oriented frame bundles}

An important example of $G$-structure is the reduction to the special linear subgroup. 

\vspace{5mm}
\begin{framed}
	\begin{center}
		\textbf{The many faces of oriented manifolds}
	\end{center}
	
	\noindent
	Let $V$ be a vector space of the same dimension as a manifold $M$. Then the following notions are equivalent:
	
	\begin{enumerate}
		\item an oriented manifold $M$.
		
		\item an $SL(V)$-structure on $M$, which is called an \textbf{orientation}.
		
		\item a nowhere-vanishing top-form on $M$, which will be called a \textbf{volume form}.
		
		\item a $GL^+(V)$-structure on $M$, which will be called the \textbf{oriented frame bundle}.
	\end{enumerate}
	\vspace{3mm}\end{framed}

A line bundle is trivial if and only if it admits a nowhere-vanishing global section.
In this sense, a volume form can be thought as a trivialisation of the top-form bundle $\wedge^n T^*M$ where $n=\text{dim}\,M$. A manifold is orientable if and only if its top-form bundle is trivial.

 \subsubsection{Cartan geometries as absolute parallelisms}

Another important (though somewhat degenerate) example of $G$-structure is the extreme case where the group $G$ reduces to the sole identity: an $\{e\}$-structure is called an \textbf{absolute parallelism} of $M$, \textit{i.e.} a section $s:M\hookrightarrow FM$. Equivalently, an absolute parallelism of $M$ is a global moving frame of the tangent frame bundle $FM$. 

\vspace{5mm}
\begin{framed}
	\begin{center}
		\textbf{Equivalent formulations of parallelised manifolds}
	\end{center}
	
	\noindent
	The following notions are equivalent:
	
	\begin{enumerate}
		\item a parallelised manifold $M$.
		
		\item an absolute parallelism of $M$. 
		
		\item a global moving frame of the tangent frame bundle $FM$.
	\end{enumerate}
	\vspace{3mm}\end{framed}
\vspace{5mm}

A manifold admitting an absolute parallelism is called \textbf{parallelisable}. A manifold $M$ is parallelisable if and only if the tangent frame bundle $FM$ is trivial (since it is a principal bundle) which, in turn, is true if and only if the tangent bundle $TM$ is trivial. For instance, any Lie goup is parallelisable (since the Maurer-Cartan one-form provides a trivialisation of the tangent bundle $TG$). Any parallelisable manifold is orientable, since an absolute parallelism provides a canonical orientation via the determinant of the global moving frame as volume form.

\vspace{5mm}
\begin{framed}\noindent
	\begin{center}
		\textbf{Equivalent formulations of parallelisable manifolds}
	\end{center}
	
	\noindent
	Let $M$ be a manifold and $V$ be a vector space such that they have the same dimension.
	The following notions are equivalent:
	\begin{enumerate}
		
		\item a parallelisable manifold $M$.
		
		\item a manifold with trivial tangent bundle $TM\cong M\times V$\,.
		
		\item a manifold with trivial tangent frame bundle $FM\cong M\times GL(V)$\,.
	\end{enumerate}
	\vspace{3mm}\end{framed}
\vspace{5mm}

A regular Lie action algebroid $M\rtimes\mathfrak{g}$ over a manifold $M$ is equivalent to a regular action $\#$ of a Lie algebra $\mathfrak{g}$ on $M$ and corresponds to the case of a Lie action algebroid with a bijective anchor 
\be
\#_\bullet\,:\,M\rtimes\mathfrak{g}\stackrel{\sim}{\to} TM\,:\,(m,y)\mapsto y^\#|_m\,.
\ee
The latter defines an absolute parallelism on $M$, \textit{i.e.} a global moving frame 
\be
s\,:\,M\hookrightarrow FM\,:\,m\mapsto\#|_m\,,
\ee
where $\#|_m:\mathfrak{g}\stackrel{\sim}{\to} T_mM$ is a tangent frame at $m$. 
In particular, this applies to a Maurer-Cartan geometry for a Lie group $G$ ($=M$) on which the corresponding Lie algebra $\mathfrak{g}$ acts regularly.
In the curved case, a Cartan parallelism \eqref{Crampindef*} is equivalent to an $H$-equivariant absolute parallelism $D:P\hookrightarrow FP:p\mapsto D_p$ on a principal $H$-bundle $P$, where the Cartan intertwiner $D$ is seen as defining a field of tangent frames
$D_p:\mathfrak{g}\stackrel{\sim}{\to} T_pP$ (with $V=\mathfrak{g}$) on $P$
whose restriction to the isotropy subalgebra is the fundamental action $D_p|_{\mathfrak{h}}=\#_p$.
\vspace{3mm}

A Koszul connection $\nabla$ on the tangent bundle $TM$ is often called an \textbf{affine connection} on $M$. 
This is a source of terminological confusion since an \textit{affine connection} on $M$ can actually also be seen as a \textit{linear} connection  
on $TM$.
In the coordinate basis, the components of an affine connection read as $\Gamma_\mu{}^\rho{}_\nu(x)$.
The \textbf{torsion tensor} is a tangent-valued differential two-form $\textsc{T}:\wedge^2 TM\to TM$ defined by
\be
\text{\textsc{T}}({X},{Y})=\nabla_{{X}}{Y}-\nabla_{{Y}}{X}-[{X},{Y}]\,.
\ee
An affine connection $\nabla$ is called \textbf{torsionless} if the torsion tensor vanishes, $\nabla_{{X}}{Y}-\nabla_{{Y}}{X}=[{X},{Y}]$. Consequently, a torsionless affine connection on $M$ is also sometimes called a \textbf{symmetric (affine) connection} on $M$, because its components are symmetric in the coordinate basis, $\Gamma_\mu{}^\rho{}_\nu=\Gamma_\nu{}^\rho{}_\mu$.
A vector field ${X}$ on $M$ such that $\nabla_{{X}}{X}=0$ is often called an \textbf{affine geodesic vector field for the affine connection} $\nabla$ on $M$. A symmetric connection is entirely determined by its affine geodesic vector fields.

\section{Densities and weights}\label{densitiesweights}

The simplest possible examples of associated vector bundles are line bundles $P\times_{r_H}{\mathbb R}$ over $P/H$ associated to a principal $H$-bundle $P$ through a one-dimensional representation $r$ of $H$ (on the vector space ${\mathbb R}$). 

A particularly important example is the line bundle $FM\times_{|\det|_{GL(V)}}{\mathbb R}$ associated to the tangent frame bundle $FM$ of the manifold $M$ through the absolute value of the determinant representation, $\det:GL(V)\to GL({\mathbb R})$, of the general linear group $GL(V)$. 
This trivial line bundle is called the \textbf{density bundle} and will be denoted $|\det|M$.
Its sections, \textit{i.e.} $|\det|_{GL(V)}$-covariant fields on $M$, are called \textbf{densities} because they are the ${\mathbb R}$-valued fields which can be integrated over the manifold $M$. In fact, the integration $\int_M$ is a linear form on the associated sheaf, which will be called the \textbf{density sheaf} and denoted 
\be
|\det|(M)\,:=\,\Gamma(FM\times_{|\det|_{GL(V)}}{\mathbb R})\,\cong\,{C}^\infty(FM)\otimes_{|\det|_{GL(V)}}{\mathbb R}\,.
\ee
An affine connection $\Gamma$ on $M$ is equivalent to a principal connection on the frame bundle, hence it induces an associated linear connection $A$ on the density bundle $|\det|M$. Its components are the contraction of the ones of the affine connection. More explicitly, the connection one-form reads $A=\omega^a{}_a$ and its components $A_\mu=\Gamma^\mu_{\nu\mu}$ in a coordinate basis. 

 \subsection{Integration \textit{vs} orientation}\label{integrationvsoritentation}

A nowhere-vanishing density $\nu$ on $M$ is sometimes called a \textbf{volume density} (or \textbf{integration measure}). 
They can be thought as sections of a ray bundle, which will be denoted $|\det|_0M$ and called the \textbf{volume density bundle}.
Accordingly, a point $v\in|\det|_0M$ of the volume density bundle is called a \textbf{volume element at the point} $m=\pi(v)\in M$. By definition, the value $v=\nu|_m$ of a volume density at a point $m\in M$ is a volume element at this point.

A volume density on $M$ is equivalent to a trivialisation of the volume density bundle $|\det|_0M$. Accordingly, any volume density $\nu$ defines a principal connection on the volume density bundle $|\det|_0M$, whose corresponding exact connection one-form reads $A=d(\ln\nu)$.
Given a metric, the square-root $\nu=\sqrt{\det\,g_{\alpha\beta}}$ of its determinant provides the \textbf{canonical volume density on a Riemannian manifold}.

\vspace{3mm}

\noindent\textbf{Remark:} Note that integration over manifolds is always well-defined since volume densities always exist (because the density bundle is trivial). It should be emphasised that the density bundle is canonically defined on any manifold. In particular, it is not necessary above to assume the manifold to be orientable. 
However, on an oriented manifold $M$ the densities can be canonically identified with the top forms, \textit{i.e.} $|\det|(M)\cong\Omega^n(M)$ in such case. Therefore, top-forms can be integrated on orientable manifolds, which explains the frequent (slightly abusive) terminology ``volume forms'' for nowhere-vanishing top-forms.

 \subsection{Density bundles}\label{densitybundle}

More generally, the line bundle $FM\times_{|\det|^{\texttt{w}}_{GL(V)}}{\mathbb R}$ associated to the tangent frame bundle $FM$ through the $\texttt{w}$th power $|\det|^{\texttt{w}}$ of (the absolute value of) the determinant representation of the general linear group $GL(n)$ will be called the \textbf{density bundle of weight} ${\texttt{w}}$ and will be denoted $|\det|^{\texttt{w}}M$. The $|\det|^{\texttt{w}}_{GL(V)}$-covariant fields on $M$ are called \textbf{densities of weight} ${\texttt{w}}$, or ${\texttt{w}}$-densities for short. The previous densities, \textit{i.e.} elements of $|\det|(M)$, are densities of weight ${\texttt{w}}=1$. 
A ${\texttt{w}}$-density $\varphi_{\texttt{w}}\in |\det|^{\texttt{w}}(M)$ can be thought equivalently as a $|\det|^{\texttt{w}}_{GL(V)}$-equivariant map $f$ on the tangent frame bundle $FM$, \textit{i.e.} $f\in{C}^\infty(FM)$ such that $f(ph)=|\det(h)|^{-{\texttt{w}}}\cdot f(p)$ for all $p\in FM$ and $h\in GL(V)$. Accordingly, the corresponding transformation law for the density reads in coordinates as 
\be\label{transfolawdensity}
\varphi'_{\texttt{w}}(x')\,=\,\Big|\,\det\left(\frac{\partial x^{\prime\mu}}{\partial x^\nu}\right)\,\Big|^{-{\texttt{w}}}\,\varphi_{\texttt{w}}(x)\,.
\ee
Infinitesimally, it corresponds to the Lie derivative of a ${\texttt{w}}$-density, which reads in coordinates as:
\be\label{Liederdensity}
{\mathcal L}_{{X}}\varphi_{\texttt{w}}\,=\,X^\mu\partial_\mu\varphi_{\texttt{w}}\,+\,{\texttt{w}}\, \partial_\mu X^\mu\,\varphi_{\texttt{w}}\,.
\ee
The space of densities forms a commutative algebra such that $|\det|^{\texttt{w}_1}|\det|^{\texttt{w}_2}\subset|\det|^{\texttt{w}_1+\texttt{w}_2}$\,.

\vspace{2mm}

\noindent{\small\textbf{Remark:} The equation \eqref{Liederdensity} for ${\texttt{w}}=1$ leads to ${\mathcal L}_{{X}}\varphi\,=\,\partial_\mu(X^\mu\varphi)$, which shows that the integral over $M$ of the Lie derivative of a density on $M$ is a boundary term. This reflects the fact that the integral of densities are diffeomorphism invariant, as can be seen from the fact that the extra factor in \eqref{transfolawdensity} will cancels the Jacobian factor in the integral.}

\vspace{2mm}
One can of course also define tensor densities of weight ${\texttt{w}}$ by tensoring the representation $|\det|^{\texttt{w}}$ with the corresponding tensorial representation.

 \subsection{Scale bundle}\label{scalebundle}

Let $n$ be the dimension of $M$. An everywhere-positive density whose weight is equal to the inverse $1/n$ of the dimension of the manifold will be called a \textbf{conversion scale} $s(x)>0$. 
A conversion scale is equivalent to a trivialisation of the ray bundle $|\det|_0^{1/n}M$. Obviously, a conversion scale multiplied by any positive real number $\Omega>0$ defines a new conversion scale $s'=\Omega\,s$\,. Therefore, conversion scales are sections of a principal $\mathbb{R}$-bundle, which will be called the \textbf{conversion scale bundle} and denoted $\Sigma$. This principal bundle $\mathbb{R}\lefttorightarrow\Sigma\twoheadrightarrow M$ is a trivial bundle: $\Sigma\cong M\times\mathbb{R}$. 
Concretely, one can take the positive numbers $s$ as defining a coordinate along the fibre. 

Let $L$ be a one-dimensional vector space.
The \textbf{bundle of conformal densities of conformal weight} $w$ is the line bundle ${\mathcal E}[w]:=\Sigma\times_{r^w_{\mathbb R}}L$ associated to the conversion scale bundle with respect to the representation 
\be\label{rw}
r^w:\mathbb{R}\to GL(L):t\mapsto\exp^{-wt}
\ee
of the Lie group $\mathbb{R}$ on the one-dimensional vector space $L$. 
An $r^w_{\mathbb R}$-covariant field on $M$ with respect to the latter one-dimensional representation will be called a \textbf{Weyl-covariant field of conformal weight} $w$. (This choice of terminology will become clear later.)
For instance, a Weyl-covariant field of conformal weight zero is merely a function on $M$. In fact, the line bundle ${\mathcal E}[0]$ is isomorphic to the trivial bundle $M\times\mathbb R$.
The transformation law for a Weyl-covariant field $\phi_w\in\Gamma\big(\,{\mathcal E}[w]\,\big)|$ of conformal weight $w$ reads in coordinates as 
\be\label{transfolawdensityconf}
\phi_w'(x)\,=\,\Omega^{w}(x)\,\phi_w(x)
\ee
when the conversion scale is changed as $s'=\Omega\,s$.
Of course, one can tensor these line bundles ${\mathcal E}[w]$ with each other (note that ${\mathcal E}[w_1]\otimes{\mathcal E}[w_2]\cong{\mathcal E}[w_1+w_2]$) or with other vector bundles.

A \textbf{scale} is a Weyl-covariant field of conformal weight one. The line bundle of scales is ${\mathcal E}[1]$.
A \textbf{positive scale} is a scale which is everywhere positive. Obviously, a positive scale multiplied by any positive real number $\Omega>0$ defines a new positive scale $\sigma'=\Omega\,\sigma$\,. Therefore, positive scales are sections of a ray bundle called the \textbf{positive scale bundle}, sometimes denoted ${\mathcal E}_+[1]$. This principal bundle $\mathbb{R}\lefttorightarrow{\mathcal E}_+[1]\twoheadrightarrow M$
is a trivial bundle: ${\mathcal E}_+[1]\cong M\times\mathbb{R}_+$. A positive scale is nothing but a trivialisation of this principal $\mathbb{R}$-bundle. Concretely, one can take the positive numbers $\sigma$ as defining a coordinate on the fibre. 
The fundamental action of the group $\mathbb R$ is by multiplication $\sigma\mapsto\sigma'=\Omega\,\sigma$ by positive factors $\Omega=\exp t>0$ where $t\in\mathbb R$. The fundamental vector field $V=\sigma\frac{\partial}{\partial\sigma}$ on the ray bundle ${\mathcal E}_+[1]$ will be called the \textbf{homothety vector field}.
A vertical automorphism of the positive scale bundle $\Sigma$,
\be\label{verticalWeyl}
\sigma'\,=\,\Omega(x)\,\sigma\,,\qquad x'=x\,,
\ee
is called a \textbf{Weyl transformation} where $\Omega$ is an everywhere-positive function on $M$ called a \textbf{conformal factor} in this context.
The corresponding change of positive scales is obviously
\be\label{transfolawdensityconfscale}
\sigma'(x)\,=\,\Omega(x)\,\sigma(x)\,.
\ee

\vspace{3mm}
\noindent\textbf{Remark:} Although their fibres are both one-dimensional, the bundle $\cal E$ of scales is a \textit{vector} bundle (more precisely, a line bundle) while the bundles $\Sigma$ and ${\mathcal E}_+[1]$ of conversion and of positive scales are \textit{principal} bundles (more precisely,  ray bundles). 
\vspace{3mm}

Retrospectively, the line bundle ${\mathcal E}[w]\cong{\mathcal E}_+[1]\times_{r^w_{\mathbb R}}L$ is isomorphic to the vector bundle associated to the positive scale bundle with respect to the representation \eqref{rw} of the Lie group $\mathbb{R}$ on the one-dimensional vector space $L$.
Moreover, the conversion scale bundle is isomorphic to the fibrewise product of two ray bundles: the conversion scale bundle and the positive scale bundle, $\Sigma\cong|\det|_0^{1/n}M\times{\mathcal E}_+[1]$ in the sense that (by construction) its sections are simultaneously densities of weight $1/n$ and Weyl-covariant fields of conformal weight $1$.
In fact, picking a conversion scale $s$ (or, equivalently, a volume density $\nu=s^n$) seen as an element of $|\det|^{1/n}(M)\otimes\Gamma({\mathcal E}[1])$ defines, for each $w\in\mathbb R$, an isomorphism of line bundles between ${\mathcal E}[w]$ and $|\det|^{-w/n}M$ which reads,
\be
{\mathcal E}[w]\,\stackrel{\sim}{\to}\,|\det|^{-w/n}M
\,:\,\phi_w\mapsto\varphi_{_{-w/n}}:=s^{-w}\phi_w\,,
\ee
since $s^{-w}$ is a nowhere-vanishing element of $|\det|^{-w/n}(M)\otimes\Gamma({\mathcal E}[-w])$ and since ${\mathcal E}[-w]\otimes\,{\mathcal E}[+w]\cong{\mathcal E}[0]=M\times\mathbb R$.
The inverse of this isomorphism is obviously
\be
|\det|^{\texttt{w}}M\,\stackrel{\sim}{\to}\,{\mathcal E}[-n\texttt{w}]
\,:\,\varphi_{\texttt{w}}\mapsto\phi_{-n\texttt{w}}:=s^{-n\texttt{w}}\varphi_{\texttt{w}}\,.
\ee

A \textbf{conformal density of (conformal) weight} $w$ is an $r^w$-equivariant map $\pmb{\phi}_w\in {C}^\infty(\Sigma)\otimes_{r^w_{\mathbb R}}L$ from the conversion scale bundle $\Sigma$ to the one-dimensional $\mathbb R$-module $L$. 
In down-to-earth terms, a conformal density of weight $h$ is a homogeneous function on the positive scale bundle $\Sigma$ of homogeneity degree $w$ with respect to the homothety vector field: ${\mathcal L}_V\pmb{\phi}_w=w\,\pmb{\phi}_w$.
In coordinates, a conformal density of conformal weight $w$ reads
\be\label{confdens}
{\pmb{\phi}}_w(x,\sigma)\,=\,\sigma^{w}\,f(x)
\ee
where $f(x)$ is a function on the base manifold $M$.
A choice of positive scale provides a concrete isomorphism between the vector spaces of Weyl-covariant field and the space of conformal densities (of the same conformal weight $w$). Explicitly, a Weyl-covariant field $\phi_w$ of conformal weight $w$ is obtained from a conformal density $\pmb{\phi}_w$ of weight $w$ via pullback by the positive scale:
\be
\phi_w(x)\,:=\,\pmb{\phi}_w\big(\,x\,,\,\sigma(x)\,\big)\,=\,\sigma^{w}(x)\,f(x)\,,
\ee
where \eqref{confdens} was used in the last equality. This makes manifest the transformation law \eqref{transfolawdensityconf} under Weyl transformation \eqref{transfolawdensityconfscale}.

\vspace{3mm}
\noindent\textbf{Remark on terminology:} Although they are distinct objects, conformal densities of weight $w$ and Weyl-covariant fields of conformal weight $w$ are most often not distinguished via a different vocabulary in the literature.

 \subsection{Conformal metric}\label{confdensitybundle}

Note that, despite the terminology, conformal densities (\textit{etc}) as defined above can be introduced without specifying a conformal structure (\textit{cf.} the comment in \cite[Subsection 2.4.1]{Curry:2014yoa}). However, a choice of representative metric $g$ in a conformal class leads to a canonical volume density $\nu=\sqrt{\det\,g_{\alpha\beta}}$ and, consequently, to a canonical conversion scale
$s=\nu^{1/n}=|\det\,g_{\alpha\beta}|^{1/2n}$ which is simultaneously a density of weight $1/n$ and a Weyl-covariant fields of conformal weight $1$.
For such a choice, the conformal metric becomes a Weyl-covariant metric field, \textit{i.e.} a conformal class $[g]$ of metrics $g$, of conformal weight two, which means that it transforms under a Weyl transformation via
\be
g'_{\mu\nu}(x)\,=\,\Omega^2(x)\,g_{\mu\nu}(x)\,,
\ee
as it should. Equivalently, a conformal metric is an everywhere positive-definite section $\pmb{g}$ of the bundle ${\mathcal E}[2]\,\otimes\,\odot^2 T^*M$ over $M$. Indeed, a conformal metric can be thought either as a Weyl-covariant metric field denoted by $[g]$, or as a conformal density denoted by $\pmb{g}$.

A \textbf{Carrollian metric} is a symmetric rank-two tensor on a ray bundle which is horizontal, semi-positive-definite and whose radical is spanned by the fundamental vector field.
A conformal metric is equivalent to a Carrollian metric $\pmb{g}$ on the positive scale bundle ${\mathcal E}_+[1]$ which is homogeneous of degree two with respect to the homothety vector field (\textit{i.e.} $\mathcal{L}_V\pmb{g}=2\pmb{g}$). Given a representatives $g$ of the conformal class $[g]$ and a choice of positive scale, it reads in coordinates as
\be
ds^2_{\Sigma}=\pmb{g}_{\mu\nu}(x,\sigma)\,dx^\mu dx^\nu=\sigma^2 g_{\mu\nu}(x)\,dx^\mu dx^\nu
\ee

 \subsection{Lost in translation}

Unfortunately, technical terms like ``weight'' and ``density'' have slightly distinct meaning for different specialists. As we have seen, conformal geometers use different terminology and normalisation\footnote{See \textit{e.g.} the lectures notes on conformal geometry  \cite{Eastwood:1996,Curry:2014yoa}.} for densities and weights. Let us compare three communitites of specialists: differential geometers, conformal geometers and conformal field theorists.

\paragraph{Differential geometers:} \textit{cf.} Subsection \ref{densitybundle} for their conventions on densities and weights.

\paragraph{Conformal geometers:} \textit{cf.} Subsection \ref{confdensitybundle} for their conventions on conformal densities and their weight. Given a choice of metric representative, there is a way to map densities of weight $\texttt{w}=\frac{\Delta}{n}$ (in the sense of differential geometers) to conformal densities of weight $w=-\Delta$ (in the sense of conformal geometers) via multiplication by the suitable power of the volume density, and vice versa. Nevertheless, stricty speaking being a density of weight $\texttt{w}$ and/or being a conformal density of weight $w$ are simply two unrelated distinct properties.

\paragraph{Conformal Field Theorists:} \textit{cf.} Subsection \ref{parabolicKleingeom} for their conventions on conformal primary fields on the celestial sphere and their scaling dimension.
The conformal metric of the celestial sphere is fixed since it defines the background geometry, hence it is of scaling dimension zero.
A conformal and tensorial density on the celestial sphere which is of weight $\texttt{w}$ and of conformal weight $w$, say
$$\varphi_{\texttt{w}}\in\Gamma\Big(\,|\det|^{\texttt{w}}M\,\otimes\,{\mathcal E}[w]\,\otimes\,(\otimes^s T^*S^n)\,\otimes\,(\otimes^t TS^n)\,\Big)\,,$$
defines a conformal primary field of scaling dimension $\Delta=n\texttt{w}-w+s-t$ and rank $s+t$ (see \textit{e.g.} \cite[Section 2.2]{Bekaert:2022ipg}).

\begin{table}
	\begin{center}
		\begin{tabular}{
				|c|c|c|c|}
			\hline
			Context & Bundles & Sections & Label \\
			\hline\hline
			Differential & Line bundle & Density & Weight  \\
			geometry & $|\det|M=FM\times_{|\det|_{GL(V)}}{\mathbb R}$ & & $1$ \\
			\hline
			& Line bundle & $\texttt{w}$-density & Weight  \\
			& $|\det|^{\texttt{w}}M$ & & $\texttt{w}$ \\
			\hline
			& Ray bundle & Volume & Weight  \\
			& $|\det|_0M=FM\times_{|\det|_{GL(V)}}{\mathbb R}_0$ & density & $1$ \\
			\hline\hline
			Conformal & Ray bundle & Positive & Conformal  \\
			geometry & $\mathcal{E}_+[1]$ & scale & weight $1$ \\
			\hline
			& Line bundle & Scale & Conformal  \\
			& $\mathcal{E}[1]=\mathcal{E}_+[1]\times_{r_{\mathbb R}}L$ & & weight $1$ \\
			\hline
			& Line bundle & Conformal & Conformal  \\
			& $\mathcal{E}[w]=\mathcal{E}_+[1]\times_{r^w_{\mathbb R}}L$ & density & weight $w$ \\
			\hline\hline
			Conformal & Homogeneous bundle & Conformal & Scaling \\
			field theory & $SO(n+1,1)\times_{r^{-\Delta}_{ICO(n)}}L$ & primary field & dimension $\Delta$ \\
			\hline
		\end{tabular}
	\end{center}
	\caption{Summary on scalar densities and their weights}
\end{table}

\section{Soldering}\label{solderring}

\subsection{Solder form}

Let $P$ be a principal $H$-bundle over $M:=P/H$.
Consider a vector space $V$ of the same dimension as the base manifold, \textit{i.e.} dim\,$V$ = dim\,$M$ = dim\,$P$ - dim\,$H\,$, and let $V$ carry a linear representation $r$ of the Lie group $H$. Then the associated vector bundle $\mathbb{V}:=P \times_{r_H} V$ has the same base $M$ and the same rank as the tangent bundle $TM$\,. 

A \textbf{solder form, associated to the representation} $r$\textbf{, on a principal} $H$-\textbf{bundle} $P$, is a $V$-valued $r_H$-equivariant horizontal differential one-form $\theta\,\in\,\Omega^1(P)\,\otimes_{r_H}V$ on $P$ inducing a vector bundle isomorphism $\Theta:T\frac{P}{H}\stackrel{\sim}{\to} P \times_{r_H} V$ from the base tangent bundle $T\frac{P}{H}$ to the associated vector bundle 
$\mathbb{V}$. 
A solder form is an $r_H$-equivariant map $\theta:TP\to V$, thus it induces a map $\theta:\frac{TP}{H}\to \mathbb{V}$ from the Atiyah algebroid to the associated vector bundle.
Consider the following commutative diagram
\be
\begin{array}
	[c]{ccccc}%
	P\times_{Ad_H}\mathfrak{h} & \stackrel{\#}{\hookrightarrow} & \frac{TP}{H} & \stackrel{\pi_*}{\twoheadrightarrow} & T\frac{P}{H} \\
	& & & & \\
	& & & \theta\searrow\quad & \downarrow \Theta \\
	& & & & \\
	& & & & P \times_{r_H} V\quad
\end{array}
\ee
where the first row is the Atiyah sequence of the principal $H$-bundle $P$. 
The solder form $\theta$ induces the isomorphism $\Theta$ (in the sense that $\theta=\Theta\circ\pi_*$), from which one can see that it is horizontal (in the sense that $\theta\circ\#=0$).\footnote{
Another way to understand that the latter property is necessary, is to observe that, by definition, a solder form implements the following isomorphisms $T\frac{P}{H}\cong \mathbb V \cong\,\frac{\text{Im\,$\theta$}}{H}$ of vector bundles. Compatibility with the isomorphisms  of vector bundles, $T\frac{P}{H}\cong \frac{TP}{H}\,/\,\frac{VP}{H}$ and  $\text{Im}\,\theta\cong TP\,/\,\text{Ker}\,\theta$, implies that the kernel of the solder form is the vertical distribution: $\text{Ker}\,\theta=VP$ (which explains why solder forms must be horizontal).}

The projection $\omega^{\mathfrak{g}/\mathfrak{h}}:\frac{TP}{H}\twoheadrightarrow P \times_{Ad_H} \mathfrak{g}/\mathfrak{h}$ of the codomain of a Cartan connection one-form $\omega^\mathfrak{g}$ to the transvection module $\mathfrak{g}/\mathfrak{h}$ of a Klein pair is a solder form, associated to the adjoint representation, on the principal $H$-bundle $P$, providing a vector bundle isomorphism $\Theta:T\frac{P}{H}\stackrel{\sim}{\to} P \times_{Ad_H} \mathfrak{g}/\mathfrak{h}$
from the tangent bundle of the base to the transvection bundle, as in \eqref{salsiciettatrac}. When the Klein pair is reductive, the vector bundle isomorphism $\theta:\frac{TP}{H}\stackrel{\sim}{\to} P \times_{Ad_H} \mathfrak{p}$ is nothing but the inverse of the Cartan covariant derivative.

Let $\{\texttt{T}_i\}$ denote a basis of $\mathfrak{h}$ and let $\{\texttt{e}_a\}$ denote a moving frame of $\mathbb{V}$.
The $H$-equivariance of the $V$-valued horizontal one-form on $P$ 
\be\label{soldform}
\theta=\theta_\mu^{a}(x,y)\,dx^\mu\otimes \texttt{e}_a
\ee
reads in components as:
\be
{\mathcal L}_{\texttt{T}_i^\#}\theta_\mu^a\,=\,-\,(\textsc{T}_i)^a{}_{b}\,\theta_\mu^b\,,
\label{equivtheta}
\ee
where 
\be
\textsc{T}_i=r(\texttt{T}_i)\,,\qquad\textsc{T}_i\texttt{e}_b=(\textsc{T}_i)^a{}_{b}\texttt{e}_a\,.
\ee
The isomorphism property requires det$(\,\theta_\mu^{a})\neq0$ everywhere.
Given any gauge $s:M\hookrightarrow P$ of the principal $H$-bundle $P$ with a coordinate expression $y^i(x^\mu)$, the pullback of the solder form is a $V$-valued one-form $s^*\theta=\theta_\mu^{a}\big(x,y(x)\big)\,dx^\mu\otimes \texttt{e}_a$ on $M$.

The solder form is a map $\theta:TP\twoheadrightarrow V$ such that $\text{Ker}\,\theta=VP$. Therefore, the field of linear surjections $\theta_p:T_pP\twoheadrightarrow V$ induces a field of linear frames $\Theta_m:T_mM\stackrel{\sim}{\to} \mathbb{V}_m$ of the associated bundle $\mathbb{V}$. In this sense, $\theta_p\in F_{\pi(p)}\mathbb{V}$. In fact, if $r$ is faithful, then the map $\theta_\bullet:P\hookrightarrow F\mathbb{V}:p\mapsto\theta_p$ is a reduction of the principal $GL(V)$-bundle $F\mathbb{V}$ to the $H$-bundle $P$. Moreover, the solder form induces the isomorphism $TM\cong \mathbb{V}$ therefore the tangent frame bundle of $M$ is isomorphic to the linear frame bundle of the vector bundle $\mathbb{V}$ over $M$ (\textit{i.e.} $FM=F\,TM\cong F\mathbb{V}$). Consequently, when the representation $r$ of $H$ is faithful,  the existence of a solder form associated to the representation $r$, on a principal $H$-bundle $P$, implies that $P$ is an $H$-structure on $P/H$.

The tangent frame bundle ${\mathcal F}_M:FM\twoheadrightarrow M$ is a principal $GL(V)$-bundle over $M$ endowed with a canonical solder form associated to the fundamental representation of the general linear group. Indeed, a point $p\in F\,M$ of the tangent frame bundle of $M$ is described by a tangent frame $\texttt{e}|_m:V\stackrel{\sim}{\to} T_mM$ at the base point $m=\pi(p)$. Consider the associated vector bundle $\mathbb{V}:=FM \times_{GL(V)}\, V$ via the fundamental representation of $GL(V)$ on $V$. 
The \textbf{fundamental form on the tangent frame bundle} is the solder form $\theta:T\,FM\to \mathbb{V}$ on the principal $GL(V)$-bundle $FM$ defined by $\theta|_p=(\texttt{e}|_m)^{-1}\circ({\mathcal F}_M)_{*p}:T_pFM\to V$ where $(\texttt{e}|_m)^{-1}:T_mM\stackrel{\sim}{\to} V$ is the tangent coframe and $({\mathcal F}_M)_{*p}:T_pFM\to T_mM$ is the pushforward of the projection of the tangent frame bundle.\footnote{See \textit{e.g.} \cite[Proposition 2.1]{Kob63} for a concise presentation.} In coordinates, the fibre coordinates of the tangent frame bundle are $e^\mu_a$ of which the
components $\theta_\mu^{a}$ of the solder form \eqref{soldform} are the entries of the inverse matrix. The fundamental form on  the tangent frame bundle $FM$ provides a canonical isomorphism $\Theta:TM\stackrel{\sim}{\to} FM \times_{GL(V)}\, V$ of vector bundles from the tangent bundle $TM$ to the associated vector bundle $\mathbb V$.
For an $H$-structure $P\subseteq FM$ the pullback of the fundamental form along the reduction provides a solder form on the principal $H$-bundle $P$. Therefore, a solder form is precisely the extra datum that turns a principal bundle over $M$ into a reduction of the frame bundle of $M$. In other words, 
a principal $H$-bundle $P$ is an $H$-structure on $P/H$ if and only if it is endowed with a solder form (associated to a faithful representation of $H$ on a vector space such that dim\,$V$ = dim\,$P$ - dim\,$H\,$) \cite[Subsection 1.3.6]{Cap}.

\vspace{3mm}
\noindent\textbf{Example (Tangent bundle as associated vector bundle)\,:} Let $M$ be a manifold of dimension $n$ with local coordinates $x^\mu$. The elements of the vector space $V={\mathbb R}^n$ will be denoted with an arrow on top of the letter. Tangent frames are written as $\texttt{e}_a=e_a^\mu\partial_\mu$, thus $(x^\mu,e_a^\nu)$ are coordinates on the tangent frame bundle $FM$.
The fundamental action of $\mathfrak{gl}(n)$ on the tangent frame bundle $FM$ corresponds to the fundamental vector fields 
$(\texttt{T}{}_a{}^b)^\#=\,e^\mu_a\frac{\partial}{\partial e_b^\mu}$ on $FM$.

Let $\rho:h\mapsto\textsl{h}$ be the defining representation of the general group $GL(n)$ on ${\mathbb R}^n$.
Via the fundamental form, the tangent bundle $TM$ can be identified with the vector bundle $FM\times_{GL(n)}{\mathbb R}^n$ associated to this linear representation. In this way, a tangent vector $X^\mu\,\partial_\mu=X^a\texttt{e}_a$ is identified with an equivalence class 
$(\texttt{e},\overrightarrow{X})\sim(\texttt{e}',\overrightarrow{X}')$ of elements in $FM\times{\mathbb R}^n$, where the two tangent frames are related by $\texttt{e}'=\texttt{e}\textsl{h}$ while the two vectors are $\overrightarrow{X}=X^a\vec{e}_a$ and $\overrightarrow{X}'=X'{}^a\vec{e}_a=\textsl{h}^{-1}\vec X$. Their components are related by the relation $\texttt{e}'_a=\texttt{e}_b\,\textsl{h}^b{}_a$ and $X'{}^a=(\textsl{h}^{-1})^a{}_b\,X^b$.
Therefore, in local coordinates, this equivalence relation reads $(e_a^\mu,X^a)\sim(e_a^\mu{}',X'{}^a)$ in agreement with $X^a\texttt{e}_a=X'{}^a\texttt{e}'_a$ and $e_a^\mu{}'=e_b^\mu\,\textsl{h}^b{}_a$.

The fundamental form on $FM$ is the ${\mathbb R}^n$-valued $GL(n)$-equivariant horizontal one-form 
\be\label{theta1form}
\theta=\theta^a_\mu \,dx^\mu\otimes\vec{e}_a=\theta^a\otimes\vec{e}_a\,,
\ee
where $\theta^a_\mu$ are the coframe components, \textit{i.e.} the inverse of the tangent frame components $e^\mu_a$.
At the level of sheaves, it induces the isomorphism
\ba
\Theta&:& \Gamma(TM)\stackrel{\sim}{\to}{C}^\infty(FM)\otimes_{{}_{GL(n)}}{\mathbb R}^n\nonumber\\
&:& X^\mu(x)\frac{\partial}{\partial x^\mu}=X^a(x)\,\texttt{e}_a\mapsto X^a(x)\,\vec{e}_a\,,
\label{solderisoTM}
\ea
where there is an implicit dependence on the vertical coordinate in the components $X^a(x):=\theta_\mu^a X^\mu(x)$. This dependence is responsible for the equivariance of the vector-valued field.
The isomorphism 
\be
 \Gamma(TM)\cong\Gamma\big(FM\times_{GL(n)}{\mathbb R}^n\big)\cong {C}^\infty(FM)\otimes_{GL(n)}{\mathbb R}^n
\ee
relates a tangent vector field ${X}=X^\mu(x)\,\partial_\mu=X^a(x)\,\texttt{e}_a$ in $ \Gamma(TM)$ to an equivariant vector-valued field $\overrightarrow{X}(x)=X^a(x)\,\vec{e}_a$ in ${C}^\infty(FM)\otimes_{GL(n)}{\mathbb R}^n$.

Tensoring and dualising the former relations defines the contravariant and covariant tensor bundles $\otimes\, TM$ and $\otimes\, T^*M$ as associated vector bundles to the representation $\rho^\otimes$ on $\otimes{\mathbb R}^n$ and $\otimes{\mathbb R}^{n*}$.
For instance, the isomorphism 
\be
\Gamma\Big(\bigotimes\,TM\Big)\,\cong\,{C}^\infty(FM)\,\otimes_{_{GL(n)}}\bigotimes{\mathbb R}^n
\ee
between the spaces of contravariant tensor fields on $M$ and of $GL(n)$-equivariant maps from $FM$ to $\bigotimes{\mathbb R}^n$, 
relies in coordinates on the relation 
\be
T^{\mu_1\cdots\,\mu_r}(x)\,\partial_{\mu_1}\otimes\cdots\otimes\partial_{\mu_r}=T^{a_1\cdots\, a_r}(x)\,\texttt{e}_{a_1}\otimes\cdots\otimes\texttt{e}_{a_r}
\ee
where the components 
\be
T^{a_1\cdots \,a_r}(x)\,:=\,\theta^{a_1}_{\mu_1}\cdots\theta^{a_r}_{\mu_r}\,
T^{\mu_1\cdots\,\mu_r}(x)
\ee
have an implicit dependence on the vertical coordinates. 
Similarly, the isomorphism 
\be
\Gamma\Big(\bigotimes\,T^*M\Big)\,\cong\,{C}^\infty(FM)\,\otimes_{_{GL(n)}}\bigotimes{\mathbb R}^{n*}
\ee
between the spaces of covariant tensor fields on $M$ and of $GL(n)$-equivariant maps from $FM$ to $\bigotimes{\mathbb R}^{n*}$, 
rests in coordinates on the equality
\be\label{tensorcovcomponents}
T_{\mu_1\cdots\,\mu_r}(x)\,dx^{\mu_1}\otimes\cdots\otimes dx^{\mu_r}=T_{a_1\cdots\, a_r}(x)\,\texttt{e}^{a_1*}\otimes\cdots\otimes\texttt{e}^{a_r*}
\ee
where the components in the coframe basis read 
\be
T_{a_1\cdots \,a_r}(x)\,:=\,e_{a_1}^{\mu_1}\cdots e_{a_r}^{\mu_r}\,
T_{\mu_1\cdots\,\mu_r}(x)\,.
\ee

The subset of the automorphisms of the frame bundle which preserve the solder form \eqref{theta1form}  are generated by the invariant vector fields $\hat{\Xi}:=\xi^\mu(x)\,\frac{\partial}{\partial x^\mu}+\partial_\nu \xi^\mu(x)\,e_a^\nu\,\frac{\partial}{\partial e_a^\mu}$. Their action on tensor-valued $GL(n)$-equivariant maps on $FM$ reproduces the standard transformation rule of $GL(n)$-covariant tensor fields on $M$. In other words, the Lie derivative of tensor-valued $GL(n)$-equivariant maps on $FM$ with respect to the vector field $\hat{\Xi}$
is intertwined with the Lie derivative of $GL(n)$-covariant tensor fields on $M$ with respect to the base vector field $\tau_{M*}\hat{\Xi}=:\hat{\xi}=\xi^\mu(x)\,\frac{\partial}{\partial x^\mu}$.

Consider the orthogonal subgroup $O(n)\subset GL(n)$ of the general linear group and an $O(n)$-structure $OFM\subset FM$. The diagonal metric $\delta=\delta_{ab}\,\vec{e}^{\,a*}\odot\vec{e}^{\,b*}$ on ${\mathbb R}^n$ is an $O(n)$-invariant tensor.
Therefore the $O(n)$-equivariant map $\delta_{ab}\,\texttt{e}^{a*}\odot\texttt{e}^{b*}\in{C}^\infty(OFM)\,\otimes_{_{O(n)}}\odot^2{\mathbb R}^{n*}$ from the orthogonal frame bundle $OFM$ to $\odot^2{\mathbb R}^{n*}$ is constant. The corresponding $GL(n)$-equivariant map from the tangent frame bundle $FM$ to $\odot^2{\mathbb R}^{n*}$ defines the metric tensor field $g\in\odot^2\mathcal{T}^{*}(M)$ via the equality, in coordinates,
\be\label{metricomponents}
g_{\mu\nu}(x)\,dx^{\mu}\odot dx^{\nu}=\delta_{ab}\,\texttt{e}^{a*}\odot\texttt{e}^{b*}\,,
\ee
so that its components are given by $g_{\mu\nu}(x)=\delta_{ab}\,\theta_\mu^a\theta_\nu^b$ in the coordinate basis and $\delta_{ab}=g_{\mu\nu}(x)\,e^\mu_a e^\nu_b$ in the coframe basis, as usual.

 \subsection{Reductive Cartan geometry}

In the case of a reductive Klein pair, the short exact sequences of $\mathfrak{h}$-module morphisms in \eqref{salsiciettatracred}  splits, which has several consequences.
On the one hand, the restriction $\omega^\mathfrak{h}$ of the codomain of a Cartan connection one-form $\omega^\mathfrak{g}$ to the isotropy subalgebra $\mathfrak{h}$ of a reductive Klein pair is a principal connection one-form $\omega^\mathfrak{h}:\frac{TP}{H}\twoheadrightarrow P \times_{Ad_H} \mathfrak{h}$ on the principal $H$-bundle $P$.
On the other hand, the restriction $\omega^\mathfrak{p}$ of the codomain of a Cartan connection one-form $\omega^\mathfrak{g}$ to the transvection submodule $\mathfrak{p}$ of a reductive Klein pair is a solder form $\eta:=\omega^\mathfrak{p}:\frac{TP}{H}\twoheadrightarrow P \times_{Ad_H} \mathfrak{p}$ on the principal $H$-bundle $P$ providing a vector bundle isomorphism $\Theta:T\frac{P}{H}\stackrel{\sim}{\to} P \times_{Ad_H} \mathfrak{p}$
from the tangent bundle $T\frac{P}{H}$ of the base to the associated vector bundle $P \times_{Ad_H} \mathfrak{p}$.
Composing this isomorphism with the Cartan covariant derivative, \textit{i.e.} the restriction of the Cartan intertwiner $D:P \times_{Ad_H} \mathfrak{g}\stackrel{\sim}{\to}\frac{TP}{H}$ to the transvection submodule $\mathfrak{p}$, defines the horizontal lift $\gamma=D\circ \Theta:T\frac{P}{H}\hookrightarrow\frac{TP}{H}$ of the principal connection $\omega^\mathfrak{h}$.
The solder form $\omega^\mathfrak{p}$, together with the principal connection $\omega^\mathfrak{h}$, implies a fibrewise-bijective morphism of vector bundles over $P$ from the horizontal distribution $HP$ to the vector bundle
$P\times\mathfrak{p}$. 

A Cartan geometry such that the linear representation $Ad:H\to GL(\mathfrak{p})$ of $H$ on the transvection module $\mathfrak{p}=\mathfrak{g}/\mathfrak{h}$ is faithful, \textit{i.e.} injective, is called a \textbf{first order Cartan geometry} \cite[Definition 5.3.20]{Sharpe}. Otherwise, it is called a \textbf{higher order Cartan geometry}. For a first-order Cartan geometry modeled on a reductive Klein pair $\mathfrak{g}=\mathfrak{h}\inplus\mathfrak{p}$, the principal $H$-bundle $P$ can be interpreted as an $H$-structure on $P/H$ with fundamental form $\omega^\mathfrak{p}$ and principal connection $\omega^\mathfrak{h}$ \cite[Section A.2]{Sharpe}.

\vspace{3mm}
\noindent\textbf{Example (Affine Cartan geometry)\,:} The paradigmatic example of first order Cartan geometry  modeled on a reductive Klein pair is the tangent frame bundle endowed with a Cartan connection modeled on the affine Klein geometry.

An affine Cartan geometry is an isomorphism 
\be\label{isoCartanaff}
\omega^{\mathfrak{igl}(n)}\,:\,\frac{T\,FM}{GL(n)}\,\stackrel{\sim}{\longrightarrow}\, FM\times_{{}_{Ad_{GL(n)}}}\mathfrak{igl}(n)
\ee
of Lie algebroids between the Atiyah algebroid of the principal $GL(n)$-bundle of tangent frames and the adjoint tractor bundle for the affine algebra $\mathfrak{igl}(n)$.
More explicitly, a Cartan connection one-form on the principal $GL(n)$-bundle $FM$ modeled on the Klein pair $\mathfrak{gl}(n)\subset\mathfrak{igl}(n)$
is an $\mathfrak{igl}(n)$-valued one-form on $FM$ reading in coordinates as
\ba
\omega^{\mathfrak{igl}(n)}
&=&\theta^a_\mu(dx^\mu\otimes \texttt{P}_a+de_b^\mu\otimes \texttt{T}_a{}^b)+\Gamma_\mu{}^a{}_b(x)\, dx^\mu\otimes \texttt{T}_a{}^b\nonumber\\
&=&\theta^a_\mu \,dx^\mu\otimes \texttt{P}_a+\big(\,\Gamma_\mu{}^a{}_b(x)\, dx^\mu+\theta^a_\mu \,de_b^\mu\big)\otimes \texttt{T}_a{}^b
\label{affineCartanconnexion}
\ea
where $\theta^a_\mu$ is the inverse of $e_b^\nu$. The implicit dependence on the vertical coordinates for
$\Gamma_\mu{}^a{}_b(x):=\theta_\nu^a\,e^\rho_b\,\Gamma_\mu\,{}^\nu{}_\rho(x)$ is such that this one-form is $GL(n)$-equivariant.
We can rewrite the Cartan one-form as
\be\label{affCartconn}
\omega^{\mathfrak{igl}(n)}
=\theta^a\otimes \texttt{P}_a+\big(\theta^a{}_b+\Gamma_c{}^a{}_b(x)\, \theta^c\big)\otimes \texttt{T}_a{}^b
\ee
by introducing the basis $\theta^a:=\theta^a_\mu dx^\mu$ of base one-forms and $\theta^a{}_b:=\theta^a_\mu de_b^\mu$ of vertical one-forms, as well as the $GL(n)$-equivariant field $\Gamma_c{}^a{}_b(x):=e_c^\mu \Gamma_\mu{}^a{}_b(x)$.

On the one hand, the restriction of the codomain of the Cartan connection one-form to the translation ideal defines a solder form on $FM$, \textit{i.e.} the ${\mathbb R}^n$-valued $GL(n)$-equivariant horizontal one-form on $FM$ 
\be\label{Cartsold}
\omega^{{\mathbb R}^n}=\theta^a_\mu \,dx^\mu\otimes\texttt{P}_a=\theta^a\otimes\texttt{P}_a\,,
\ee
which is nothing but the fundamental form on the tangent frame bundle $FM$ with $V={\mathbb R}^n$.
The corresponding isomorphism
\be
\Theta: \Gamma(TM)\stackrel{\sim}{\to}{C}^\infty(FM)\otimes_{GL(n)}{\mathbb R}^n
\ee
of vector spaces takes the same form in components as the right-hand-side of \eqref{Cartsold}.
On the other hand, the restriction of the codomain of the Cartan connection one-form to the general linear subalgebra defines a principal connection one-form on $FM$, \textit{i.e.} the $\mathfrak{gl}(n)$-valued $GL(n)$-equivariant one-form on the principal $GL(n)$-bundle $FM$
\be
\omega^{\mathfrak{gl}(n)}=\theta^a_\nu\,(\,de_b^\nu\,+\,\Gamma_\mu\,{}^\nu{}_\rho(x)\,e^\rho_b\,dx^\mu)\otimes \texttt{T}_a{}^b
=\omega{}^a{}_b\otimes \texttt{T}_a{}^b\,.
\ee
The corresponding invariant horizontal lift
\be\label{invhorliftTM}
\gamma\,:\, \Gamma(TM)\hookrightarrow\Gamma(T\,FM)^{GL(n)}\,:\,X^\mu(x)\,\partial_\mu\mapsto X^\mu(x)\nabla_\mu
\ee
maps the coordinate basis vector fields $\partial_\mu$ on $M$ to the right $GL(n)$-invariant horizontal vector fields
\be\label{affinehorizontalvect field}
\nabla_\mu=\frac{\partial}{\partial x^\mu}-\,\Gamma_\mu{}^b{}_c(x)\,(\texttt{T}{}_b{}^c)^\#=\frac{\partial}{\partial x^\mu}-\Gamma_\mu\,{}^\nu{}_\rho(x)\,e_a^\rho\,\frac{\partial}{\partial e_a^\nu}
\ee
on $FM$.
The torsion two-form of the Cartan connection one-form $\omega^{\mathfrak{igl}(n)}$ is the ${\mathbb R}^n$-valued $GL(n)$-equivariant horizontal two-form on $FM$ given by
\ba
\Omega^{{\mathbb R}^n}&=&(d\theta^a+\omega{}^a{}_b\,\theta^b)\otimes\texttt{P}_a\nonumber\\
&=&(d\theta^a+\theta{}^a{}_b\,\theta^b+\Gamma_b\,{}^a{}_c(x)\, \theta^b\theta^c)\otimes\texttt{P}_a\,,
\ea
where
\be
\Gamma_b\,{}^a{}_c(x)\, \theta^b\theta^c=\theta_\rho^a\,\Gamma_\mu\,{}^\rho{}_\nu(x)\, dx^\mu\,dx^\nu
\ee
The curvature two-form of the principal connection identifies with the restriction of the codomain of the Cartan curvature two-form to the general linear subalgebra
and is the $\mathfrak{gl}(n)$-valued $GL(n)$-equivariant horizontal two-form on $FM$ given by
\ba
\Omega^{\mathfrak{gl}(n)}&=&(d\omega{}^a{}_b+\omega{}^a{}_c\,\omega{}^c{}_b)\otimes\texttt{T}_a{}^b\nonumber\\
&=&\left[\,d\theta{}^a{}_b\,+\,\theta_\rho^a\,\big(\partial_\mu\Gamma_\nu\,{}^\rho{}_\sigma(x)+\Gamma_\mu\,{}^\rho{}_\tau(x)\,\Gamma_\nu\,{}^\tau{}_\sigma(x)\,\big)\,e^\sigma_b\,dx^\mu\,dx^\nu\right]\otimes\texttt{T}_a{}^b\,.\nonumber
\ea 

The Cartan intertwiner of this affine Cartan geometry is the inverse of \eqref{isoCartanaff}, \textit{i.e.} it is the isomorphism
\be
D\,:\,FM\times_{{}_{Ad_{GL(n)}}}\mathfrak{igl}(n)\,\stackrel{\sim}{\longrightarrow}\, \frac{T\,FM}{GL(n)}
\ee
of Lie algebroids between the adjoint tractor bundle for the affine algebra and the Atiyah algebroid of the tangent frame bundle.
The Cartan covariant derivative is defined by the Cartan fundamental vector fields for the translation ideal, which are the following vector fields on $FM$
\be
D_a=e_a^\mu\,\left(\frac{\partial}{\partial x^\mu}-\,\Gamma_\mu{}^b{}_c(x)\,(\texttt{T}{}_b{}^c)^\#\right)=e_a^\mu\,\frac{\partial}{\partial x^\mu}-\,\Gamma_\mu\,{}^\nu{}_\rho(x)\,e_a^\mu\,e_b^\rho\,\frac{\partial}{\partial e_b^\nu}\,.
\ee
At the level of sheaves, the Cartan intertwiner is an isomorphism of ${C}^\infty(M)$-modules
\be
D:{C}^\infty(FM)\otimes_{{}_{Ad_{GL(n)}}}\mathfrak{igl}(n)\,\stackrel{\sim}{\longrightarrow}\, \Gamma(T\,FM)^{GL(n)}\,.
\ee
Since it is modeled on a reductive Klein geometry, it decomposes into the fundamental action
\ba
\#&:&{C}^\infty(FM)\otimes_{{}_{Ad_{GL(n)}}}\mathfrak{gl}(n)\,\stackrel{\sim}{\longrightarrow}\, {\mathcal V}_{GL(n)}(FM)\nonumber\\
&:&\varepsilon{}^a{}_b(x)\,\texttt{T}{}_a{}^b\mapsto \varepsilon{}^\mu{}_\nu(x)\, e_a^\nu\,\frac{\partial}{\partial e_a^\mu}
\ea
where $\varepsilon{}^a{}_b(x):=\varepsilon{}^\mu{}_\nu(x)\,e_\mu^a\,\theta^\nu_b$ has an implicit dependence on the vertical coordinates,
and into the Cartan covariant derivative
\ba
D&:&{C}^\infty(FM)\otimes_{{}_{Ad_{GL(n)}}}{\mathbb R}^n\,\longrightarrow\, \Gamma(T\,FM)^{GL(n)}\nonumber\\
&:&X^a(x)\,\texttt{P}_a\mapsto X^\mu(x)\nabla_\mu\,.\label{Cartcovder}
\ea
The relation $\gamma=D\circ\Theta$ between the Cartan covariant derivative \eqref{Cartcovder}, the fundamental form \eqref{solderisoTM} and the horizontal lift \eqref{invhorliftTM} is quite manifest in components since we made use of \eqref{affinehorizontalvect field}.

Consider the defining representation of the general linear group ${GL(n)}$ on ${\mathbb R}^n$. The associated vector bundle is the tangent bundle $TM\cong FM\times_{GL(n)}{\mathbb R}^n$ and the Koszul connection $\nabla_\bullet:=\Theta^{-1}\circ (\gamma\bullet\otimes\, id_{{\mathbb R}^n})\circ\, \Theta$ on $TM$ is an affine connection on $M$ with components $\Gamma_\mu\,{}^\rho{}_\nu(x)$. More generally, the vector representation, tensor representations and their dual representations define the usual expression for the Koszul connection on $\otimes\, TM$ and $\otimes\, T^*M$, as can be checked from the action of the invariant horizontal lift
\eqref{invhorliftTM}-\eqref{affinehorizontalvect field} on the explicit form of tensors such as
\eqref{tensorcovcomponents}. As one can see, an affine connection in the sense of Koszul (on $TM$) is equivalent to a connection in the sense of Cartan (on $FM$, modeled on the affine geometry). This is due to the fact that the tangent bundle is endowed with a canonical solder form $\theta$ so this extra ingredient of a Cartan connection with respect to a linear connection is provided for free on the tangent bundle.
For a generic moving frame $e_a^\mu(x)$, the principal gauge connection is a one-form on $M$ reading $\omega^{\mathfrak{gl}(n)}=\omega_\mu{}^a{}_b(x)\, dx^\mu\otimes \texttt{T}_a{}^b$
whose components are given by the usual expression
\be
\omega_\mu{}^a{}_b(x)= \,\theta^a_\nu(x) \big(\partial_\mu e_b^\nu(x)\,+\,\Gamma_\mu\,{}^\nu{}_\rho(x)\,e^\rho_b(x)\big)\,.
\ee

The subset of the automorphisms of the frame bundle which preserve the solder form \eqref{Cartsold}  are generated by the invariant vector fields $\hat{\Xi}:=\xi^\mu(x)\,\frac{\partial}{\partial x^\mu}+\partial_\nu \xi^\mu(x)\,e_a^\nu\,\frac{\partial}{\partial e_a^\mu}$.
The transformation of the affine Cartan connection under such diffeomorphisms reproduces the standard transformation rules of an affine connection under a diffeomorphism of the base. For infinitesimal transformations, the expression in components can be obtained by computing the Lie derivative of the vector field \eqref{affinehorizontalvect field} along $\hat{\Xi}$.

\subsection{Soldering}

Consider a reduction \eqref{reductionprincbundl} of the principal $G$-bundle $P$ to the $H$-bundle $Q$. The associated Klein geometry bundle is $E= P\times_G G/H$. The base space of all those bundles is $M=P/G=Q/H$.
The adjoint representation of the principal group $G$ on the transvection module $\mathfrak{g}/\mathfrak{h}$ defines the associated vector bundle $\mathbb{V}:=P\times_{Ad_G}\mathfrak{g}/\mathfrak{h}$. 
A reduction $i:Q\hookrightarrow P$ is equivalent to a section $\sigma:M\hookrightarrow E$. 
Let $V_{\sigma(M)}E:=\sigma^*VE=\{(m,v)\in M\times VE\,|\,\sigma(m)=\pi(v)\}$ denote the pullback of the vertical distribution $VE$ along the section $\sigma$. There is a canonical isomorphism of vector bundles over $P/G$: $V_{\sigma(M)}E\cong \mathbb{V}$. A solder form on the principal $G$-bundle $P$ associated to the adjoint representation of the principal group $G$ on the transvection module $\mathfrak{g}/\mathfrak{h}$ is an isomorphism $\Theta:TM\stackrel{\sim}{\to} \mathbb{V}\cong V_{\sigma(M)}E$ of vector bundles over $M$.

To summarise, let $\pi:E\twoheadrightarrow M$ be a Klein geometry bundle with fibres $E_m\cong G/H\,$. 
The \textbf{soldering of a Klein geometry bundle} $E$ to $M$ \textbf{through a distinguished section} $\sigma:M\hookrightarrow E$ is a linear isomorphism $\Theta:TM\stackrel{\sim}{\to} V_{\sigma(M)}E$ of vector bundles, called the \textbf{solder form of the soldering}.
More concretely, this means that we have a field on $M$ of linear frames $\Theta|_m:T_mM\stackrel{\sim}{\to} V_{\sigma(m)}E$ of vector spaces.
In other words, a soldering is a manner of attaching the fibres of $E$ in such a way that they can be regarded as tangent to $M$.
Notice that a soldering of $E$ to $M$ requires dim$(G/H)$=\,dim$(E_m)$=\,dim$M$ thus dim\,$E=2\,$dim$M$.
A soldering is therefore composed of two levels of identifications: at zeroth order, the distinguished section $\sigma$ specifies the point of attachment $\sigma(m)$ of $G/H$ to each point $m$ of the base $M$ while, at first order, the solder form identifies $\mathfrak{g}/\mathfrak{h}$ with the tangent space $T_mM$.

Let $H_2=H_1\ltimes I$ be, on the one hand, a semidirect product of Lie groups and, on the other hand, a subgroup $H_2\subset G$ of the Lie group $G$. Consider a principal $H_2$-bundle $P_2$ over $M:=P_2/H_2$. 
\be
\begin{array}
	[c]{ccccccc}%
	I&\trianglelefteqslant & H_2 &  \lefttorightarrow & P_2 & &\\
	&&  & & & \searrow & \\ 
	&&\downarrow & & \downarrow & & M\cong \frac{P_1}{H_1}\cong\frac{P_2}{H_2} \\ 
	&& & & & \nearrow & \\ 
	&&H_1\cong \frac{H_2}{I} & \lefttorightarrow & P_1\cong \frac{P_2}{I} & & \\
	&& & & & & 
\end{array}
\label{projprincipgeom}
\ee
A moving frame of the principal $I$-bundle $P_2$ over $P_1:=P_2/I$ is a reduction of the principal $H_2$-bundle $P_2$ to the principal $H_1$-bundle $P_1$. Consider an $H_2$-equivariant \textbf{solder form} $\theta^\mathfrak{g}:TP_2\to \mathfrak{g}$ on $P_2$ (seen as an $I$-bundle over $P_1$) associated to the adjoint representation of $I$ on $\mathfrak{g}$ inducing an $H_1$-equivariant vector bundle isomorphism $\Theta^\mathfrak{g}:TP_1\stackrel{\sim}{\to} P_2 \times_{Ad_I}\mathfrak{g}$ from the tangent bundle $TP_1$ to the adjoint tractor bundle of the $I$-principal bundle $P_2$ over $P_1$. Such an $H_2$-equivariant solder form $\theta^\mathfrak{g}:TP_2\to \mathfrak{g}$ will be said \textbf{reducible to a Cartan connection one-form} $\omega^\mathfrak{g}:TP_1\stackrel{\sim}{\to} P_1 \times\mathfrak{g}$ on the reduced $H_1$-bundle $P_1$ modeled on the Klein pair $\mathfrak{h}_1\subseteq\mathfrak{g}$
if there exists a trivialisation of the adjoint tractor bundle $\tau:P_2 \times_{Ad_I}\mathfrak{g}\stackrel{\sim}{\to} P_1\times\mathfrak{g}$ such that the canonical isomorphism between the vertical sub-bundle and the adjoint bundle is reproduced as $\tau\circ\Theta^\mathfrak{g}|_{VP_1}=\iota:VP_1\stackrel{\sim}{\to} P_1\rtimes\mathfrak{h}_1$. In fact, $\omega^\mathfrak{g}=\tau\circ\Theta^\mathfrak{g}$ defines a Cartan connection one-form on the principal $H_1$-bundle $P_1$.

\clearpage



\begin{thebibliography}{99}

\bibitem{Cartan}
\'E.~Cartan, \textit{La m\'ethode du rep\`ere mobile, la th\'eorie des groupes continus et les espaces g\'en\'eralis\'es} 
(Hermann, 1935).

\bibitem{Ehresmann:1950}
C.~Ehresmann, ``Les connexions infinit\'esimales dans un espace fibr\'e diff\'erentiable'' in G.~Thone (ed), \textit{Colloque de topologie (espaces fibrés) Bruxelles, 1950} (Masson, 1951) 29; 
S\'eminaire Bourbaki {\bf 1} (1952) 153.

\bibitem{Kob63}
S.~Kobayashi and K.~Nomizu, \textit{Foundations of Differential Geometry: Vol. 1 \& 2} (Wiley, 1963 \& 1969).

\bibitem{Kob72}
S.~Kobayashi, \textit{Transformation Groups in Differential Geometry} (Springer-Verlag, 1972).

\bibitem{Sharpe}
R.~W.~Sharpe, \textit{Differential Geometry: Cartan's Generalization of Klein's Erlangen Program} (Springer, 1997).

\bibitem{Wise:2006sm}
D.~K.~Wise,
``MacDowell-Mansouri gravity and Cartan geometry,''
Class. Quant. Grav. \textbf{27} (2010) 155010 [arXiv:gr-qc/0611154 [gr-qc]].

\bibitem{CrampinSaunders}
M.~Crampin and D.~Saunders, \textit{Cartan geometries and their symmetries: a Lie algebroid approach}  (Atlantis Press, 2016).

\bibitem{Bekaert:2023cmi}
X.~Bekaert, ``Geometric tool kit for higher-spin gravity (Part I): An introduction to the geometry of differential operators,'' Int. J. Mod. Phys. A \textbf{38} (2023) no.8, 2330003 [arXiv:2301.08069 [hep-th]].

\bibitem{Bekaert:2022dlx}
X.~Bekaert, N.~Kowalzig and P.~Saracco,
``Universal enveloping algebras of Lie-Rinehart algebras: crossed products, connections, and curvature,'' Lett.\ Math.\ Phys.\ {\bf 114} (2024) 140
[arXiv:2208.00266 [math.RA]].

\bibitem{Reviews}
	M.~A.~Vasiliev,
	``Higher spin gauge theories in four-dimensions, three-dimensions, and two-dimensions,''
  Int.\ J.\ Mod.\ Phys.\ D {\bf 5} (1996) 763
  [hep-th/9611024];
  ``Higher spin gauge theories in various dimensions,''
  Fortsch.\ Phys.\  {\bf 52} (2004) 702
  [hep-th/0401177];
	``Higher spin gauge theories in any dimension,''
  Comptes Rendus Physique {\bf 5} (2004) 1101
  [hep-th/0409260];\\
	X.~Bekaert, S.~Cnockaert, C.~Iazeolla and M.~A.~Vasiliev,
  ``Nonlinear higher spin theories in various dimensions,''
  hep-th/0503128;\\
	V.~E.~Didenko and E.~D.~Skvortsov,
  ``Elements of Vasiliev theory,'' 
  Lect. Notes Phys. \textbf{1028} (2024) 269
  [arXiv:1401.2975 [hep-th]];\\
	M.~A.~Vasiliev,
	``Higher-spin theory and space-time metamorphoses,''
  Lect.\ Notes Phys.\ {\bf 892} (2015) 227
  [arXiv:1404.1948 [hep-th]];\\
	D.~Ponomarev,
	``Basic introduction to higher-spin theories,''
	Int. J. Theor. Phys. \textbf{62} (2023) 146
	[arXiv:2206.15385 [hep-th]];\\
	A.~Campoleoni and S.~Fredenhagen,
	``Higher-spin gauge theories in three spacetime dimensions,''
	Lect. Notes Phys. \textbf{1028} (2024) 121 [arXiv:2403.16567 [hep-th]].

\bibitem{Introductions}
	D.~Sorokin,
  ``Introduction to the classical theory of higher spins,''
  AIP Conf.\ Proc.\  {\bf 767} (2005) 172
  [hep-th/0405069];\\
  X.~Bekaert, N.~Boulanger and P.~Sundell,
  ``How higher-spin gravity surpasses the spin two barrier: no-go theorems versus yes-go examples,''
  Rev.\ Mod.\ Phys.\  {\bf 84} (2012) 987
  [arXiv:1007.0435 [hep-th]];\\
	R.~Rahman,
	``Higher Spin Theory - Part I,''
	PoS \textbf{ModaveVIII} (2012) 004
	[arXiv:1307.3199 [hep-th]];\\
	R.~Rahman and M.~Taronna,
  ``From Higher Spins to Strings: A Primer,'' 
  Lect. Notes Phys. \textbf{1028} (2024) 1
  [arXiv:1512.07932 [hep-th]];\\
	P.~Kessel,
  ``The Very Basics of Higher-Spin Theory,''
  PoS \textbf{Modave2016} (2017) 001
  [arXiv:1702.03694 [hep-th]];\\
  S.~Pekar, ``Introduction to higher-spin theories,''
  PoS \textbf{Modave2022} (2023) 004;\\
	X.~Bekaert, N.~Boulanger, A.~Campoleoni, M.~Chiodaroli, D.~Francia, M.~Grigoriev, E.~Sezgin and E.~Skvortsov,
	``Snowmass White Paper: Higher Spin Gravity and Higher Spin Symmetry,''
	in J.~N.~Butler, R.~S.~Chivukula, M.~E.~Peskin (eds)
	\textit{Proceedings: 2021 US Community Study on the Future of Particle Physics: Snowmass 2021} (SLAC, 2023)
	[arXiv:2205.01567 [hep-th]];\\
	A.~Bengtsson, \textit{Higher Spin Field Theory (Concepts, Methods and History) Volume 1: Free Theory} (De Gruyter, 2020); 
	\textit{Volume 2: Interactions} (De Gruyter, 2023);\\
	X.~Bekaert, ``Higher spin gravity'' in \textit{Encyclopedia of Mathematical Physics (2nd edition) Volume 1} (Elsevier, 2025) 425 [hal-04509045].
	
\bibitem{Proceedings}
	R.~Argurio, G.~Barnich, G.~Bonelli and M.~Grigoriev (eds), \textit{Higher Spin Gauge Theories} (International Solvay Institutes, 2004);\\
	L.~Brink, M.~Henneaux and M.~A.~Vasiliev (eds), \textit{Higher Spin Gauge Theories} (World Scientific, 2017).
	
\bibitem{Bekaert:2023jvl}
X.~Bekaert,
``Geometric tool kit for higher spin gravity (Part II): An introduction to Lie algebroids and their enveloping algebras,''
Int. J. Mod. Phys. A \textbf{38} (2023) no.25, 2330013
[arXiv:2308.00724 [hep-th]].

\bibitem{Sardanashvily2}
G.~Sardanashvily, \textit{Lectures on differential geometry of modules and rings: Application to Quantum Theory} (Lambert Academic Publishing, 2012) [arXiv:0910.1515 [math-ph]].

\bibitem{Mackenzie2}
K.~Mackenzie, \textit{General theory of Lie groupoids and Lie algebroids} (Cambridge University Press, 2005).
	
\bibitem{Saunders}
D.~J.~Saunders, \textit{The geometry of jet bundles}  (Cambridge University Press, 1989).

\bibitem{Dubrovin}
B.~A.~Dubrovin, S.~P.~Novikov, A.~T.~Fomenko, \textit{Modern Geometry: Methods and Applications} (Springer, 1985).

\bibitem{Hamilton}
M.~J.~D.~Hamilton, \textit{Mathematical Gauge Theory: With Applications to the Standard Model of Particle Physics} (Springer, 2018).

\bibitem{Mackenzie}
K.~Mackenzie, \textit{Lie groupoids and Lie algebroids in differential geometry} (Cambridge University Press, 1987).

\bibitem{Dieudonne1970} J.~Dieudonn\'e, \textit{\'{E}l\'ements d'analyse, tome {III}}
({G}authier-{V}illars, 1970).

\bibitem{Kosmann:1979}
Y.~Kosmann-Schwarzbach, ``Infinitesimal conditions for the equivariance of morphisms of fibered manifolds,'' Proc. Amer. Math. Soc. \textbf{77}  (1979) 374.

\bibitem{Attard:2019}
J.~Attard, J.~Fran\c{c}ois, S.~Lazzarini and T.~Masson,
``Cartan Connections and Atiyah Lie Algebroids,''
J. Geom. Phys. \textbf{148}  (2020) 103541 [arXiv:1904.04915 [math-ph]].

\bibitem{Benito} 
P.~Benito, J.~Rold\'an-L\'opez, ``Examples and Patterns on Quadratic Lie Algebras,'' in H.~Albuquerque, J.~Brox, C.~Martínez, P.~Saraiva (eds), \textit{Non-Associative Algebras and Related Topics. NAART 2020} (Springer, 2023) [arXiv:2210.08257 [math.RA]].

\bibitem{BEG}
T.~N.~Bailey, M.~G.~Eastwood and A.~R.~Gover, ``Thomas's structure bundle for conformal, projective and related structures,'' 
Rocky Mountain J.\ Math.\ {\bf 24}  (1994) 1191.

\bibitem{Crampin}
M.~Crampin, ``Cartan connections and Lie algebroids,'' 
SIGMA {\bf 5}  (2009) 061 [arXiv:0906.2554 [math.DG]].

\bibitem{Bernshtein}
I. N. Bernshtein and B. I. Rozenfel’d, ``Homogeneous spaces of infinite-dimensional
Lie algebras and the characteristic classes of foliation,'' Uspehi Mat.\ Nauk {\bf 28}  (1973) 103.

\bibitem{Hall}
B.~C.~Hall, \textit{Lie Groups, Lie Algebras, and Representations: An Elementary Introduction}, Graduate texts in Mathematics \textbf{222}  (Springer, 2015).

\bibitem{Campoleoni:2010zq}
A.~Campoleoni, S.~Fredenhagen, S.~Pfenninger and S.~Theisen,
``Asymptotic symmetries of three-dimensional gravity coupled to higher-spin fields,''
JHEP \textbf{11}   (2010) 007
[arXiv:1008.4744 [hep-th]];\\
C.~Bunster, M.~Henneaux, A.~Perez, D.~Tempo and R.~Troncoso,
``Generalized Black Holes in Three-dimensional Spacetime,''
JHEP \textbf{05}  (2014) 031
[arXiv:1404.3305 [hep-th]].

\bibitem{Lee2}
J.~M.~Lee, \textit{Introduction to Smooth Manifolds}  (2nd edition), Graduate Texts in Mathematics {\bf 218}  (Springer, 2012).

\bibitem{Dieck} T.~t.~Dieck, \textit{Algebraic topology}
(European Mathematical Society, 2008).

\bibitem{Koszul}
J.-L.~Koszul, ``Homologie et cohomologie des algèbres de Lie,'' Bull. Soc. Math. France {\bf 78} (1950) 65.

\bibitem{Steenrod}
N.~Steenrod, \textit{The Topology of Fibre Bundles} (Princeton University Press, 1951).

\bibitem{Michor}
P.~W.~Michor, \textit{Topics in Differential Geometry}, Graduate Studies in Mathematics {\bf 93} (American Mathematical Society, 2008)

\bibitem{Green:2014}
M. Green, P. Griffiths and M. Kerr, ``Special Values of Automorphic Cohomology Classes,'' Memoirs of the American Mathematical Society
{\bf 231} (American Mathematical Society, 2015).


\bibitem{Kosmann-Schwarzbach}
Y.~Kosmann-Schwarzbach and K.~C.~H.~Mackenzie, ``Differential operators and actions of Lie algebroids,'' in T.~Voronov  (Ed.), \textit{Quantization, Poisson Brackets and Beyond}, Contemporary Mathematics {\bf 315}  (American Mathematical Society, 2002) 213 [arXiv:math/0209337 [math.DG]].

\bibitem{Eastwood:1987ki}
M.~G.~Eastwood and J.~W.~Rice, ``Conformally invariant differential operators on Minkowski space and their curved analogues,'' Commun.\ Math.\ Phys.\  {\bf 109}  (1987) 207.

\bibitem{Eastwood:1996}
M.~G.~Eastwood, ``Notes on conformal differential geometry'' in the Proceedings of the 15th Winter School `Geometry and Physics' (Circolo Matematico di Palermo, 1996).

\bibitem{Joseph:1974}
A.~Joseph, ``Minimal realizations and spectrum generating algebras,''
Commun. Math. Phys. \textbf{36} (974) 325.

\bibitem{Vogan}
D. A. Vogan, ``Gelfand-Kirillov dimension for Harish-Chandra modules,'' Inventiones
Mathematicae {\bf 48} (1978) 75.

\bibitem{Joseph:1976}
A.~Joseph, ``The minimal orbit in a simple Lie algebra and its associated maximal ideal,''
Annales Sci.\ \'Ecole Norm.\ Sup.\ {\bf 9} (1976) 1.

\bibitem{Joung:2014qya}
E.~Joung and K.~Mkrtchyan, ``Notes on higher-spin algebras: minimal representations and structure constants,''
JHEP {\bf 1405}  (2014) 103 [arXiv:1401.7977 [hep-th]].

\bibitem{Cap}
A.~Cap and J.~Slovak, \textit{Parabolic Geometries} (American Mathematical Society, 2009).

\bibitem{Beachy} 
J.~A.~Beachy, \textit{Introductory Lectures on Rings and Modules}, London Mathematical Society Student Texts \textbf{47} (Cambridge University Press, 1999).

\bibitem{Dobrev}
V.~K.~Dobrev, \textit{Invariant Differential Operators: Noncompact Semisimple Lie Algebras and Groups}  (De Gruyter, 2016).

\bibitem{Serre:2001}
J.-P.~Serre, \textit{Complex Semisimple Lie Algebras} (Springer, 2001).

\bibitem{Humphreys}
J.~E.~Humphreys, \textit{Introduction to Lie Algebras and Representation Theory}, Graduate Texts in Mathematics \textbf{9} (Springer, 1972).

\bibitem{Humphreys2}
J.~E.~Humphreys, \textit{Representations of Semisimple Lie Algebras in the BGG Category $\mathcal O$}, Graduate Studies in Mathematics \textbf{94} (American Mathematical Society, 2008).

\bibitem{Duflo}
M.~Duflo, ``Sur la classification des id\'eaux primitifs dans l'alg\`ebre enveloppante d'une alg\`ebre de Lie semi-simple,'' Ann.\ of Math.\ {\bf 105} (1977) 107.

\bibitem{Ovsienko}
V.~Ovsienko and S.~Tabachnikov, \textit{Projective Differential Geometry Old and New: From the Schwarzian Derivative to Cohomology of Diffeomorphism Groups}  (Cambridge University Press, 2005).

\bibitem{Feigin1988}
B.~L. Feigin, ``The Lie algebras $\mathfrak{gl} (\lambda)$ and cohomologies of Lie algebras of differential operators,'' Russian Math.\ Surveys {\bf 43}  (1988) 169.

\bibitem{Mazorchuk}
V. Mazorchuk, \textit{Lectures on $\mathfrak{sl}(2,\mathbb{C})$-modules} (Imperial College Press, 2010).

\bibitem{Catren:2014vza}
G.~Catren, ``Geometric foundations of Cartan gauge gravity,'' Int.\ J.\ Geom.\ Meth.\ Mod.\ Phys.\  {\bf 12}   (2015) 1530002 [arXiv:1407.7814 [gr-qc]].

\bibitem{Curry:2014yoa}
S.~Curry and A.~R.~Gover, ``An introduction to conformal geometry and tractor calculus, with a view to applications in general relativity,'' in T.~Daud\'e, D.~H\"afner, J.-P.~Nicolas, \textit{Asymptotic Analysis in General Relativity}, London Mathematical Society Lecture Note Series {\bf 443} (Cambridge University Press, 2018) [arXiv:1412.7559 [math.DG]].

\bibitem{Bekaert:2022ipg}
X.~Bekaert and B.~Oblak,
``Massless scalars and higher-spin BMS in any dimension,''
JHEP \textbf{11} (2022) 022
[arXiv:2209.02253 [hep-th]].

\end{thebibliography}
\end{document}